\documentclass[letterpaper,twocolumn,10pt]{article}

\usepackage{usenix}

\usepackage{textcase}

\usepackage{bbding}

\usepackage{booktabs} 
\usepackage{epsfig,endnotes}
\usepackage{subfigure}
\usepackage{mathtools}
\usepackage{amsmath,amsfonts,dsfont,amssymb}
\usepackage{subfigure}
\usepackage{graphicx,xcolor}
\usepackage{epstopdf}
\graphicspath{{figures/}}
\usepackage{array}
\usepackage{color}
\usepackage{amssymb}
\usepackage[english, nocase]{babel}
\usepackage[utf8]{inputenc}
\usepackage[colorinlistoftodos]{todonotes}
\usepackage{algpseudocode}
\usepackage{multirow}
\usepackage{tabularx}
\usepackage{makecell}
\usepackage{array}
\usepackage{setspace}
\usepackage{textcomp}
\usepackage{bm}
\usepackage{xspace}
\usepackage{comment}
\usepackage{caption}
\usepackage[ruled,linesnumbered,vlined]{algorithm2e}
\usepackage{amsmath}

\usepackage{wasysym}
\usepackage{colortbl}
\usepackage{threeparttable}
\usepackage{tablefootnote}
\usepackage{hhline}

\newcommand{\sys}{{\sc Trident}\xspace}
\newcommand{\sysB}{{\sc Trident}}

\definecolor{backcolor}{HTML}{EBEFEA}

\begin{document}

\title{\Large \bf Expose Your Disguise: Recovering Source Speaker Identity From Voice Conversion}

\author{
{\rm Hanlei Zhang}\\
Zhejiang University\\
Hangzhou, China\\
{zhanghanlei@zju.edu.cn}
\and
{\rm Zhongming Ma}\\
Zhejiang University\\
Hangzhou, China\\
{zmma@zju.edu.cn}
\and
{\rm Mingyang Zhang}\\
Ant Group\\
Shanghai, China\\
{zhangmingyang.zmy@antgroup.com}
\and
{\rm Tengfei Liu}\\
Ant Group\\
Shanghai, China\\
{aaron.ltf@antgroup.com}
\and
{\rm Yushi Cheng}\\
Zhejiang University\\
Hangzhou, China\\
{yushicheng@zju.edu.cn}
\and
{\rm Yanjiao Chen}\\
Zhejiang University\\
Hangzhou, China\\
{chenyj.thu@gmail.com}
}

\maketitle

\begin{abstract}
Voice conversion (VC) poses a significant threat to biometric security by allowing attackers to impersonate target speakers. In forensic contexts, recovering the source speaker's identity from converted audio is vital for narrowing the field of suspects. To address this, we propose \sys, a retracing framework designed to restore a source speaker's original identity from a converted audio sample. \sys utilizes a three-pronged architecture consisting of a primary extractor and two auxiliary branches. The first auxiliary branch identifies the underlying voice conversion mechanism. This design acknowledges that even if the exact conversion strategy is unknown, a high-performance model adopted by the attacker is typically a derivative or variant of established mainstream ones. The second auxiliary branch extracts a latent representation of the target speaker, facilitating the isolation of target-specific traits from the composite converted audio sample. Finally, the main extractor leverages insights from both auxiliary branches to decouple confounding factors and distill a highly discriminative representation of the source speaker's identity. Experimental results demonstrate that \sys achieves an accuracy as high as $90.99\%$ against 7 state-of-the-art voice conversion methods. Furthermore, \sys maintains robust performance under challenging conditions, including telephony channels, unseen languages, and adaptive scenarios.

\end{abstract}

\begin{figure}[t]
\centering     
    \includegraphics[width=0.45\textwidth]{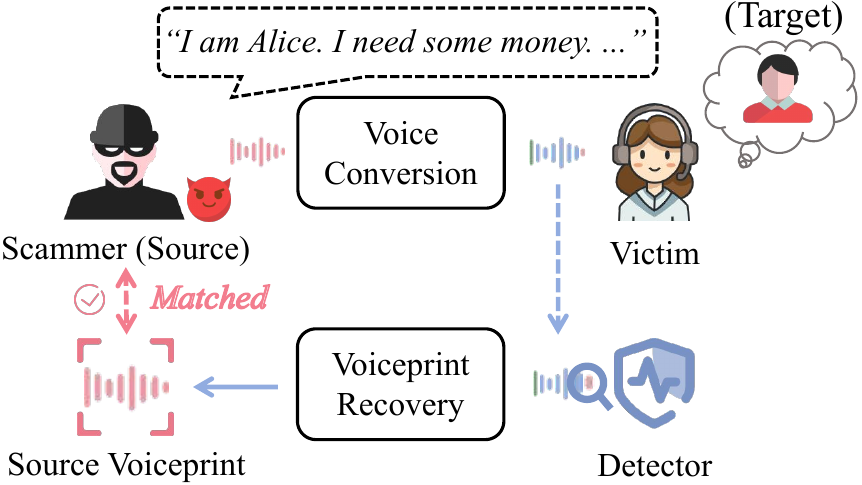}
    \caption{The application scenario of voiceprint recovery.}
    \label{fig.scenarios}
\end{figure}

\section{Introduction}

By enabling the creation of high-fidelity vocal impersonations, speech deepfake technology has become a potent tool for social engineering, allowing attackers to manipulate human trust on an unprecedented scale. Voice conversion and text-to-speech (TTS) synthesis are two representative techniques of speech deepfakes. TTS generates vocal output from raw text but typically incurs higher latency due to its dependency on complete text input and additional style conditioning. Furthermore, due to the lack of fine-grained prosodic details in raw text, TTS produces less nuanced prosody than voice conversion, rendering it unsuitable for real-time fraudulent interactions. Conversely, voice conversion maps the acoustic features of a source speaker onto a target identity, allowing the attacker to project their own emotional nuances and natural rhythm onto the converted audio. Therefore, voice conversion can generate high-fidelity speech that preserves fine-grained prosodic attributes, with real-time streaming systems now achieving sub-100 ms latencies \cite{yang2024streamvc}.

In this manner, voice conversion provides a viable mechanism for real-time impersonation and live-call spoofing \cite{lorenzo2018voice, yi2020voice, wang2020asvspoof, fraud, fraud_2, fraud_3} as shown in Figure \ref{fig.scenarios}. A notable real-world manifestation occurred in 2024, when an employee at Arup transferred approximately HK\$200 million to scammers who impersonated the company’s CFO and other senior executives during a live video conference \cite{fraud_3}. In this full-on video call, real-time voice conversion enabled interactions between scammers and the victim, who made money transfers across 15 transactions following the scammers' orders \cite{10.1145/3772374}. Prior research mainly focused on deepfake detection, a binary task that differentiates genuine audio from fake audio \cite{conti2022deepfake, li2022comparative, mostaani2022breathing, doan2023bts, kulangareth2024investigation, liu2023betray, kumarivoiceradar,  kawa2022specrnet, yadav2024compression, doan2023gan, jung2022aasist}, which falls short in forensic attribution. Unlike TTS outputs, which are entirely synthetic, voice conversion produces composite signals that inherently retain underlying acoustic traces of the source speaker alongside the target speaker. This composition presents a unique opportunity for source identity recovery, a critical capability for narrowing down suspect pools and unmasking malicious actors.

\begin{figure}[t]
\centering     
    \includegraphics[width=0.49\textwidth]{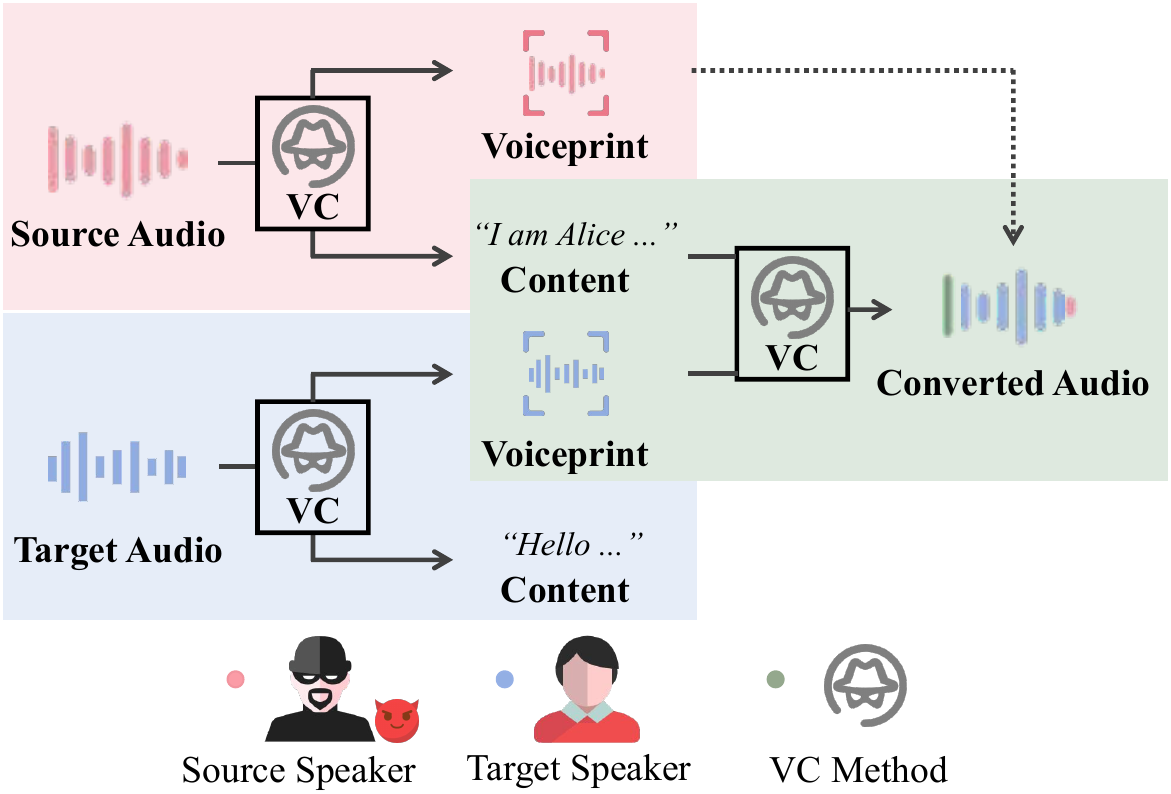}
    \caption{The workflow of voice conversion. Different colors represent voiceprint parts that carry information targeting different speakers or methods. Variations in audio waveforms correspond to differences in speech content.}

    \label{fig.voice}
\end{figure}

As illustrated in Figure \ref{fig.voice}, voice conversion works in a representation-mapping-reconstruction pipeline \cite{sisman2020overview}. The source audio sample is first decomposed into representations of the speech and the speaker. These representations are then transformed by a mapping module to match the target speaker's characteristics, and finally reconstructed as time-domain signals. While early voice conversion methods rely primarily on statistical modeling such as gaussian mixture models \cite{aihara2012gmm, hwang2013incorporating}, modern approaches have shifted toward deep learning, specifically the encoder-decoder architecture. In this framework, an encoder performs representation extraction, and a decoder handles the combined mapping and reconstruction process \cite{sisman2020overview}. Because the mapping-reconstruction stage is critical for ensuring that the converted audio captures the natural characteristics of the target speaker, significant research has focused on integrating advanced generative models and architectures into the decoder to enhance synthesis quality, such as variational autoencoders (VAEs) \cite{chou2019one, wu2020one}, U-Nets \cite{wu2020vqvc+, chen2021again}, sequence-to-sequence (Seq2Seq) models \cite{hayashi2020voice, liu2021any}, generative adversarial networks (GANs) \cite{li2021starganv2, li2023freevc}, and diffusion models \cite{choi2023diff, choi2024dddm}. However, despite extensive efforts to achieve seamless voice conversion, it is shown that source speaker leakage is inevitable in converted audio samples \cite{wellington2024quantifying, panariello2024voiceprivacy}. Furthermore, voice conversion also introduces method-specific \emph{fingerprints}, e.g., unique spectral signatures or prosodic artifacts, into converted audios.

While pioneering research \cite{deng2023catch} and subsequent studies \cite{cai2023identifying, ma2024distillation, zhang2024target} have achieved satisfactory performance in source speaker recovery, they often struggle to generalize across advanced or previously unseen voice conversion methods. This limitation stems from the complex coupling of heterogeneous information embedded within converted audio, which presents two primary challenges. First, different voice conversion techniques obfuscate the source speaker's voiceprint in distinctive ways. Consequently, a naive extraction approach that overlooks these methodological variations fails to generalize. Second, the target speaker’s characteristics, which are the dominant component of the converted audio, further mask the residual source voiceprint, severely hindering direct extraction. To overcome these challenges, we propose \sys, a robust framework equipped with two specialized modules. These modules disentangle the two confounding components, namely the features of the voice conversion method and the target speaker, from the converted samples, thereby facilitating the recovery of the source speaker's voiceprint.

\begin{itemize}
\item \textit{How can the recovery process be customized to diversified voice conversion paradigms?}
\end{itemize}

Under the black-box threat model, existing works \cite{deng2023catch, cai2023identifying, ma2024distillation, zhang2024target} largely overlook the impact of specific voice conversion architecture on the recovery process. The common practice is to train a recovery model on converted audio samples labeled with the source speaker in a supervised way. While effective for voice conversion methods seen during training, these models often suffer from poor generalization. We observe recovery accuracy plummeting to below 10\% when encountering unseen conversion techniques \cite{deng2023catch}. Despite architectural nuances, we find that high-performance voice conversion methods generally derive from a few mainstream paradigms featuring different decoders, e.g., VAEs \cite{chou2019one, wu2020one}, U-Nets \cite{wu2020vqvc+, chen2021again}, Seq2Seq models \cite{hayashi2020voice, liu2021any}, GANs \cite{li2021starganv2, li2023freevc}, and diffusion models \cite{choi2023diff, choi2024dddm}. Leveraging this insight, we equip \sys with the auxiliary branch to capture method-specific information. Specifically, a classifier categorizes the input into different voice conversion families. The latent representation of the classifier is provided to guide the primary extractor in a more informed recovery process.

\begin{itemize}
\item \textit{How can the recovery pipeline better isolate source identities from target-dominated audio?}
\end{itemize}

The influence of the target speaker on source voiceprint recovery is notable. While existing frameworks \cite{deng2023catch} attempt to mitigate this by providing the target voiceprint as an input, this often leads to \emph{information dilution} as the signal propagates through deep layers. In contrast, \sys shifts target information to the output via a dedicated auxiliary branch designed to explicitly extract the target speaker identity from the converted audio. To enforce distinctness, we employ a cosine similarity (COS) loss to maximize the margin between the source and target representations. Specifically, for converted audio (e.g., Source: Bob, Target: Alice), we maximize the dissimilarity between the primary extractor and this auxiliary branch. For authentic speech (e.g., Alice), we minimize the distance. This contrastive approach ensures the primary extractor effectively filters out target interference to recover a pure source identity.

Extensive experiments demonstrate that \sys outperforms existing methods across 7 voice conversion methods. It achieves over 90\% source identification accuracy for unseen speakers and is effective even when audio is transmitted over real-world telephony and voice over internet protocol (VoIP) channels. Furthermore, the framework exhibits robust cross-lingual generalization. A model trained exclusively on English successfully generalizes to Spanish, French, German, and Amharic. Finally, \sys maintains high accuracy even when attackers use pre-processing or adaptive techniques.

Our main contributions are summarized as follows.

\begin{itemize}
\item We present \sys, a recovery framework that enhances forensic accountability by isolating source speaker identities across diverse voice conversion paradigms.

\item We establish a three-pronged architecture to effectively tap into the information of the conversion method and the target speaker to facilitate better source speaker identity recovery. 

\item We validate the performance of \sys through extensive experiments and verify its robustness across gender subgroups, telephony scenarios, adaptive adversaries, unseen voice conversion methods, and unseen languages. 
\end{itemize}

\section{Preliminaries}

\subsection{Speaker Recognition Systems}
Voiceprint is the unique characteristic of audio that can be associated with the identity of the speaker. To enable speaker recognition, voiceprint representations are extracted from audio samples as
\begin{equation}
v=\mathcal{V}(x),
\label{eq:voiceprint_extractor}
\end{equation}
where $v$ is the voiceprint representation of an audio sample $x$ and $\mathcal{V}$ denotes a particular speaker recognition system. 

There are two phases in speaker recognition systems: enrollment and verification. In the enrollment phase, the voiceprint representation of a speaker $i\in I$ is extracted from several reference audio samples $D^{r}=\{(x^{r},i)\}$ and aggregated as a template $v^{r}_{i}=\mathop{\mathrm{Average}}\limits_{(x^{r},i)\in D^{r}}\mathcal{V}(x^{r})$. Reference audio samples are indispensable for speaker recognition, providing the biometric templates for comparison. In the verification phase, the voiceprint representation of an audio sample is extracted and compared against all recorded templates to find the most similar match $\mathcal{O}(x)=\mathop{\mathrm{arg\,max}}\limits_{i\in I}{\mathcal{S}(\mathcal{V}(x), v^{r}_{i})}$. $S$ typically employs the cosine similarity function. ECAPA-TDNN \cite{desplanques2020ecapa} is one of the mainstream speaker recognition systems. ECAPA-TDNN enhances the time delay neural network (TDNN) architecture with SE-Res2Blocks and attentive stat pooling, obtaining remarkable performance on VoxCeleb \cite{nagrani2017voxceleb, chung2018voxceleb2}, the widely-used speaker recognition dataset.

\begin{figure}[t] 
\centering     
    \includegraphics[width=0.45\textwidth]{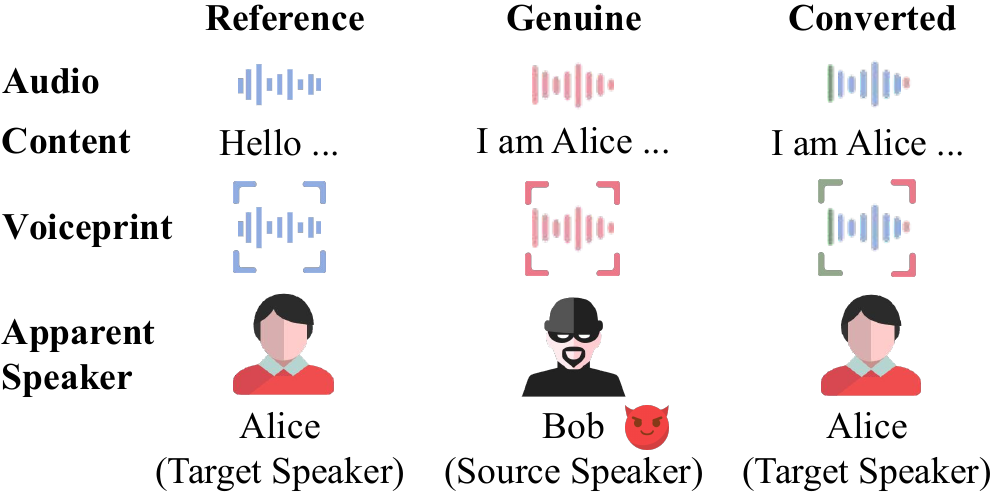}
    \caption{Comparison of information contained in reference, genuine, and converted audio samples. The reference audio is the enrollment audio of Alice, which has significant differences from the genuine audio of Bob, i.e., the attacker, in both voiceprint and content. However, Bob can change the voiceprint to that of Alice by voice conversion.}
    \label{fig.compare}
\end{figure}

\subsection{Voice Conversion}
Voice conversion aims to transform the style of an audio sample while preserving its linguistic content (i.e., the spoken words). As shown in Figure \ref{fig.compare}, most voice conversion methods convert the source speaker’s identity into that of a target speaker for various applications, e.g., fraud. To achieve this, voice conversion typically follows a representation-mapping-reconstruction pipeline. It first extracts representations corresponding to speaker identity and linguistic content. After mapping the source speaker representation to the target speaker, the converted audio is finally reconstructed from the transformed representations.

Early voice conversion methods focus on mapping source-to-target speaker features, including spectral envelopes \cite{stylianou1998continuous}, spectrum parameters \cite{abe1990voice}, and spectral parameter trajectory \cite{toda2007voice}. Beyond classical statistical models \cite{aihara2012gmm, hwang2013incorporating}, methods also utilize deep neural networks for feature extraction, e.g., Deep Belief Nets \cite{nakashika2013voice}, DBLSTM-RNNs \cite{sun2015voice}, and GANs \cite{kaneko2018cyclegan, kaneko2019cyclegan}. However, these traditional mapping approaches often suffer from limited performance and a failure to generalize to unseen speakers. To overcome these limitations, contemporary voice conversion systems resort to encoder-decoder architectures. In this framework, the encoder disentangles the audio sample into voiceprint/identity representation and linguistic content representation. The decoder then facilitates the mapping-reconstruction stage by transforming the source voiceprint to match the target speaker and synthesizing the final audio through the integration of this converted voiceprint representation with the original linguistic content. The decoder, handling the mapping-reconstruction stage, is a critical component that determines the quality of converted audio.

Voice conversion has undergone a significant evolution driven by advancements in generative models. Early research leveraged VAEs \cite{kingma2014autoencoding} to learn a structured latent representation of speech. While these models facilitate conversion by swapping speaker embeddings, as in AdaIN-VC \cite{chou2019one} and VQVC \cite{wu2020one}, they often suffer from loss of critical acoustic details. To address this, later frameworks integrated U-Net \cite{ronneberger2015u} architectures by employing skip connections to propagate low-level features directly to the decoder. Models such as AGAIN-VC \cite{chen2021again} and VQVC+ \cite{wu2020vqvc+} enhanced reconstruction fidelity. To better model prosody, Seq2Seq frameworks such as the Voice Transformer Network \cite{hayashi2020voice} and BNE-Seq2seqMoL \cite{liu2021any} were introduced to learn explicit temporal alignments. Building on these structural improvements, the integration of GANs \cite{goodfellow2014generative} shifted the research focus toward enhancing perceptual naturalness. Models like StarGANv2-VC \cite{li2021starganv2} and FreeVC \cite{li2023freevc} utilize adversarial objectives to generate high-fidelity speech, often integrating VAEs or U-Nets within the GAN framework. Most recently, diffusion models \cite{sohl2015deep} employ an iterative denoising process to recover acoustic features from gaussian noise. By modeling the complex distribution of speech data, systems such as DPM-VC \cite{popovdiffusion}, Diff-HierVC \cite{choi2023diff}, and DDDM-VC \cite{choi2024dddm} capture fine-grained speaker characteristics and generate high-fidelity audio. 

\subsection{Voice Conversion Countermeasures}

In the case of malicious use of voice conversion, countermeasures have been proposed to either detect voice conversion or recover the source speaker identity from converted audio.

Voice conversion detection aims to make use of various information to detect whether an audio sample is a converted one or not, such as spectral \cite{tak2021end2}, emotional \cite{conti2022deepfake}, physical \cite{li2022comparative}, perceptual \cite{li2022comparative}, breathing \cite{mostaani2022breathing}, silence \cite{doan2023bts}, pause \cite{kulangareth2024investigation} features, dual-channel stereo information \cite{liu2023betray}, frequency dynamics and oscillations \cite{kumarivoiceradar}. The detection model may take the form of convolutional neural networks \cite{ge2021raw, hua2021towards, tak2021end, zhang2021multi}, residual networks \cite{ma2021rw, kawa2022specrnet}, transformers \cite{yadav2024compression}, generative adversarial networks \cite{doan2023gan}, and graph attention networks \cite{tak2021end2, tak2021graph, jung2022aasist}. However, voice conversion detection methods are not able to recover the identity of the source speaker.

\begin{figure*}[t] 
\centering     
    \includegraphics[width=1.0\textwidth]{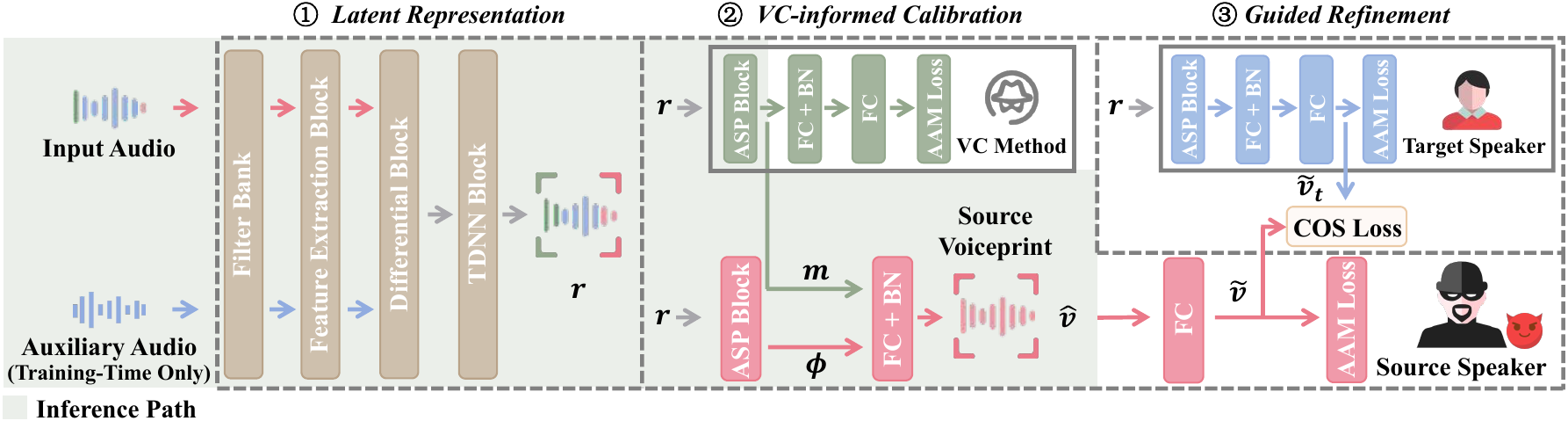}
    \vspace{-0.6cm}
    \caption{The overview of \sys.}
    \label{fig.overview}
\end{figure*}

Voiceprint recovery tries to achieve the more challenging goal of restoring the identity information of the source speaker from converted audio. Revelio \cite{deng2023catch} is the state-of-the-art  voiceprint recovery method, which introduces a differential rectification algorithm to recover the source voiceprint from converted audio. However, Revelio exhibits limited generalization across diverse voice conversion methods. Another work \cite{cai2023identifying} employs a residual convolutional network \cite{he2016deep} for recovering source identity, yet its performance is considerably inferior to that of Revelio. Subsequent efforts \cite{ma2024distillation, zhang2024target} have introduced incremental modifications to Revelio, such as incorporating a mask estimation \cite{zhang2024target} or a distillation algorithm \cite{ma2024distillation}, yielding results that are either worse or, at best, on par with Revelio. To address this limitation, \sys eliminates interference from both method-specific distortion and target speaker characteristics to recover the source speaker voiceprint robustly across diverse voice conversion methods.

Another related line of work lies in speaker de-anonymization, which has thus far exclusively tackled traditional speaker anonymization techniques that modify speaker-specific characteristics while preserving baseline acoustic features \cite{miao2025benchmark}. Unlike these conventional anonymization methods, voice conversion explicitly alters the source audio to mimic a specific target identity. Existing speaker de-anonymization mainly exploits conventional acoustic features (e.g., phoneme timing patterns \cite{tomashenko2025analysis, tomashenko2025exploiting} and accent cues \cite{bakari2026identity}) and linguistic content (e.g., semantic keyword patterns \cite{gaznepoglu2025you} and stylometric fingerprinting \cite{aggazzotti2026content}) for speaker recognition. Some methods \cite{seungmin2026evaluating, zhang2025attacking} identify speakers based on acoustic embedding similarity between anonymized and original speech, with embeddings extracted using pre-trained speaker recognition models. To the best of our knowledge, the unique challenges posed by voice conversion remain unaddressed within the existing speaker de-anonymization literature.

\section{System Model}

This section defines the goals, capabilities, and knowledge of the attacker and the defender.

\subsection{Attacker}

\textbf{Goal}. The attacker intends to use voice conversion to produce audio samples that sound similar to the target speaker.

\noindent\textbf{Knowledge}. We assume that the attacker has knowledge of any widely-used voice conversion methods.

\noindent\textbf{Capability}. We assume that the attacker has sufficient computational resources to train and implement any voice conversion model, has access to enough audio samples of the target speaker to ensure a well-performing model, and has the ability to arbitrarily modify audio samples to hinder voiceprint recovery.

\subsection{Defender}

\textbf{Goal}. The defender aims to extract the source voiceprint from a converted audio or the sole voiceprint from a genuine audio.

\noindent\textbf{Knowledge}. We assume that the defender does not know whether the input audio is converted or genuine. The defender does not know the specific voice conversion method used by the attacker, but has knowledge of commonly-used voice conversion methods.

\noindent\textbf{Capability}. The defender can collect public voice datasets for training the voiceprint recovery model. During inference, the model requires only a single input audio to extract the source speaker voiceprint, with no additional data and no prior exposure to the source or target speaker during training. The extracted voiceprint can then be used for speaker recognition. We assume a voiceprint database is available for this purpose, which can be constructed from only a few reference audio samples per speaker. This assumption is reasonable given that the mass collection of voices has become increasingly common across various domains, such as social media, banking, and prisons \cite{tiktok2021, bank2019}. If the true identity of the source speaker is already enrolled in the defender's database, the speaker recognition system yields a ``hit,'' successfully mapping the audio to a specific identity. Conversely, if the speaker is not present in the database, the system registers a ``miss,'' and the extracted voiceprint can be archived to facilitate future cross-referencing as the database expands. We discuss these two cases in terms of forensic settings in Section \ref{sec.deployment_scenario}.

\emph{Disclaimer.} Note that TTS is out of our scope since TTS leaves no source voiceprint to be recovered. Nevertheless, \sys can be used to trace synthesized audio back to the specific TTS tools.

\section{Methodology}

As shown in Figure \ref{fig.overview}, \sys consists of three main modules, i.e., latent representation, VC-informed calibration, and guided refinement.

{\begin{itemize}
    \item \textbf{Latent representation.} The latent representation module applies to extract the basic latent representation of the input audio, which encodes information including the source speaker voiceprint, the target speaker voiceprint, and the method-specific fingerprints. The output latent representation requires further purification to obtain a highly discriminative representation of the source speaker's identity.
    \item \textbf{VC-informed calibration.} The VC-informed calibration module identifies the specific voice conversion method and guides the model to refine the extracted latent representation, steering it towards the source speaker voiceprint. This module consists of the primary feature extractor and the first auxiliary branch dedicated to VC method identification.
    \item \textbf{Guided refinement.} The guided refinement module extracts a latent representation of the target speaker, facilitating the isolation of target-specific traits from the composite converted audio. This module incorporates the second auxiliary branch, which biases the primary extractor toward extracting the source speaker voiceprint and away from the target speaker voiceprint during training.
\end{itemize}}

\subsection{Latent Representation}
\label{section.method_shared}

The latent representation module is applied to extract basic representation of the audio sample. As shown in Figure \ref{fig.overview}, we adapt the internal architecture of widely-used speaker recognition models as the backbone of this module. 

The latent representation module is composed of a filter bank layer, TDNN blocks, SERes blocks, and a differential block. The filter bank layer is used to extract Mel-frequency-domain features, which are widely used for speaker recognition. Then a feature extraction block, namely a TDNN block and three SERes blocks, helps extract time-domain features and global channel interdependence. The TDNN block consists of one one-dimensional convolution (Conv1D) layer, one rectified linear unit (ReLU) activation layer, and one batch normalization (BN) layer. An SERes block consists of one TDNN block, one Res2-Dilated-TDNN layer, another TDNN block, and one squeeze-excitation layer. The outputs of three SERes blocks are concatenated and fed into a differential block. 

The differential block is specially designed to assist in guided refinement. In addition to the outputs of the SERes blocks, the differential block accepts an auxiliary input. Specifically, during training, if the input audio sample $x$ is converted, a genuine audio sample from the target speaker serves as this auxiliary input; if $x$ is genuine, the auxiliary input is set to a zero embedding. Internally, the differential block utilizes a built-in TDNN block to process the difference between the two inputs, which is then added back to the original input. This combined representation is subsequently fed into another TDNN block for further feature extraction.

During inference, the auxiliary input of the differential block is fixed as a zero embedding regardless of the input type. This design strictly conforms to our threat model, which assumes the defender has no prior knowledge of whether the test audio is converted or genuine. Notably, using a genuine target speaker sample during training but a zero embedding during inference for a converted input $x$ might raise concerns regarding a training-inference mismatch. However, our ablation study confirms that incorporating genuine target speaker audio as the auxiliary input during training yields superior performance compared to using zero embeddings across both phases. Detailed analysis is provided in Section \ref{section.ablation_ti}.

\subsection{VC-informed Calibration}
\label{section.method_vcspecific}

The VC-informed calibration module introduces the information of the voice conversion method to guide the model to refine the extracted embedding towards the source speaker voiceprint. To achieve this objective, we equip this module with three submodules, i.e., VC method inference, attention-based refinement, and feature aggregation. Let $r$ denote the output of the latent representation module.  

\begin{figure}[t]
    \centering
    	\begin{minipage}[b]{0.65\linewidth}
    		\centering
    		\includegraphics[trim=0mm 0mm 0mm 0mm, clip, width=\textwidth]{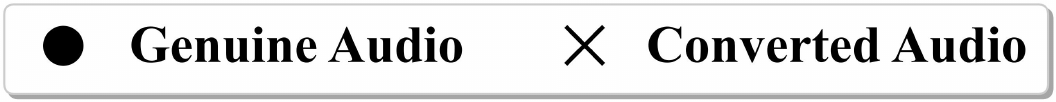}
    	\end{minipage} \\
    \subfigure[O]{
    	\begin{minipage}[b]{0.42\linewidth} 
    		\centering
    		\includegraphics[trim=0mm 0mm 0mm 0mm, clip, width=0.95\textwidth]{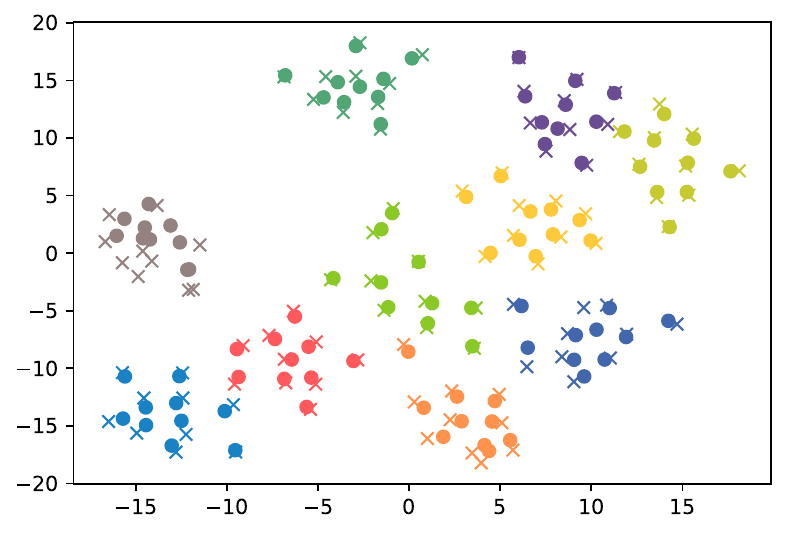}
    	\end{minipage}
    }
     \vspace{-0.1cm}
    \subfigure[O+M]{
    	\begin{minipage}[b]{0.42\linewidth}
    		\centering
    		\includegraphics[trim=0mm 0mm 0mm 0mm, clip, width=0.95\textwidth]{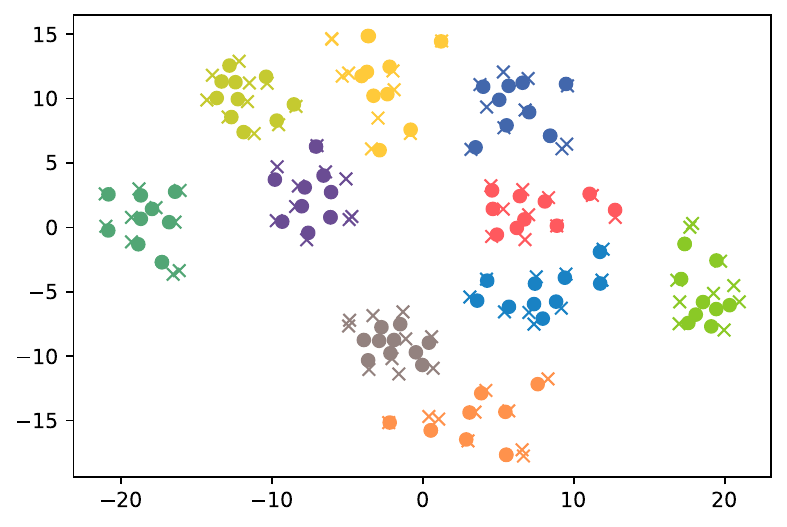}
    	\end{minipage}
    }
    \subfigure[O+G]{
    	\begin{minipage}[b]{0.42\linewidth}
    		\centering
    		\includegraphics[trim=0mm 0mm 0mm 0mm, clip, width=0.95\textwidth]{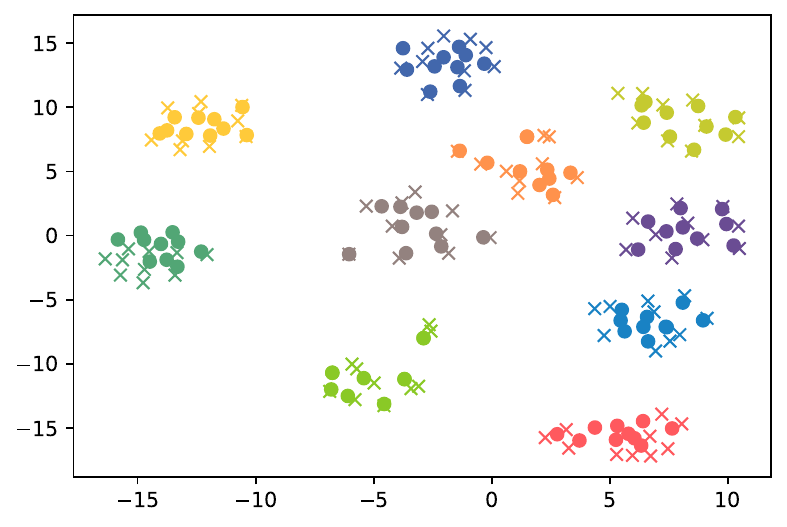}
    	\end{minipage}
    }
    \hspace{-0.1cm}
    \vspace{0.2cm}
    \subfigure[O+M+G]{
    	\begin{minipage}[b]{0.42\linewidth}
    		\centering
    		\includegraphics[trim=0mm 0mm 0mm 0mm, clip, width=0.95\textwidth]{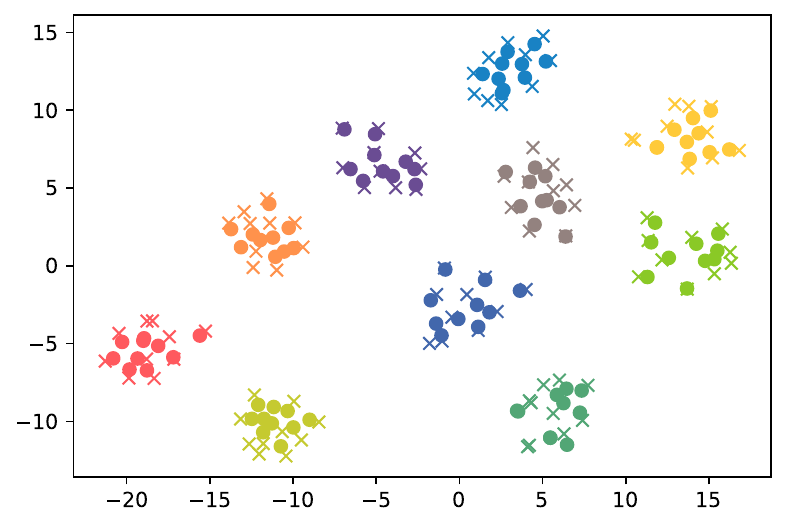}
    	\end{minipage}
    }
    \resizebox{0.75\linewidth}{!}{
    \begin{tabular}{ccccc}
    \Xhline{1pt}
        \multirow{2}{*}{\textbf{Distance}} & \multicolumn{4}{c}{\textbf{Branch Combination}} \\
        & \textbf{O}  & \textbf{O+M}   & \textbf{O+G} & \textbf{O+M+G}\\
      \Xhline{1pt}
      Inter-Speaker ($\uparrow$) & 0.84 & 0.86 & 0.92 & 0.96  \\
      Intra-Speaker ($\downarrow$) & 0.59 & 0.54 & 0.49 & 0.48  \\
    \Xhline{1pt}
    \end{tabular}
    }
	\caption{Distribution of voiceprints extracted by models with different auxiliary branch combinations, which is visualized by t-SNE \cite{van2008visualizing}. M and G are the abbreviations for the VC method inference and the guided refinement. O denotes a variant of \sys that excludes the two auxiliary branches. Different colors represent different ground-truth speakers. Distances are computed using cosine distance between the extracted voiceprints, which subtracts the cosine similarity from unity.}
	\label{fig:ablation}
\end{figure}

\subsubsection{VC method inference} The VC method inference submodule aims to infer the specific VC method used to convert the input audio samples. More specifically, during training, the defender constructs converted samples using various widely-used VC methods so that the method used to convert a training sample is known. Let $\mathcal{M}$ denote the set of VC methods used during the training phase. A label $0$ is added to indicate that the audio sample is genuine.

Note that we assume that the defender does not know the specific VC method adopted to convert an input sample during inference. The ground-truth VC method utilized by the attacker may belong to $\mathcal{M}$ or not. Our extensive empirical study demonstrates that even if the ground-truth VC method is not in $\mathcal{M}$, the information from the VC method inference submodule is still instrumental. This is because a well-performed VC method is usually a variant of one or multiple mainstream VC methods. The VC method inference submodule will find the most \emph{similar} method to the ground-truth one, narrowing down the searching scope. Since different families of VC methods shape the converted audio samples in different ways, being aware of the VC method or its proxy provides valuable guidelines for \sys to further process the embedding to distill the source voiceprint. Figure \ref{fig:ablation} illustrates the t-SNE \cite{van2008visualizing} visualization of extracted voiceprint space under different auxiliary branch combinations. It is shown that the extracted voiceprint embeddings from converted samples cluster more closely to those of genuine samples in the presence of the VC method inference branch. This indicates that this auxiliary branch provides crucial guidance for \sys to accurately distill the residual source voiceprint.

The VC method inference submodule processes input $r$ through 2 blocks, i.e., an ASP block and a classification block. The ASP block features attentive stat pooling (ASP). The classification block comprises a fully-connected (FC) layer and a batch normalization (BN) layer. Note that during training, the ASP block and the classification block will be jointly trained, but only the ASP block is used during inference. Let $m$ denote the output of the ASP block of the VC method inference submodule.  The additive angular margin (AAM) loss is employed as the loss function. We will describe the details of the ASP block as we introduce the attention-based refinement submodule and the AAM loss as we introduce the feature aggregation submodule. 

\begin{figure}[t] 
\centering     
    \includegraphics[width=0.48\textwidth]{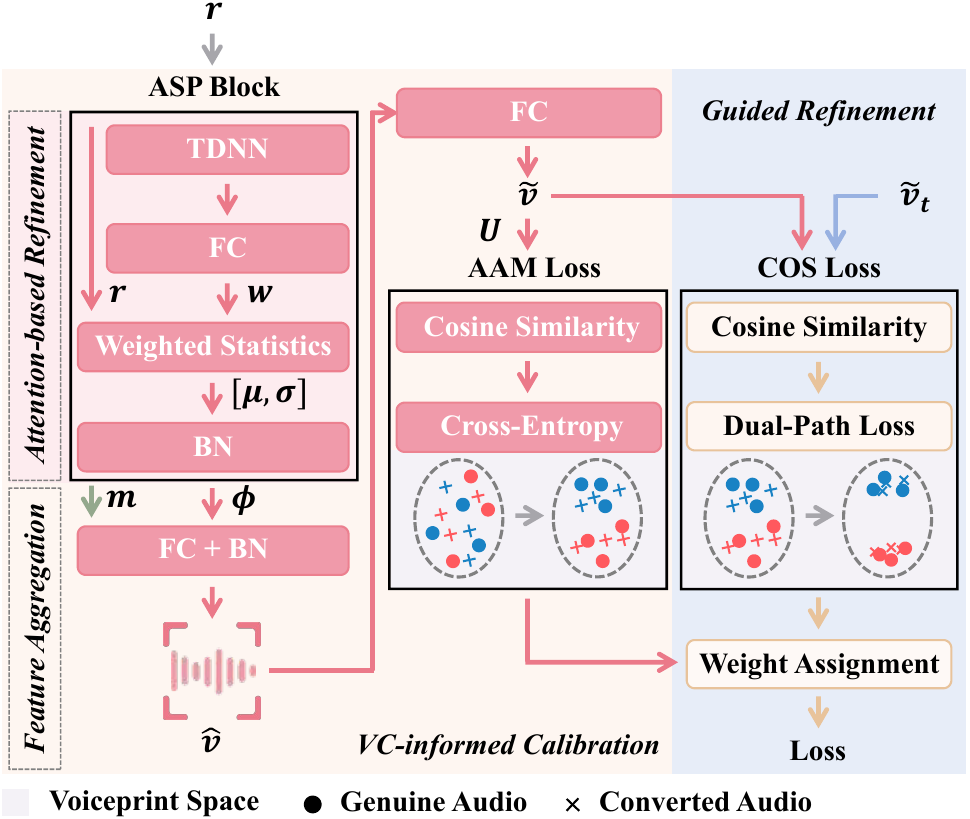}
    \caption{The workflow of VC-informed calibration and guided refinement. The changes of voiceprint distributions visualized by t-SNE \cite{van2008visualizing} are schematized to illustrate the effects of VC-informed calibration and guided refinement.}
    \label{fig.extractor}
\end{figure}

\subsubsection{Attention-based refinement}

The attention-based refinement submodule is composed of one ASP block, which is designed to refine the feature representations to be more discriminative.

The internal structure of ASP is shown in Figure \ref{fig.extractor}. The input $v$ is first processed by one TDNN block and one FC layer. The output represents inter-frame attention weights, which reflect each frame's importance to the source speaker identity. We have
\begin{equation}
w_{{j}}=\frac{FC(TDNN(r_{j}))}{\sum_{j=1}^{T}FC(TDNN(r_{j}))}.
\label{eq:diffcbackend1}
\end{equation}
where $r_{j}$ is the $j$-th frame of $r$, $w_{{j}}$ is the attentive weight of $r_j$, and $T$ is the number of frames.

Combining $r_{j}$ with $w_{{j}}$ can capture the speaker-related features, effectively filtering out irrelevant information. The weighted statistics, namely mean and standard deviation, are calculated to model the speaker-related distribution patterns, which are associated with the identity information of speakers. We have
\begin{equation}
\begin{split}
\mu_r&=\sum_{j=1}^{T}r_{j}w_{{j}},\\ \sigma_r&=\sqrt{\sum_{j=1}^{T}w_{{j}}(r_{j}-\mu)^{2}}.
\end{split}
\end{equation}

Then a BN layer is used to standardize the concatenated statistic features, aligning their numerical scales to avoid biased weighting in the following FC layers.
\begin{equation}
\phi=BN([\mu, \sigma]).
\end{equation}

\subsubsection{Feature aggregation}

The feature aggregation submodule takes the output of the VC method inference submodule $m$ and the output of the attention-based refinement submodule $\phi$, and outputs the final extracted voiceprint $\hat{v}$. The feature aggregation submodule contains one FC layer and one BN layer. 

During training, the design of the loss function is important since it guides the learning process of the model. Naturally, the optimization goal is to minimize the distance between the extracted voiceprint $\hat{v}$ and the ground-truth voiceprint $v$. Note that the ground-truth voiceprint $v$ is the source speaker's voiceprint of a converted audio sample or the sole speaker's voiceprint of a genuine audio sample.

We employ the additive angular margin (AAM) loss. The AAM loss forces the extracted voiceprint of the same category to have high similarity, thereby enabling the model to generalize to unseen speakers. The AAM loss learns a weight matrix $U$ that represents corresponding embeddings for each speaker, which can be used to calculate and enhance intra-class compactness and inter-class distinctiveness in the angular space.

One FC layer is utilized to perform feature transformation that enables more training-efficient loss computation.
\begin{equation}
\tilde{v}=FC(\hat{v}).
\label{eq:diffcbackend1}
\end{equation}

Then the cosine similarity between the output embedding $\tilde{v}$ and the weighted matrix is calculated as
\begin{equation}
\cos \theta_{i}=\frac{\tilde{v}^{T}u_{i}}{||\tilde{v}||_{2}||u_i||_{2}}.
\end{equation}
where $u_i$ is the representative embedding of speaker $i$. $||\cdot||_{2}$ represents the $L_2$ norm. $\cos \theta_{i}$ computes the similarity score between the output embedding $\tilde{v}$ and the representative embedding $u_{i}$ of speaker $i$.

As for the source speaker $y_s$ in the label, AAM loss will add a predefined angular margin $\delta$, which enforces tighter angular clustering for positive pairs while pushing negative pairs beyond a fixed angular boundary.
\begin{equation}
\cos(\theta_{y_{s}}+\delta)=\cos \theta_{y_{s}}\cos \delta - \sin \theta_{y_{s}}\sin \delta.
\end{equation}

To prevent sluggish gradient updates, we introduce a scaling factor $\beta$ to amplify the cosine similarity scores. 
\begin{equation}
s_{i} = 
\begin{cases}
\beta\cos \theta_{i}, & i \neq y_{s},\\
\beta \cos (\theta_{y_{s}}+\delta), & i = y_{s}.
\end{cases}
\end{equation}

Consequently, the final AAM loss function, an improved softmax cross-entropy loss with additive angular margin, can be formulated as

\begin{equation}
\mathcal{L}_{AAM}=-\frac{1}{N}\sum_{(x_i,y_i)\in \mathcal{D}}\ln \frac{e^{s_{y_{s}}}}{\sum_{i}e^{s_{i}}}.
\end{equation}
where $N$ is the total number of samples in dataset $\mathcal{D}$.

\subsection{Guided Refinement}
\label{section.loss}

During training, we teach the model to learn to simultaneously master two tasks: (1) extract the source speaker voiceprint from converted audio samples, and (2) extract the sole speaker voiceprint from genuine audio samples. However, the first task is obviously more challenging than the second one. Our extensive empirical study shows that the existing models fail to yield satisfactory results in the face of advanced voice conversion methods. To address this problem, we further improve the model with a guided refinement module. This module elicits more information to aid the more complicated task of extracting the source speaker voiceprint from converted audio samples. 

The guided refinement module is an additional subnetwork, which will be used for training but not for inference. The input of the subnetwork is the output of the latent representation module $r$ and the output of the subnetwork is the target speaker of converted audios or the sole speaker of genuine audios. Let $\tilde{v}_{t}$ denote the output voiceprint of the guided refinement module. 

During training, we leverage the extracted target speaker voiceprint $\tilde{v}_{t}$ to further guide the refinement of the extracted source speaker voiceprint $\tilde{v}$ via an additional loss term. The loss term aims to reduce the similarity between $\tilde{v}_{t}$ and $\tilde{v}$ of converted audio samples, while increasing the similarity between $\tilde{v}_{t}$ and $\tilde{v}$ of genuine audio samples.

\begin{equation}
\mathcal{L}_{COS}=\mathcal{L}_{COS_{c}}-\mathcal{L}_{COS_{g}}.
\label{eq:diffbackendloss2}
\end{equation}
\begin{equation}
\mathcal{L}_{COS_{c}}=\frac{1}{N_{c}}\sum_{x_{c}\in \mathcal{D}_{c}}\frac{\tilde{v}^{T}\tilde{v}_t}{||\tilde{v}||_{2}||\tilde{v}_t||_{2}}.
\label{eq:diffbackendloss2}
\end{equation}
\begin{equation}
\mathcal{L}_{COS_{g}}=\frac{1}{N_{g}}\sum_{x_{g}\in \mathcal{D}_{g}}\frac{\tilde{v}^{T}\tilde{v}_t}{||\tilde{v}||_{2}||\tilde{v}_t||_{2}}.
\label{eq:diffbackendloss2}
\end{equation}
in which $\mathcal{D}_c$ is the subset of converted audio samples and $\mathcal{D}_g$ is the subset of genuine audio samples in the training dataset. $N_{c}$ is the number of converted audios in $\mathcal{D}_{c}$ and $N_{g}$ is the number of genuine audios in $\mathcal{D}_{g}$. As shown in Figure \ref{fig:ablation}, the guided refinement branch further improves the performance of source speaker recovery, increasing the inter-speaker distance to 0.96.

The overall loss function for training is 
\begin{equation}
    \mathcal{L} = \mathcal{L}_{AAM}+\lambda \mathcal{L}_{COS},
\end{equation}
in which $\lambda$ is a hyperparameter.

During the whole training process, we also implement commonly-used data augmentation techniques to boost model performance, including modifying speed, adding noise, adding reverberation, and randomly dropping frames.

\section{Evaluation}
\label{section.evaluation}

We conduct extensive experiments to evaluate the effectiveness of \sys. The experimental setup is detailed in Section \ref{section.setup}. In this section, we take different voice conversion methods (Section \ref{section.overall}), gender subgroups (Section \ref{section.gender}), telephony codecs (Section \ref{section.telephony}), real-world scenarios (Section \ref{section.real_world}), large-scale settings (Section \ref{section.large_scale}), preprocessing obfuscation (Section \ref{section.preprocess}), subjective study (Section \ref{section.study}), adaptive scenarios (Section \ref{section.adaptive}), unseen voice conversion methods (Section \ref{section.unseenvc}), and unseen languages (Section \ref{section.language}) into consideration. We conduct two ablation studies to demonstrate the effect of auxiliary branches and the impact of training-inference inconsistency in Section \ref{section.ablation} and \ref{section.ablation_ti}. We also explore the impact of some parameters on the method performance in Sections \ref{section.imapct_number}-\ref{section.impact_loss_weight}. The comprehensive experimental results demonstrate the superiority of \sys.

\subsection{Setup}
\label{section.setup}

\subsubsection{Datasets}

We conduct experiments on three basic speaker recognition datasets: VoxCeleb1 \cite{nagrani2017voxceleb}, VoxCeleb2 \cite{chung2018voxceleb2}, and LibriSpeech \cite{panayotov2015librispeech}. These datasets are collected from open resources such as interview videos and audiobooks. Due to their high quality, diversity, and open accessibility, these datasets are widely used in speaker identification and speech recognition. Moreover, these three datasets contain extensive speakers and audio samples. We have annotated the number of speakers and audio samples from each dataset in the Appendix (Table \ref{tab.datasets}). The number of speakers achieves 9623, which are split into 9583 speakers for training and the remaining 40 speakers for testing. The detailed information on converted dataset generation is presented in the Appendix \ref{section.dataset}.

Genuine samples are randomly sampled from audios of each speaker. For the training datasets, the total number of genuine samples is consistent with the converted audio number of one voice conversion method. For the testing dataset, we randomly sample 50 genuine audios from each speaker to form the genuine test set. As for reference audios, three reference audios are randomly sampled from the remaining genuine audios to form the voiceprint templates for each speaker. There is no overlap between reference audios and test genuine audios.

As for voice preprocessing, we resample the audios to 16kHz in both the training phase and testing phase. The length of audios is randomly trimmed to 6 seconds due to the memory limitation of GPUs in the training phase and trimmed or padded to 20 seconds for high-precision authentication in the testing phase.

\subsubsection{Voice Conversion Methods}

We evaluate our method against seven representative voice conversion methods, providing systematic coverage of all mainstream families. They are AGAIN-VC \cite{chen2021again}, VQVC \cite{wu2020one}, VQVC+ \cite{wu2020vqvc+}, BNE-Seq2seqMoL \cite{liu2021any}, FreeVC \cite{li2023freevc}, Diff-HierVC \cite{choi2023diff}, and DDDM-VC \cite{choi2024dddm}. For brevity and clarity, we use the concise prefixes of compound method names (e.g., AGAIN for AGAIN-VC, BNE for BNE-Seq2seqMoL, Diff for Diff-HierVC, and DDDM for DDDM-VC) where no ambiguity arises.

\subsubsection{Evaluation Metrics}

We employ the Equal Error Rate and Top-$k$ Accuracy for performance evaluation.
\begin{itemize}
\item \textbf{Equal Error Rate (EER)} is a performance metric widely used in speaker recognition, defined as the error rate at the point where the false positive rate equals the false negative rate. A lower EER corresponds to higher discriminative capability in voiceprint recognition.

\item \textbf{Top-$k$ Accuracy (Top-$k$ ACC)} is widely used in multi-label classification tasks, by calculating the proportion of correct labels in the top $k$ predicted results. However, in order to verify the model performance in unseen speakers, we use Top-$k$ ACC to measure the cosine similarity between $\hat{v}$. It means that the Top-$k$ ACC depends on the rank of cosine similarity between the extracted voiceprints and the reference voiceprint templates. We assign $k$ the values of 1, 3, and 5 in experiments, which can comprehensively reflect the model performance.
\end{itemize}

\begin{table}[t]
  \centering
  \footnotesize
  \caption{Performance of voiceprint recovery methods against various voice conversion methods.}
  \vspace{-0.2cm}
  \label{tab.overall}
  \resizebox{1.00\linewidth}{!}{
  \begin{threeparttable}
    \begin{tabular}{crccccc}
    \Xhline{1pt}
      \multirow{2}{*}{\textbf{VC}} & \multicolumn{1}{c}{\multirow{2}{*}{\textbf{Metric}}} & \multicolumn{4}{c}{\textbf{Voiceprint Recovery Method}} \\
      & & \textbf{MFA} & \textbf{ECAPA} & \textbf{Revelio} & \multicolumn{1}{c}{\textbf{\sysB} }\\
      \Xhline{1pt}
      \multirow{4}{*}{Clean}
      & EER ($\downarrow$)
      & 15.93\% & 4.24\% & 2.38\% & \cellcolor{backcolor!100}\textbf{1.11\%}\\
      \addlinespace[0.2pt]
      \cline{2-6}
      \addlinespace[1pt]
      & Top-1 ACC ($\uparrow$)
      & 69.14\% & 83.77\% & 93.06\% & \cellcolor{backcolor!100}\textbf{98.49\%}\\
      & Top-3 ACC ($\uparrow$)
      & 82.24\% & 88.58\% & 97.26\% & \cellcolor{backcolor!100}\textbf{99.86\%}\\
      & Top-5 ACC ($\uparrow$)
      & 82.76\% & 90.50\% & 98.60\% & \cellcolor{backcolor!100}\textbf{99.97\%}\\
      \addlinespace[0.2pt]
      \cline{1-6}
      \addlinespace[1pt]
      \multirow{4}{*}{AGAIN}
      & EER ($\downarrow$)
      & 35.87\% & 31.15\% & 4.66\% & \cellcolor{backcolor!100}\textbf{1.34\%}\\
      \addlinespace[0.2pt]
      \cline{2-6}
      \addlinespace[1pt]
      & Top-1 ACC ($\uparrow$)
      & 22.44\% & 10.87\% & 83.88\% & \cellcolor{backcolor!100}\textbf{97.79\%} \\
      & Top-3 ACC ($\uparrow$)
      & 38.46\% & 21.86\% & 92.88\% & \cellcolor{backcolor!100}\textbf{99.81\%}\\
      & Top-5 ACC ($\uparrow$)
      & 50.45\% & 29.49\% & 95.74\% & \cellcolor{backcolor!100}\textbf{99.94\%}\\
      \addlinespace[0.2pt]
      \cline{1-6}
      \addlinespace[1pt]
      \multirow{4}{*}{VQVC}
      & EER ($\downarrow$)
      & 45.91\% & 40.81\% & 8.48\% & \cellcolor{backcolor!100}\textbf{6.83\%}\\
      \addlinespace[0.2pt]
      \cline{2-6}
      \addlinespace[1pt]
      & Top-1 ACC ($\uparrow$)
      & 10.03\% & 4.81\% & 64.54\% & \cellcolor{backcolor!100}\textbf{83.39\%}\\
      & Top-3 ACC ($\uparrow$)
      & 22.76\% & 12.92\% & 86.53\% & \cellcolor{backcolor!100}\textbf{94.58\%}\\
      & Top-5 ACC ($\uparrow$)
      & 30.48\% & 19.97\% & 92.95\% & \cellcolor{backcolor!100}\textbf{97.05\%} \\
      \addlinespace[0.2pt]
      \cline{1-6}
      \addlinespace[1pt]
      \multirow{4}{*}{VQVC+}
      & EER ($\downarrow$)
      & 43.37\% & 36.69\% & 5.54\% & \cellcolor{backcolor!100}\textbf{3.85\%}\\
      \addlinespace[0.2pt]
      \cline{2-6}
      \addlinespace[1pt]
      & Top-1 ACC ($\uparrow$)
      & 21.60\% & 11.15\% & 77.92\% & \cellcolor{backcolor!100}\textbf{94.36\%}\\
      & Top-3 ACC ($\uparrow$)
      & 40.42\% & 24.14\% & 91.41\% & \cellcolor{backcolor!100}\textbf{99.14\%} \\
      & Top-5 ACC ($\uparrow$)
      & 47.95\% & 32.98\% & 94.58\% & \cellcolor{backcolor!100}\textbf{99.71\%} \\
      \addlinespace[0.2pt]
      \cline{1-6}
      \addlinespace[1pt]
      \multirow{4}{*}{BNE}
      & EER ($\downarrow$)
      & 43.48\% & 46.17\% & 10.46\% & \cellcolor{backcolor!100}\textbf{7.92\%}\\
      \addlinespace[0.2pt]
      \cline{2-6}
      \addlinespace[1pt]
      & Top-1 ACC ($\uparrow$)
      & 13.05\% & 1.73\% & 60.96\% & \cellcolor{backcolor!100}\textbf{77.40\%}\\
      & Top-3 ACC ($\uparrow$)
      & 21.64\% & 8.88\% & 84.46\% & \cellcolor{backcolor!100}\textbf{90.03\%}\\
      & Top-5 ACC ($\uparrow$)
      & 30.22\% & 15.96\% & 91.03\% & \cellcolor{backcolor!100}\textbf{95.29\%}\\
      \addlinespace[0.2pt]
      \cline{1-6}
      \addlinespace[1pt]
      \multirow{4}{*}{FreeVC} 
      & EER ($\downarrow$)
      & 33.91\% & 39.99\% & 5.38\% & \cellcolor{backcolor!100}\textbf{3.87\%}\\
      \addlinespace[0.2pt]
      \cline{2-6}
      \addlinespace[1pt]
      & Top-1 ACC ($\uparrow$)
      & 22.12\% & 6.28\% & 86.19\% & \cellcolor{backcolor!100}\textbf{93.69\%}\\
      & Top-3 ACC ($\uparrow$)
      & 35.87\% & 17.02\% & 94.01\% & \cellcolor{backcolor!100}\textbf{98.59\%}\\
      & Top-5 ACC ($\uparrow$)
      & 43.78\% & 25.00\% & 96.64\% & \cellcolor{backcolor!100}\textbf{99.58\%} \\
      \addlinespace[0.2pt]
      \cline{1-6}
      \addlinespace[1pt]
      \multirow{4}{*}{Diff}
      & EER ($\downarrow$)
      & 36.93\% & 32.61\% & 4.77\% & \cellcolor{backcolor!100}\textbf{2.61\%}\\
      \addlinespace[0.2pt]
      \cline{2-6}
      \addlinespace[1pt]
      & Top-1 ACC ($\uparrow$)
      & 19.26\% & 9.52\% & 90.58\% & \cellcolor{backcolor!100}\textbf{95.35\%}  \\
      & Top-3 ACC ($\uparrow$)
      & 28.27\% & 29.30\% & 98.11\% & \cellcolor{backcolor!100}\textbf{99.62\%}  \\
      & Top-5 ACC ($\uparrow$)
      & 37.08\% & 39.90\% & 99.18\% & \cellcolor{backcolor!100}\textbf{99.90\%}  \\
      \addlinespace[0.2pt]
      \cline{1-6}
      \addlinespace[1pt]
      \multirow{4}{*}{DDDM}
      & EER ($\downarrow$)
      & 41.43\% & 35.12\% & 4.26\% & \cellcolor{backcolor!100}\textbf{3.02\%}\\
      \addlinespace[0.2pt]
      \cline{2-6}
      \addlinespace[1pt]
      & Top-1 ACC ($\uparrow$)
      & 22.53\% & 6.47\% & 88.33\% & \cellcolor{backcolor!100}\textbf{94.98\%} \\
      & Top-3 ACC ($\uparrow$)
      & 33.78\% & 22.82\% & 95.45\% & \cellcolor{backcolor!100}\textbf{99.26\%} \\
      & Top-5 ACC ($\uparrow$)
      & 42.50\% & 33.01\% & 97.37\% & \cellcolor{backcolor!100}\textbf{99.84\%}  \\
      \Xhline{1pt}
    \end{tabular}
         \end{threeparttable}
    }
\end{table}

In addition, we also take clean performance into consideration by calculating the EER and Top-$k$ ACC on genuine audios. The clean performance reflects the model's ability to correctly identify the true speaker rather than fabricating a non-existent source speaker for non-converted audio samples. All experiments are repeated five times and the average values are taken as the experimental results.

\subsubsection{Baselines and Method Settings}
We compare our method against representative baselines, including the powerful speaker recognition models MFA-Conformer \cite{zhang2022mfa} and ECAPA-TDNN \cite{desplanques2020ecapa}, as well as the state-of-the-art voice recovery method Revelio \cite{deng2023catch}. As for \sys, we set the default configuration as $\delta=0.2$, $\beta=30$, and $\lambda=0.1$. The parameters of the model are detailed in the Appendix (Table \ref{tab.model}). In the training phase, we initialize the latent representation module of both Revelio and \sys with the parameters of an ECAPA-TDNN pretrained on VoxCeleb1\&2. We utilize the Adam \cite{kingma2014adam} optimizer to train models for 4 epochs with a learning rate set as 1e-3, a weight decay rate set as 2e-6, and the batch size set as 24 until the loss no longer decreases. Each model is trained on a mixture of genuine audios and converted audio samples processed by these seven VC methods, considering that a robust model should learn to restore voiceprints converted by different methods. All models are tested with the auxiliary input fixed as a zero embedding.

\subsection{Overall Effectiveness}
\label{section.overall}

Here we compare MFA-Conformer, ECAPA-TDNN, Revelio, and our method \sys across seven different voice conversion methods. As shown in Table \ref{tab.overall}, MFA-Conformer and ECAPA-TDNN exhibit significant vulnerability to converted audios, which underscores the necessity of voiceprint recovery for voice authentication security. However, the existing method Revelio only achieves 6.22\% EER and 78.91\% Top-1 ACC on converted audios. As for \sys, our method substantially improves voiceprint recovery performance, attaining 4.21\% EER, 90.99\% Top-1 ACC, 97.29\% Top-3 ACC, and 98.76\% Top-5 ACC on converted audios. These experimental results demonstrate that the three-pronged structure enables the extraction of highly distinctive source voiceprints from complex converted audios. Furthermore, \sys also achieves a high clean performance on genuine audio samples, yielding a 1.11\% EER and a 98.49\% Top-1 ACC. This not only demonstrates that \sys correctly identifies genuine audio as non-converted with a low false positive rate, but also highlights its capability to accurately recognize the true speaker of a genuine audio sample.

\begin{figure}[t]
    \centering
    	\begin{minipage}[b]{0.65\linewidth}
    		\centering
    		\includegraphics[trim=0mm 0mm 0mm 0mm, clip, width=\textwidth]{Section/Pictures/Draw/Point/Legend.pdf}
    	\end{minipage} \\
    \subfigure[MFA-Conformer]{
    	\begin{minipage}[b]{0.42\linewidth}
    		\centering
    		\includegraphics[trim=0mm 0mm 0mm 0mm, clip, width=0.95\textwidth]{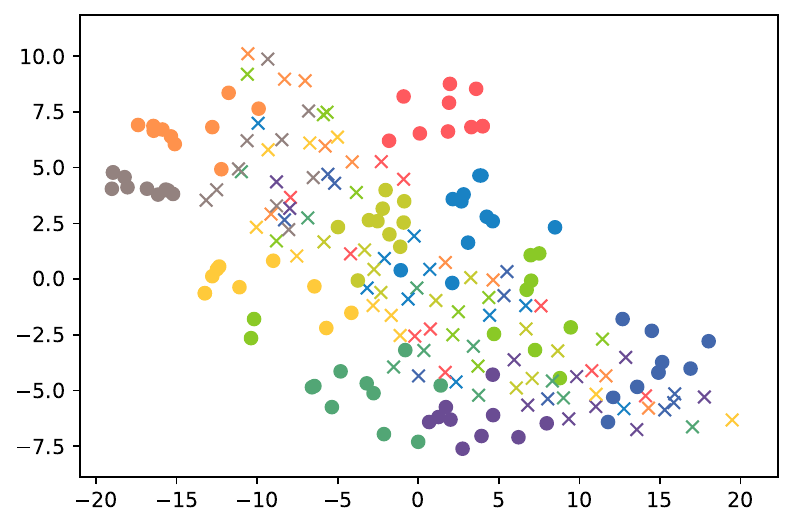}
    	\end{minipage}
    }
    \vspace{-0.1cm}
    \subfigure[ECAPA-TDNN]{
    	\begin{minipage}[b]{0.42\linewidth}
    		\centering
    		\includegraphics[trim=0mm 0mm 0mm 0mm, clip, width=0.95\textwidth]{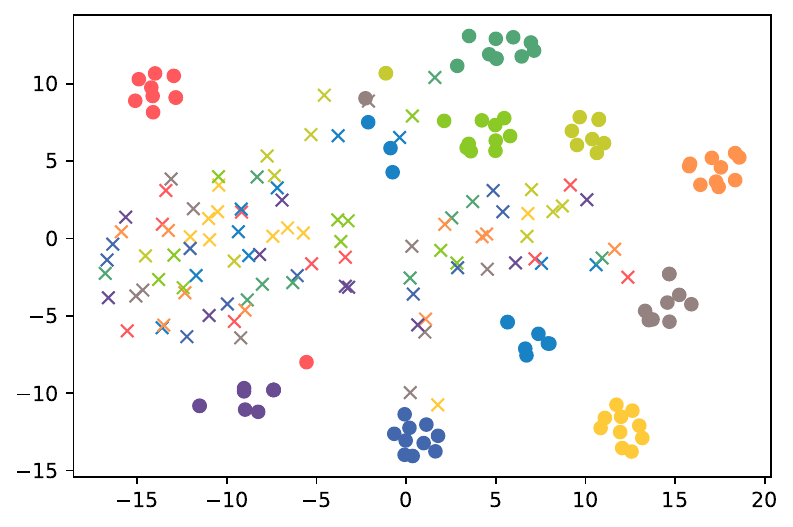}
    	\end{minipage}
    }
    \subfigure[Revelio]{
    	\begin{minipage}[b]{0.42\linewidth}
    		\centering
    		\includegraphics[trim=0mm 0mm 0mm 0mm, clip, width=0.95\textwidth]{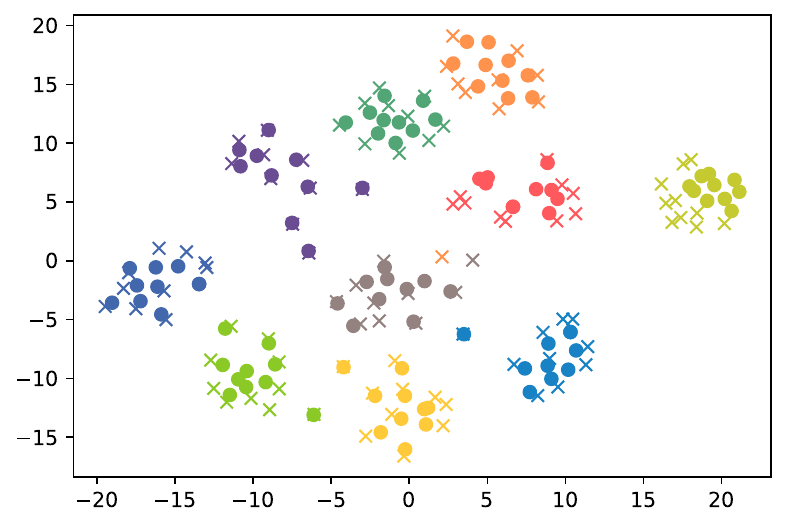}
    	\end{minipage}
    }
    \vspace{-0.2cm}
    \subfigure[\sys]{
    	\begin{minipage}[b]{0.42\linewidth}
    		\centering
    		\includegraphics[trim=0mm 0mm 0mm 0mm, clip, width=0.95\textwidth]{Section/Pictures/Draw/Point/Nformat.pdf}
    	\end{minipage}
    }
	\caption{Distribution of voiceprints extracted by MFA-Conformer, ECAPA-TDNN, Revelio, and \sys, which is visualized by t-SNE \cite{van2008visualizing}. Different colors represent different ground-truth speakers.}
	\label{fig:tsne}
\end{figure}

To further illustrate the effectiveness, we visualize the cumulative distribution function (CDF) of cosine similarity between extracted voiceprints and reference voiceprint templates. And as we can see in the Appendix (Figure \ref{fig:cdf_n}), the cosine similarity between an extracted voiceprint and the reference voiceprint of the ground-truth speaker is much higher than that of other speakers, which intuitively demonstrates the strong discriminative ability of \sys. The CDFs of other methods are also provided in the Appendix (Figures \ref{fig:cdf_mfa}-\ref{fig:cdf_r}). Additionally, we employ t-SNE \cite{van2008visualizing} to visualize the discriminative ability of different voiceprint recovery methods. Figure \ref{fig:tsne} shows that voiceprints of the same ground-truth speaker are well clustered by \sys.

\subsection{Performance by Gender Subgroups}
\label{section.gender}

Audio samples exhibit inherent differences in pitch, harmonics, and speaking rate across genders, which may influence the quality of converted audios. To investigate the impact of gender, we divide the converted audios into four groups, male-to-male (M$\rightarrow$M), female-to-female (F$\rightarrow$F), male-to-female (M$\rightarrow$F), and female-to-male (F$\rightarrow$M). The first refers to the gender of the source speaker, i.e., the attacker, and the second refers to the gender of the target speaker. As shown in the Appendix (Table \ref{tab.gender}), the performance of \sys varies in different gender subgroups, but within an acceptable range. \sys achieves an average of 91.16\% (M$\rightarrow$M), 89.46\% (F$\rightarrow$F), 90.18\% (M$\rightarrow$F) and 90.44\% (F$\rightarrow$M) on Top-1 ACC. Based on the empirical evidence presented, gender differences do not appear to exert a significant impact on the model performance of \sys.

\subsection{Performance under Telephony Codecs}
\label{section.telephony}
Telephony is a canonical domain within voice communication technologies. We adopt four commonly-used audio codecs to simulate the voice quality degradation in real telephony transmissions.

\begin{itemize}
\item \textbf{G.711} \cite{recommendation1988pulse} is a widely adopted pulse-code modulation standard for digital telephony. It employs companding algorithms ($\mu$-law and A-law) to enhance the dynamic range of analog signals during digitization. The $\mu$-law variant is predominantly deployed in North America and Japan, whereas A-law is the standard in Europe and other international regions. Both operate at 8-bit resolution and an 8 kHz sampling rate, making them fundamental to traditional PSTN and VoIP systems.

\item \textbf{GSM-FR} \cite{etsi2000digital} processes audio signals as 13-bit linear PCM sampled at 8 kHz. As the inaugural speech coding standard for GSM mobile networks, it balances acceptable voice quality with the bandwidth constraints of early cellular systems.

\item \textbf{AMR-NB} \cite{ekudden1999adaptive} is a versatile narrow-band speech codec designed for GSM and UMTS networks. Its multi-rate capability dynamically adjusts bit rates based on network conditions, improving efficiency without compromising voice clarity.

\end{itemize}

We adopt four audio codecs to process audio converted by these seven different VC methods. As shown in the Appendix (Figure \ref{fig:tel}), we compare the performance of MFA-Conformer, ECAPA-TDNN, Revelio, and \sys. In these experiments, \sys maintains a 7.50\% EER and a 94.75\% Top-5 ACC on converted audio. As for genuine audio, \sys achieves a 2.62\% EER and a 95.63\% Top-1 ACC. It demonstrates the resilience of \sys to distortion, quantization noise, and information loss inherent in telephony codecs.

\subsection{Performance under Real-World Scenarios}
\label{section.real_world}

To assess real-world performance, we evaluate \sys across both telephony and VoIP channels. To simulate a realistic telephony scenario, the audio is played via a studio-grade loudspeaker, transmitted over the air, captured by an iPhone 13, and subsequently routed through a commercial cellular network to an iPhone 11. For the VoIP scenario, the speech is transmitted through a widely adopted commercial platform (Discord) via a virtual audio device, configured at encoding bitrates of 32 kbps and 64 kbps. The evaluation results across four voiceprint recovery methods are summarized in Table \ref{tab.real_trident} and the Appendix (Tables \ref{tab.real_revelio}--\ref{tab.real_mfa}). In the telephony scenario, all voiceprint recovery methods exhibit a performance degradation when dealing with audio manipulated by certain VC techniques, such as VQVC, VQVC+, and BNE. This drop in accuracy occurs because the telephone channel distortion exacerbates the artifacts and noise inherently introduced by these specific VC methods, thereby severely hindering downstream speaker recognition. Despite these severe real-world channel distortions, \sys consistently sustains high performance on genuine audio. For converted audio, \sys maintains robust performance, achieving an EER of 7.14\%, alongside Top-1, Top-3, and Top-5 accuracies of 86.01\%, 94.84\%, and 97.54\%, respectively. These results underscore the strong practical applicability and resilience of \sys in real-world deployment scenarios.

\begin{table}[t]
  \begin{center}
    \caption{Performance of \sys under real-world scenarios.}
    \label{tab.real_trident}
    \vspace{-0.2cm}
    \resizebox{1.0\linewidth}{!}{
    \begin{tabular}{cr c c c}
      \Xhline{1pt}
      \multicolumn{1}{c}{\multirow{2}{*}{\textbf{VC}}} & \multicolumn{1}{c}{\multirow{2}{*}{\textbf{Metric}}} &  \multicolumn{3}{c}{\textbf{Voiceprint Recovery Method}} \\
      \multicolumn{1}{c}{}& & \textbf{Telephone} & \textbf{VoIP-32kbps} & \textbf{VoIP-64kbps} \\
      \Xhline{1pt}
      \multirow{4}{*}{Clean}
      & EER ($\downarrow$)
      & 4.39\% & 2.71\%  & 2.22\% \\
      \addlinespace[0.2pt]
      \cline{2-5}
      \addlinespace[1pt]
      & Top-1 ACC ($\uparrow$)
      & 94.98\% & 96.78\% & 97.25\%  \\
      & Top-3 ACC ($\uparrow$)
      & 98.80\% & 98.89\% & 99.21\%  \\
      & Top-5 ACC ($\uparrow$)
      & 99.76\% & 99.78\% & 99.93\%  \\
      \addlinespace[0.2pt]
      \cline{1-5}
      \addlinespace[1pt]
      \multirow{4}{*}{AGAIN}
      & EER ($\downarrow$)
      & 6.22\% & 5.27\%  & 4.61\%  \\
      \addlinespace[0.2pt]
      \cline{2-5}
      \addlinespace[1pt]
      & Top-1 ACC ($\uparrow$)
      & 93.95\% & 94.74\% & 96.32\%  \\
      & Top-3 ACC ($\uparrow$)
      & 98.82\% & 98.95\% & 99.08\%  \\
      & Top-5 ACC ($\uparrow$)
      & 99.21\% & 99.61\% & 99.74\%  \\
      \addlinespace[0.2pt]
      \cline{1-5}
      \addlinespace[1pt]
      \multirow{4}{*}{VQVC}
      & EER ($\downarrow$)
      & 14.74\% & 8.63\%  & 7.89\%  \\
      \addlinespace[0.2pt]
      \cline{2-5}
      \addlinespace[1pt]
      & Top-1 ACC ($\uparrow$)
      & 63.68\% & 81.05\% & 83.29\%  \\
      & Top-3 ACC ($\uparrow$)
      & 85.66\% & 93.95\% & 94.37\%  \\
      & Top-5 ACC ($\uparrow$)
      & 92.53\% & 96.95\% & 97.16\%  \\
      \cline{1-5}
      \addlinespace[1pt]
      \multirow{4}{*}{VQVC+}
      & EER ($\downarrow$)
      & 9.40\% & 5.66\%  & 5.14\%  \\
      \addlinespace[0.2pt]
      \cline{2-5}
      \addlinespace[1pt]
      & Top-1 ACC ($\uparrow$)
      & 79.57\% & 92.24\% & 93.29\%  \\
      & Top-3 ACC ($\uparrow$)
      & 90.27\% & 97.90\% & 98.42\%  \\
      & Top-5 ACC ($\uparrow$)
      & 95.52\% & 98.82\% & 99.45\%  \\
      \cline{1-5}
      \addlinespace[1pt]
      \multirow{4}{*}{BNE}
      & EER ($\downarrow$)
      & 15.53\% & 8.68\%  & 8.32\%  \\
      \addlinespace[0.2pt]
      \cline{2-5}
      \addlinespace[1pt]
      & Top-1 ACC ($\uparrow$)
      & 60.79\% & 76.18\% & 77.37\%  \\
      & Top-3 ACC ($\uparrow$)
      & 82.90\% & 89.93\% & 90.32\%  \\
      & Top-5 ACC ($\uparrow$)
      & 91.87\% & 93.95\% & 94.97\%  \\
      \cline{1-5}
      \addlinespace[1pt]
      \multirow{4}{*}{FreeVC}
      & EER ($\downarrow$)
      & 8.44\% & 6.71\%  & 4.22\%  \\
      \addlinespace[0.2pt]
      \cline{2-5}
      \addlinespace[1pt]
      & Top-1 ACC ($\uparrow$)
      & 85.08\% & 90.53\% & 92.58\%  \\
      & Top-3 ACC ($\uparrow$)
      & 93.21\% & 94.08\% & 97.54\%  \\
      & Top-5 ACC ($\uparrow$)
      & 96.45\% & 98.68\% & 99.43\%  \\
      \addlinespace[0.2pt]
      \cline{1-5}
      \addlinespace[1pt]
      \multirow{4}{*}{Diff}
      & EER ($\downarrow$)
      & 6.35\% & 3.82\%  & 3.27\%  \\
      \addlinespace[0.2pt]
      \cline{2-5}
      \addlinespace[1pt]
      & Top-1 ACC ($\uparrow$)
      & 89.37\% &  93.29\% & 94.08\%  \\
      & Top-3 ACC ($\uparrow$)
      & 97.90\% & 98.79\% & 99.34\%  \\
      & Top-5 ACC ($\uparrow$)
      & 98.74\% & 99.20\% & 99.87\%  \\
      \addlinespace[0.2pt]
      \cline{1-5}
      \addlinespace[1pt]
      \multirow{4}{*}{DDDM}
      & EER ($\downarrow$)
      & 8.97\% & 4.17\%  & 3.80\%  \\
      \addlinespace[0.2pt]
      \cline{2-5}
      \addlinespace[1pt]
      & Top-1 ACC ($\uparrow$)
      & 81.45\% & 93.05\% & 94.40\%  \\
      & Top-3 ACC ($\uparrow$)
      & 94.21\% & 97.68\% & 98.42\%  \\
      & Top-5 ACC ($\uparrow$)
      & 97.24\% & 99.34\% & 99.61\%  \\
      \Xhline{1pt}
    \end{tabular}
    }
  \end{center}
\end{table}

\subsection{Performance under Large-Scale Settings}
\label{section.large_scale}

To evaluate the robustness of \sys on a larger scale, we expand the test set from 40 to 100 speakers, and ultimately include all test speakers from LibriSpeech and VoxCeleb1\&2. For VC methods, we adopt AGAIN, FreeVC, Diff, and DDDM. Table \ref{tab.large_scale} and Table \ref{tab.large_scale_all} in Appendix \ref{section.all_test} present the evaluation results on 100 speakers and all test speakers, respectively. While baselines suffer from degraded performance on genuine audio, \sys consistently maintains strong performance. As shown in Table \ref{tab.large_scale}, on converted audio, Revelio experiences a severe performance drop, with its Top-1 ACC falling to 67.57\%. In contrast, \sys sustains a notably higher Top-1 ACC of 81.54\%. These findings validate the scalability and robustness of \sys on large-scale datasets.

\begin{table}[t]
  \begin{center}
    \caption{Performance of voiceprint recovery methods under large-scale settings.}
    \label{tab.large_scale}
    \vspace{-0.2cm}
    \resizebox{1.0\linewidth}{!}{
    \begin{tabular}{cr c c c c}
      \Xhline{1pt}
      \multicolumn{1}{c}{\multirow{2}{*}{\textbf{VC}}} & \multicolumn{1}{c}{\multirow{2}{*}{\textbf{Metric}}} &  \multicolumn{4}{c}{\textbf{Voiceprint Recovery Method}} \\
      \multicolumn{1}{c}{}& & \textbf{MFA} & \textbf{ECAPA} & \textbf{Revelio} & \textbf{\sys} \\
      \Xhline{1pt}
      \multirow{4}{*}{Clean}
      & EER ($\downarrow$)
      & 17.82\% & 7.89\%  & 5.09\% & \cellcolor{backcolor!100}\textbf{2.32\%} \\
      \addlinespace[0.2pt]
      \cline{2-6}
      \addlinespace[1pt]
      & Top-1 ACC ($\uparrow$)
      & 64.97\% & 79.70\% & 86.78\% & \cellcolor{backcolor!100}\textbf{95.30\%}\\
      & Top-3 ACC ($\uparrow$)
      & 74.69\% & 85.29\% & 92.33\% & \cellcolor{backcolor!100}\textbf{99.36\%} \\
      & Top-5 ACC ($\uparrow$)
      & 79.44\% & 88.58\% & 95.59\% & \cellcolor{backcolor!100}\textbf{99.83\%} \\
      \addlinespace[0.2pt]
      \cline{1-6}
      \addlinespace[1pt]
      \multirow{4}{*}{AGAIN}
      & EER ($\downarrow$)
      & 35.57\% & 32.45\%  & 9.38\% & \cellcolor{backcolor!100}\textbf{5.47\%} \\
      \addlinespace[0.2pt]
      \cline{2-6}
      \addlinespace[1pt]
      & Top-1 ACC ($\uparrow$)
      & 15.13\% & 10.72\% & 66.31\% & \cellcolor{backcolor!100}\textbf{82.88\%}\\
      & Top-3 ACC ($\uparrow$)
      & 31.31\% & 23.35\% & 85.16\% & \cellcolor{backcolor!100}\textbf{94.90\%} \\
      & Top-5 ACC ($\uparrow$)
      & 42.16\% & 32.61\% & 90.06\% & \cellcolor{backcolor!100}\textbf{97.25\%} \\
      \addlinespace[0.2pt]
      \cline{1-6}
      \addlinespace[1pt]
      \multirow{4}{*}{FreeVC}
      & EER ($\downarrow$)
      & 23.81\% & 37.28\%  & 10.35\% & \cellcolor{backcolor!100}\textbf{5.52\%} \\
      \addlinespace[0.2pt]
      \cline{2-6}
      \addlinespace[1pt]
      & Top-1 ACC ($\uparrow$)
      & 20.78\% & 7.71\% & 65.38\% & \cellcolor{backcolor!100}\textbf{72.98\%} \\
      & Top-3 ACC ($\uparrow$)
      & 39.10\% & 17.78\% & 84.99\% & \cellcolor{backcolor!100}\textbf{90.38\%} \\
      & Top-5 ACC ($\uparrow$)
      & 51.37\% & 25.45\% & 89.92\% & \cellcolor{backcolor!100}\textbf{94.82\%}\\
      \addlinespace[0.2pt]
      \cline{1-6}
      \addlinespace[1pt]
      \multirow{4}{*}{Diff}
      & EER ($\downarrow$)
      & 34.95\% & 30.99\%  & 7.69\% & \cellcolor{backcolor!100}\textbf{4.35\%} \\
      \addlinespace[0.2pt]
      \cline{2-6}
      \addlinespace[1pt]
      & Top-1 ACC ($\uparrow$)
      & 20.22\% &  13.47\% & 69.89\% & \cellcolor{backcolor!100}\textbf{88.40\%} \\
      & Top-3 ACC ($\uparrow$)
      & 33.56\% & 29.95\% & 87.79\% & \cellcolor{backcolor!100}\textbf{97.04\%} \\
      & Top-5 ACC ($\uparrow$)
      & 41.37\% & 39.29\% & 92.77\% & \cellcolor{backcolor!100}\textbf{98.93\%} \\
      \addlinespace[0.2pt]
      \cline{1-6}
      \addlinespace[1pt]
      \multirow{4}{*}{DDDM}
      & EER ($\downarrow$)
      & 35.76\% & 33.59\%  & 8.75\% & \cellcolor{backcolor!100}\textbf{5.32\%} \\
      \addlinespace[0.2pt]
      \cline{2-6}
      \addlinespace[1pt]
      & Top-1 ACC ($\uparrow$)
      & 16.79\% &  10.03\% & 68.69\% & \cellcolor{backcolor!100}\textbf{81.89\%} \\
      & Top-3 ACC ($\uparrow$)
      & 32.64\% & 21.82\% & 85.81\% & \cellcolor{backcolor!100}\textbf{94.31\%} \\
      & Top-5 ACC ($\uparrow$)
      & 42.71\% & 30.68\% & 91.55\% & \cellcolor{backcolor!100}\textbf{97.11\%} \\
      \Xhline{1pt}
    \end{tabular}
    }
  \end{center}
\end{table}

\begin{figure*}[t]
    \centering
    	\begin{minipage}[b]{0.6\linewidth}
    		\centering
    		\includegraphics[trim=0mm 0mm 0mm 0mm, clip, width=\textwidth]{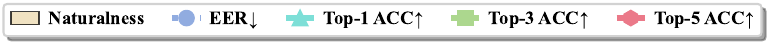}
    	\end{minipage} \\
    \subfigure[VQVC+]{
    	\begin{minipage}[b]{0.28\linewidth}
    		\centering
    		\includegraphics[trim=0mm 0mm 0mm 0mm, clip, width=0.95\textwidth]{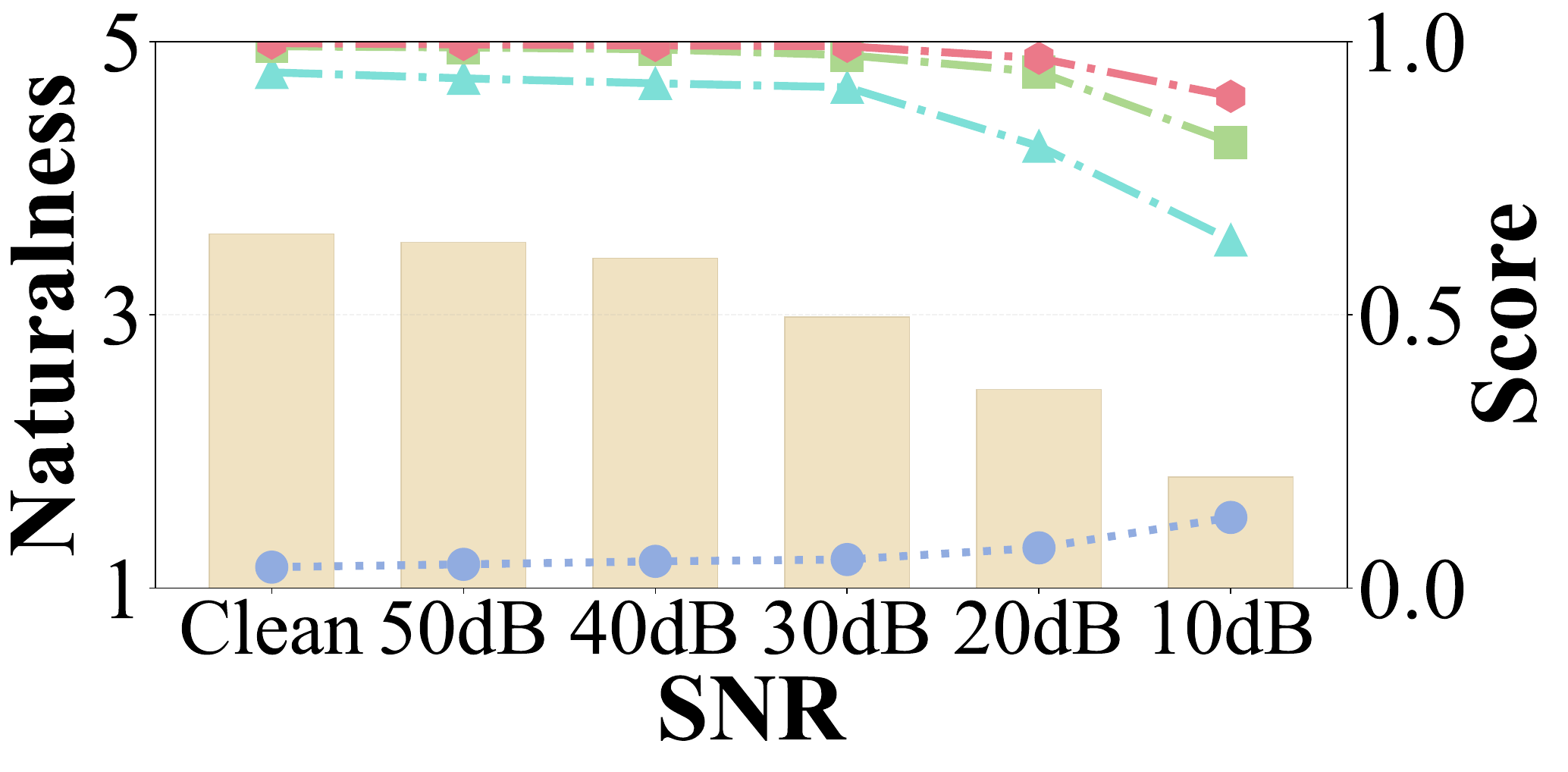}
    	\end{minipage}
    }
    \subfigure[FreeVC]{
    	\begin{minipage}[b]{0.28\linewidth}
    		\centering
    		\includegraphics[trim=0mm 0mm 0mm 0mm, clip, width=0.95\textwidth]{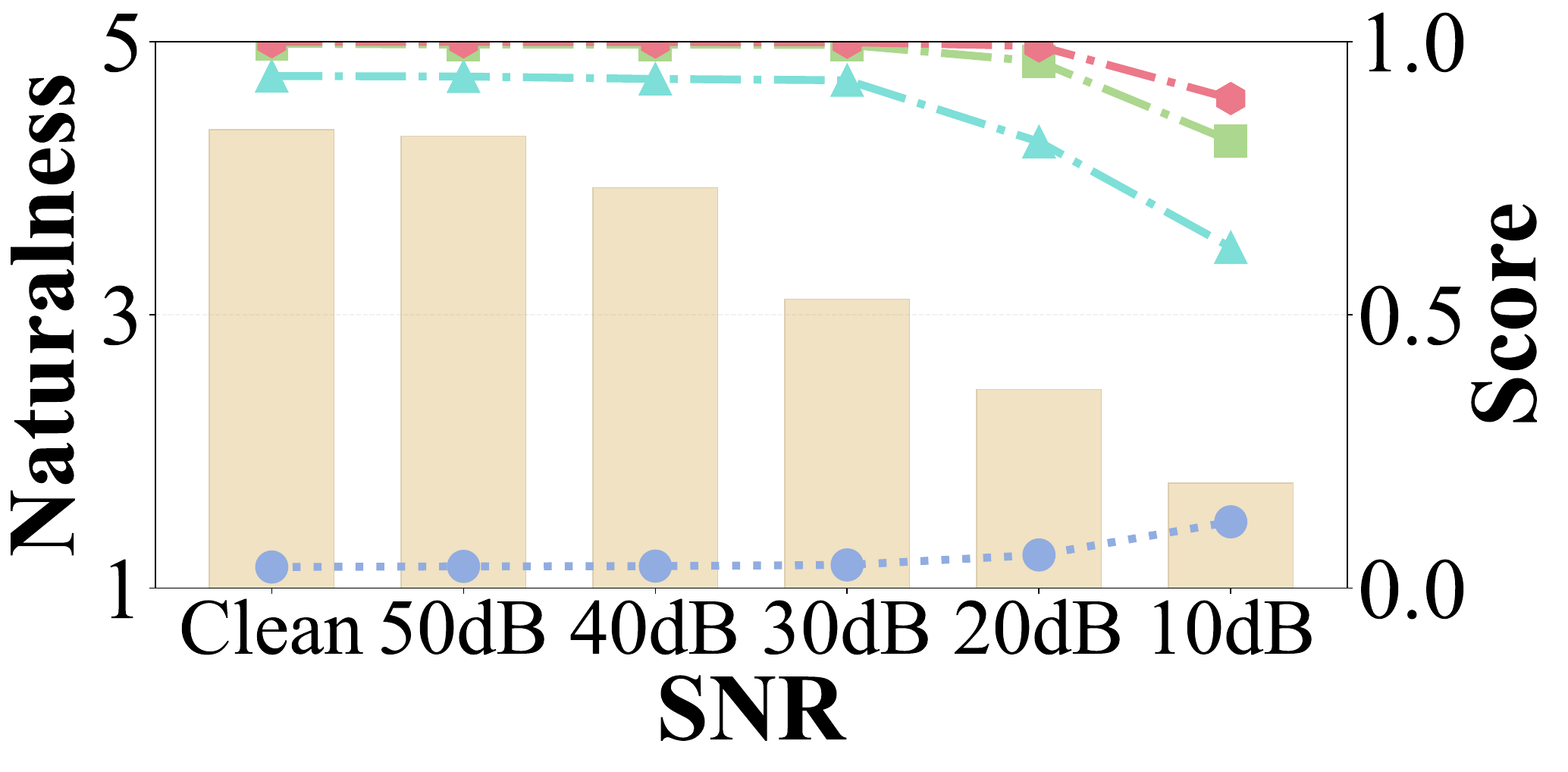}
    	\end{minipage}
    }
    \vspace{-0.2cm}
    \subfigure[Diff]{
    	\begin{minipage}[b]{0.28\linewidth}
    		\centering
    		\includegraphics[trim=0mm 0mm 0mm 0mm, clip, width=0.95\textwidth]{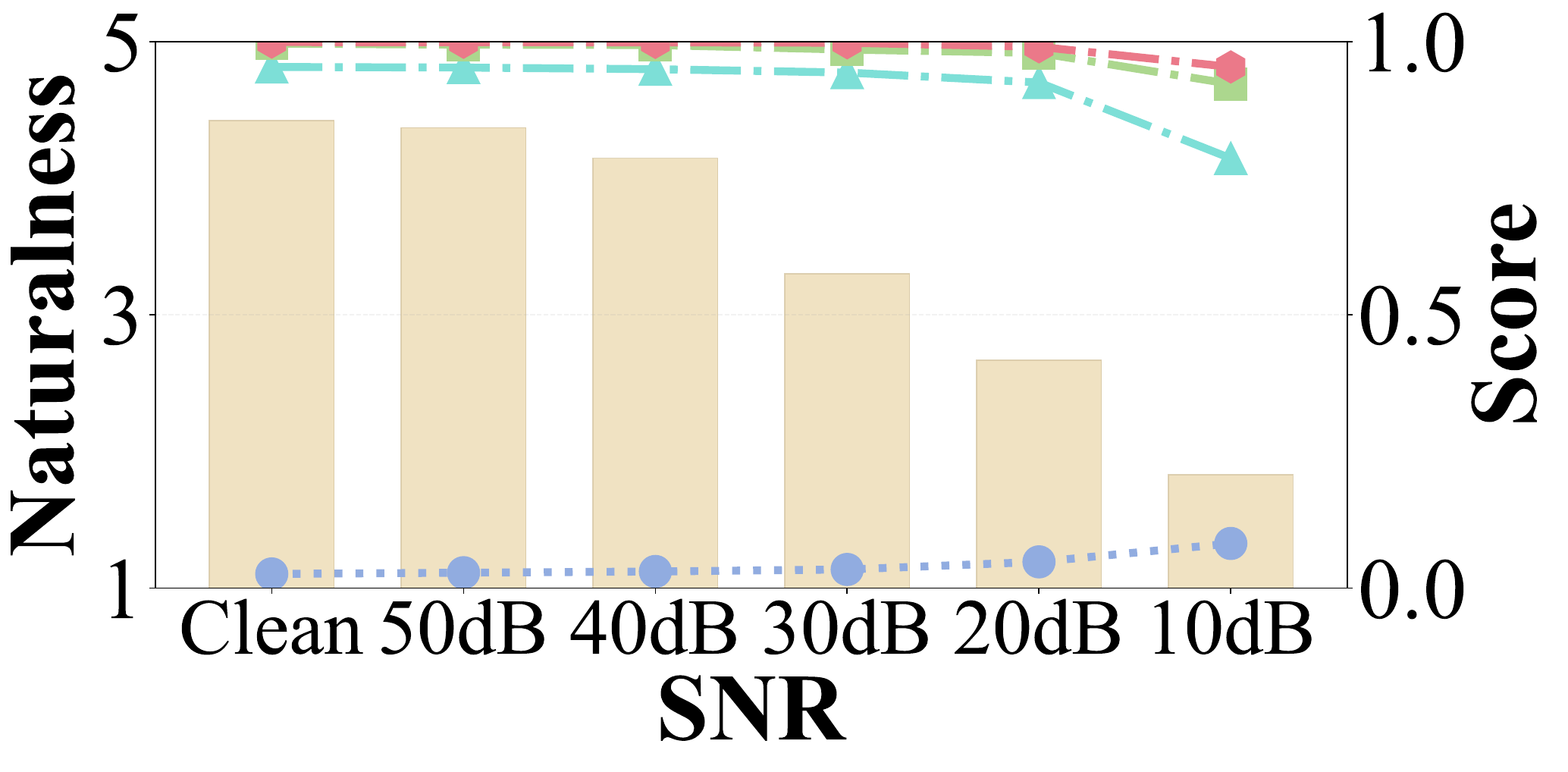}
    	\end{minipage}
    }
	\caption{Results of the subjective study.}
	\label{fig:study}
\end{figure*}

\subsection{Robustness to Preprocessing Obfuscation}
\label{section.preprocess}

To avoid being identified, an attacker may employ audio pre-processing techniques to obfuscate the source features in converted audios. We evaluate Revelio and \sys on audio samples subjected to three commonly-used audio pre-processing techniques: (1) \textbf{Additive White Gaussian Noise} with signal-to-noise ratios (SNR) of 30, 25, 20, 15, and 10 dB; (2) \textbf{Low-pass Filtering} via a Butterworth filter with cutoff frequencies at 8, 6, 4, 3, and 2 kHz; (3) \textbf{MP3 Compression} at bit rates of 128, 96, 64, 48, and 32 kbit/s. Results are shown in the Appendix (Figures \ref{fig:line_noise_r}-\ref{fig:line_compression}). The performance of \sys has a noticeable degradation only in the case of severe signal degradation, e.g., with additive white Gaussian noise at 10 dB, where the subjective naturalness rating falls to only 1.81 out of 5 in Section \ref{section.study}. Even in this extreme case, \sys maintains a Top‑5 accuracy of 91.52\%.

\subsection{Subjective Study}
\label{section.study}

We conduct a formal subjective user study to empirically correlate the trends of machine failure and human suspicion, where 30 participants are recruited to evaluate the naturalness of the converted audios. For each of VQVC+, FreeVC, and Diff, we randomly select 5 converted audio and obfuscate each audio with Additive White Gaussian Noise at five SNR levels (50, 40, 30, 20, and 10 dB), resulting in 90 audios. Each participant is asked to rate the naturalness of each audio sample on a scale from 1 to 5 (1 for completely unnatural and 5 for completely natural). As shown in Figure \ref{fig:study}, when the model's performance begins to decline under white noise at an SNR of 20 dB, the corresponding mean human naturalness rating has already dropped to $2.53$. This conspicuous impairment of sound quality would likely alert human victims, rendering these VC techniques impractical for real-time fraud under such noisy conditions.

\subsection{More Advanced Adaptive Adversary}
\label{section.adaptive}

If an adaptive attacker attempts to circumvent \sys, one potential strategy is to perform voice conversion iteratively to disrupt the two auxiliary branches. We consider two potential scenarios.  In the first scenario, the attacker applies two different voice conversion methods to convert audios with the same target speaker. By combining two different methods, the attacker tries to hinder the first auxiliary branch from identifying the specific VC method. In the second scenario, the attacker applies the same voice conversion method twice with different target speakers. The audio is first converted to match a randomly selected speaker, and then processed a second time to match the final target speaker. The repetitive voice conversion renders the target voiceprint heterogeneous, attempting to mislead the model trained with the second auxiliary branch. These repeatedly converted audios present a significant challenge to voiceprint recovery.

The experimental results are shown in the Appendix (Tables \ref{tab.adaptive_first}-\ref{tab.adaptive_second}). In the first scenario, we adopt VQVC+, FreeVC, Diff, and DDDM to further disguise audios converted first by AGAIN. Revelio can only maintain an 8.26\% EER and a 72.69\% Top-1 ACC against the composite methods. In contrast, \sys can still achieve a 6.67\% EER, an 82.34\% Top-1 ACC, a 94.22\% Top-3 ACC, and a 97.04\% Top-5 ACC on average, which proves the robustness of the first auxiliary branch in complex scenarios. In the second scenario, we adopt AGAIN, VQVC+, FreeVC, Diff, and DDDM to convert the same audio samples twice with two different target speakers. Despite the added complexity of multiple target speakers, \sys achieves a 5.20\% EER, an 89.03\% Top-1 ACC, a 97.31\% Top-3 ACC, and a 98.82\% Top-5 ACC on average. These results demonstrate the robustness of \sys, which consistently outperforms existing approaches against more advanced adaptive attackers.

\subsection{Generalization to unseen VC}
\label{section.unseenvc}

To evaluate model performance on unseen VC methods, we utilize AGAIN, VQVC, BNE, FreeVC, and Diff to train different voiceprint recovery models and measure them with AdaIN-VC \cite{chou2019one}, VQVC+, StarGANv2-VC \cite{li2021starganv2}, and DDDM. These unseen VC methods employ different types of generative models, which can provide a reasonable assessment. The experimental results are shown in the Appendix (Table \ref{tab.unseenvc}). Compared to Revelio, \sys exhibits a stable advantage. Despite never having been exposed to these VC methods during training, \sys achieves competitive performance, attaining an average Top-5 ACC of 93.07\%. These results demonstrate that \sys can generalize to other unseen VC methods.

\subsection{Cross-lingual Evaluation}
\label{section.language}

Language can greatly impact model performance in automatic speech recognition and speaker identification tasks. To evaluate this effect, we consider six typologically diverse languages: Spanish, French, German, Chinese, Amharic, and Finnish. Spanning four distinct language families, this selection enables a comprehensive and reliable generalization analysis. Additionally, we include VCTK \cite{yamagishi2019cstr}, an out-of-distribution English dataset. Detailed descriptions of these datasets are provided in Table \ref{tab:unseen_language} in the Appendix. The voiceprint recovery methods, all trained exclusively on English data, are evaluated across these out-of-scope test sets considering AGAIN, FreeVC, Diff, and DDDM as the VC methods. As illustrated in Figure \ref{fig:lang} in the Appendix, \sys achieves an EER of 2.69\% and a Top-1 accuracy of 96.65\% on genuine audio, while attaining a 97.70\% average Top-5 ACC on converted audio. These results substantially outperform competing methods, demonstrating the strong cross-domain generalization capability of \sys.

\subsection{Decoupling Effects of Auxiliary Branches}
\label{section.ablation}

\begin{table}[t]
  \centering
  \footnotesize
  \caption{Effect of each auxiliary branch.}
  \vspace{-0.2cm}
  \label{tab:ablation}
  \resizebox{1.00\linewidth}{!}{
  \begin{threeparttable}
    \begin{tabular}{crcccc}
    \Xhline{1pt}
      \multirow{2}{*}{\textbf{VC}} & \multicolumn{1}{c}{\multirow{2}{*}{\textbf{Metric}}} & \multicolumn{4}{c}{\textbf{Voiceprint Recovery Method\tnote{$\dagger$}}} \\
      & & \textbf{O} & \textbf{O+M} & \multicolumn{1}{c}{\textbf{O+G}} & \textbf{O+M+G} \\
      \Xhline{1pt}
      \multirow{4}{*}{Clean}
      & EER ($\downarrow$)
      & 2.45\% & 1.84\% & 1.87\% & \cellcolor{backcolor!100}\textbf{1.11\%} \\
      \addlinespace[0.2pt]
      \cline{2-6}
      \addlinespace[1pt]
      & Top-1 ACC ($\uparrow$)
      & 95.10\% & 97.63\% & 97.21\% & \cellcolor{backcolor!100}\textbf{98.49\%} \\
      & Top-3 ACC ($\uparrow$)
      & 97.44\% & 99.68\% & 99.78\% & \cellcolor{backcolor!100}\textbf{99.86\%} \\
      & Top-5 ACC ($\uparrow$)
      & 98.37\% & 99.96\% & 99.94\% & \cellcolor{backcolor!100}\textbf{99.97\%} \\
      \addlinespace[0.2pt]
      \cline{1-6}
      \addlinespace[1pt]
      \multirow{4}{*}{AGAIN}
      & EER ($\downarrow$)
      & 2.61\% & 1.48\% & 2.34\% & \cellcolor{backcolor!100}\textbf{1.34\%} \\
      \addlinespace[0.2pt]
      \cline{2-6}
      \addlinespace[1pt]
      & Top-1 ACC ($\uparrow$)
      & 88.85\% & 96.12\% & 92.50\% & \cellcolor{backcolor!100}\textbf{97.79\%} \\
      & Top-3 ACC ($\uparrow$)
      & 95.77\% & 99.04\% & 98.08\% & \cellcolor{backcolor!100}\textbf{99.81\%} \\
      & Top-5 ACC ($\uparrow$)
      & 97.02\% & 99.68\% & 99.42\% & \cellcolor{backcolor!100}\textbf{99.94\%} \\
      \addlinespace[0.2pt]
      \cline{1-6}
      \addlinespace[1pt]
      \multirow{4}{*}{VQVC}
      & EER ($\downarrow$)
      & 7.74\% & 7.73\% & 7.09\% & \cellcolor{backcolor!100}\textbf{6.83\%} \\
      \addlinespace[0.2pt]
      \cline{2-6}
      \addlinespace[1pt]
      & Top-1 ACC ($\uparrow$)
      & 76.80\% & 78.59\% & 80.10\% & \cellcolor{backcolor!100}\textbf{83.39\%} \\
      & Top-3 ACC ($\uparrow$)
      & 90.51\% & 93.08\% & 91.73\% & \cellcolor{backcolor!100}\textbf{94.58\%} \\
      & Top-5 ACC ($\uparrow$)
      & 94.97\% & 96.67\% & 96.25\% & \cellcolor{backcolor!100}\textbf{97.05\%} \\
      \addlinespace[0.2pt]
      \cline{1-6}
      \addlinespace[1pt]
      \multirow{4}{*}{VQVC+}
      & EER ($\downarrow$)
      & 5.47\% & 4.34\% & 4.50\% & \cellcolor{backcolor!100}\textbf{3.85\%} \\
      \addlinespace[0.2pt]
      \cline{2-6}
      \addlinespace[1pt]
      & Top-1 ACC ($\uparrow$)
      & 86.47\% & 88.97\% & 87.28\% & \cellcolor{backcolor!100}\textbf{94.36\%} \\
      & Top-3 ACC ($\uparrow$)
      & 94.84\% & 97.31\% & 95.03\% & \cellcolor{backcolor!100}\textbf{99.14\%} \\
      & Top-5 ACC ($\uparrow$)
      & 96.67\% & 99.20\% & 97.53\% & \cellcolor{backcolor!100}\textbf{99.71\%} \\
      \addlinespace[0.2pt]
      \cline{1-6}
      \addlinespace[1pt]
      \multirow{4}{*}{BNE}
      & EER ($\downarrow$)
      & 10.72\% & 8.85\% & 8.05\% & \cellcolor{backcolor!100}\textbf{7.92\%} \\
      \addlinespace[0.2pt]
      \cline{2-6}
      \addlinespace[1pt]
      & Top-1 ACC ($\uparrow$)
      & 69.20\% & 72.56\% & 72.24\% & \cellcolor{backcolor!100}\textbf{77.40\%} \\
      & Top-3 ACC ($\uparrow$)
      & 87.47\% & 90.00\% & 89.84\% & \cellcolor{backcolor!100}\textbf{90.03\%} \\
      & Top-5 ACC ($\uparrow$)
      & 93.08\% & 95.16\% & 94.49\% & \cellcolor{backcolor!100}\textbf{95.29\%} \\
      \addlinespace[0.2pt]
      \cline{1-6}
      \addlinespace[1pt]
      \multirow{4}{*}{FreeVC}
      & EER ($\downarrow$)
      & 4.56\% & 3.92\% & 4.17\% & \cellcolor{backcolor!100}\textbf{3.87\%} \\
      \addlinespace[0.2pt]
      \cline{2-6}
      \addlinespace[1pt]
      & Top-1 ACC ($\uparrow$)
      & 86.54\% & 93.37\% & 92.69\% & \cellcolor{backcolor!100}\textbf{93.69\%} \\
      & Top-3 ACC ($\uparrow$)
      & 95.51\% & 98.24\% & 97.73\% & \cellcolor{backcolor!100}\textbf{98.59\%} \\
      & Top-5 ACC ($\uparrow$)
      & 97.85\% & 99.55\% & 98.38\% & \cellcolor{backcolor!100}\textbf{99.58\%} \\
      \addlinespace[0.2pt]
      \cline{1-6}
      \addlinespace[1pt]
      \multirow{4}{*}{Diff}
      & EER ($\downarrow$)
      & 3.05\% & 2.85\% & 2.88\% & \cellcolor{backcolor!100}\textbf{2.61\%} \\
      \addlinespace[0.2pt]
      \cline{2-6}
      \addlinespace[1pt]
      & Top-1 ACC ($\uparrow$)
      & 93.62\% & 94.90\% & 94.46\% & \cellcolor{backcolor!100}\textbf{95.35\%} \\
      & Top-3 ACC ($\uparrow$)
      & 98.78\% & 99.46\% & 99.42\% & \cellcolor{backcolor!100}\textbf{99.62\%} \\
      & Top-5 ACC ($\uparrow$)
      & 99.52\% & 99.81\% & 99.87\% & \cellcolor{backcolor!100}\textbf{99.90\%} \\
      \addlinespace[0.2pt]
      \cline{1-6}
      \addlinespace[1pt]
      \multirow{4}{*}{DDDM}
      & EER 
      & 4.94\% & 4.37\% & 4.72\% & \cellcolor{backcolor!100}\textbf{3.02\%} \\
      \addlinespace[0.2pt]
      \cline{2-6}
      \addlinespace[1pt]
      & Top-1 ACC ($\uparrow$)
      & 91.92\% & 94.33\% & 92.66\% & \cellcolor{backcolor!100}\textbf{94.98\%} \\
      & Top-3 ACC ($\uparrow$)
      & 97.89\% & 99.11\% & 98.75\% & \cellcolor{backcolor!100}\textbf{99.26\%} \\
      & Top-5 ACC ($\uparrow$)
      & 99.04\% & 99.74\% & 99.65\% & \cellcolor{backcolor!100}\textbf{99.84\%} \\
      \Xhline{1pt}
    \end{tabular}%
        \begin{tablenotes}  
        \footnotesize         
        \item[$\dagger$] M and G are the abbreviations for the VC method inference and the guided refinement. O denotes a variant of \sys that excludes the two auxiliary branches.
      \end{tablenotes}  
         \end{threeparttable}
    }
\end{table}

\begin{table}[t]
  \centering
  \footnotesize
    \caption{Impact of training-inference mismatch.}
  \vspace{-0.2cm}
  \label{tab:inconsistency}
  \resizebox{1.00\linewidth}{!}{
  \begin{threeparttable}
    \begin{tabular}{crccc}
    \Xhline{1pt}
      \multirow{2}{*}{\textbf{VC}} & \multicolumn{1}{c}{\multirow{2}{*}{\textbf{Metric}}} & \multicolumn{3}{c}{\textbf{Voiceprint Recovery Method\tnote{$\dagger$}}} \\
      & & \textbf{\sys$^{\bm{-}}$} & \textbf{\sys} & \multicolumn{1}{c}{\textbf{\sys$^{\bm{+}}$}} \\
      \Xhline{1pt}
      \multirow{4}{*}{Clean}
      & EER ($\downarrow$)
      & 1.78\% & 1.11\% & \cellcolor{backcolor!100}\textbf{1.06} \%\\
      \addlinespace[0.2pt]
      \cline{2-5}
      \addlinespace[1pt]
      & Top-1 ACC ($\uparrow$)
      & 98.13\% & 98.49\% & \cellcolor{backcolor!100}\textbf{99.07\%} \\
      & Top-3 ACC ($\uparrow$)
      & 99.81\% & 99.86\% & \cellcolor{backcolor!100}\textbf{100.00\%} \\
      & Top-5 ACC ($\uparrow$)
      & 99.95\% & 99.97\% & \cellcolor{backcolor!100}\textbf{100.00\%} \\
      \addlinespace[0.2pt]
      \cline{1-5}
      \addlinespace[1pt]
      \multirow{4}{*}{AGAIN}
      & EER ($\downarrow$)
      & 2.45\% & 1.34\% & \cellcolor{backcolor!100}\textbf{1.28\%} \\
      \addlinespace[0.2pt]
      \cline{2-5}
      \addlinespace[1pt]
      & Top-1 ACC ($\uparrow$)
      & 96.51\% & 97.79\% & \cellcolor{backcolor!100}\textbf{98.88\%} \\
      & Top-3 ACC ($\uparrow$)
      & 98.53\% & 99.81\% & \cellcolor{backcolor!100}\textbf{99.94\%} \\
      & Top-5 ACC ($\uparrow$)
      & 99.81\% & 99.94\% & \cellcolor{backcolor!100}\textbf{99.97\%} \\
      \addlinespace[0.2pt]
      \cline{1-5}
      \addlinespace[1pt]
      \multirow{4}{*}{VQVC}
      & EER ($\downarrow$)
      & 8.25\% & 6.83\% & \cellcolor{backcolor!100}\textbf{6.06\%} \\
      \addlinespace[0.2pt]
      \cline{2-5}
      \addlinespace[1pt]
      & Top-1 ACC ($\uparrow$)
      & 80.71\% & 83.39\% & \cellcolor{backcolor!100}\textbf{88.59\%} \\
      & Top-3 ACC ($\uparrow$)
      & 93.74\% & 94.58\% & \cellcolor{backcolor!100}\textbf{97.82\%} \\
      & Top-5 ACC ($\uparrow$)
      & 96.56\% & 97.05\% & \cellcolor{backcolor!100}\textbf{98.94\%} \\
      \addlinespace[0.2pt]
      \cline{1-5}
      \addlinespace[1pt]
      \multirow{4}{*}{VQVC+}
      & EER ($\downarrow$)
      & 4.05\% & 3.85\% & \cellcolor{backcolor!100}\textbf{3.35\%} \\
      \addlinespace[0.2pt]
      \cline{2-5}
      \addlinespace[1pt]
      & Top-1 ACC ($\uparrow$)
      & 93.77\% & 94.36\% & \cellcolor{backcolor!100}\textbf{95.69\%} \\
      & Top-3 ACC ($\uparrow$)
      & 98.86\% & 99.14\% & \cellcolor{backcolor!100}\textbf{99.65\%} \\
      & Top-5 ACC ($\uparrow$)
      & 99.54\% & 99.71\% & \cellcolor{backcolor!100}\textbf{99.96\%} \\
      \addlinespace[0.2pt]
      \cline{1-5}
      \addlinespace[1pt]
      \multirow{4}{*}{BNE} 
      & EER ($\downarrow$)
      & 8.31\% & 7.92\% & \cellcolor{backcolor!100}\textbf{6.31\%} \\
      \addlinespace[0.2pt]
      \cline{2-5}
      \addlinespace[1pt]
      & Top-1 ACC ($\uparrow$)
      & 77.15\% & 77.40\% & \cellcolor{backcolor!100}\textbf{79.15\%} \\
      & Top-3 ACC ($\uparrow$)
      & 90.24\% & 90.03\% & \cellcolor{backcolor!100}\textbf{92.24\%} \\
      & Top-5 ACC ($\uparrow$)
      & 94.65\% & 95.29\% & \cellcolor{backcolor!100}\textbf{95.87\%} \\
      \addlinespace[0.2pt]
      \cline{1-5}
      \addlinespace[1pt]
      \multirow{4}{*}{FreeVC} 
      & EER ($\downarrow$)
      & 4.50\% & 3.87\% & \cellcolor{backcolor!100}\textbf{3.02\%} \\
      \addlinespace[0.2pt]
      \cline{2-5}
      \addlinespace[1pt]
      & Top-1 ACC ($\uparrow$)
      & 91.15\% & 93.69\% & \cellcolor{backcolor!100}\textbf{95.55\%} \\
      & Top-3 ACC ($\uparrow$)
      & 98.42\% & 98.59\% & \cellcolor{backcolor!100}\textbf{99.33\%} \\
      & Top-5 ACC ($\uparrow$)
      & 99.35\% & 99.58\% & \cellcolor{backcolor!100}\textbf{99.81\%} \\
      \addlinespace[0.2pt]
      \cline{1-5}
      \addlinespace[1pt]
      \multirow{4}{*}{Diff}
      & EER ($\downarrow$)
      & 3.94\% & 2.61\% & \cellcolor{backcolor!100}\textbf{1.93\%} \\
      \addlinespace[0.2pt]
      \cline{2-5}
      \addlinespace[1pt]
      & Top-1 ACC ($\uparrow$)
      & 94.74\% & 95.35\% & \cellcolor{backcolor!100}\textbf{97.52\%} \\
      & Top-3 ACC ($\uparrow$)
      & 98.88\% & 99.62\% & \cellcolor{backcolor!100}\textbf{99.73\%} \\
      & Top-5 ACC ($\uparrow$)
      & 99.46\% & 99.90\% & \cellcolor{backcolor!100}\textbf{100.00\%} \\
      \addlinespace[0.2pt]
      \cline{1-5}
      \addlinespace[1pt]
      \multirow{4}{*}{DDDM} 
      & EER 
      & 3.19\% & 3.02\% & \cellcolor{backcolor!100}\textbf{2.47\%} \\
      \addlinespace[0.2pt]
      \cline{2-5}
      \addlinespace[1pt]
      & Top-1 ACC ($\uparrow$)
      & 94.08\% & 94.98\% & \cellcolor{backcolor!100}\textbf{96.03\%} \\
      & Top-3 ACC ($\uparrow$)
      & 98.87\% & 99.26\% & \cellcolor{backcolor!100}\textbf{99.68\%} \\
      & Top-5 ACC ($\uparrow$)
      & 99.58\% & 99.84\% & \cellcolor{backcolor!100}\textbf{100.00\%} \\
      \Xhline{1pt}
    \end{tabular}%
        \begin{tablenotes}  
        \footnotesize         
        \item[$\dagger$] \sys$^{\bm{-}}$ and \sys$^{\bm{+}}$ use zero embeddings and genuine target speaker samples respectively as auxiliary inputs for converted samples during both training and inference.
      \end{tablenotes}  
         \end{threeparttable}
    }
\end{table}

To evaluate the contribution of each extracted representation to voiceprint recovery, we conduct ablation experiments on \sys by removing its two auxiliary branches: VC method inference (which extracts method-specific information) and guided refinement (which extracts target speaker voiceprint components). The variants include: (1) \sys without both branches, (2) \sys without guided refinement, (3) \sys without VC method inference, and (4) the complete \sys. As shown in Table \ref{tab:ablation}, \sys achieves the best performance when both auxiliary branches are present. We further employ t-SNE \cite{van2008visualizing} to visualize the discriminative ability of different branch combinations in Figure \ref{fig:ablation}. Notably, these two auxiliary branches significantly enhance the discriminative ability of \sys.

To quantitatively assess the disentanglement performance of \sys, we use three metrics: Adjusted Rand Index (ARI), Normalized Mutual Information (NMI), and Correlation Coefficient Matrix. As shown in the Appendix (Table \ref{tab.distengle}), \sys achieves mean ARI and NMI scores of 0.906 and 0.949, indicating strong alignment between the extracted voiceprint components and their corresponding labels (i.e., VC method, target speaker, source speaker). Additionally, all off-diagonal correlations are below 0.2, confirming their mutual independence. These results demonstrate that intra-branch correlations are high and inter-branch correlations are low, indicating an effective decoupling of different components (i.e., VC method, target speaker, source speaker) by each branch.

\subsection{Ablation Study on Differential Block}
\label{section.ablation_ti}

To evaluate the impact of training-inference mismatch, we conduct an ablation study where the auxiliary inputs for converted samples during training and testing are different or the same. In the former case, the auxiliary input is a genuine target speaker sample during training and a zero embedding during inference for a converted audio sample, i.e., the design of \sys. In the latter case, we consider zero embeddings for both training and inference (denoted as \sys$^-$) and genuine target speaker samples for both training and inference (denoted as \sys$^+$). Note that \sys$^+$ is an ideal case but cannot be implemented since the threat model assumes that the defender does not know whether the input is converted or not. As demonstrated in Table \ref{tab:inconsistency}, the inclusion of the genuine target speaker sample as an auxiliary does improve the performance of \sys over the baseline strategy, namely \sys$^-$, in voiceprint recovery tasks. Furthermore, these results demonstrate that \sys does not rely on potential label leakage as a training shortcut that could impair inference performance.

\subsection{Impact of Reference Audio Number}
\label{section.imapct_number}

We explore the impact of the reference audio number. We separately set the number of reference audios per speaker to 1, 5, 10, 15, and 20. The experimental results are shown in the Appendix (Figure \ref{fig:line_ref}). By increasing the reference audio number, the performance of \sys is significantly improved. The EER decreases from 7.81\% to 2.85\%. The accuracy increases from 79.82\% to 95.87\% (Top-1), from 91.01\% to 99.14\% (Top-3) and from 94.71\% to 99.68\% (Top-5) in converted audios. These experimental results validate the importance of reference audios. \sys exhibits superior performance across different reference audio numbers, where a single reference audio per speaker suffices to match the performance of Revelio employing three reference audio samples per speaker.

\subsection{Impact of Audio Length}
\label{section.impact_length}

We explore the impact of the input audio length. We separately set the audio length to 1, 5, 10, 15, and 20 s. According to the specified audio length, the input audios are randomly trimmed. The experimental results are shown in the Appendix (Figure \ref{fig:line_len}). When the input audios are trimmed into 1 s, the most useful information has been pruned, which severely limits the effectiveness of methods. However, significant changes occur when the audios are trimmed into 5 s. \sys can preliminarily recover the voiceprint of converted audios with the audio length of 5 s. As the audio length increases, the method performance becomes better. These experimental results demonstrate the superiority of \sys.

\subsection{Impact of Sampling Rate}
\label{section.impact_sampling_rate}

We explore the impact of the sampling rate to simulate inconsistent sampling rates of input audios. We separately set the sampling rate to 12, 14, 16, 18, and 20 kHz. Experiments are conducted under two conditions, one without resampling and the other with resampling. The experimental results without resampling are shown in the Appendix (Figure \ref{fig:line_sr}). Different sampling rates can cause distortion of the voiceprint, resulting in timbre changes and information loss. Therefore, the different sampling rates in these experiments pose a significant challenge for voiceprint recovery. \sys maintains good performance within the 14–18 kHz sampling rate range but exhibits significant performance fluctuations outside this range. However, simple resampling can handle most of the issues as shown in the Appendix (Figure \ref{fig:line_sr_c}).

\subsection{Impact of Learning Rate}
\label{section.impact_learning_rate}

We explore the impact of the learning rate. We separately set the learning rate as 0.0005, 0.0008, 0.0010, 0.0012, and 0.0015. The experimental results are shown in the Appendix (Figure \ref{fig:line_lr}). A learning rate of 0.0005 leads to premature convergence to suboptimal local minima, while rates above 0.0012 cause unstable training and non-convergence. The intermediate value of 0.0010 achieves the highest accuracy, demonstrating a robust optimum. These experiments provide empirical justification for our parameter choice.

\subsection{Impact of Loss Weight}
\label{section.impact_loss_weight}

We explore the impact of the loss weight $\lambda$. We separately set the loss weight $\lambda$ as 0.05, 0.08, 0.10, 0.12, and 0.15. The experimental results are shown in the Appendix (Figure \ref{fig:line_lw}). A high $\lambda$ of 0.15 compromises main-task learning and degrades voiceprint recovery, whereas a low $\lambda$ of 0.05 provides insufficient guidance, leading to suboptimal performance. The intermediate value of 0.10 achieves the best trade-off between the main and auxiliary tasks, yielding the highest recovery accuracy. These experiments justify our choice of $\lambda$.

\section{Discussion}

\subsection{Data dependence}

An inherent limitation of speaker recognition systems is their dependence on a comprehensive voiceprint database. In our experiments, we mitigate this dependency by synthesizing  voiceprint enrollment templates using only three reference audio samples per speaker, a requirement that is feasible in most practical scenarios. To further stress-test this dependency, we evaluate an extreme condition using only a single reference audio sample per speaker, as described in Section \ref{section.imapct_number}. Even under this severe data limitation, \sys achieves an average Top-5 accuracy of 94.71\%. These empirical findings underscore that our framework remains robust and effective even under extreme data-scarcity conditions.

\subsection{Cross-lingual generalization}

In cross-lingual experiments, our method achieves the best performance on all languages tested. However, on Chinese, which is sharply divergent from English, our method suffers a notable performance drop. This drop in accuracy suggests that pronounced acoustic and phonetic discrepancies stemming from cross-lingual variations can impede model generalization. To achieve consistent generalization across markedly divergent languages remains a critical and promising direction for future work.

\subsection{Computational Resources}

\sys is computationally efficient and suitable for real-time deployment. For inference, it processes a 20-second audio input with 30 ms latency while maintaining a memory consumption below 880 MB. In terms of training, it only requires 80 hours on four NVIDIA 3090 GPUs in Section \ref{section.evaluation}.

\subsection{Deployment Scenario}
\label{sec.deployment_scenario}

\sys holds significant potential for forensic applications, leveraging the fact that a voiceprint serves as a powerful biometric modality combining an individual’s unique vocal tract physiology with learned neuromuscular habits \cite{hollien2002forensic}. In investigative workflows, voiceprints are evaluated using distinct candidate set paradigms depending on the availability of case-specific intelligence. In a closed candidate set scenario, investigators isolate a bounded pool of suspects based on external leads (e.g., phone numbers or financial transactions). \sys can then be deployed to extract the source speaker voiceprint, execute a $1:N$ comparative ranking, and generate similarity scores to prioritize leads. Conversely, in an open candidate set scenario, investigators do not assume the true speaker is present in the database. Consequently, after querying the database with the voiceprint extracted by \sys, an absolute verification threshold is required to determine whether the true speaker matches an existing entry or is entirely absent \cite{hansen2015speaker}. If the similarity score falls below this threshold, the system triggers a rejection. In such scenarios, the extracted voiceprint can be securely archived. If a new suspect emerges later, this stored voiceprint allows investigators to look back and make a retrospective match without needing to start the investigation from scratch. This verification threshold can be calibrated using cumulative distribution function analysis.

\section{Conclusion}

This paper introduces \sys, a novel method to reconstruct the source speaker's original identity from converted audios. We propose a three-pronged structure featuring a primary extractor supported by two auxiliary branches. One auxiliary branch identifies the underlying VC mechanism, which helps distill the source voiceprint. The other auxiliary branch extracts a latent representation of the target speaker, facilitating the isolation of target-specific traits from the composite converted audio. The primary extractor recovers the source voiceprint with the assistance of the other two auxiliary branches, which distills a highly discriminative representation of the source speaker's identity. Extensive experiments prove that \sys can achieve robust and high performance to recover the voiceprint of the source speaker even in complex scenarios, such as telephony scenarios, unseen languages scenarios, and adaptive scenarios.

\section*{Acknowledgment}

This work was supported by the National Natural Science Foundation of China (Grant No. 62271280), the Ant Group through CCF-Ant Research Fund (Grant No. CCF-AFSG RF20230301), and the Zhejiang Key Laboratory of Electrical Technology and System on Renewable Energy. Yanjiao Chen and Yushi Cheng are the corresponding authors.

\bibliographystyle{plain}
\bibliography{reference}

@article{sisman2020overview,
  title={An overview of voice conversion and its challenges: From statistical modeling to deep learning},
  author={Sisman, Berrak and Yamagishi, Junichi and King, Simon and Li, Haizhou},
  journal={IEEE/ACM Transactions on Audio, Speech, and Language Processing},
  volume={29},
  pages={132--157},
  year={2020},
  publisher={IEEE}
}

@article{aihara2012gmm,
  title={GMM-based emotional voice conversion using spectrum and prosody features},
  author={Aihara, Ryo and Takashima, Ryoichi and Takiguchi, Tetsuya and Ariki, Yasuo},
  journal={American Journal of Signal Processing},
  volume={2},
  number={5},
  pages={134--138},
  year={2012},
  publisher={Scientific and Academic Publishing}
}

@inproceedings{hwang2013incorporating,
  title={Incorporating global variance in the training phase of GMM-based voice conversion},
  author={Hwang, Hsin-Te and Tsao, Yu and Wang, Hsin-Min and Wang, Yih-Ru and Chen, Sin-Horng},
  booktitle={2013 Asia-Pacific Signal and Information Processing Association Annual Summit and Conference},
  pages={1--6},
  year={2013},
  organization={IEEE}
}

@article{zhang2024target,
  title={Target speaker filtration by mask estimation for source speaker traceability in voice conversion},
  author={Zhang, Junfei and Zhang, Xiongwei and Sun, Meng and Zou, Xia and Jia, Chong and Li, Yihao},
  journal={Engineering Applications of Artificial Intelligence},
  volume={136},
  pages={109071},
  year={2024},
  publisher={Elsevier}
}

@inproceedings{wellington2024quantifying,
  title={Quantifying Source Speaker Leakage in One-to-One Voice Conversion},
  author={Wellington, Scott and Liu, Xuechen and Yamagishi, Junichi},
  booktitle={2024 International Conference of the Biometrics Special Interest Group (BIOSIG)},
  pages={1--6},
  year={2024},
  organization={IEEE}
}

@misc{tiktok2021,
  author = {Megan McCluskey},
  title = {TikTok Has Started Collecting Your Faceprints and Voiceprints. Here's What It Could Do With Them},
  howpublished = {\url{https://time.com/6071773/tiktok-faceprints-voiceprints-privacy/}},
  year = {2021},
  note = {Accessed: 2025-12-27}
}

@misc{bank2019,
  author = {Cassandra Deskus and Joshua R. Fattal},
  title = {A Fourth Amendment Framework for Voiceprint Database Searches},
  howpublished = {\url{https://www.justsecurity.org/66571/a-fourth-amendment-framework-for-voiceprint-database-searches/}},
  year = {2019},
  note = {Accessed: 2025-12-27}
}

@misc{fraud,
    title = {{Fraudsters Used AI to Mimic CEO’s Voice in Unusual Cybercrime Case}},
    howpublished = {\url{https://www.wsj.com/articles/fraudsters-use-ai-to-mimic-ceos-voice-in-unusual-cybercrime-case-11567157402}},
    year = {2019}
}

@misc{fraud_2,
    title = {{Fraudsters Cloned Company Director’s Voice In \$35 Million Heist, Police Find}},
    howpublished = {\url{https://www.forbes.com/sites/thomasbrewster/2021/10/ 14/huge-bank-fraud-uses-deep-fake-voice-tech-to-steal-millions/}},
    year = {2021}
}

@misc{fraud_3,
    title = {{Finance worker pays out \$25 million after video call with deepfake ‘chief financial officer’}},
    howpublished = {\url{https://edition.cnn.com/2024/02/04/asia/deepfake-cfo-scam-hong-kong-intl-hnk}},
    year = {2024}
}

@article{10.1145/3772374,
author = {Belani, Gaurav},
title = {AI Rewrites the Rules of Phishing, Cybercrime},
year = {2025},
volume = {69},
number = {1},
journal = {Communications of the ACM},
pages = {8–9},
}

@inproceedings{lorenzo2018voice,
  title={The Voice Conversion Challenge 2018: Promoting Development of Parallel and Nonparallel Methods},
  author={Lorenzo-Trueba, Jaime and Yamagishi, Junichi and Toda, Tomoki and Saito, Daisuke and Villavicencio, Fernando and Kinnunen, Tomi and Ling, Zhenhua},
  booktitle={The Speaker and Language Recognition Workshop (Odyssey 2018)},
  pages={195},
  year={2018},
  organization={ISCA}
}

@inproceedings{yi2020voice,
  title={Voice conversion challenge 2020: Intra-lingual semi-parallel and cross-lingual voice conversion},
  author={Yi, Zhao and Huang, Wen-Chin and Tian, Xiaohai and Yamagishi, Junichi and Das, Rohan Kumar and Kinnunen, Tomi and Ling, Zhen-Hua and Toda, Tomoki},
  booktitle={Joint Workshop for the Blizzard Challenge and Voice Conversion Challenge 2020},
  year={2020},
  organization={ISCA}
}

@article{wang2020asvspoof,
  title={ASVspoof 2019: A large-scale public database of synthesized, converted and replayed speech},
  author={Wang, Xin and Yamagishi, Junichi and Todisco, Massimiliano and Delgado, H{\'e}ctor and Nautsch, Andreas and Evans, Nicholas and Sahidullah, Md and Vestman, Ville and Kinnunen, Tomi and Lee, Kong Aik and others},
  journal={Computer Speech \& Language},
  volume={64},
  pages={101114},
  year={2020},
  publisher={Elsevier}
}

@article{abe1990voice,
  title={Voice conversion through vector quantization},
  author={Abe, Masanobu and Nakamura, Satoshi and Shikano, Kiyohiro and Kuwabara, Hisao},
  journal={Journal of the Acoustical Society of Japan (E)},
  volume={11},
  number={2},
  pages={71--76},
  year={1990},
  publisher={Acoustical Society of Japan}
}

@article{stylianou1998continuous,
  title={Continuous probabilistic transform for voice conversion},
  author={Stylianou, Yannis and Capp{\'e}, Olivier and Moulines, Eric},
  journal={IEEE Transactions on speech and audio processing},
  volume={6},
  number={2},
  pages={131--142},
  year={1998},
  publisher={IEEE}
}

@article{toda2007voice,
  title={Voice conversion based on maximum-likelihood estimation of spectral parameter trajectory},
  author={Toda, Tomoki and Black, Alan W and Tokuda, Keiichi},
  journal={IEEE Transactions on Audio, Speech, and Language Processing},
  volume={15},
  number={8},
  pages={2222--2235},
  year={2007},
  publisher={IEEE}
}

@inproceedings{nakashika2013voice,
  title={Voice conversion in high-order eigen space using deep belief nets.},
  author={Nakashika, Toru and Takashima, Ryoichi and Takiguchi, Tetsuya and Ariki, Yasuo},
  booktitle={Interspeech},
  pages={369--372},
  year={2013}
}

@inproceedings{sun2015voice,
  title={Voice conversion using deep bidirectional long short-term memory based recurrent neural networks},
  author={Sun, Lifa and Kang, Shiyin and Li, Kun and Meng, Helen},
  booktitle={2015 IEEE international conference on acoustics, speech and signal processing (ICASSP)},
  pages={4869--4873},
  year={2015},
  organization={IEEE}
}

@inproceedings{chou2019one,
  title={One-Shot Voice Conversion by Separating Speaker and Content Representations with Instance Normalization},
  author={Chou, Ju-chieh and Lee, Hung-Yi},
  booktitle={Proc. Interspeech 2019},
  pages={664--668},
  year={2019}
}

@inproceedings{wu2020one,
  title={One-shot voice conversion by vector quantization},
  author={Wu, Da-Yi and Lee, Hung-yi},
  booktitle={ICASSP 2020-2020 IEEE International Conference on Acoustics, Speech and Signal Processing (ICASSP)},
  pages={7734--7738},
  year={2020},
  organization={IEEE}
}

@article{wu2020vqvc+,
  title={VQVC+: One-Shot Voice Conversion by Vector Quantization and U-Net Architecture},
  author={Wu, Da-Yi and Chen, Yen-Hao and Lee, Hung-yi},
  journal={Interspeech 2020},
  year={2020},
  publisher={ISCA}
}

@inproceedings{chen2021again,
  title={Again-vc: A one-shot voice conversion using activation guidance and adaptive instance normalization},
  author={Chen, Yen-Hao and Wu, Da-Yi and Wu, Tsung-Han and Lee, Hung-yi},
  booktitle={ICASSP 2021-2021 IEEE International Conference on Acoustics, Speech and Signal Processing (ICASSP)},
  pages={5954--5958},
  year={2021},
  organization={IEEE}
}

@article{liu2021any,
  title={Any-to-many voice conversion with location-relative sequence-to-sequence modeling},
  author={Liu, Songxiang and Cao, Yuewen and Wang, Disong and Wu, Xixin and Liu, Xunying and Meng, Helen},
  journal={IEEE/ACM Transactions on Audio, Speech, and Language Processing},
  volume={29},
  pages={1717--1728},
  year={2021},
  publisher={IEEE}
}

@inproceedings{li2021starganv2,
  title={StarGANv2-VC: A Diverse, Unsupervised, Non-Parallel Framework for Natural-Sounding Voice Conversion},
  author={Li, Yinghao Aaron and Zare, Ali and Mesgarani, Nima},
  booktitle={Proc. Interspeech 2021},
  pages={1349--1353},
  year={2021}
}

@inproceedings{kaneko2018cyclegan,
  title={Cyclegan-vc: Non-parallel voice conversion using cycle-consistent adversarial networks},
  author={Kaneko, Takuhiro and Kameoka, Hirokazu},
  booktitle={2018 26th European Signal Processing Conference (EUSIPCO)},
  pages={2100--2104},
  year={2018},
  organization={IEEE}
}

@inproceedings{kaneko2019cyclegan,
  title={Cyclegan-vc2: Improved cyclegan-based non-parallel voice conversion},
  author={Kaneko, Takuhiro and Kameoka, Hirokazu and Tanaka, Kou and Hojo, Nobukatsu},
  booktitle={ICASSP 2019-2019 IEEE International Conference on Acoustics, Speech and Signal Processing (ICASSP)},
  pages={6820--6824},
  year={2019},
  organization={IEEE}
}

@inproceedings{choi2023diff,
  title={Diff-HierVC: Diffusion-based Hierarchical Voice Conversion with Robust Pitch Generation and Masked Prior for Zero-shot Speaker Adaptation},
  author={Choi, Ha-Yeong and Lee, Sang-Hoon and Lee, Seong-Whan},
  booktitle={Proc. Interspeech 2023},
  pages={2283--2287},
  year={2023}
}

@inproceedings{popovdiffusion,
  title={Diffusion-Based Voice Conversion with Fast Maximum Likelihood Sampling Scheme},
  author={Popov, Vadim and Vovk, Ivan and Gogoryan, Vladimir and Sadekova, Tasnima and Kudinov, Mikhail Sergeevich and Wei, Jiansheng},
  booktitle={International Conference on Learning Representations},
  year={2022}
}

@article{goodfellow2014generative,
  title={Generative adversarial nets},
  author={Goodfellow, Ian J and Pouget-Abadie, Jean and Mirza, Mehdi and Xu, Bing and Warde-Farley, David and Ozair, Sherjil and Courville, Aaron and Bengio, Yoshua},
  journal={Advances in neural information processing systems},
  volume={27},
  year={2014}
}

@inproceedings{sohl2015deep,
  title={Deep unsupervised learning using nonequilibrium thermodynamics},
  author={Sohl-Dickstein, Jascha and Weiss, Eric and Maheswaranathan, Niru and Ganguli, Surya},
  booktitle={International conference on machine learning},
  pages={2256--2265},
  year={2015},
  organization={pmlr}
}

@inproceedings{ma2024distillation,
  title={Distillation-Based Feature Extraction Algorithm For Source Speaker Verification},
  author={Ma, Xinlei and Lu, Wenhuan and Zhang, Ruiteng and Xu, Junhai and Lu, Xugang and Wei, Jianguo},
  booktitle={2024 IEEE Spoken Language Technology Workshop (SLT)},
  pages={1240--1246},
  year={2024},
  organization={IEEE}
}

@inproceedings{choi2024dddm,
  title={Dddm-vc: Decoupled denoising diffusion models with disentangled representation and prior mixup for verified robust voice conversion},
  author={Choi, Ha Yeong and Lee, Sang Hoon and Lee, Seong Whan},
  booktitle={Proceedings of the AAAI Conference on Artificial Intelligence},
  volume={38},
  number={16},
  pages={17862--17870},
  year={2024}
}

@inproceedings{tak2021end2,
  title={End-to-End Spectro-Temporal Graph Attention Networks for Speaker Verification Anti-Spoofing and Speech Deepfake Detection},
  author={Tak, Hemlata and Jung, Jee-Weon and Patino, Jose and Kamble, Madhu and Todisco, Massimiliano and Evans, Nicholas},
  booktitle={ASVSPOOF 2021, Automatic Speaker Verification and Spoofing Countermeasures Challenge},
  pages={1--8},
  year={2021},
  organization={ISCA}
}

@inproceedings{ge2021raw,
  title={Raw Differentiable Architecture Search for Speech Deepfake and Spoofing Detection},
  author={Ge, Wanying and Patino, Jose and Todisco, Massimiliano and Evans, Nicholas},
  booktitle={ASVSPOOF 2021, Automatic Speaker Verification and Spoofing Countermeasures Challenge},
  pages={22--28},
  year={2021},
  organization={ISCA}
}

@inproceedings{zhang2021multi,
  title={Multi-task Learning in Utterance-level and Segmental-level Spoof Detection},
  author={Zhang, Lin and Wang, Xin and Cooper, Erica and Yamagishi, Junichi},
  booktitle={Proc. ASVSPOOF 2021},
  pages={9--15},
  year={2021}
}

@article{hua2021towards,
  title={Towards end-to-end synthetic speech detection},
  author={Hua, Guang and Teoh, Andrew Beng Jin and Zhang, Haijian},
  journal={IEEE Signal Processing Letters},
  volume={28},
  pages={1265--1269},
  year={2021},
  publisher={IEEE}
}

@inproceedings{ma2021rw,
  title={RW-Resnet: A Novel Speech Anti-Spoofing Model Using Raw Waveform},
  author={Ma, Youxuan and Ren, Zongze and Xu, Shugong},
  booktitle={Proc. Interspeech 2021},
  pages={4144--4148},
  year={2021}
}

@inproceedings{conti2022deepfake,
  title={Deepfake speech detection through emotion recognition: a semantic approach},
  author={Conti, Emanuele and Salvi, Davide and Borrelli, Clara and Hosler, Brian and Bestagini, Paolo and Antonacci, Fabio and Sarti, Augusto and Stamm, Matthew C and Tubaro, Stefano},
  booktitle={ICASSP 2022-2022 IEEE international conference on acoustics, speech and signal processing (ICASSP)},
  pages={8962--8966},
  year={2022},
  organization={IEEE}
}

@inproceedings{li2022comparative,
  title={A comparative study on physical and perceptual features for deepfake audio detection},
  author={Li, Menglu and Ahmadiadli, Yasaman and Zhang, Xiao-Ping},
  booktitle={Proceedings of the 1st International Workshop on Deepfake Detection for Audio Multimedia},
  pages={35--41},
  year={2022}
}

@inproceedings{doan2023bts,
  title={BTS-E: Audio deepfake detection using breathing-talking-silence encoder},
  author={Doan, Thien-Phuc and Nguyen-Vu, Long and Jung, Souhwan and Hong, Kihun},
  booktitle={ICASSP 2023-2023 IEEE International Conference on Acoustics, Speech and Signal Processing (ICASSP)},
  pages={1--5},
  year={2023},
  organization={IEEE}
}

@inproceedings{mostaani2022breathing,
  title={On Breathing Pattern Information in Synthetic Speech},
  author={Mostaani, Zohreh and Doss, Mathew Magimai},
  booktitle={Proc. Interspeech 2022},
  pages={2768--2772},
  year={2022}
}

@inproceedings{tak2021graph,
  title={Graph Attention Networks for Anti-Spoofing},
  author={Tak, Hemlata and Jung, Jee-weon and Patino, Jose and Todisco, Massimiliano and Evans, Nicholas},
  booktitle={Proc. Interspeech 2021},
  pages={2356--2360},
  year={2021}
}

@inproceedings{tak2021end,
  title={End-to-end anti-spoofing with rawnet2},
  author={Tak, Hemlata and Patino, Jose and Todisco, Massimiliano and Nautsch, Andreas and Evans, Nicholas and Larcher, Anthony},
  booktitle={ICASSP 2021-2021 IEEE International Conference on Acoustics, Speech and Signal Processing (ICASSP)},
  pages={6369--6373},
  year={2021},
  organization={IEEE}
}

@inproceedings{jung2022aasist,
  title={Aasist: Audio anti-spoofing using integrated spectro-temporal graph attention networks},
  author={Jung, Jee-weon and Heo, Hee-Soo and Tak, Hemlata and Shim, Hye-jin and Chung, Joon Son and Lee, Bong-Jin and Yu, Ha-Jin and Evans, Nicholas},
  booktitle={ICASSP 2022-2022 IEEE international conference on acoustics, speech and signal processing (ICASSP)},
  pages={6367--6371},
  year={2022},
  organization={IEEE}
}

@inproceedings{kawa2022specrnet,
  title={Specrnet: Towards faster and more accessible audio deepfake detection},
  author={Kawa, Piotr and Plata, Marcin and Syga, Piotr},
  booktitle={2022 IEEE International Conference on Trust, Security and Privacy in Computing and Communications (TrustCom)},
  pages={792--799},
  year={2022},
  organization={IEEE}
}

@inproceedings{liu2023betray,
  title={Betray Oneself: A Novel Audio DeepFake Detection Model via Mono-to-Stereo Conversion},
  author={Liu, Rui and Zhang, Jinhua and Gao, Guanglai and Li, Haizhou},
  booktitle={Proc. Interspeech 2023},
  pages={3999--4003},
  year={2023}
}

@article{yadav2024compression,
  title={Compression Robust Synthetic Speech Detection Using Patched Spectrogram Transformer},
  author={Yadav, Amit Kumar Singh and Xiang, Ziyue and Bhagtani, Kratika and Bestagini, Paolo and Tubaro, Stefano and Delp, Edward J},
  journal={CoRR},
  year={2024}
}

@article{kulangareth2024investigation,
  title={Investigation of deepfake voice detection using speech pause patterns: Algorithm development and validation},
  author={Kulangareth, Nikhil Valsan and Kaufman, Jaycee and Oreskovic, Jessica and Fossat, Yan},
  journal={JMIR biomedical engineering},
  volume={9},
  pages={e56245},
  year={2024},
  publisher={JMIR Publications Toronto, Canada}
}

@inproceedings{doan2023gan,
  title={Gan discriminator based audio deepfake detection},
  author={Doan, Thien Phuc and Hong, Kihun and Jung, Souhwan},
  booktitle={Proceedings of the 2nd Workshop on Security Implications of Deepfakes and Cheapfakes},
  pages={29--32},
  year={2023}
}

@inproceedings{kumarivoiceradar,
  title={VoiceRadar: Voice Deepfake Detection using Micro-Frequency and Compositional Analysis},
  author={Kumari, Kavita and Abbasihafshejani, Maryam and Pegoraro, Alessandro and Rieger, Phillip and Arshi, Kamyar and Jadliwala, Murtuza and Sadeghi, Ahmad-Reza},
  booktitle={NDSS},
  year={2025}
}

@inproceedings{desplanques2020ecapa,
  title={Ecapa-tdnn: Emphasized channel attention, propagation and aggregation in tdnn based speaker verification},
  author={Desplanques, Brecht and Thienpondt, Jenthe and Demuynck, Kris},
  booktitle={21st Annual conference of the International Speech Communication Association (INTERSPEECH 2020)},
  pages={3830--3834},
  year={2020},
  organization={International Speech Communication Association (ISCA)}
}

@inproceedings{deng2023catch,
  title={Catch you and i can: Revealing source voiceprint against voice conversion},
  author={Deng, Jiangyi and Chen, Yanjiao and Zhong, Yinan and Miao, Qianhao and Gong, Xueluan and Xu, Wenyuan},
  booktitle={32nd USENIX Security Symposium (USENIX Security 23)},
  pages={5163--5180},
  year={2023}
}

@inproceedings{nagrani2017voxceleb,
  title={VoxCeleb: A Large-Scale Speaker Identification Dataset},
  author={Nagrani, Arsha and Chung, Joon Son and Zisserman, Andrew},
  booktitle={Proc. Interspeech 2017},
  pages={2616--2620},
  year={2017}
}

@inproceedings{chung2018voxceleb2,
  title={VoxCeleb2: Deep Speaker Recognition},
  author={Chung, Joon Son and Nagrani, Arsha and Zisserman, Andrew},
  booktitle={Proc. Interspeech 2018},
  pages={1086--1090},
  year={2018}
}

@inproceedings{panayotov2015librispeech,
  title={Librispeech: an asr corpus based on public domain audio books},
  author={Panayotov, Vassil and Chen, Guoguo and Povey, Daniel and Khudanpur, Sanjeev},
  booktitle={2015 IEEE international conference on acoustics, speech and signal processing (ICASSP)},
  pages={5206--5210},
  year={2015},
  organization={IEEE}
}

@inproceedings{kingma2014adam,
  title={Adam: A method for stochastic optimization},
  author={Kingma, Diederik P and Ba, Jimmy},
  booktitle={International Conference on Learning Representations},
  year={2015}
}

@article{van2008visualizing,
  title={Visualizing data using t-SNE.},
  author={Van der Maaten, Laurens and Hinton, Geoffrey},
  journal={Journal of machine learning research},
  volume={9},
  number={11},
  year={2008}
}

@inproceedings{recommendation1988pulse,
  title={Pulse code modulation (PCM) of voice frequencies},
  author={Recommendation, CCITT},
  booktitle={ITU},
  year={1988}
}

@inproceedings{etsi2000digital,
  title={Digital cellular telecommunications system (Phase 2+)(GSM); Full rate speech; Transcoding, GSM 06.10 version 8.1. 1 Release 1999},
  author={ETSI},
  booktitle={Standard ETSI EN 300 961, European Telecommunications Standards Institute},
  year={2000},
}

@inproceedings{ekudden1999adaptive,
  title={The adaptive multi-rate speech coder},
  author={Ekudden, E and Hagen, R and Johansson, I and Svedberg, J},
  booktitle={1999 IEEE workshop on speech coding proceedings. model, coders, and error criteria (cat. no. 99EX351)},
  pages={117--119},
  year={1999},
  organization={IEEE}
}

@inproceedings{li2023freevc,
  title={Freevc: Towards high-quality text-free one-shot voice conversion},
  author={Li, Jingyi and Tu, Weiping and Xiao, Li},
  booktitle={ICASSP 2023-2023 IEEE International Conference on Acoustics, Speech and Signal Processing (ICASSP)},
  pages={1--5},
  year={2023},
  organization={IEEE}
}

@inproceedings{cai2023identifying,
  title={Identifying source speakers for voice conversion based spoofing attacks on speaker verification systems},
  author={Cai, Danwei and Cai, Zexin and Li, Ming},
  booktitle={ICASSP 2023-2023 IEEE International Conference on Acoustics, Speech and Signal Processing (ICASSP)},
  pages={1--5},
  year={2023},
  organization={IEEE}
}

@inproceedings{zhang2022mfa,
  title={MFA-Conformer: Multi-scale Feature Aggregation Conformer for Automatic Speaker Verification},
  author={Zhang, Yang and Lv, Zhiqiang and Wu, Haibin and Zhang, Shanshan and Hu, Pengfei and Wu, Zhiyong and Lee, Hung-yi and Meng, Helen},
  booktitle={Proc. Interspeech 2022},
  pages={306--310},
  year={2022}
}

@article{panariello2024voiceprivacy,
  title={The voiceprivacy 2022 challenge: Progress and perspectives in voice anonymisation},
  author={Panariello, Michele and Tomashenko, Natalia and Wang, Xin and Miao, Xiaoxiao and Champion, Pierre and Nourtel, Hubert and Todisco, Massimiliano and Evans, Nicholas and Vincent, Emmanuel and Yamagishi, Junichi},
  journal={IEEE/ACM Transactions on Audio, Speech, and Language Processing},
  year={2024},
  publisher={IEEE}
}

@inproceedings{he2016deep,
  title={Deep residual learning for image recognition},
  author={He, Kaiming and Zhang, Xiangyu and Ren, Shaoqing and Sun, Jian},
  booktitle={Proceedings of the IEEE conference on computer vision and pattern recognition},
  pages={770--778},
  year={2016}
}

@inproceedings{ronneberger2015u,
  title={U-net: Convolutional networks for biomedical image segmentation},
  author={Ronneberger, Olaf and Fischer, Philipp and Brox, Thomas},
  booktitle={International Conference on Medical image computing and computer-assisted intervention},
  pages={234--241},
  year={2015},
  organization={Springer}
}

@inproceedings{kingma2014autoencoding,
  title={Auto-Encoding Variational Bayes},
  author={Kingma, Diederik P and Welling, Max},
  booktitle={International Conference on Learning Representations (ICLR)},
  year={2014},
}

@article{hayashi2020voice,
  title={Voice Transformer Network: Sequence-to-Sequence Voice Conversion Using Transformer with Text-to-Speech Pretraining},
  author={Hayashi, Tomoki and Kameoka, Hirokazu and Wu, Yi-Chiao and Huang, Wen-Chin and Toda, Tomoki},
  journal={Interspeech 2020},
  year={2020},
  publisher={ISCA}
}

@article{tachbelie2014, 
  Author = {Martha Tachbelie and Solomon Teferra Abate and Laurent Besacier}, 
  Journal = {Speech Communication}, 
  Publisher = {Elsevier}, 
  Title = {Using different acoustic, lexical and language modeling units for ASR of an under-resourced language - Amharic}, 
  Volume = {56}, 
  Year = {2014}
 }

@misc{eduskuntav1.5,
  title = {Plenary Sessions of the Parliament of Finland (eduskunta)},
  author = {{Kielipankki}},
  year = {2023},
  howpublished = {The Language Bank of Finland},
  url = {https://www.kielipankki.fi/download/eduskunta/v1.5/},
}

@misc{yamagishi2019cstr,
  author = {Yamagishi, Junichi and Veaux, Christophe and MacDonald, Kirsten},
  title = {CSTR VCTK Corpus: English Multi-speaker Corpus for CSTR Voice Cloning Toolkit (version 0.92)},
  year = {2019},
  publisher = {University of Edinburgh. The Centre for Speech Technology Research (CSTR)},
  doi = {10.7488/ds/2645},
  url = {https://doi.org/10.7488/ds/2645}
}

@inproceedings{pratap2020mls,
  title={MLS: A Large-Scale Multilingual Dataset for Speech Research},
  author={Pratap, Vineel and Xu, Qiantong and Sriram, Anuroop and Synnaeve, Gabriel and Collobert, Ronan},
  booktitle={Proc. Interspeech 2020},
  pages={2757--2761},
  year={2020}
}

@inproceedings{bu2017aishell,
  title={Aishell-1: An open-source mandarin speech corpus and a speech recognition baseline},
  author={Bu, Hui and Du, Jiayu and Na, Xingyu and Wu, Bengu and Zheng, Hao},
  booktitle={2017 20th conference of the oriental chapter of the international coordinating committee on speech databases and speech I/O systems and assessment (O-COCOSDA)},
  pages={1--5},
  year={2017},
  organization={IEEE}
}

@inproceedings{yang2024streamvc,
  title={Streamvc: Real-time low-latency voice conversion},
  author={Yang, Yang and Kartynnik, Yury and Li, Yunpeng and Tang, Jiuqiang and Li, Xing and Sung, George and Grundmann, Matthias},
  booktitle={ICASSP 2024-2024 IEEE International Conference on Acoustics, Speech and Signal Processing (ICASSP)},
  pages={11016--11020},
  year={2024},
  organization={IEEE}
}

@inproceedings{gaznepoglu2025you,
  title={You Are What You Say: Exploiting Linguistic Content for VoicePrivacy Attacks},
  author={Gaznepoglu, {\"U}nal Ege and Leschanowsky, Anna and Aloradi, Ahmad and Singh, Prachi and Tenbrinck, Daniel and Habets, Emanu{\"e}l AP and Peters, Nils},
  booktitle={Proc. Interspeech 2025},
  pages={4238--4242},
  year={2025}
}

@inproceedings{tomashenko2025analysis,
  title={Analysis of speech temporal dynamics in the context of speaker verification and voice anonymization},
  author={Tomashenko, Natalia and Vincent, Emmanuel and Tommasi, Marc},
  booktitle={ICASSP 2025-2025 IEEE International Conference on Acoustics, Speech and Signal Processing (ICASSP)},
  pages={1--5},
  year={2025},
  organization={IEEE}
}

@article{miao2025benchmark,
  title={A benchmark for multi-speaker anonymization},
  author={Miao, Xiaoxiao and Tao, Ruijie and Zeng, Chang and Wang, Xin},
  journal={IEEE Transactions on Information Forensics and Security},
  year={2025},
  publisher={IEEE}
}

@inproceedings{tomashenko2025exploiting,
  title={Exploiting Context-dependent Duration Features for Voice Anonymization Attack Systems},
  author={Tomashenko, Natalia and Vincent, Emmanuel and Tommasi, Marc},
  booktitle={Proc. Interspeech 2025},
  pages={5128--5132},
  year={2025}
}

@inproceedings{bakari2026identity,
  title={Identity leakage through accent cues in voice anonymisation},
  author={Bakari, Rayane and Le Blouch, Olivier and Gengembre, Nicolas and Evans, Nicholas and Panariello, Michele},
  booktitle={ICASSP 2026-2026 IEEE International Conference on Acoustics, Speech and Signal Processing (ICASSP)},
  pages={13657--13661},
  year={2026},
  organization={IEEE}
}

@inproceedings{aggazzotti2026content,
  title={Content Anonymization for Privacy in Long-Form Audio},
  author={Aggazzotti, Cristina and Garg, Ashi and Cai, Zexin and Andrews, Nicholas},
  booktitle={ICASSP 2026-2026 IEEE International Conference on Acoustics, Speech and Signal Processing (ICASSP)},
  pages={13412--13416},
  year={2026},
  organization={IEEE}
}

@inproceedings{seungmin2026evaluating,
  title={EVALUATING IDENTITY LEAKAGE IN SPEAKER DE-IDENTIFICATION SYSTEMS},
  author={Seungmin Seo and Oleg Aulov and Afzal Godil and Kevin Mangold},
  booktitle={ICASSP 2026-2026 IEEE International Conference on Acoustics, Speech and Signal Processing (ICASSP)},
  year={2026},
  organization={IEEE}
}

@inproceedings{zhang2025attacking,
  title={Attacking voice anonymization systems with augmented feature and speaker identity difference},
  author={Zhang, Yanzhe and Bi, Zhonghao and Xiao, Feiyang and Yang, Xuefeng and Zhu, Qiaoxi and Guan, Jian},
  booktitle={ICASSP 2025-2025 IEEE International Conference on Acoustics, Speech and Signal Processing (ICASSP)},
  pages={1--2},
  year={2025},
  organization={IEEE}
}

@book{hollien2002forensic,
  title={Forensic voice identification},
  author={Hollien, Harry Francis},
  year={2002},
  publisher={Academic Press}
}

@article{hansen2015speaker,
  title={Speaker recognition by machines and humans: A tutorial review},
  author={Hansen, John HL and Hasan, Taufiq},
  journal={IEEE Signal processing magazine},
  volume={32},
  number={6},
  pages={74--99},
  year={2015},
  publisher={IEEE}
}

\appendix

\section{Converted Audio Generation}
\label{section.dataset}
Following the description below, we process each dataset with every voice conversion method to generate adequate voice conversion samples.

\begin{table}[h]
  \centering
  \footnotesize
  \caption{Description of the datasets.}
  \label{tab.datasets}
  \resizebox{1.0\linewidth}{!}{
  \begin{threeparttable}[b]
    \begin{tabular}{cccc}
    \Xhline{1pt}
        \textbf{Dataset\tnote{$\dagger$}} &\textbf{Language}  & \textbf{Speakers}   & \textbf{Converted Audios\tnote{$\ddagger$}}\\
      \Xhline{1pt}
      \textit{train-clean-100} & \multirow{5}{*}{English} & 251 & 25,100  \\
      \textit{train-clean-360} &  & 921 & 92,100  \\
      \textit{train-other-500} &  & 1,166 & 116,600 \\
      VoxCeleb1    &    & 1,251 & 125,100  \\
      VoxCeleb2    &    & 5,994 & 239,760  \\
      \cline{1-4}
      \textit{test-clean}  & English    & 40  & 31,200 \\
    \Xhline{1pt}
    \end{tabular}%
        \begin{tablenotes}  
        \footnotesize         
        \item[$\dagger$] \textit{train-clean-100}, \textit{train-clean-360} and \textit{train-other-500} are training sets in LibriSpeech. \textit{test-clean} is the testing set in LibriSpeech, which we utilize to evaluate the performance of different voiceprint recovery methods.
        \item[$\ddagger$] It refers to the converted audio number of each voice conversion method.
      \end{tablenotes}  
         \end{threeparttable}
    }
\end{table}

\begin{itemize}
\item For training sets, we generate one sample for each source-target pair. For each speaker in \textit{train-clean-100}, \textit{train-clean-360}, \textit{train-other-500}, and VoxCeleb1, we randomly sample another 100 speakers for each source speaker to form source-target pairs and generate converted audios. For VoxCeleb2, we randomly select another 40 speakers as target speakers to convert audios of each source speaker.

\item For testing sets, we generate twenty samples for each source-target pair. For each source speaker in \textit{test-clean}, we select the remaining 39 speakers as target speakers to generate converted audios. As for datasets of unseen languages in Section \ref{section.language}, we also exhaustively pair all speakers to generate twenty converted audios for each source-target pair.
\end{itemize}

\section{Performance on all test speakers}
\label{section.all_test}

For large-scale evaluation, we include all 231 test speakers from the test sets of LibriSpeech (40 speakers in \textit{test-clean} and 33 in \textit{test-other}, with no overlap), VoxCeleb1 (40 speakers), and VoxCeleb2 (118 speakers). All other settings follow those described in Section \ref{section.setup}. As shown in Table \ref{tab.large_scale_all}, scaling up the test set generally degrades the performance of all voiceprint recovery methods. However, unlike MFA-Conformer, which suffers from a notable increase in EER, the other methods benefit from the cross-dataset variability, yielding EERs that are lower than or comparable to those reported in Section \ref{section.large_scale}. Overall, \sys achieves the best performance, yielding a 4.57\% EER and a 90.83\% Top-5 ACC on converted audio, thereby confirming its robustness in large-scale scenarios.

\begin{table}[t]
  \begin{center}
    \caption{Performance of voiceprint recovery methods on all test speakers across the three datasets.}
    \label{tab.large_scale_all}
    \resizebox{1.0\linewidth}{!}{
    \begin{tabular}{cr c c c c}
      \Xhline{1pt}
      \multicolumn{1}{c}{\multirow{2}{*}{\textbf{VC}}} & \multicolumn{1}{c}{\multirow{2}{*}{\textbf{Metric}}} &  \multicolumn{4}{c}{\textbf{Voiceprint Recovery Method}} \\
      \multicolumn{1}{c}{}& & \textbf{MFA} & \textbf{ECAPA} & \textbf{Revelio} & \textbf{\sys} \\
      \Xhline{1pt}
      \multirow{4}{*}{Clean}
      & EER ($\downarrow$)
      & 18.53\% & 6.94\%  & 4.22\% & \cellcolor{backcolor!100}\textbf{1.64\%} \\
      \addlinespace[0.2pt]
      \cline{2-6}
      \addlinespace[1pt]
      & Top-1 ACC ($\uparrow$)
      & 38.47\% & 73.56\% & 81.27\% & \cellcolor{backcolor!100}\textbf{92.03\%}\\
      & Top-3 ACC ($\uparrow$)
      & 52.18\% & 83.40\% & 93.46\% & \cellcolor{backcolor!100}\textbf{96.82\%} \\
      & Top-5 ACC ($\uparrow$)
      & 59.19\% & 87.17\% & 96.22\% & \cellcolor{backcolor!100}\textbf{97.81\%} \\
      \addlinespace[0.2pt]
      \cline{1-6}
      \addlinespace[1pt]
      \multirow{4}{*}{AGAIN}
      & EER ($\downarrow$)
      & 44.83\% & 32.02\%  & 8.48\% & \cellcolor{backcolor!100}\textbf{4.25\%} \\
      \addlinespace[0.2pt]
      \cline{2-6}
      \addlinespace[1pt]
      & Top-1 ACC ($\uparrow$)
      & 4.17\% & 5.70\% & 62.91\% & \cellcolor{backcolor!100}\textbf{75.18\%}\\
      & Top-3 ACC ($\uparrow$)
      & 7.78\% & 11.40\% & 79.01\% & \cellcolor{backcolor!100}\textbf{89.03\%} \\
      & Top-5 ACC ($\uparrow$)
      & 10.08\% & 15.23\% & 84.91\% & \cellcolor{backcolor!100}\textbf{92.59\%} \\
      \addlinespace[0.2pt]
      \cline{1-6}
      \addlinespace[1pt]
      \multirow{4}{*}{FreeVC}
      & EER ($\downarrow$)
      & 42.44\% & 36.19\%  & 9.31\% & \cellcolor{backcolor!100}\textbf{5.32\%} \\
      \addlinespace[0.2pt]
      \cline{2-6}
      \addlinespace[1pt]
      & Top-1 ACC ($\uparrow$)
      & 4.35\% & 3.61\% & 50.63\% & \cellcolor{backcolor!100}\textbf{64.99\%} \\
      & Top-3 ACC ($\uparrow$)
      & 9.11\% & 8.24\% & 71.65\% & \cellcolor{backcolor!100}\textbf{82.30\%} \\
      & Top-5 ACC ($\uparrow$)
      & 12.77\% & 11.73\% & 79.64\% & \cellcolor{backcolor!100}\textbf{87.64\%}\\
      \addlinespace[0.2pt]
      \cline{1-6}
      \addlinespace[1pt]
      \multirow{4}{*}{Diff}
      & EER ($\downarrow$)
      & 46.54\% & 31.63\%  & 7.20\% & \cellcolor{backcolor!100}\textbf{4.15\%} \\
      \addlinespace[0.2pt]
      \cline{2-6}
      \addlinespace[1pt]
      & Top-1 ACC ($\uparrow$)
      & 6.60\% & 5.61\% & 61.59\% & \cellcolor{backcolor!100}\textbf{72.18\%} \\
      & Top-3 ACC ($\uparrow$)
      & 11.67\% & 12.90\% & 78.02\% & \cellcolor{backcolor!100}\textbf{88.28\%} \\
      & Top-5 ACC ($\uparrow$)
      & 14.24\% & 17.61\% & 84.37\% & \cellcolor{backcolor!100}\textbf{93.21\%} \\
      \addlinespace[0.2pt]
      \cline{1-6}
      \addlinespace[1pt]
      \multirow{4}{*}{DDDM}
      & EER ($\downarrow$)
      & 47.09\% & 35.96\%  & 8.08\% & \cellcolor{backcolor!100}\textbf{4.56\%} \\
      \addlinespace[0.2pt]
      \cline{2-6}
      \addlinespace[1pt]
      & Top-1 ACC ($\uparrow$)
      & 4.91\% & 4.14\% & 53.15\% & \cellcolor{backcolor!100}\textbf{68.05\%} \\
      & Top-3 ACC ($\uparrow$)
      & 8.49\% & 9.88\% & 74.06\% & \cellcolor{backcolor!100}\textbf{83.91\%} \\
      & Top-5 ACC ($\uparrow$)
      & 10.94\% & 13.93\% & 81.67\% & \cellcolor{backcolor!100}\textbf{89.87\%} \\
      \Xhline{1pt}
    \end{tabular}
    }
  \end{center}
\end{table}

\section{Ethical Considerations}

The authors confirm their strict adherence to relevant ethical guidelines throughout this research. Ethical considerations have been critically examined and upheld during all stages of the work.

\textbf{Motivation.} This study is motivated by the need to defend against malicious applications of voice conversion technology. Such technology can be exploited to impersonate target speakers and manipulate human trust on an unprecedented scale, leading to serious consequences such as widespread misinformation, substantial financial loss, and general societal harm. Defending against voice conversion-based spoofing is therefore essential for reliable speaker recognition.

\textbf{Privacy.} All experiments utilize publicly available benchmark datasets (VoxCeleb1 \cite{nagrani2017voxceleb}, VoxCeleb2 \cite{chung2018voxceleb2}, and LibriSpeech \cite{panayotov2015librispeech}), which are widely adopted in the speech research community under compliant licenses. Our use of the data aligns with ethical standards and the respective license agreements. Moreover, the voiceprint vectors generated by \sys do not directly reveal personal speaker attributes such as age or gender. Speaker identity could only be inferred if an attacker already possesses a sufficient set of reference audios to build a voiceprint database. Given these safeguards, we believe the research benefits of this work substantially outweigh any potential risks.

\section{Open Science}

The authors are committed to open science policy and publicly share research outputs. In accordance with this policy, we provide artifacts at https://github.com/Forliqr/TRIDENT, which include all materials necessary to evaluate the paper's core contributions, i.e., code, datasets, models, and evaluation pipelines.

\begin{table}[h]
  \centering
  \footnotesize
  \caption{Detailed description of \sys.}
  \label{tab.model}
  \resizebox{1.0\linewidth}{!}{
  \begin{threeparttable}[b]
    \begin{tabular}{cccc}
    \Xhline{1pt}
        \textbf{Module} & \textbf{Block}  & \textbf{Output Size}   & \textbf{Params}\\
      \Xhline{1pt}
      \textbf{Input} & \multicolumn{3}{c}{\makecell{Input Audio Size: (1, T)\\Auxiliary Audio Size: (1, T)}}\\
      \cline{1-4}
      \multirow{7}{*}{\textbf{\makecell{Latent\\Representation}}} & Filter Bank & (80, T)& -\\
      & Conv1D+ReLU+BN & (1024, T) & 412,672\\
      & SE-Res2Bloack & (1024, T) & 2,713,344\\
      & SE-Res2Bloack & (1024, T) & 2,713,344\\
      & SE-Res2Bloack & (1024, T) & 2,713,344\\
      & Differential Block & (3072, T) & 9,446,400\\
      & Conv1D+ReLU+BN & (3072, T) & 9,446,400\\
      \cline{1-4}
      \multirow{3}{*}{\textbf{\makecell{VC Method\\Inference}}} & ASP & (6144, 1) & 1,588,608 \\
      & FC+BN & (32, 1) & 196,640\\
      & FC & (9583\tnote{$\dagger$}, 1) & 306,656\\
      \cline{1-4}
      \multirow{2}{*}{\textbf{\makecell{Attention-based\\Refinement}}} & \multirow{2}{*}{ASP} & \multirow{2}{*}{(6144, 1)} & \multirow{2}{*}{1,588,608} \\
      & & & \\
      \cline{1-4}
      \multirow{2}{*}{\textbf{\makecell{Feature\\Aggregation}}} & FC+BN & (192, 1) & 2,359,488 \\
      & FC & (9583, 1) & 1,839,936\\
      \cline{1-4}
      \multirow{3}{*}{\textbf{\makecell{Guided\\Refinement}}} & ASP & (6144, 1) & 1,588,608\\
      & FC+BN & (192, 1) & 1,179,840\\
      & FC & (9583, 1) & 1,839,936\\
      \cline{1-4}
      \textbf{Output} & \multicolumn{3}{c}{\makecell{Output Voiceprint Size: (192, 1)\\Output Size: (9853, 1)}} \\
      \Xhline{1pt}
    \end{tabular}
        \begin{tablenotes}  
        \footnotesize         
        \item[$\dagger$] We have reserved sufficient labels here for more voice conversion methods.
      \end{tablenotes}  
         \end{threeparttable}
    }
\end{table}

\begin{table}[h]
  \begin{center}
    \caption{Performance of \sys by gender subgroups.}
    \label{tab.gender}
    \resizebox{1.0\linewidth}{!}{
    \begin{tabular}{c r c c c c}
      \Xhline{1pt}
      \textbf{VC} & \multicolumn{1}{c}{\textbf{Metric}} & \textbf{M$\rightarrow$M} & \textbf{F$\rightarrow$F} & \textbf{M$\rightarrow$F} & \textbf{F$\rightarrow$M} \\
      \Xhline{1pt}
      \multirow{4}{*}{AGAIN}
      & EER ($\downarrow$)
      & 1.27\% & 1.87\% & 1.26\% & 1.17\% \\
      \addlinespace[0.2pt]
      \cline{2-6}
      \addlinespace[1pt]
      & Top-1 ACC ($\uparrow$)
      & 97.76\% & 97.68\% & 97.00\% & 97.88\% \\
      & Top-3 ACC ($\uparrow$)
      & 99.87\% & 98.82\% & 99.63\% & 99.25\% \\
      & Top-5 ACC ($\uparrow$)
      & 100.00\% & 99.87\% & 100.00\% & 99.50\% \\
      \cline{1-6}
      \multirow{4}{*}{VQVC}
      & EER ($\downarrow$)
      & 7.26\% & 7.18\% & 6.78\% & 6.67\% \\
      \addlinespace[0.2pt]
      \cline{2-6}
      \addlinespace[1pt]
      & Top-1 ACC ($\uparrow$)
      & 82.66\% & 81.50\% & 83.50\% & 83.75\% \\
      & Top-3 ACC ($\uparrow$)
      & 94.47\% & 93.34\% & 96.13\% & 94.25\% \\
      & Top-5 ACC ($\uparrow$)
      & 97.76\% & 96.08\% & 97.88\% & 96.75\% \\
      \cline{1-6}
      \multirow{4}{*}{VQVC+}
      & EER ($\downarrow$)
      & 3.64\% & 4.17\% & 3.63\% & 3.87\% \\
      \addlinespace[0.2pt]
      \cline{2-6}
      \addlinespace[1pt]
      & Top-1 ACC ($\uparrow$)
      & 94.90\% & 91.18\% & 93.88\% & 93.63\% \\
      & Top-3 ACC ($\uparrow$)
      & 99.47\% & 97.24\% & 99.12\% & 98.50\% \\
      & Top-5 ACC ($\uparrow$)
      & 100.00\% & 98.82\% & 99.63\% & 100.00\% \\
      \cline{1-6}
      \multirow{4}{*}{BNE}
      & EER ($\downarrow$)
      & 7.76\% & 8.08\% & 8.00\% & 8.26\% \\
      \addlinespace[0.2pt]
      \cline{2-6}
      \addlinespace[1pt]
      & Top-1 ACC ($\uparrow$)
      & 78.97\% & 75.05\% & 74.50\% & 74.55\% \\
      & Top-3 ACC ($\uparrow$)
      & 93.42\% & 91.74\% & 89.88\% & 88.85\% \\
      & Top-5 ACC ($\uparrow$)
      & 96.05\% & 92.42\% & 93.38\% & 93.75\% \\
      \cline{1-6}
      \multirow{4}{*}{FreeVC}
      & EER ($\downarrow$)
      & 2.70\% & 3.17\% & 2.76\% & 3.37\% \\
      \addlinespace[0.2pt]
      \cline{2-6}
      \addlinespace[1pt]
      & Top-1 ACC ($\uparrow$)
      & 93.50\% & 92.47\% & 94.25\% & 91.63\% \\
      & Top-3 ACC ($\uparrow$)
      & 98.82\% & 97.11\% & 99.38\% & 96.63\% \\
      & Top-5 ACC ($\uparrow$)
      & 99.61\% & 99.74\% & 99.88\% & 99.63\% \\
      \cline{1-6}
      \multirow{4}{*}{Diff}
      & EER ($\downarrow$)
      & 2.93\% & 2.91\% & 2.30\% & 2.76\% \\
      \addlinespace[0.2pt]
      \cline{2-6}
      \addlinespace[1pt]
      & Top-1 ACC ($\uparrow$)
      & 95.24\% & 94.61\% & 93.25\% & 97.38\% \\
      & Top-3 ACC ($\uparrow$)
      & 98.68\% & 99.21\% & 98.38\% & 99.75\% \\
      & Top-5 ACC ($\uparrow$)
      & 99.47\% & 99.74\% & 99.88\% & 100.00\% \\
      \cline{1-6}
      \multirow{4}{*}{DDDM}
      & EER ($\downarrow$)
      & 3.06\% & 3.13\% & 3.00\% & 3.64\% \\
      \addlinespace[0.2pt]
      \cline{2-6}
      \addlinespace[1pt]
      & Top-1 ACC ($\uparrow$)
      & 95.12\% & 93.75\% & 94.90\% & 94.25\% \\
      & Top-3 ACC ($\uparrow$)
      & 99.24\% & 98.37\% & 99.25\% & 99.13\% \\
      & Top-5 ACC ($\uparrow$)
      & 99.61\% & 99.95\% & 99.63\% & 99.63\% \\
      \Xhline{1pt}
    \end{tabular}
    }
  \end{center}
\end{table}

\begin{table}[h]
  \begin{center}
    \caption{Performance of voiceprint recovery methods over the composite voice conversion methods.}
    \label{tab.adaptive_first}
    \resizebox{1.0\linewidth}{!}{
    \begin{tabular}{cr c c c c}
      \Xhline{1pt}
      \multicolumn{1}{c}{\multirow{2}{*}{\textbf{VC}}} & \multicolumn{1}{c}{\multirow{2}{*}{\textbf{Metric}}} &  \multicolumn{4}{c}{\textbf{Voiceprint Recovery Method}} \\
      \multicolumn{1}{c}{}& & \textbf{MFA} & \textbf{ECAPA} & \textbf{Revelio} & \textbf{\sys} \\
      \Xhline{1pt}
      \multirow{4}{*}{VQVC+}
      & EER ($\downarrow$)
      & 44.79\% & 42.00\%  & 10.10\% & \cellcolor{backcolor!100}\textbf{8.76\%} \\
      \addlinespace[0.2pt]
      \cline{2-6}
      \addlinespace[1pt]
      & Top-1 ACC ($\uparrow$)
      & 14.88\% & 5.99\% & 58.56\% & \cellcolor{backcolor!100}\textbf{75.48\%}\\
      & Top-3 ACC ($\uparrow$)
      & 30.72\% & 15.03\% & 80.29\% & \cellcolor{backcolor!100}\textbf{91.03\%} \\
      & Top-5 ACC ($\uparrow$)
      & 42.61\% & 22.40\% & 88.21\% & \cellcolor{backcolor!100}\textbf{95.35\%} \\
      \addlinespace[0.2pt]
      \cline{1-6}
      \addlinespace[1pt]
      \multirow{4}{*}{FreeVC}
      & EER ($\downarrow$)
      & 35.19\% & 47.71\%  & 7.52\% & \cellcolor{backcolor!100}\textbf{5.86\%} \\
      \addlinespace[0.2pt]
      \cline{2-6}
      \addlinespace[1pt]
      & Top-1 ACC ($\uparrow$)
      & 27.21\% & 4.01\% & 78.16\% & \cellcolor{backcolor!100}\textbf{85.61\%} \\
      & Top-3 ACC ($\uparrow$)
      & 36.83\% & 13.46\% & 91.27\% & \cellcolor{backcolor!100}\textbf{95.22\%} \\
      & Top-5 ACC ($\uparrow$)
      & 44.23\% & 21.28\% & 95.31\% & \cellcolor{backcolor!100}\textbf{98.01\%}\\
      \addlinespace[0.2pt]
      \cline{1-6}
      \addlinespace[1pt]
      \multirow{4}{*}{Diff}
      & EER ($\downarrow$)
      & 40.30\% & 40.00\%  & 7.29\% & \cellcolor{backcolor!100}\textbf{6.19\%} \\
      \addlinespace[0.2pt]
      \cline{2-6}
      \addlinespace[1pt]
      & Top-1 ACC ($\uparrow$)
      & 17.16\% &  3.61\% & 80.62\% & \cellcolor{backcolor!100}\textbf{86.22\%} \\
      & Top-3 ACC ($\uparrow$)
      & 34.19\% & 18.61\% & 92.90\% & \cellcolor{backcolor!100}\textbf{95.67\%} \\
      & Top-5 ACC ($\uparrow$)
      & 41.12\% & 28.37\% & 95.95\% & \cellcolor{backcolor!100}\textbf{97.69\%} \\
      \addlinespace[0.2pt]
      \cline{1-6}
      \addlinespace[1pt]
      \multirow{4}{*}{DDDM}
      & EER ($\downarrow$)
      & 46.21\% & 40.00\%  & 8.12\% & \cellcolor{backcolor!100}\textbf{5.88\%} \\
      \addlinespace[0.2pt]
      \cline{2-6}
      \addlinespace[1pt]
      & Top-1 ACC ($\uparrow$)
      & 13.27\% &  3.61\% & 73.43\% & \cellcolor{backcolor!100}\textbf{82.05\%} \\
      & Top-3 ACC ($\uparrow$)
      & 29.53\% & 18.61\% & 89.62\% & \cellcolor{backcolor!100}\textbf{94.97\%} \\
      & Top-5 ACC ($\uparrow$)
      & 38.65\% & 28.37\% & 94.32\% & \cellcolor{backcolor!100}\textbf{97.12\%} \\
      \Xhline{1pt}
    \end{tabular}
    }
  \end{center}
\end{table}

\begin{table}[h]
  \begin{center}
    \caption{Performance of voiceprint recovery methods over the repetitive voice conversion methods.}
    \label{tab.adaptive_second}
    \resizebox{1.0\linewidth}{!}{
    \begin{tabular}{cr c c c c}
      \Xhline{1pt}
      \multicolumn{1}{c}{\multirow{2}{*}{\textbf{VC}}} & \multicolumn{1}{c}{\multirow{2}{*}{\textbf{Metric}}} &  \multicolumn{4}{c}{\textbf{Voiceprint Recovery Method}} \\
      \multicolumn{1}{c}{} & & \textbf{MFA} & \textbf{ECAPA} & \textbf{Revelio} & \textbf{\sys} \\
      \Xhline{1pt}
      \multirow{4}{*}{AGAIN}
      & EER ($\downarrow$)
      & 33.96\% & 38.31\%  & 5.88\% & \cellcolor{backcolor!100}\textbf{3.08\%} \\
      \addlinespace[0.2pt]
      \cline{2-6}
      \addlinespace[1pt]
      & Top-1 ACC ($\uparrow$)
      & 18.26\% & 11.35\% & 85.22\% & \cellcolor{backcolor!100}\textbf{96.47\%} \\
      & Top-3 ACC ($\uparrow$)
      & 37.54\% & 24.07\% & 94.07\% & \cellcolor{backcolor!100}\textbf{99.55\%} \\
      & Top-5 ACC ($\uparrow$)
      & 48.82\% & 33.17\% & 97.12\% & \cellcolor{backcolor!100}\textbf{99.78\%} \\
      \addlinespace[0.2pt]
      \cline{1-6}
      \addlinespace[1pt]
      \multirow{4}{*}{VQVC+}
      & EER ($\downarrow$)
      & 47.81\% & 40.80\%  & 12.11\% & \cellcolor{backcolor!100}\textbf{8.33\%}  \\
      \addlinespace[0.2pt]
      \cline{2-6}
      \addlinespace[1pt]
      & Top-1 ACC ($\uparrow$)
      & 15.35\% & 6.38\% & 56.67\% & \cellcolor{backcolor!100}\textbf{79.10\%}\\
      & Top-3 ACC ($\uparrow$)
      & 29.84\% & 16.31\% & 77.23\% & \cellcolor{backcolor!100}\textbf{93.14\%}\\
      & Top-5 ACC ($\uparrow$)
      & 38.88\% & 24.07\% & 87.16\% & \cellcolor{backcolor!100}\textbf{96.47\%}\\
      \addlinespace[0.2pt]
      \cline{1-6}
      \addlinespace[1pt]
      \multirow{4}{*}{FreeVC}
      & EER ($\downarrow$)
      & 36.10\% & 49.30\%  & 7.73\% & \cellcolor{backcolor!100}\textbf{5.29\%}  \\
      \addlinespace[0.2pt]
      \cline{2-6}
      \addlinespace[1pt]
      & Top-1 ACC ($\uparrow$)
      & 26.76\% & 3.53\% & 76.31\% & \cellcolor{backcolor!100}\textbf{89.65\%} \\
      & Top-3 ACC ($\uparrow$)
      & 39.68\% & 11.80\% & 91.92\% & \cellcolor{backcolor!100}\textbf{98.30\%}\\
      & Top-5 ACC ($\uparrow$)
      & 46.86\% & 18.27\% & 95.96\% & \cellcolor{backcolor!100}\textbf{99.46\%}\\
      \addlinespace[0.2pt]
      \cline{1-6}
      \addlinespace[1pt]
      \multirow{4}{*}{Diff}
      & EER ($\downarrow$)
      & 38.92\% & 41.50\%  & 7.59\% & \cellcolor{backcolor!100}\textbf{4.30\%} \\
      \addlinespace[0.2pt]
      \cline{2-6}
      \addlinespace[1pt]
      & Top-1 ACC ($\uparrow$)
      & 23.54\% & 2.83\% & 79.09\% & \cellcolor{backcolor!100}\textbf{92.87\%}\\
      & Top-3 ACC ($\uparrow$)
      & 37.18\% & 17.61\% & 91.89\% & \cellcolor{backcolor!100}\textbf{98.44\%} \\
      & Top-5 ACC ($\uparrow$)
      & 45.75\% & 28.85\% & 94.96\% & \cellcolor{backcolor!100}\textbf{99.50\%}\\
      \addlinespace[0.2pt]
      \cline{1-6}
      \addlinespace[1pt]
      \multirow{4}{*}{DDDM}
      & EER ($\downarrow$)
      & 41.25\% & 41.50\%  & 7.23\% & \cellcolor{backcolor!100}\textbf{5.01\%} \\
      \addlinespace[0.2pt]
      \cline{2-6}
      \addlinespace[1pt]
      & Top-1 ACC ($\uparrow$)
      & 20.51\% & 2.83\% & 73.11\% & \cellcolor{backcolor!100}\textbf{87.05\%}\\
      & Top-3 ACC ($\uparrow$)
      & 31.44\% & 17.61\% & 90.51\% & \cellcolor{backcolor!100}\textbf{97.12\%} \\
      & Top-5 ACC ($\uparrow$)
      & 39.87\% & 28.85\% & 94.94\% & \cellcolor{backcolor!100}\textbf{98.88\%}\\
      \Xhline{1pt}
    \end{tabular}
    }
  \end{center}
\end{table}

\begin{table}[h]
  \begin{center}
    \caption{Performance of voiceprint recovery methods over unseen VC.}
    \label{tab.unseenvc}
    \resizebox{1.0\linewidth}{!}{
    \begin{tabular}{c r c c c c}
      \Xhline{1pt}
    \multicolumn{1}{c}{\multirow{2}{*}{\textbf{VC}}} & \multicolumn{1}{c}{\multirow{2}{*}{\textbf{Metric}}} &  \multicolumn{4}{c}{\textbf{Voiceprint Recovery Method}} \\
      \multicolumn{1}{c}{} & & \textbf{MFA} & \textbf{ECAPA} & \textbf{Revelio} & \textbf{\sys} \\
      \Xhline{1pt}
      \multirow{4}{*}{AdaIN}
      & EER ($\downarrow$)
      & 46.16\% & 43.81\%  & 15.34\% & \cellcolor{backcolor!100}\textbf{11.70\%} \\
      \addlinespace[0.2pt]
      \cline{2-6}
      \addlinespace[1pt]
      & Top-1 ACC ($\uparrow$)
      & 3.01\% & 3.81\% & 51.54\% & \cellcolor{backcolor!100}\textbf{61.57\%} \\
      & Top-3 ACC ($\uparrow$)
      & 11.80\% & 12.15\% & 70.35\% & \cellcolor{backcolor!100}\textbf{79.68\%} \\
      & Top-5 ACC ($\uparrow$)
      & 17.95\% & 19.17\% & 83.72\% & \cellcolor{backcolor!100}\textbf{91.03\%} \\
      \addlinespace[0.2pt]
      \cline{1-6}
      \addlinespace[1pt]
      \multirow{4}{*}{VQVC+}
      & EER ($\downarrow$)
      & 43.54\% & 36.69\%  & 8.60\% & \cellcolor{backcolor!100}\textbf{7.39\%} \\
      \addlinespace[0.2pt]
      \cline{2-6}
      \addlinespace[1pt]
      & Top-1 ACC ($\uparrow$)
      & 9.84\% & 11.15\% & 69.01\% & \cellcolor{backcolor!100}\textbf{76.86\%} \\
      & Top-3 ACC ($\uparrow$)
      & 20.64\% & 24.14\% & 86.35\% & \cellcolor{backcolor!100}\textbf{90.67\%} \\
      & Top-5 ACC ($\uparrow$)
      & 25.35\% & 32.98\% & 92.47\% & \cellcolor{backcolor!100}\textbf{94.58\%} \\
      \addlinespace[0.2pt]
      \cline{1-6}
      \addlinespace[1pt]
      \multirow{4}{*}{StarGANv2}
      & EER ($\downarrow$)
      & 37.53\% & 30.92\%  & 10.71\% & \cellcolor{backcolor!100}\textbf{9.42\%} \\
      \addlinespace[0.2pt]
      \cline{2-6}
      \addlinespace[1pt]
      & Top-1 ACC ($\uparrow$)
      & 13.69\% & 20.61\% & 67.85\% & \cellcolor{backcolor!100}\textbf{71.70\%} \\
      & Top-3 ACC ($\uparrow$)
      & 23.62\% & 35.51\% & 84.19\% & \cellcolor{backcolor!100}\textbf{87.34\%} \\
      & Top-5 ACC ($\uparrow$)
      & 32.21\% & 44.39\% & 89.30\% & \cellcolor{backcolor!100}\textbf{92.15\%} \\
      \addlinespace[0.2pt]
      \cline{1-6}
      \addlinespace[1pt]
      \multirow{4}{*}{DDDM}
      & EER ($\downarrow$)
      & 43.70\% & 35.12\%  & 8.34\% & \cellcolor{backcolor!100}\textbf{7.94\%} \\
      \addlinespace[0.2pt]
      \cline{2-6}
      \addlinespace[1pt]
      & Top-1 ACC ($\uparrow$)
      & 4.58\% & 6.47\% & 70.71\% & \cellcolor{backcolor!100}\textbf{75.90\%} \\
      & Top-3 ACC ($\uparrow$)
      & 14.71\% & 22.82\% & 88.75\% & \cellcolor{backcolor!100}\textbf{90.80\%} \\
      & Top-5 ACC ($\uparrow$)
      & 21.89\% & 33.01\% & 92.98\% & \cellcolor{backcolor!100}\textbf{94.52\%} \\
      \Xhline{1pt}
    \end{tabular}
    }
  \end{center}
\end{table}

\begin{table}[h]
  \begin{center}
    \caption{Performance of Revelio under real-world scenarios.}
    \label{tab.real_revelio}
    \resizebox{1.0\linewidth}{!}{
    \begin{tabular}{cr c c c}
      \Xhline{1pt}
      \multicolumn{1}{c}{\multirow{2}{*}{\textbf{VC}}} & \multicolumn{1}{c}{\multirow{2}{*}{\textbf{Metric}}} &  \multicolumn{3}{c}{\textbf{Voiceprint Recovery Method}} \\
      \multicolumn{1}{c}{}& & \textbf{Telephone} & \textbf{VoIP-32kbps} & \textbf{VoIP-64kbps} \\
      \Xhline{1pt}
      \multirow{4}{*}{Clean}
      & EER ($\downarrow$)
      & 5.12\% & 3.78\%  & 3.43\% \\
      \addlinespace[0.2pt]
      \cline{2-5}
      \addlinespace[1pt]
      & Top-1 ACC ($\uparrow$)
      & 91.75\% & 92.13\% & 92.76\%  \\
      & Top-3 ACC ($\uparrow$)
      & 95.93\% & 96.52\% & 96.80\%  \\
      & Top-5 ACC ($\uparrow$)
      & 97.42\% & 97.86\% & 98.28\%  \\
      \addlinespace[0.2pt]
      \cline{1-5}
      \addlinespace[1pt]
      \multirow{4}{*}{AGAIN}
      & EER ($\downarrow$)
      & 9.53\% & 8.47\%  & 6.32\%  \\
      \addlinespace[0.2pt]
      \cline{2-5}
      \addlinespace[1pt]
      & Top-1 ACC ($\uparrow$)
      & 75.92\% & 80.66\% & 82.76\%  \\
      & Top-3 ACC ($\uparrow$)
      & 87.37\% & 89.87\% & 92.07\%  \\
      & Top-5 ACC ($\uparrow$)
      & 93.50\% & 94.47\% & 95.51\%  \\
      \addlinespace[0.2pt]
      \cline{1-5}
      \addlinespace[1pt]
      \multirow{4}{*}{VQVC}
      & EER ($\downarrow$)
      & 19.55\% & 9.36\%  & 8.99\%  \\
      \addlinespace[0.2pt]
      \cline{2-5}
      \addlinespace[1pt]
      & Top-1 ACC ($\uparrow$)
      & 46.18\% & 63.29\% & 64.05\%  \\
      & Top-3 ACC ($\uparrow$)
      & 72.11\% & 85.40\% & 86.26\%  \\
      & Top-5 ACC ($\uparrow$)
      & 85.79\% & 90.95\% & 91.82\%  \\
      \cline{1-5}
      \addlinespace[1pt]
      \multirow{4}{*}{VQVC+}
      & EER ($\downarrow$)
      & 12.30\% & 6.64\%  & 5.79\%  \\
      \addlinespace[0.2pt]
      \cline{2-5}
      \addlinespace[1pt]
      & Top-1 ACC ($\uparrow$)
      & 67.51\% & 75.13\% & 77.24\%  \\
      & Top-3 ACC ($\uparrow$)
      & 83.68\% & 89.11\% & 90.29\%  \\
      & Top-5 ACC ($\uparrow$)
      & 90.26\% & 92.90\% & 94.04\%  \\
      \cline{1-5}
      \addlinespace[1pt]
      \multirow{4}{*}{BNE}
      & EER ($\downarrow$)
      & 24.43\% & 11.32\%  & 10.58\%  \\
      \addlinespace[0.2pt]
      \cline{2-5}
      \addlinespace[1pt]
      & Top-1 ACC ($\uparrow$)
      & 34.21\% & 58.26\% & 60.55\%  \\
      & Top-3 ACC ($\uparrow$)
      & 65.01\% & 83.53\% & 84.37\%  \\
      & Top-5 ACC ($\uparrow$)
      & 77.37\% & 89.05\% & 90.68\%  \\
      \cline{1-5}
      \addlinespace[1pt]
      \multirow{4}{*}{FreeVC}
      & EER ($\downarrow$)
      & 11.43\% & 7.35\%  & 6.83\%  \\
      \addlinespace[0.2pt]
      \cline{2-5}
      \addlinespace[1pt]
      & Top-1 ACC ($\uparrow$)
      & 74.61\% & 82.76\% & 85.40\%  \\
      & Top-3 ACC ($\uparrow$)
      & 89.84\% & 91.65\% & 92.76\%  \\
      & Top-5 ACC ($\uparrow$)
      & 94.32\% & 95.89\% & 96.18\%  \\
      \addlinespace[0.2pt]
      \cline{1-5}
      \addlinespace[1pt]
      \multirow{4}{*}{Diff}
      & EER ($\downarrow$)
      & 10.51\% & 5.93\%  & 5.67\%  \\
      \addlinespace[0.2pt]
      \cline{2-5}
      \addlinespace[1pt]
      & Top-1 ACC ($\uparrow$)
      & 77.25\% &  88.18\% & 88.97\%  \\
      & Top-3 ACC ($\uparrow$)
      & 93.41\% & 97.49\% & 97.86\%  \\
      & Top-5 ACC ($\uparrow$)
      & 97.31\% & 98.34\% & 98.61\%  \\
      \addlinespace[0.2pt]
      \cline{1-5}
      \addlinespace[1pt]
      \multirow{4}{*}{DDDM}
      & EER ($\downarrow$)
      & 9.34\% & 6.62\%  & 5.67\%  \\
      \addlinespace[0.2pt]
      \cline{2-5}
      \addlinespace[1pt]
      & Top-1 ACC ($\uparrow$)
      & 75.79\% & 85.79\% & 88.03\%  \\
      & Top-3 ACC ($\uparrow$)
      & 90.02\% & 93.03\% & 95.26\%  \\
      & Top-5 ACC ($\uparrow$)
      & 94.61\% & 96.32\% & 96.93\%  \\
      \Xhline{1pt}
    \end{tabular}
    }
  \end{center}
\end{table}

\begin{table}[h]
  \begin{center}
    \caption{Performance of ECAPA-TDNN under real-world scenarios.}
    \label{tab.real_ecapa}
    \resizebox{1.0\linewidth}{!}{
    \begin{tabular}{cr c c c}
      \Xhline{1pt}
      \multicolumn{1}{c}{\multirow{2}{*}{\textbf{VC}}} & \multicolumn{1}{c}{\multirow{2}{*}{\textbf{Metric}}} &  \multicolumn{3}{c}{\textbf{Voiceprint Recovery Method}} \\
      \multicolumn{1}{c}{}& & \textbf{Telephone} & \textbf{VoIP-32kbps} & \textbf{VoIP-64kbps} \\
      \Xhline{1pt}
      \multirow{4}{*}{Clean}
      & EER ($\downarrow$)
      & 6.21\% & 5.87\%  & 5.50\% \\
      \addlinespace[0.2pt]
      \cline{2-5}
      \addlinespace[1pt]
      & Top-1 ACC ($\uparrow$)
      & 80.91\% & 82.73\% & 83.06\%  \\
      & Top-3 ACC ($\uparrow$)
      & 85.82\% & 86.60\% & 87.13\%  \\
      & Top-5 ACC ($\uparrow$)
      & 88.93\% & 89.13\% & 90.30\%  \\
      \addlinespace[0.2pt]
      \cline{1-5}
      \addlinespace[1pt]
      \multirow{4}{*}{AGAIN}
      & EER ($\downarrow$)
      & 34.02\% & 32.58\%  & 31.59\%  \\
      \addlinespace[0.2pt]
      \cline{2-5}
      \addlinespace[1pt]
      & Top-1 ACC ($\uparrow$)
      & 8.97\% & 9.67\% & 10.42\%  \\
      & Top-3 ACC ($\uparrow$)
      & 20.11\% & 20.21\% & 21.58\%  \\
      & Top-5 ACC ($\uparrow$)
      & 26.58\% & 27.53\% & 28.95\%  \\
      \addlinespace[0.2pt]
      \cline{1-5}
      \addlinespace[1pt]
      \multirow{4}{*}{VQVC}
      & EER ($\downarrow$)
      & 45.99\% & 42.12\%  & 41.32\%  \\
      \addlinespace[0.2pt]
      \cline{2-5}
      \addlinespace[1pt]
      & Top-1 ACC ($\uparrow$)
      & 2.63\% & 3.88\% & 4.73\%  \\
      & Top-3 ACC ($\uparrow$)
      & 9.08\% & 12.63\% & 12.76\%  \\
      & Top-5 ACC ($\uparrow$)
      & 16.05\% & 19.01\% & 19.47\%  \\
      \cline{1-5}
      \addlinespace[1pt]
      \multirow{4}{*}{VQVC+}
      & EER ($\downarrow$)
      & 39.88\% & 37.90\%  & 37.33\%  \\
      \addlinespace[0.2pt]
      \cline{2-5}
      \addlinespace[1pt]
      & Top-1 ACC ($\uparrow$)
      & 8.42\% & 10.63\% & 10.90\%  \\
      & Top-3 ACC ($\uparrow$)
      & 20.03\% & 23.92\% & 24.05\%  \\
      & Top-5 ACC ($\uparrow$)
      & 27.66\% & 32.58\% & 32.87\%  \\
      \cline{1-5}
      \addlinespace[1pt]
      \multirow{4}{*}{BNE}
      & EER ($\downarrow$)
      & 47.37\% & 47.01\%  & 46.59\%  \\
      \addlinespace[0.2pt]
      \cline{2-5}
      \addlinespace[1pt]
      & Top-1 ACC ($\uparrow$)
      & 1.37\% & 1.51\% & 1.68\%  \\
      & Top-3 ACC ($\uparrow$)
      & 8.14\% & 8.35\% & 8.68\%  \\
      & Top-5 ACC ($\uparrow$)
      & 14.24\% & 14.65\% & 15.56\%  \\
      \cline{1-5}
      \addlinespace[1pt]
      \multirow{4}{*}{FreeVC}
      & EER ($\downarrow$)
      & 42.74\% & 41.09\%  & 40.13\%  \\
      \addlinespace[0.2pt]
      \cline{2-5}
      \addlinespace[1pt]
      & Top-1 ACC ($\uparrow$)
      & 5.16\% & 5.88\% & 6.24\%  \\
      & Top-3 ACC ($\uparrow$)
      & 15.37\% & 16.17\% & 16.90\%  \\
      & Top-5 ACC ($\uparrow$)
      & 22.18\% & 23.75\% & 24.47\%  \\
      \addlinespace[0.2pt]
      \cline{1-5}
      \addlinespace[1pt]
      \multirow{4}{*}{Diff}
      & EER ($\downarrow$)
      & 35.51\% & 34.26\%  & 33.29\%  \\
      \addlinespace[0.2pt]
      \cline{2-5}
      \addlinespace[1pt]
      & Top-1 ACC ($\uparrow$)
      & 8.35\% &  9.08\% & 9.40\%  \\
      & Top-3 ACC ($\uparrow$)
      & 27.86\% & 28.16\% & 28.68\%  \\
      & Top-5 ACC ($\uparrow$)
      & 38.41\% & 38.82\% & 39.30\%  \\
      \addlinespace[0.2pt]
      \cline{1-5}
      \addlinespace[1pt]
      \multirow{4}{*}{DDDM}
      & EER ($\downarrow$)
      & 38.11\% & 37.18\%  & 36.44\%  \\
      \addlinespace[0.2pt]
      \cline{2-5}
      \addlinespace[1pt]
      & Top-1 ACC ($\uparrow$)
      & 5.95\% & 6.03\% & 6.21\%  \\
      & Top-3 ACC ($\uparrow$)
      & 20.92\% & 21.18\% & 22.37\%  \\
      & Top-5 ACC ($\uparrow$)
      & 31.97\% & 32.63\% & 32.90\%  \\
      \Xhline{1pt}
    \end{tabular}
    }
  \end{center}
\end{table}


\begin{table}[h]
  \begin{center}
    \caption{Performance of MFA-Conformer under real-world scenarios.}
    \label{tab.real_mfa}
    \resizebox{1.0\linewidth}{!}{
    \begin{tabular}{cr c c c}
      \Xhline{1pt}
      \multicolumn{1}{c}{\multirow{2}{*}{\textbf{VC}}} & \multicolumn{1}{c}{\multirow{2}{*}{\textbf{Metric}}} &  \multicolumn{3}{c}{\textbf{Voiceprint Recovery Method}} \\
      \multicolumn{1}{c}{}& & \textbf{Telephone} & \textbf{VoIP-32kbps} & \textbf{VoIP-64kbps} \\
      \Xhline{1pt}
      \multirow{4}{*}{Clean}
      & EER ($\downarrow$)
      & 21.93\% & 17.11\%  & 16.89\% \\
      \addlinespace[0.2pt]
      \cline{2-5}
      \addlinespace[1pt]
      & Top-1 ACC ($\uparrow$)
      & 62.20\% & 66.03\% & 68.62\%  \\
      & Top-3 ACC ($\uparrow$)
      & 76.84\% & 79.19\% & 79.90\%  \\
      & Top-5 ACC ($\uparrow$)
      & 79.43\% & 81.82\% & 82.30\%  \\
      \addlinespace[0.2pt]
      \cline{1-5}
      \addlinespace[1pt]
      \multirow{4}{*}{AGAIN}
      & EER ($\downarrow$)
      & 44.08\% & 37.50\%  & 36.82\%  \\
      \addlinespace[0.2pt]
      \cline{2-5}
      \addlinespace[1pt]
      & Top-1 ACC ($\uparrow$)
      & 10.13\% & 20.47\% & 22.11\%  \\
      & Top-3 ACC ($\uparrow$)
      & 25.92\% & 34.87\% & 37.37\%  \\
      & Top-5 ACC ($\uparrow$)
      & 34.08\% & 44.74\% & 46.45\%  \\
      \addlinespace[0.2pt]
      \cline{1-5}
      \addlinespace[1pt]
      \multirow{4}{*}{VQVC}
      & EER ($\downarrow$)
      & 50.53\% & 47.37\%  & 46.49\%  \\
      \addlinespace[0.2pt]
      \cline{2-5}
      \addlinespace[1pt]
      & Top-1 ACC ($\uparrow$)
      & 7.34\% & 9.42\% & 9.74\%  \\
      & Top-3 ACC ($\uparrow$)
      & 18.42\% & 21.71\% & 22.47\%  \\
      & Top-5 ACC ($\uparrow$)
      & 27.69\% & 29.47\% & 30.21\%  \\
      \cline{1-5}
      \addlinespace[1pt]
      \multirow{4}{*}{VQVC+}
      & EER ($\downarrow$)
      & 47.31\% & 44.25\%  & 43.86\%  \\
      \addlinespace[0.2pt]
      \cline{2-5}
      \addlinespace[1pt]
      & Top-1 ACC ($\uparrow$)
      & 12.56\% & 19.95\% & 20.84\%  \\
      & Top-3 ACC ($\uparrow$)
      & 24.34\% & 39.90\% & 40.11\%  \\
      & Top-5 ACC ($\uparrow$)
      & 34.93\% & 46.26\% & 46.58\%  \\
      \cline{1-5}
      \addlinespace[1pt]
      \multirow{4}{*}{BNE}
      & EER ($\downarrow$)
      & 39.87\% & 38.47\%  & 37.72\%  \\
      \addlinespace[0.2pt]
      \cline{2-5}
      \addlinespace[1pt]
      & Top-1 ACC ($\uparrow$)
      & 10.92\% & 11.82\% & 12.11\%  \\
      & Top-3 ACC ($\uparrow$)
      & 20.86\% & 21.32\% & 21.42\%  \\
      & Top-5 ACC ($\uparrow$)
      & 27.36\% & 28.42\% & 29.95\%  \\
      \cline{1-5}
      \addlinespace[1pt]
      \multirow{4}{*}{FreeVC}
      & EER ($\downarrow$)
      & 42.67\% & 38.47\%  & 35.53\%  \\
      \addlinespace[0.2pt]
      \cline{2-5}
      \addlinespace[1pt]
      & Top-1 ACC ($\uparrow$)
      & 18.61\% & 20.39\% & 21.18\%  \\
      & Top-3 ACC ($\uparrow$)
      & 29.50\% & 32.91\% & 34.58\%  \\
      & Top-5 ACC ($\uparrow$)
      & 38.55\% & 42.74\% & 42.79\%  \\
      \addlinespace[0.2pt]
      \cline{1-5}
      \addlinespace[1pt]
      \multirow{4}{*}{Diff}
      & EER ($\downarrow$)
      & 45.29\% & 42.56\%  & 41.05\%  \\
      \addlinespace[0.2pt]
      \cline{2-5}
      \addlinespace[1pt]
      & Top-1 ACC ($\uparrow$)
      & 14.34\% &  15.61\% & 16.45\%  \\
      & Top-3 ACC ($\uparrow$)
      & 25.79\% & 28.16\% & 28.68\%  \\
      & Top-5 ACC ($\uparrow$)
      & 34.34\% & 35.13\% & 36.71\%  \\
      \addlinespace[0.2pt]
      \cline{1-5}
      \addlinespace[1pt]
      \multirow{4}{*}{DDDM}
      & EER ($\downarrow$)
      & 52.32\% & 50.79\%  & 49.21\%  \\
      \addlinespace[0.2pt]
      \cline{2-5}
      \addlinespace[1pt]
      & Top-1 ACC ($\uparrow$)
      & 20.61\% & 21.58\% & 22.50\%  \\
      & Top-3 ACC ($\uparrow$)
      & 31.45\% & 33.07\% & 33.32\%  \\
      & Top-5 ACC ($\uparrow$)
      & 39.87\% & 42.24\% & 42.37\%  \\
      \Xhline{1pt}
    \end{tabular}
    }
  \end{center}
  \vspace{10cm}
\end{table}

\begin{table}[h]
  \centering
  \footnotesize
  \caption{Description of different language datasets.}
  \label{tab:unseen_language}
  \resizebox{1.00\linewidth}{!}{
  \begin{threeparttable}[b]
    \begin{tabular}{ccccc}
    \Xhline{1pt}
        \multirow{2}{*}{\textbf{Dataset}} &\multirow{2}{*}{\textbf{Language}} & \multirow{2}{*}{\textbf{Speakers}}   & \multicolumn{2}{c}{\textbf{Audio Samples}}\\
        & & & \textbf{Genuine} & \textbf{Converted} \\
      \Xhline{1pt}
      VCTK \cite{yamagishi2019cstr} & English & 20 & 1,600 & 7,600 \\
      MLS Spanish \cite{pratap2020mls} & Spanish & 20 & 1,600 & 7,600 \\
      MLS French \cite{pratap2020mls} & French & 18   & 1,440 & 6,120 \\
      MLS German \cite{pratap2020mls} & German & 30   & 2,400 & 17,400  \\
      Aishell Chinese \cite{bu2017aishell} & Chinese & 20 & 1,600 & 7,600 \\
      ALFFA Amharic \cite{tachbelie2014} & Amharic & 20 & 300 & 7,600 \\
      Eduskunta Corpus \cite{eduskuntav1.5} & Finnish & 20 & 1,600 & 7,600 \\
    \Xhline{1pt}
    \end{tabular}
         \end{threeparttable}
    }
\end{table}

\begin{table}[h]
  \begin{center}
    \caption{Disentanglement metrics for different components.}
    \label{tab.distengle}
    \resizebox{1.0\linewidth}{!}{
    \begin{tabular}{c c c c c c}
      \Xhline{1pt}
      \multicolumn{1}{c}{\multirow{2}{*}{\textbf{Feature}}} & \multicolumn{1}{c}{\multirow{2}{*}{\textbf{ARI}}} &  \multicolumn{1}{c}{\multirow{2}{*}{\textbf{NMI}}} &
      \multicolumn{3}{c}{\textbf{Correlation Coefficient Matrix}} \\
      & & & \textbf{Method} & \textbf{Target} & \textbf{Source} \\
      \Xhline{1pt}
      Method & 0.9846 & 0.9721 & 1 & -0.0559 & -0.0626\\
      Target & 0.8776 & 0.9719 & -0.0559 & 1 & 0.1867\\
      Source & 0.8557 & 0.9018 & -0.0626 & 0.1867 & 1\\
      \Xhline{1pt}
    \end{tabular}
    }
  \end{center}
\end{table}


\clearpage

\begin{figure*}[h]
    \centering
    	\begin{minipage}[b]{0.36\linewidth}
    		\centering
    		\includegraphics[trim=0mm 0mm 0mm 0mm, clip, width=\textwidth]{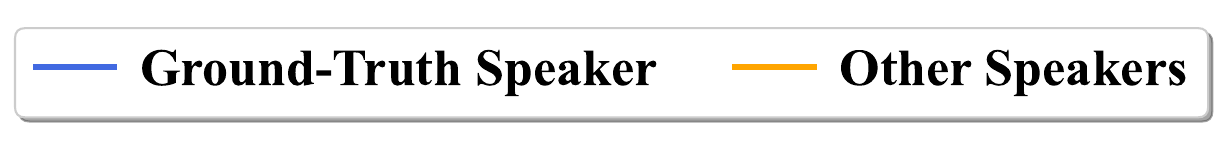}
    	\end{minipage} \\
    \vspace{-0.1cm}
    \subfigure[Clean]{
    	\begin{minipage}[b]{0.23\linewidth}
    		\centering
    		\includegraphics[trim=0mm 0mm 0mm 0mm, clip, width=0.95\textwidth]{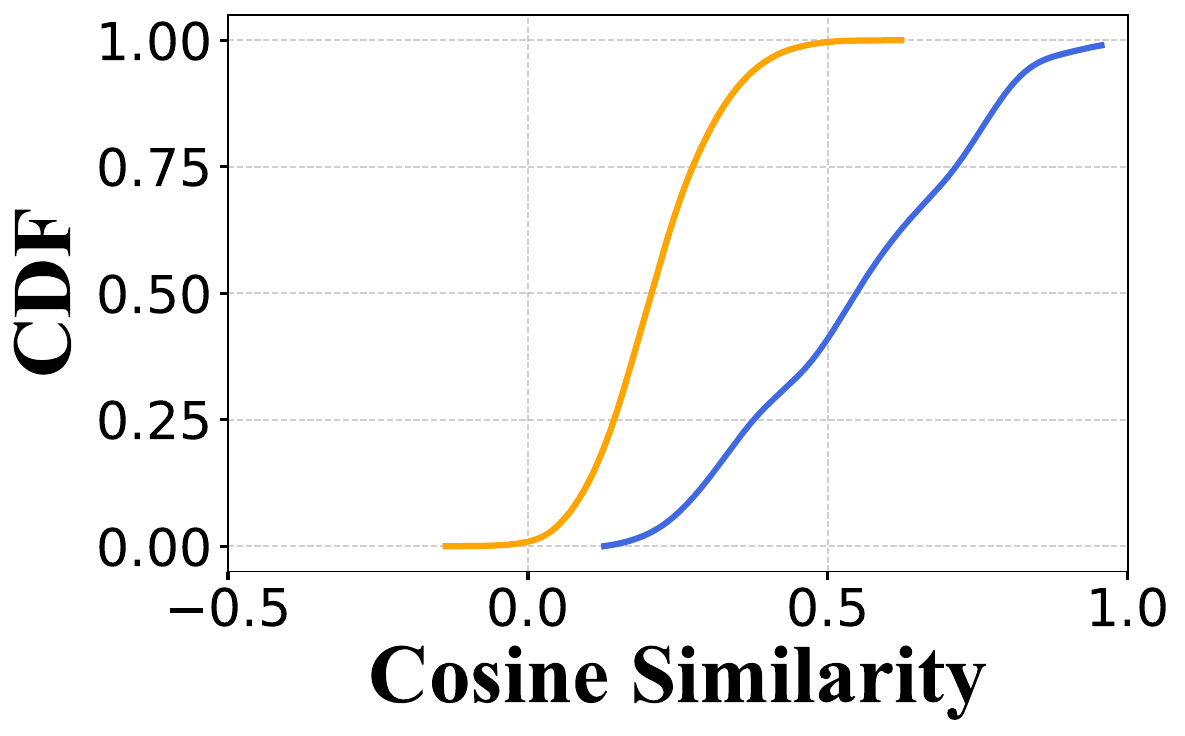}
    		\vspace{-0.1cm}
    	\end{minipage}
    }
    \subfigure[AGAIN]{
    	\begin{minipage}[b]{0.23\linewidth}
    		\centering
    		\includegraphics[trim=0mm 0mm 0mm 0mm, clip, width=0.95\textwidth]{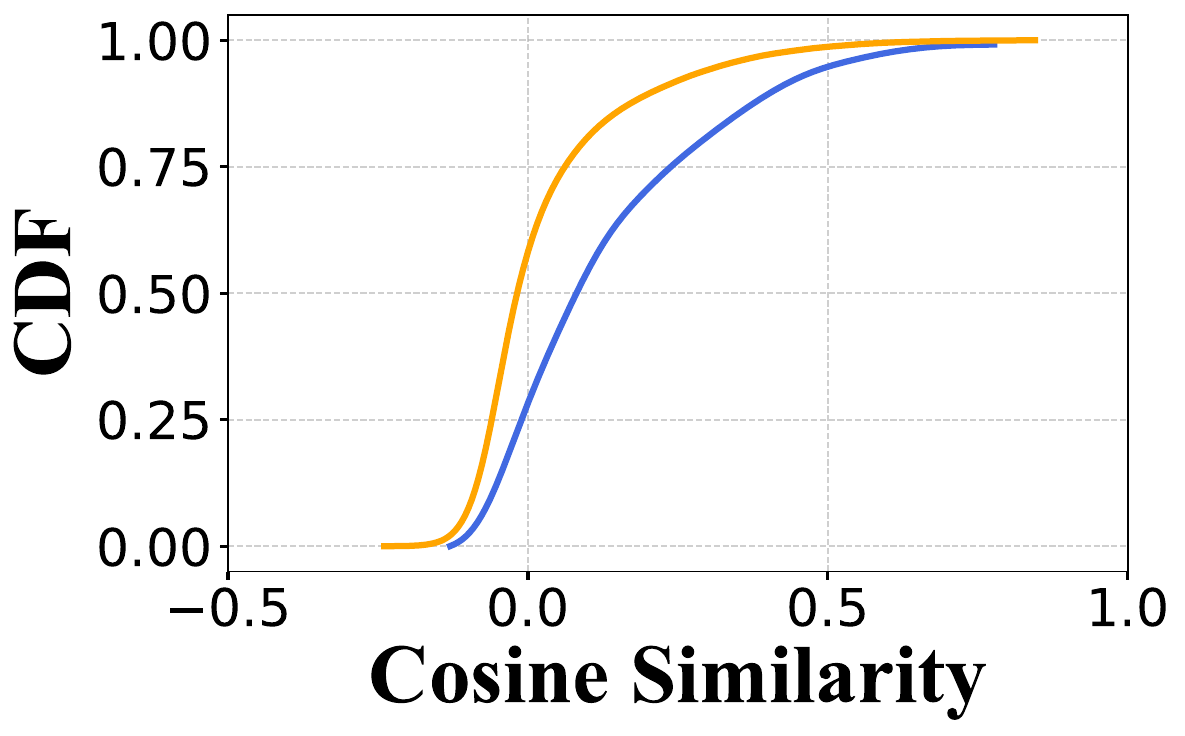}
    		\vspace{-0.1cm}
    	\end{minipage}
    }
    \subfigure[VQVC]{
    	\begin{minipage}[b]{0.23\linewidth}
    		\centering
    		\includegraphics[trim=0mm 0mm 0mm 0mm, clip, width=0.95\textwidth]{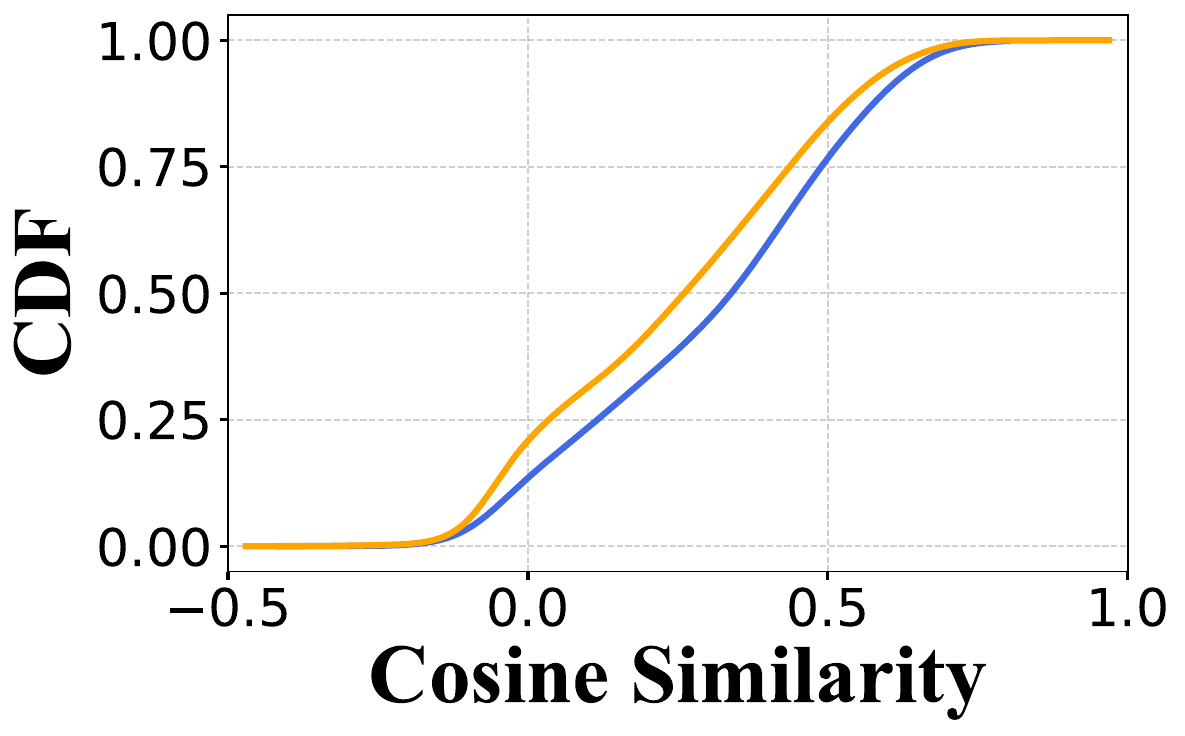}
    		\vspace{-0.1cm}
    	\end{minipage}
    }
  \vspace{-0.3cm}
    \subfigure[VQVC+]{
    	\begin{minipage}[b]{0.23\linewidth}
    		\centering
    		\includegraphics[trim=0mm 0mm 0mm 0mm, clip,width=0.95\textwidth]{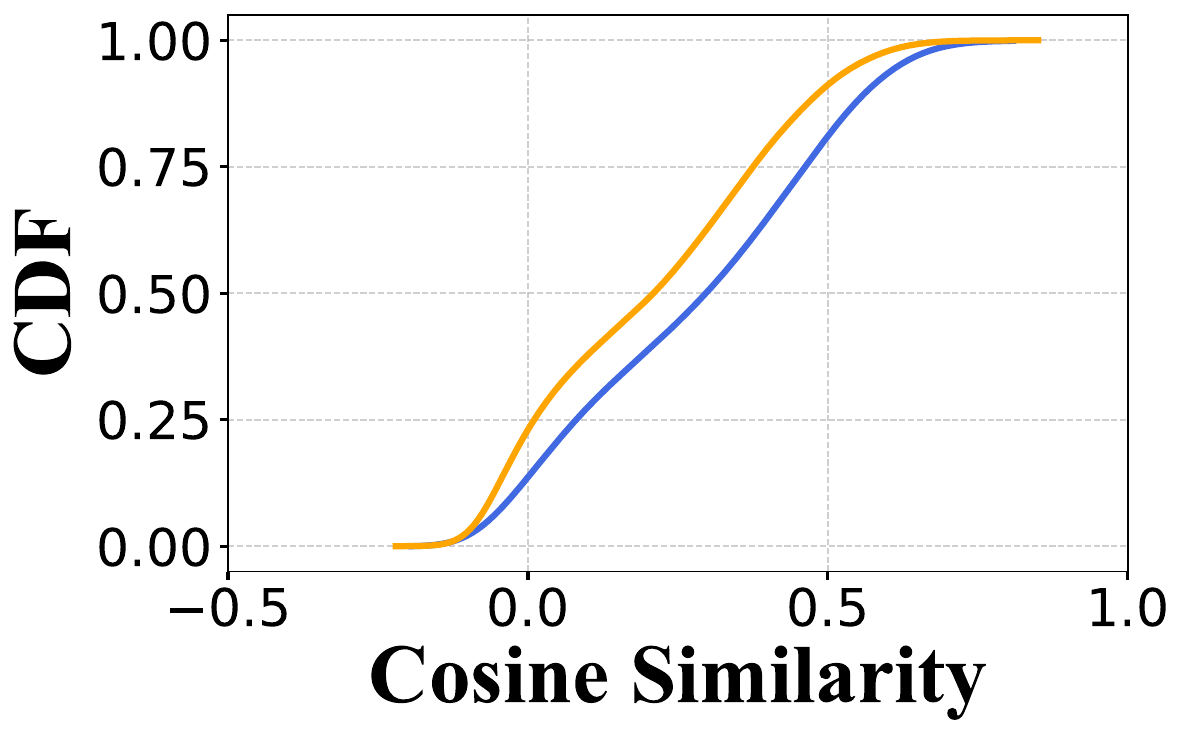}
    		\vspace{-0.1cm}
    	\end{minipage}
    }
    \subfigure[BNE]{
    	\begin{minipage}[b]{0.23\linewidth}
    		\centering
    		\includegraphics[trim=0mm 0mm 0mm 0mm, clip,width=0.95\textwidth]{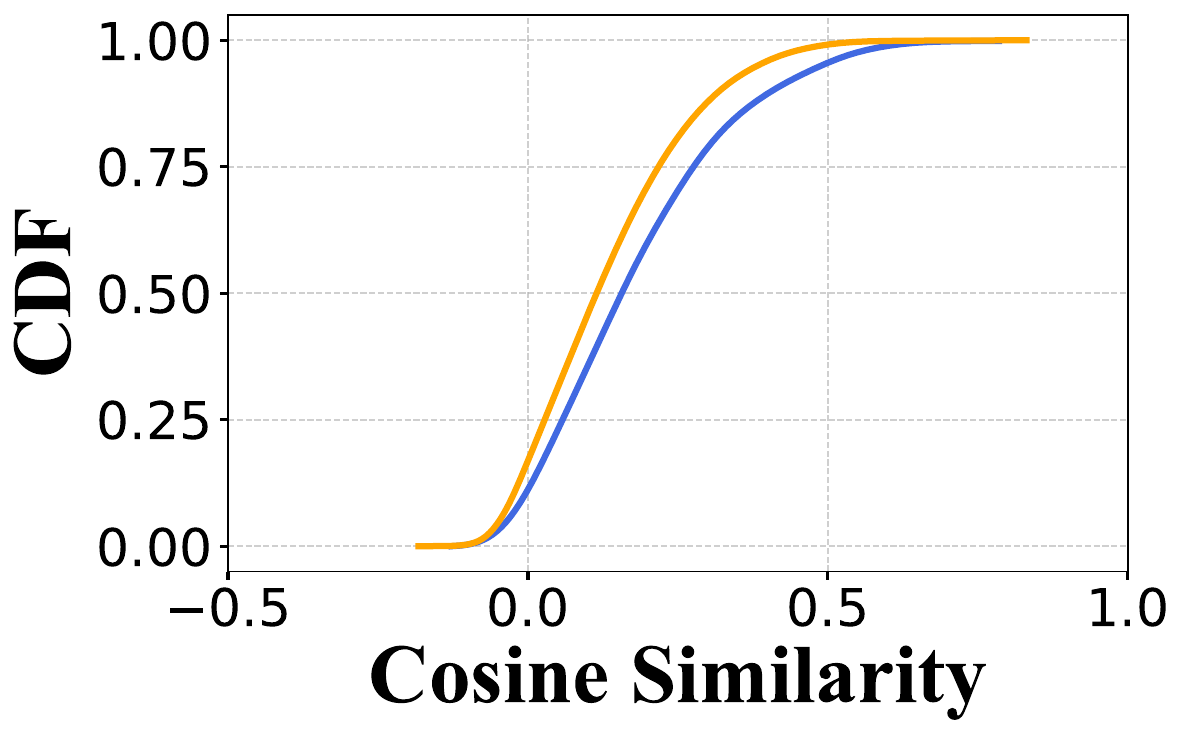}
    		\vspace{-0.1cm}
    	\end{minipage}
    }
    \subfigure[FreeVC]{
    	\begin{minipage}[b]{0.23\linewidth}
    		\centering
    		\includegraphics[trim=0mm 0mm 0mm 0mm, clip,width=0.95\textwidth]{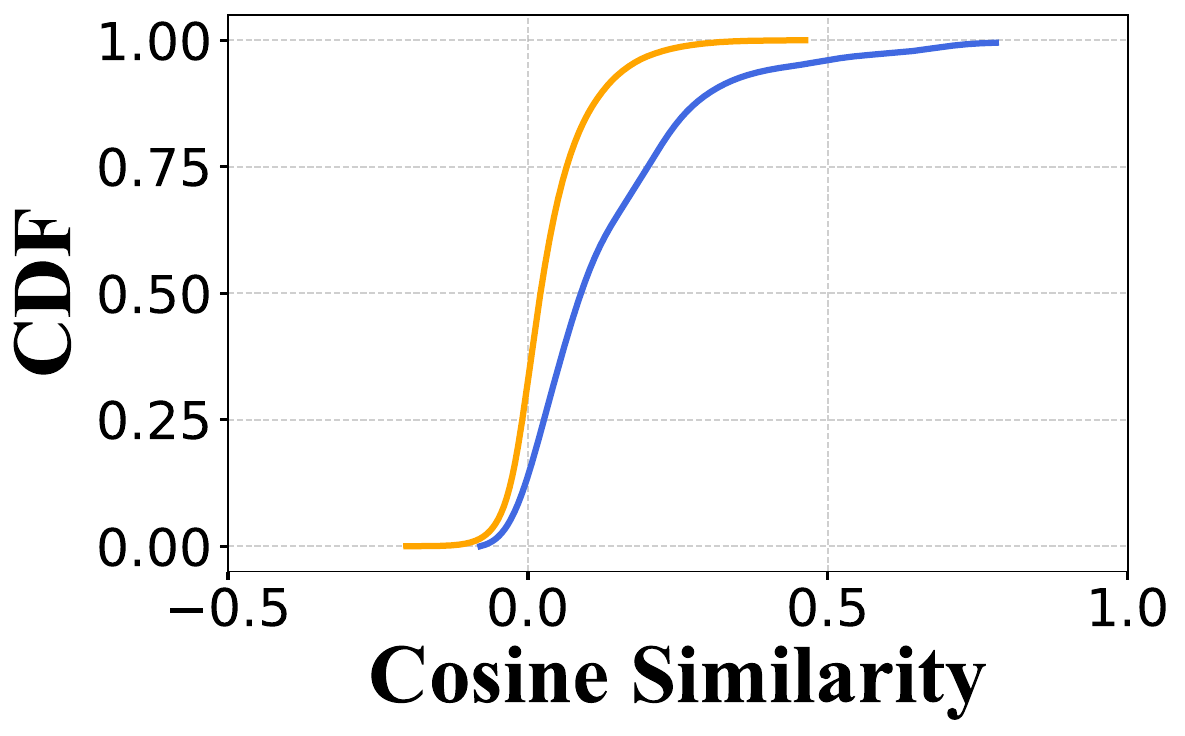}
    		\vspace{-0.1cm}
    	\end{minipage}
    }
    \subfigure[Diff]{
    	\begin{minipage}[b]{0.23\linewidth}
    		\centering
    		\includegraphics[trim=0mm 0mm 0mm 0mm, clip,width=0.95\textwidth]{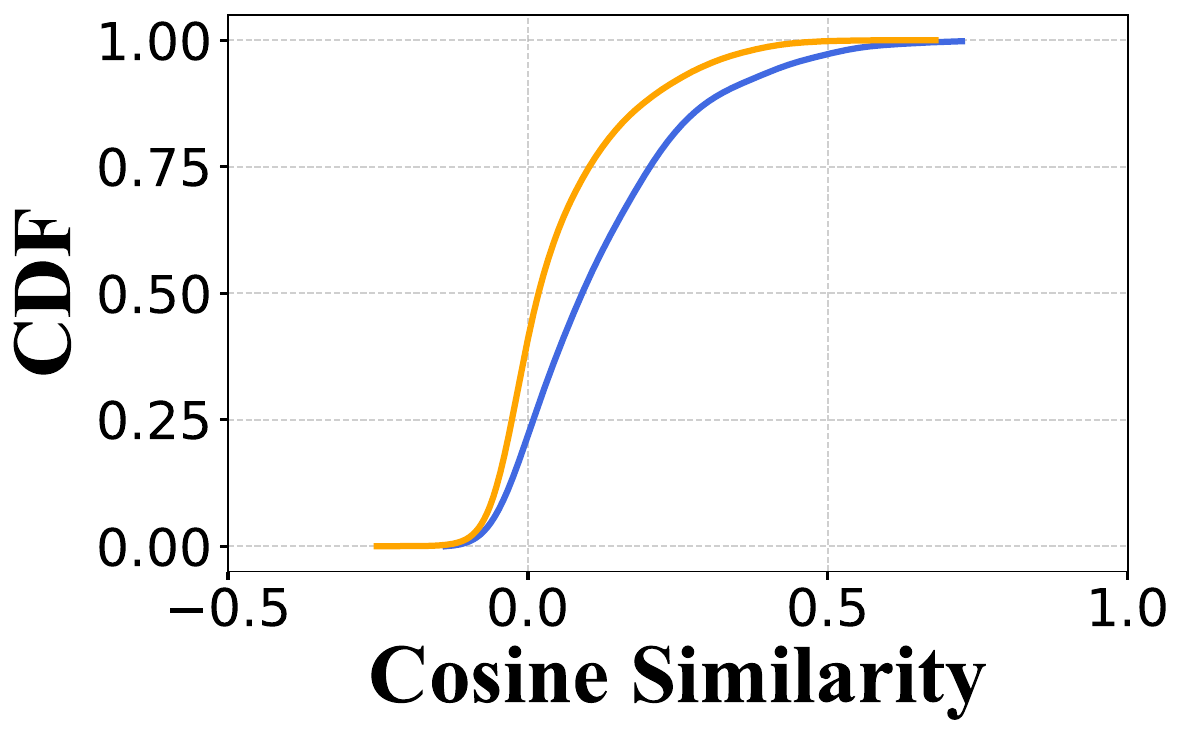}
    		\vspace{-0.1cm}
    	\end{minipage}
    }
    \subfigure[DDDM]{
    	\begin{minipage}[b]{0.23\linewidth}
    		\centering
    		\includegraphics[trim=0mm 0mm 0mm 0mm, clip,width=0.95\textwidth]{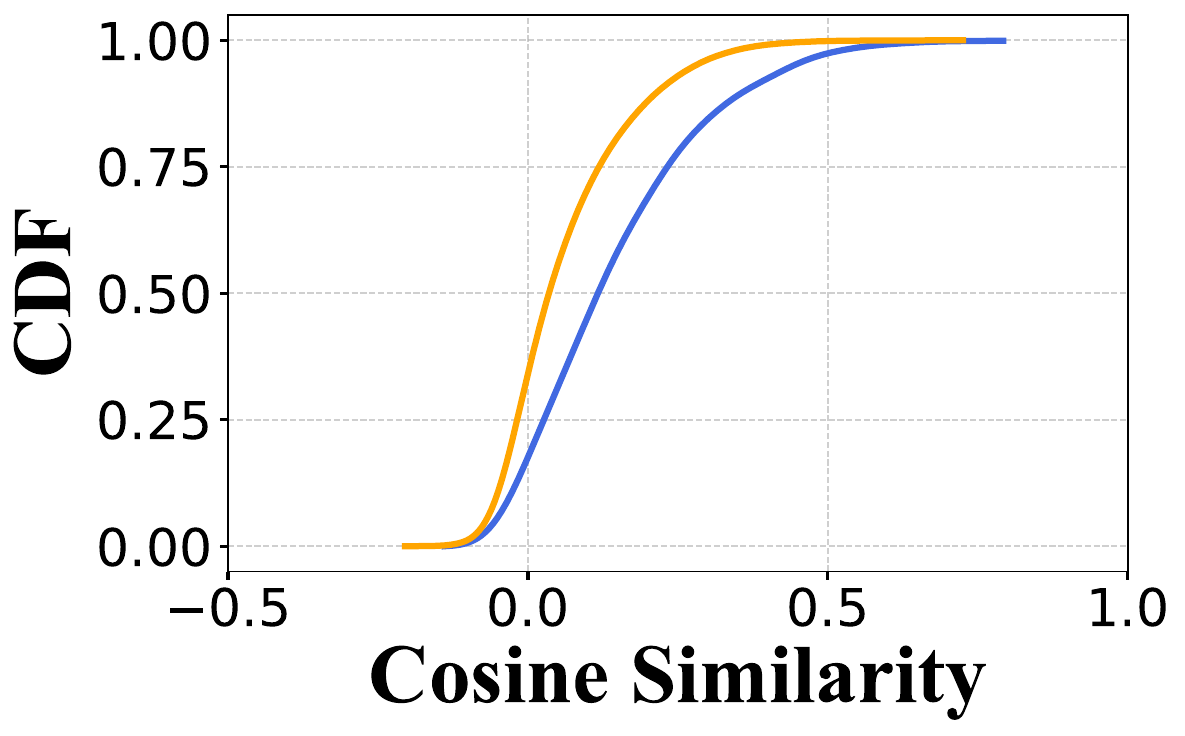}
    		\vspace{-0.1cm}
    	\end{minipage}
    }
 \vspace{-0.4cm}
	\caption{Cumulative distribution functions of MFA-Conformer against various voice conversion methods.}
	\label{fig:cdf_mfa}
\end{figure*}

\begin{figure*}[h]
    \centering
    	\begin{minipage}[b]{0.36\linewidth}
    		\centering
    		\includegraphics[trim=0mm 0mm 0mm 0mm, clip, width=\textwidth]{Section/Pictures/Draw/CDF/legend.pdf}
    	\end{minipage} \\
    \vspace{-0.1cm}
    \subfigure[Clean]{
    	\begin{minipage}[b]{0.23\linewidth}
    		\centering
    		\includegraphics[trim=0mm 0mm 0mm 0mm, clip, width=0.95\textwidth]{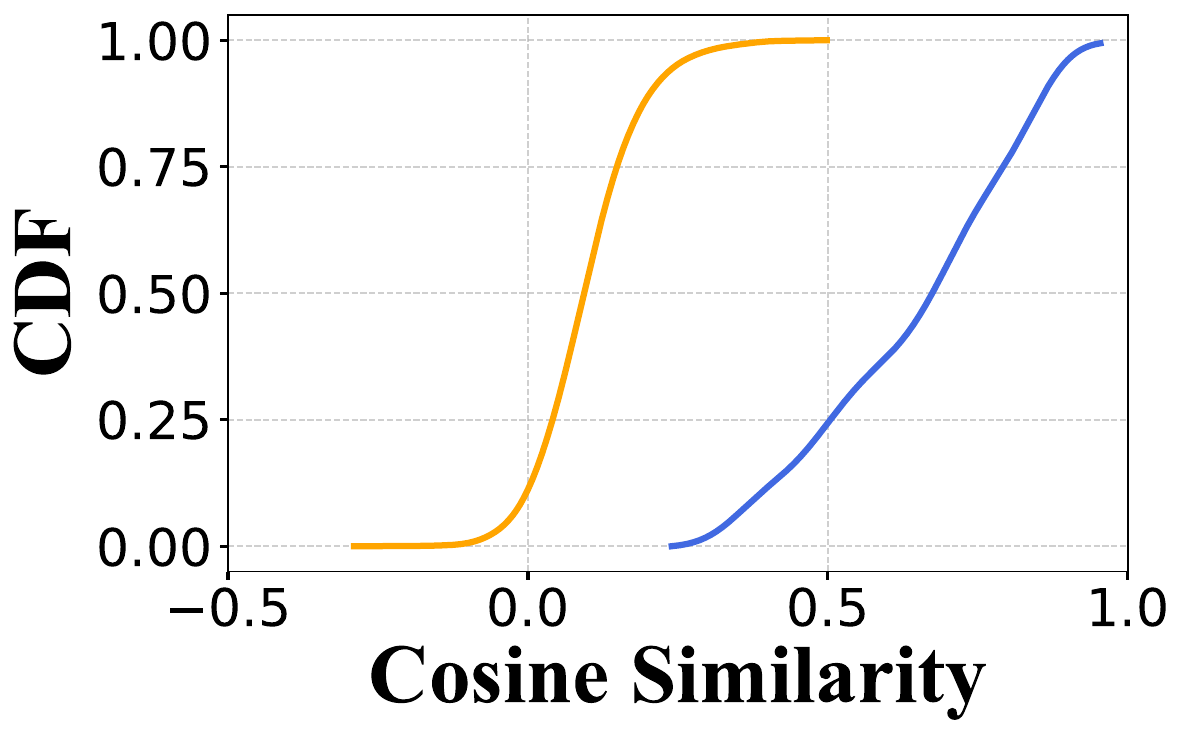}
    		\vspace{-0.1cm}
    	\end{minipage}
    }
    \subfigure[AGAIN]{
    	\begin{minipage}[b]{0.23\linewidth}
    		\centering
    		\includegraphics[trim=0mm 0mm 0mm 0mm, clip, width=0.95\textwidth]{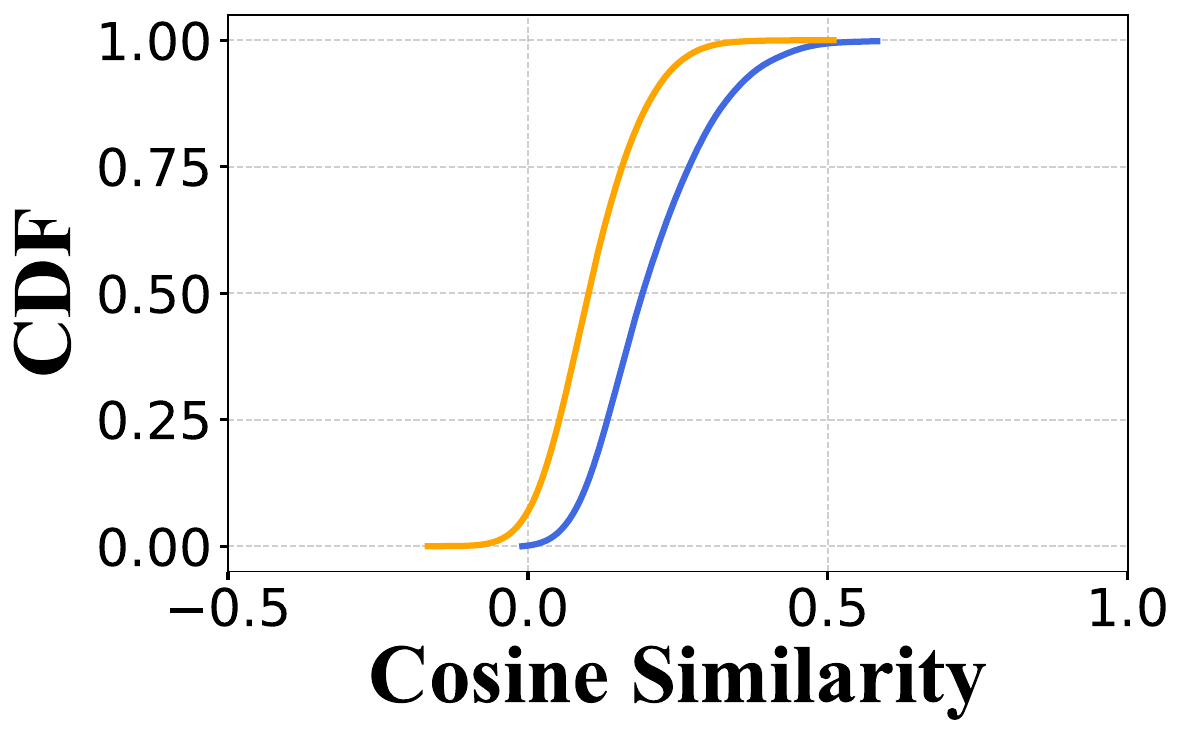}
    		\vspace{-0.1cm}
    	\end{minipage}
    }
    \subfigure[VQVC]{
    	\begin{minipage}[b]{0.23\linewidth}
    		\centering
    		\includegraphics[trim=0mm 0mm 0mm 0mm, clip, width=0.95\textwidth]{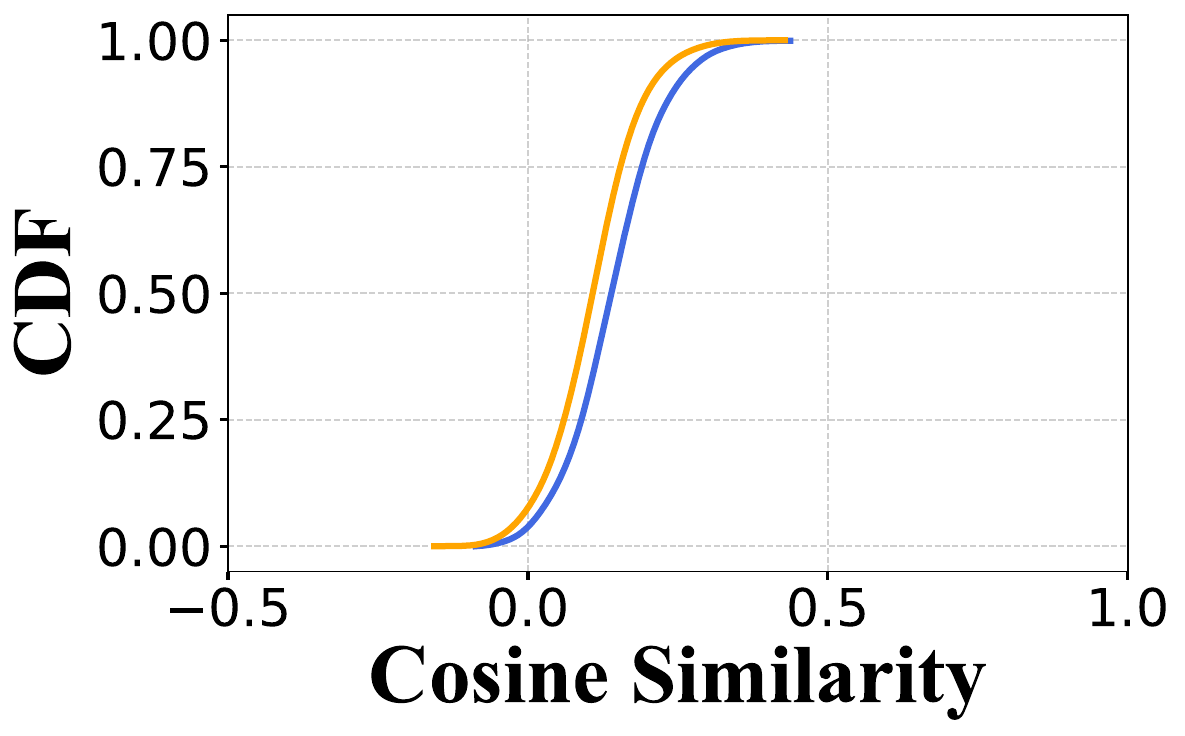}
    		\vspace{-0.1cm}
    	\end{minipage}
    }
  \vspace{-0.3cm}
    \subfigure[VQVC+]{
    	\begin{minipage}[b]{0.23\linewidth}
    		\centering
    		\includegraphics[trim=0mm 0mm 0mm 0mm, clip,width=0.95\textwidth]{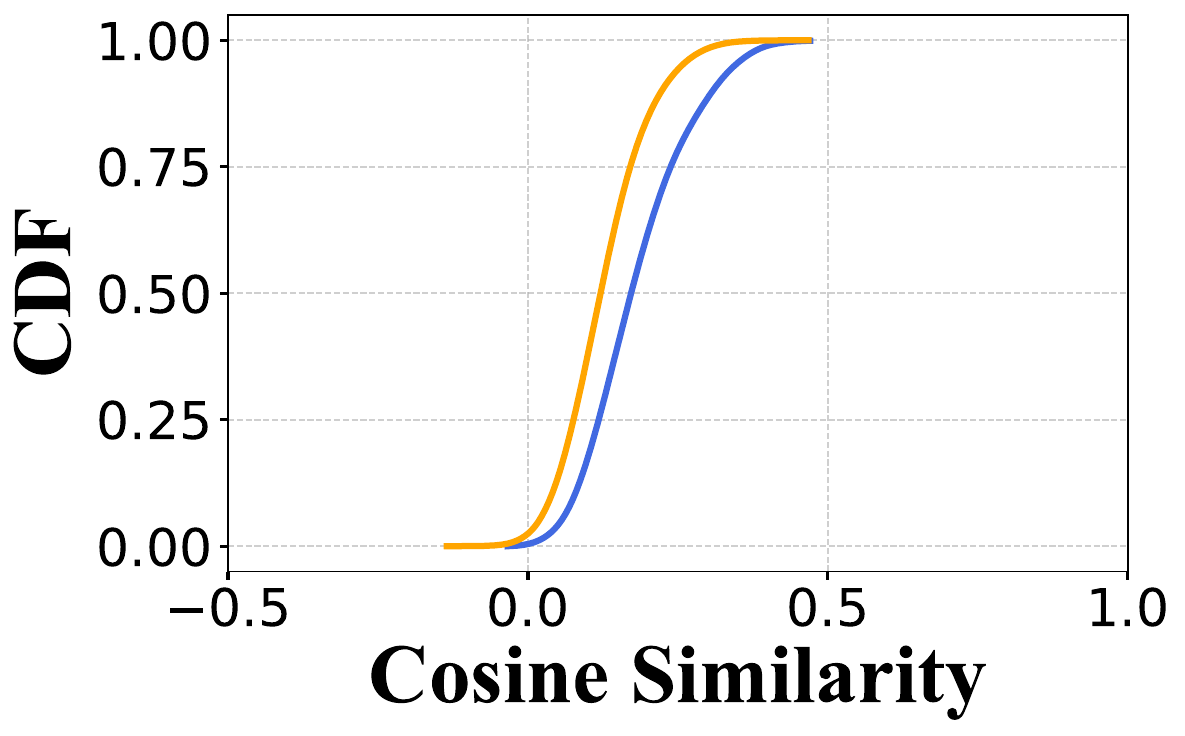}
    		\vspace{-0.1cm}
    	\end{minipage}
    }
    \subfigure[BNE]{
    	\begin{minipage}[b]{0.23\linewidth}
    		\centering
    		\includegraphics[trim=0mm 0mm 0mm 0mm, clip,width=0.95\textwidth]{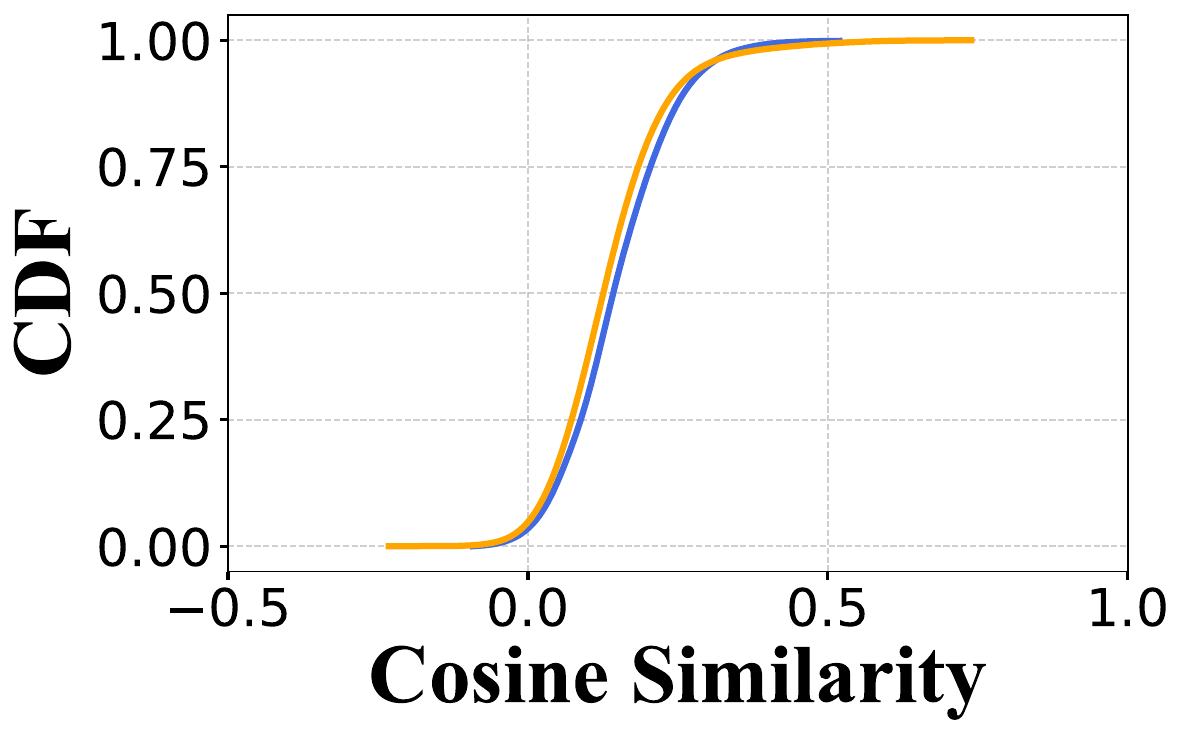}
    		\vspace{-0.1cm}
    	\end{minipage}
    }
    \subfigure[FreeVC]{
    	\begin{minipage}[b]{0.23\linewidth}
    		\centering
    		\includegraphics[trim=0mm 0mm 0mm 0mm, clip,width=0.95\textwidth]{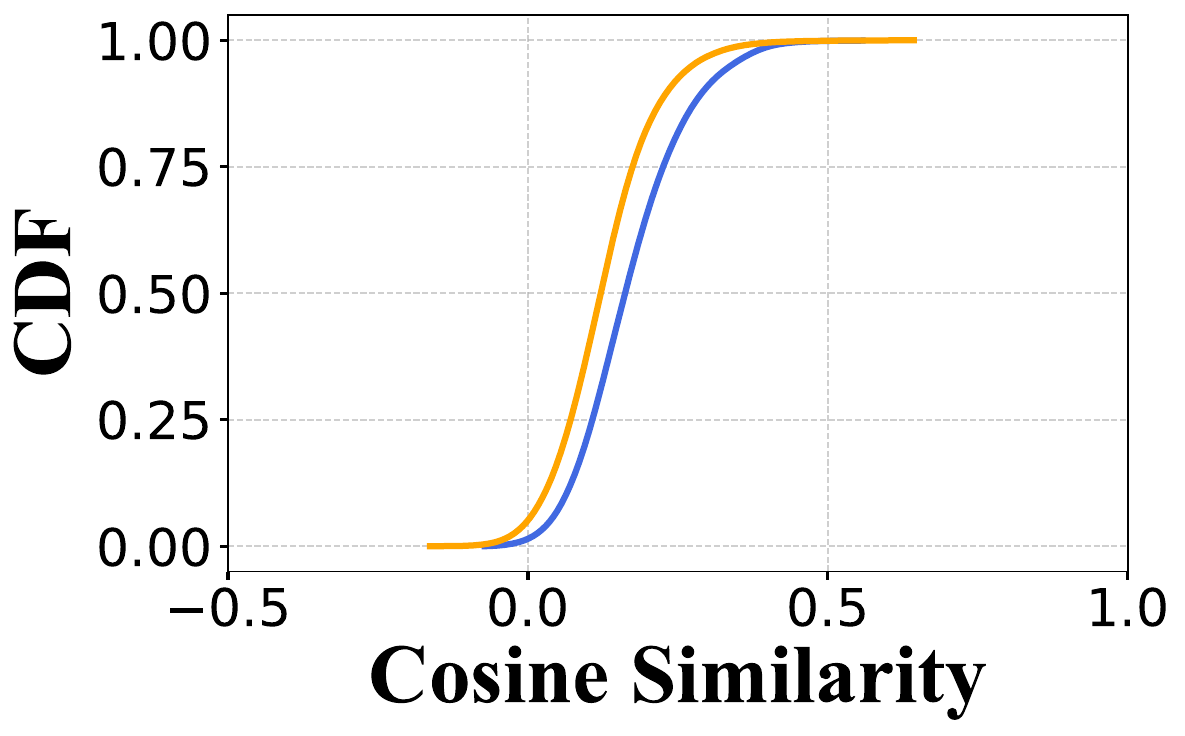}
    		\vspace{-0.1cm}
    	\end{minipage}
    }
    \subfigure[Diff]{
    	\begin{minipage}[b]{0.23\linewidth}
    		\centering
    		\includegraphics[trim=0mm 0mm 0mm 0mm, clip,width=0.95\textwidth]{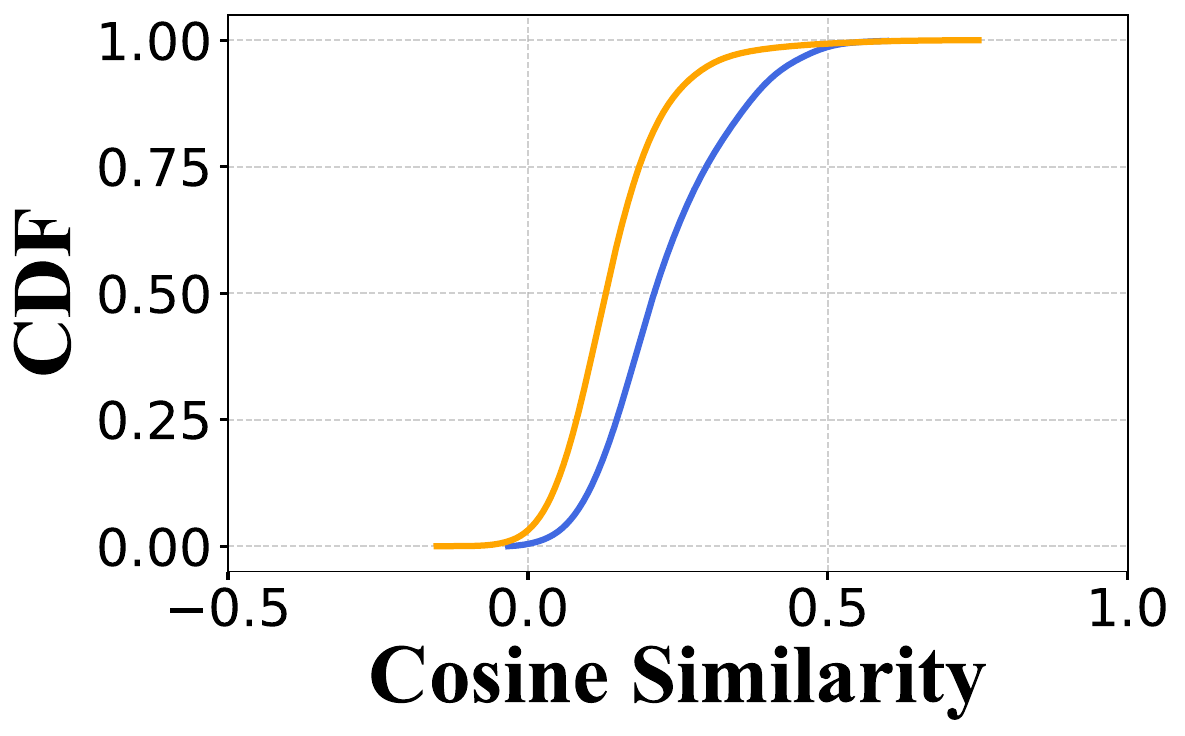}
    		\vspace{-0.1cm}
    	\end{minipage}
    }
    \subfigure[DDDM]{
    	\begin{minipage}[b]{0.23\linewidth}
    		\centering
    		\includegraphics[trim=0mm 0mm 0mm 0mm, clip,width=0.95\textwidth]{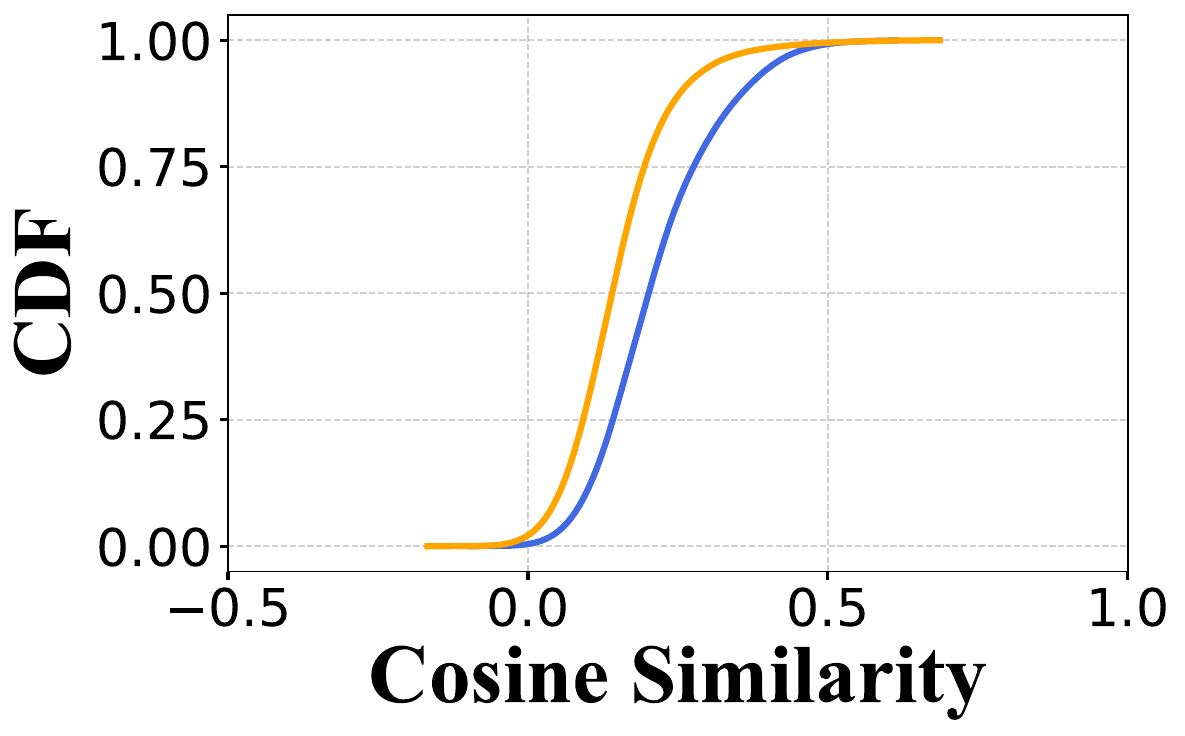}
    		\vspace{-0.1cm}
    	\end{minipage}
    }
 \vspace{-0.4cm}
	\caption{Cumulative distribution functions of ECAPA-TDNN against various voice conversion methods.}
	\label{fig:cdf_e}
\end{figure*}

\begin{figure*}[h]
    \centering
    	\begin{minipage}[b]{0.36\linewidth}
    		\centering
    		\includegraphics[trim=0mm 0mm 0mm 0mm, clip, width=\textwidth]{Section/Pictures/Draw/CDF/legend.pdf}
    	\end{minipage} \\
    \vspace{-0.1cm}
    \subfigure[Clean]{
    	\begin{minipage}[b]{0.23\linewidth}
    		\centering
    		\includegraphics[trim=0mm 0mm 0mm 0mm, clip, width=0.95\textwidth]{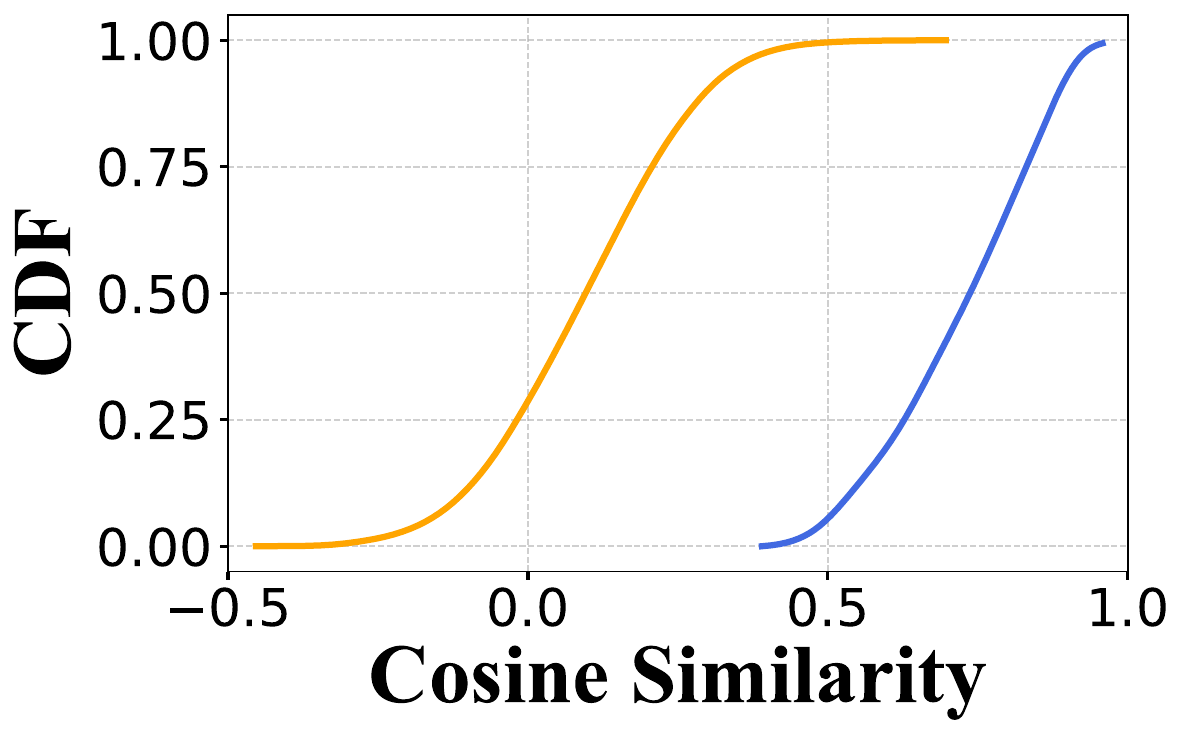}
    		\vspace{-0.1cm}
    	\end{minipage}
    }
    \subfigure[AGAIN]{
    	\begin{minipage}[b]{0.23\linewidth}
    		\centering
    		\includegraphics[trim=0mm 0mm 0mm 0mm, clip, width=0.95\textwidth]{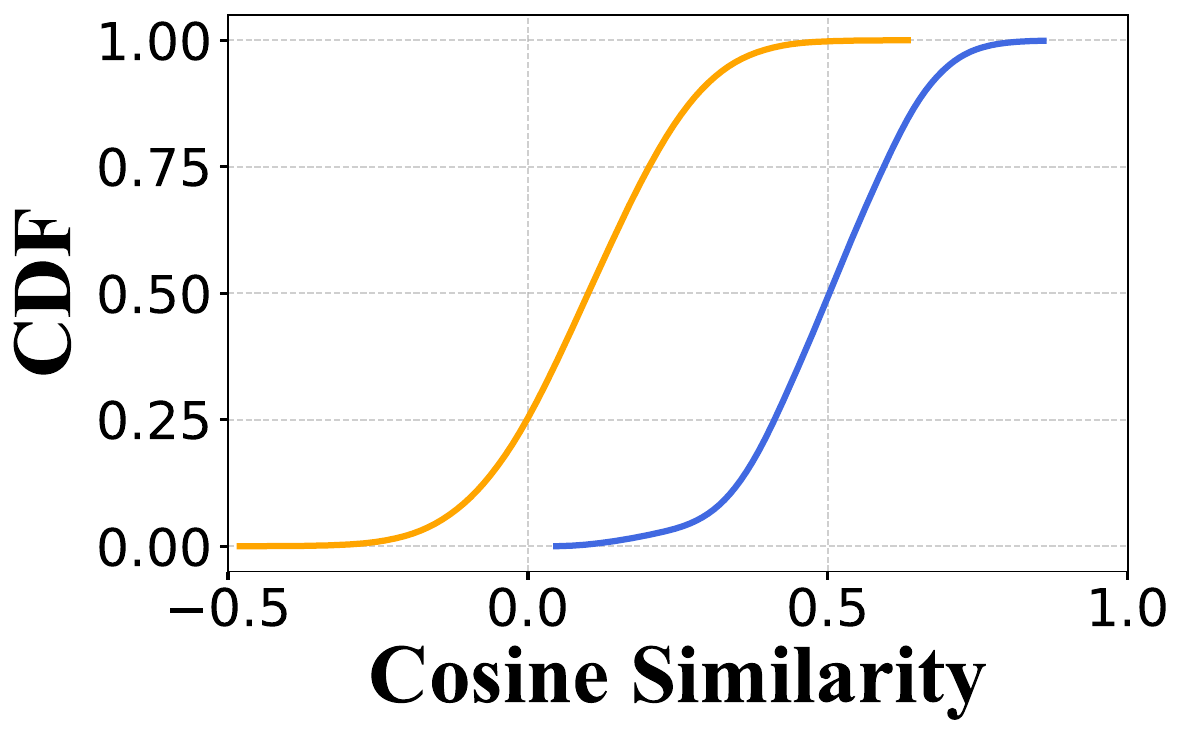}
    		\vspace{-0.1cm}
    	\end{minipage}
    }
    \subfigure[VQVC]{
    	\begin{minipage}[b]{0.23\linewidth}
    		\centering
    		\includegraphics[trim=0mm 0mm 0mm 0mm, clip, width=0.95\textwidth]{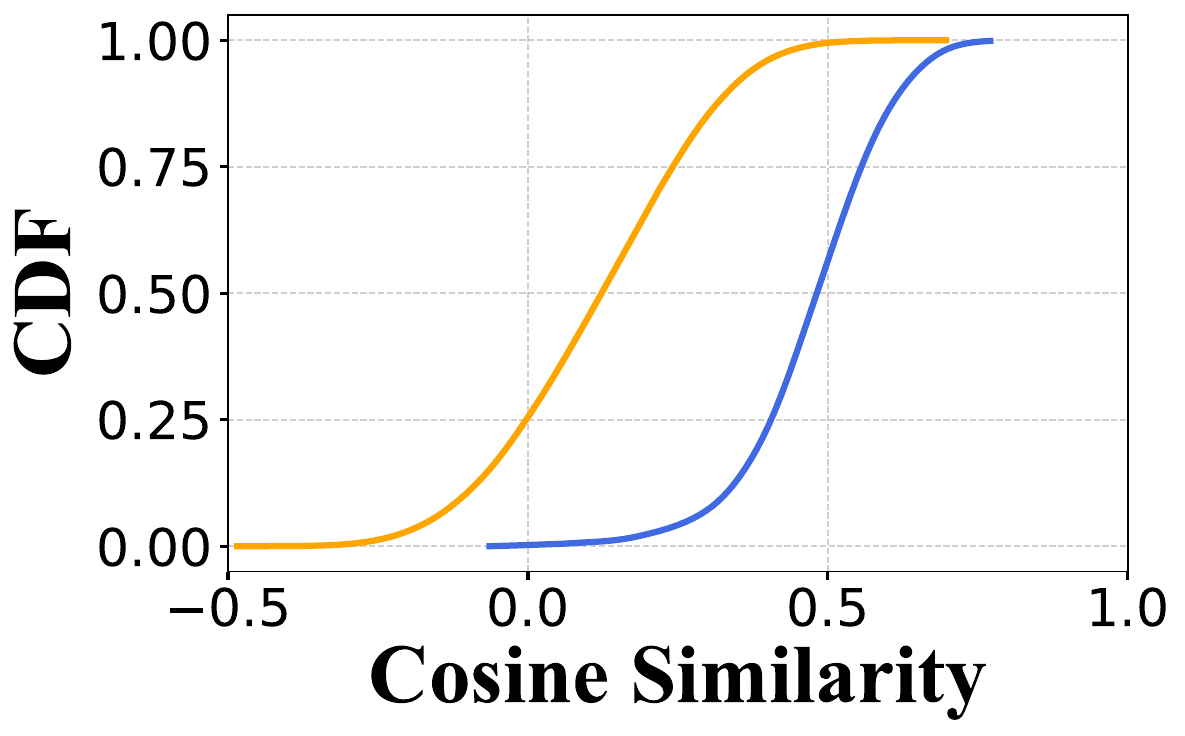}
    		\vspace{-0.1cm}
    	\end{minipage}
    }
  \vspace{-0.3cm}
    \subfigure[VQVC+]{
    	\begin{minipage}[b]{0.23\linewidth}
    		\centering
    		\includegraphics[trim=0mm 0mm 0mm 0mm, clip,width=0.95\textwidth]{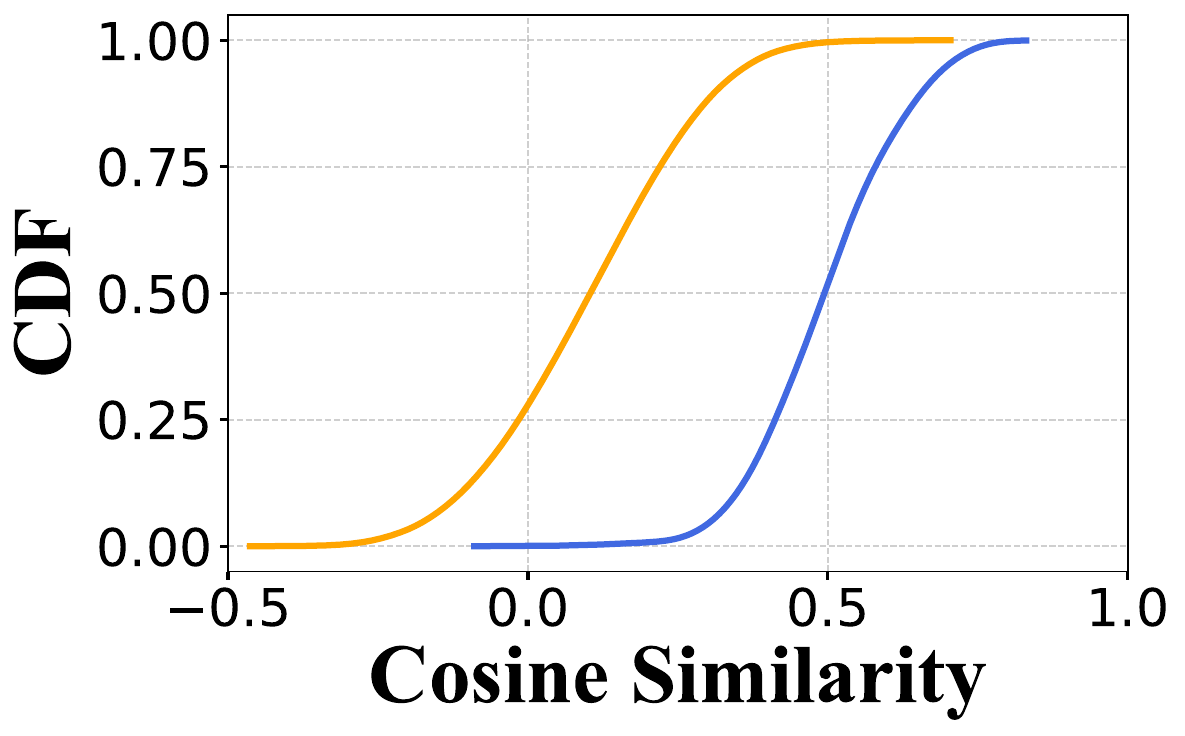}
    		\vspace{-0.1cm}
    	\end{minipage}
    }
    \subfigure[BNE]{
    	\begin{minipage}[b]{0.23\linewidth}
    		\centering
    		\includegraphics[trim=0mm 0mm 0mm 0mm, clip,width=0.95\textwidth]{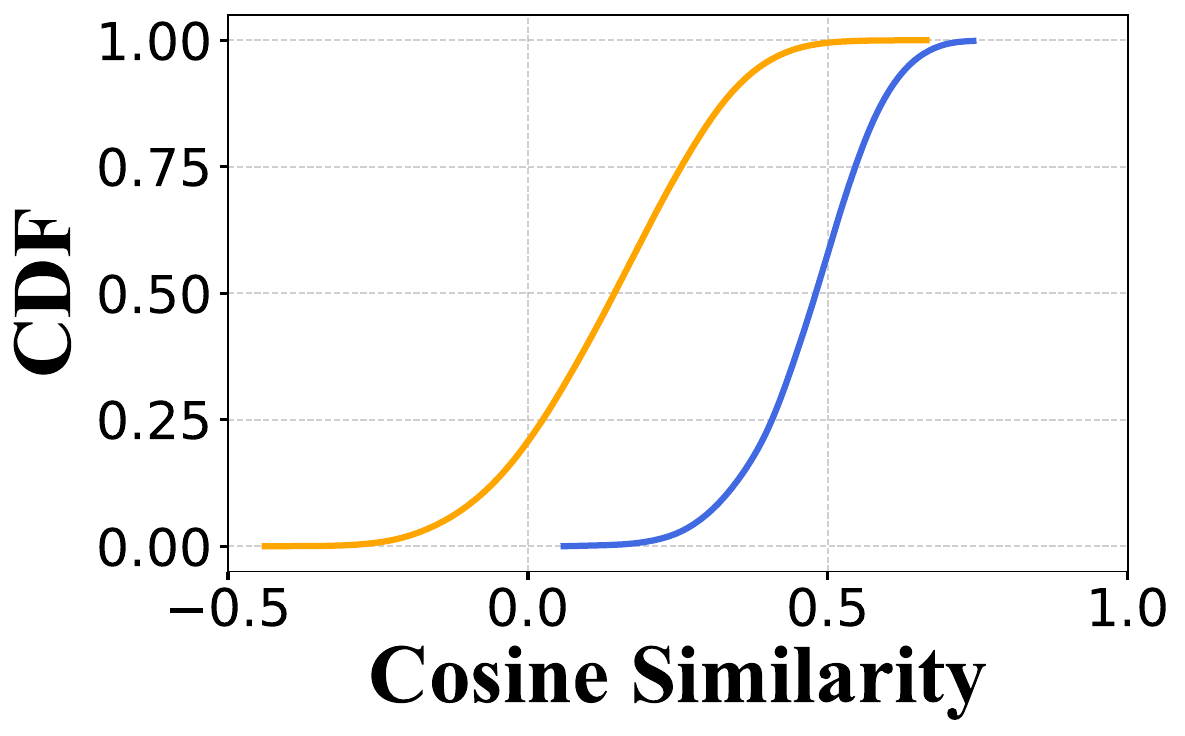}
    		\vspace{-0.1cm}
    	\end{minipage}
    }
    \subfigure[FreeVC]{
    	\begin{minipage}[b]{0.23\linewidth}
    		\centering
    		\includegraphics[trim=0mm 0mm 0mm 0mm, clip,width=0.95\textwidth]{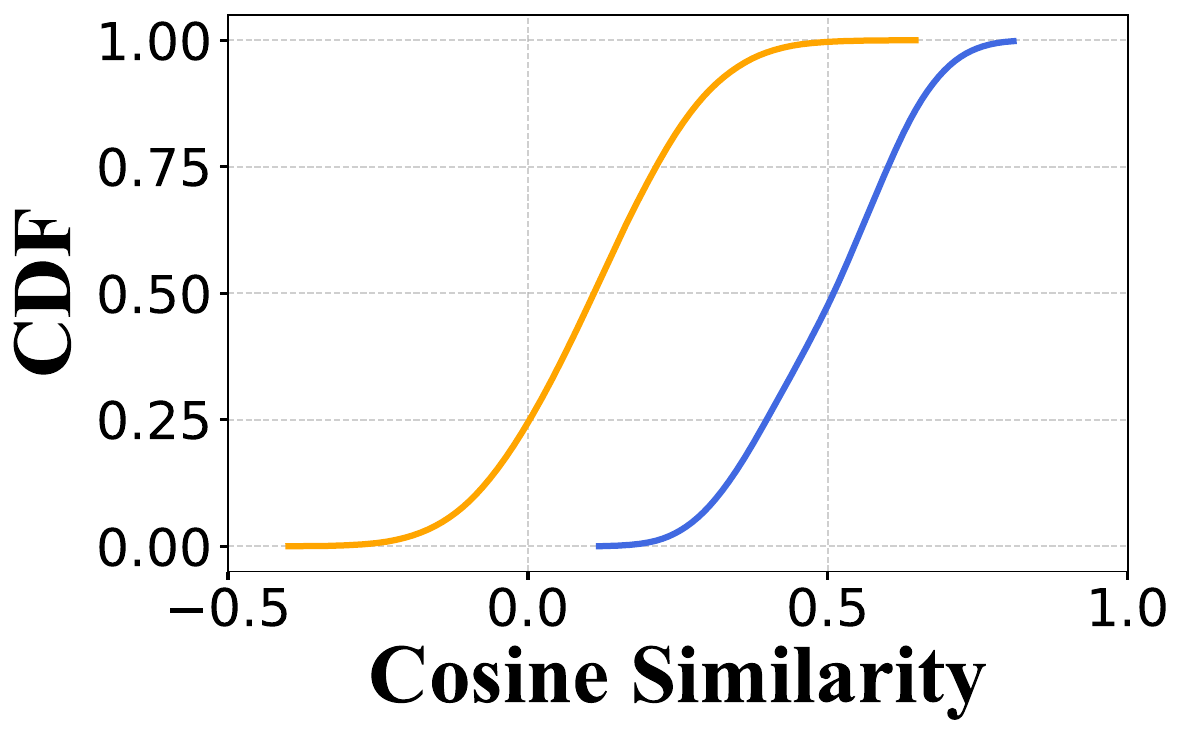}
    		\vspace{-0.1cm}
    	\end{minipage}
    }
    \subfigure[Diff]{
    	\begin{minipage}[b]{0.23\linewidth}
    		\centering
    		\includegraphics[trim=0mm 0mm 0mm 0mm, clip,width=0.95\textwidth]{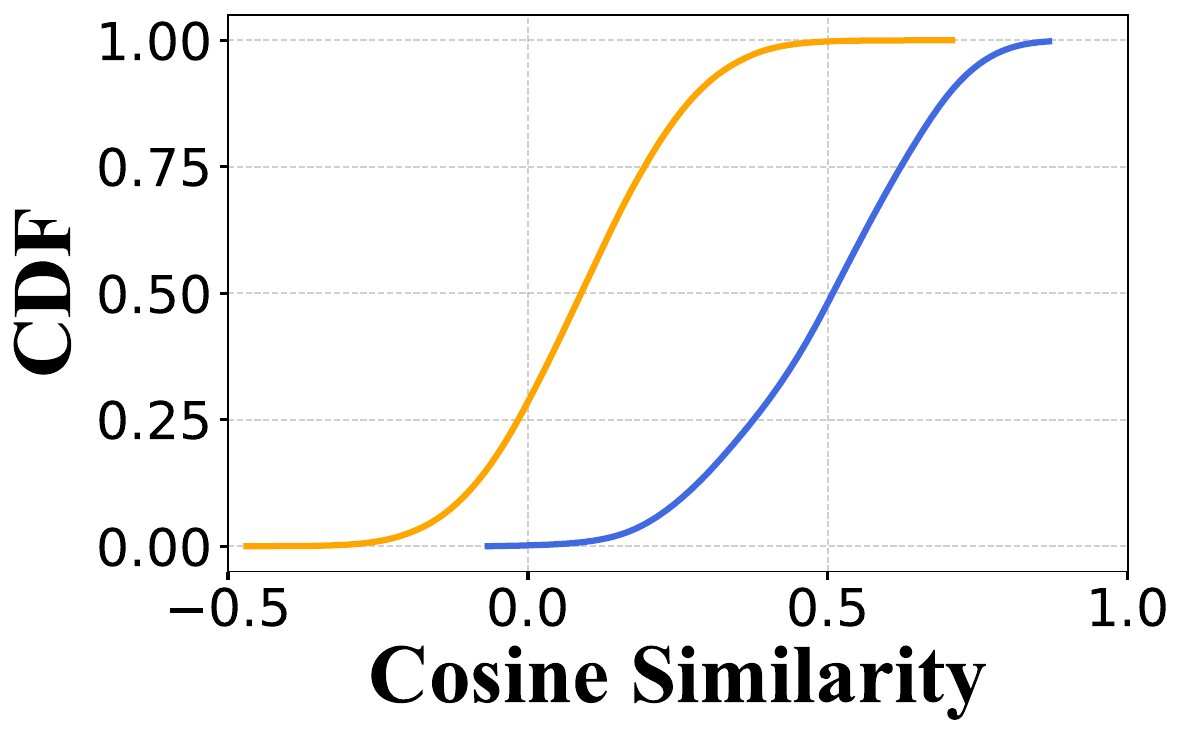}
    		\vspace{-0.1cm}
    	\end{minipage}
    }
    \subfigure[DDDM]{
    	\begin{minipage}[b]{0.23\linewidth}
    		\centering
    		\includegraphics[trim=0mm 0mm 0mm 0mm, clip,width=0.95\textwidth]{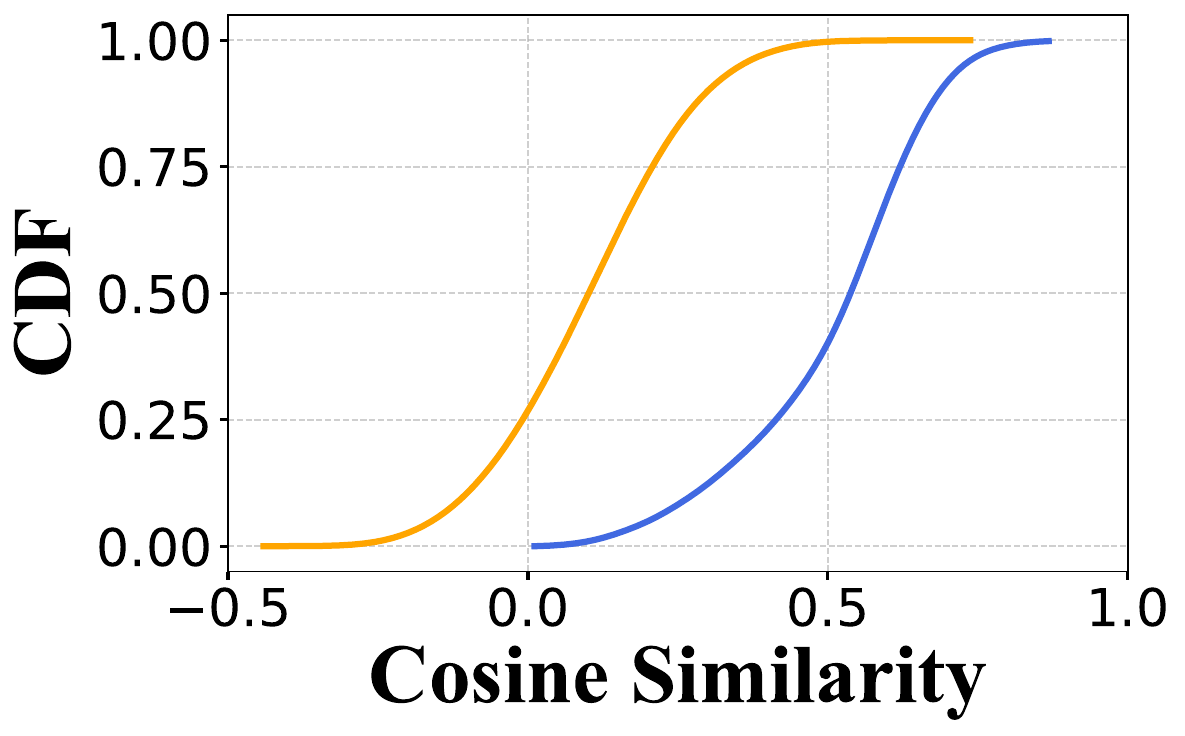}
    		\vspace{-0.1cm}
    	\end{minipage}
    }
 \vspace{-0.4cm}
	\caption{Cumulative distribution functions of Revelio against various voice conversion methods.}
	\label{fig:cdf_r}
\end{figure*}

\begin{figure*}[h]
    \centering
    	\begin{minipage}[b]{0.36\linewidth}
    		\centering
    		\includegraphics[trim=0mm 0mm 0mm 0mm, clip, width=\textwidth]{Section/Pictures/Draw/CDF/legend.pdf}
    	\end{minipage} \\
    \subfigure[Clean]{
    	\begin{minipage}[b]{0.23\linewidth}
    		\centering
    		\includegraphics[trim=0mm 0mm 0mm 0mm, clip, width=0.95\textwidth]{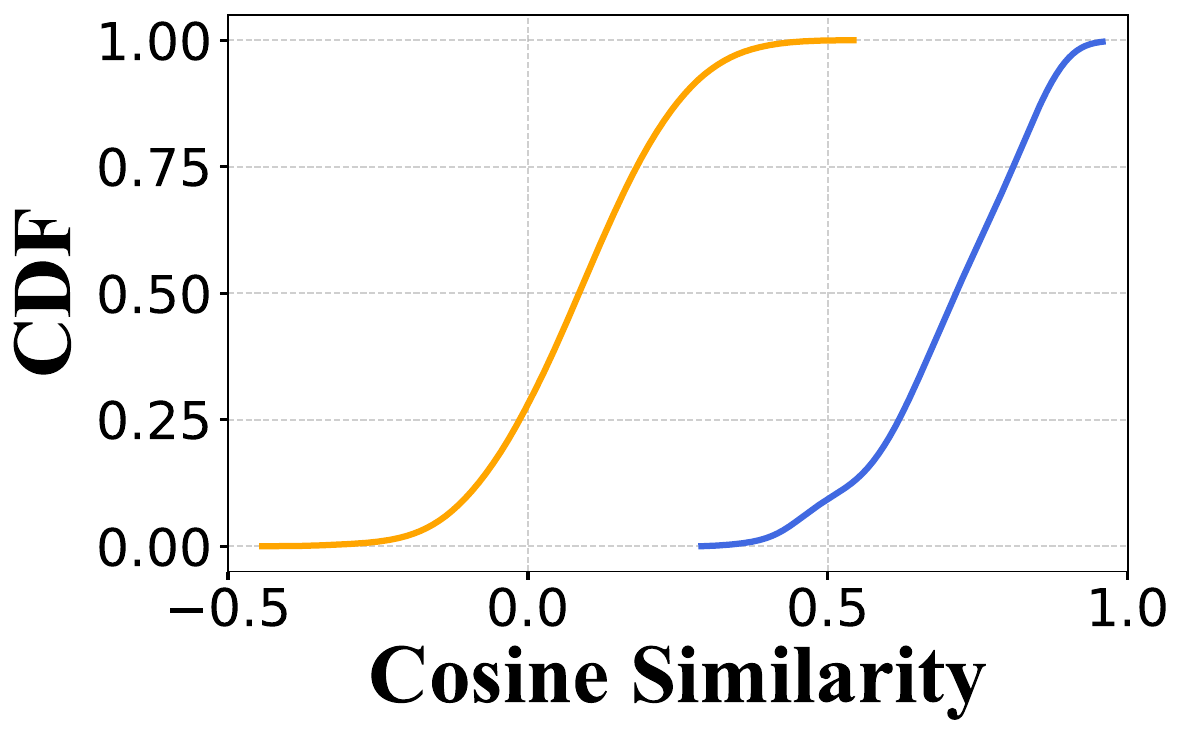}
    	\end{minipage}
    }
    \subfigure[AGAIN]{
    	\begin{minipage}[b]{0.23\linewidth}
    		\centering
    		\includegraphics[trim=0mm 0mm 0mm 0mm, clip, width=0.95\textwidth]{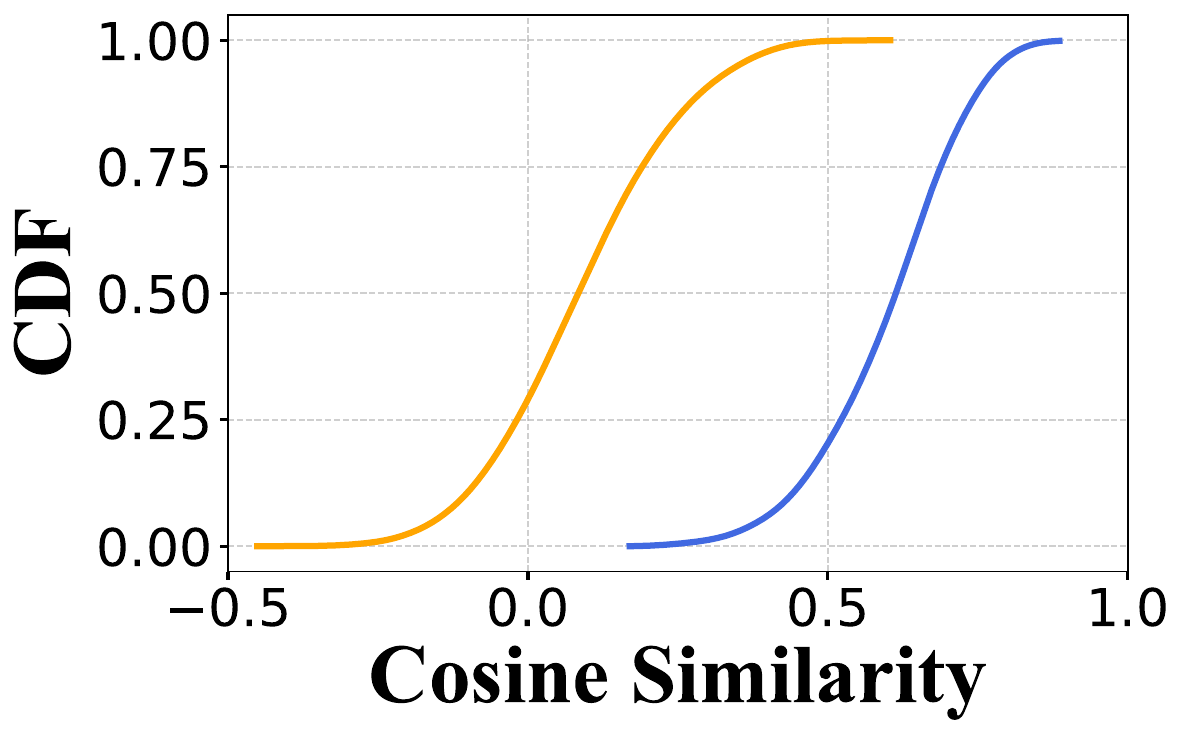}
    		\vspace{-0.1cm}
    	\end{minipage}
    }
    \subfigure[VQVC]{
    	\begin{minipage}[b]{0.23\linewidth}
    		\centering
    		\includegraphics[trim=0mm 0mm 0mm 0mm, clip, width=0.95\textwidth]{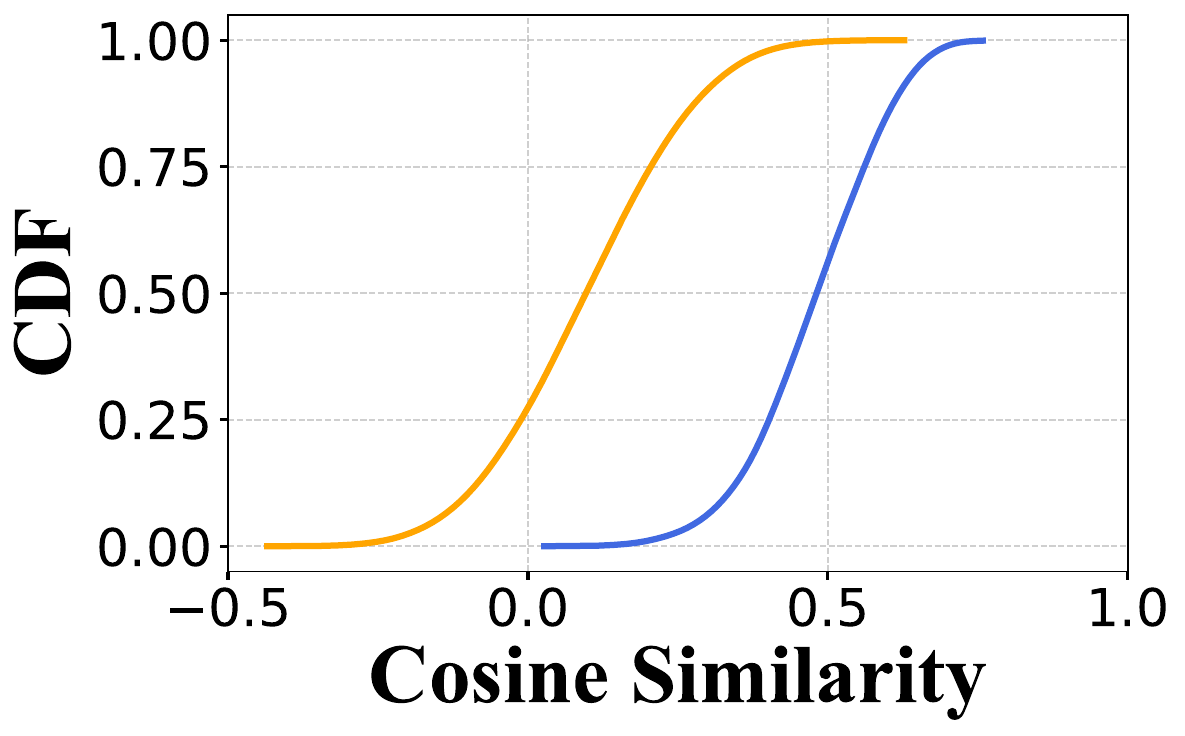}
    	\end{minipage}
    }
  \vspace{-0.3cm}
    \subfigure[VQVC+]{
    	\begin{minipage}[b]{0.23\linewidth}
    		\centering
    		\includegraphics[trim=0mm 0mm 0mm 0mm, clip,width=0.95\textwidth]{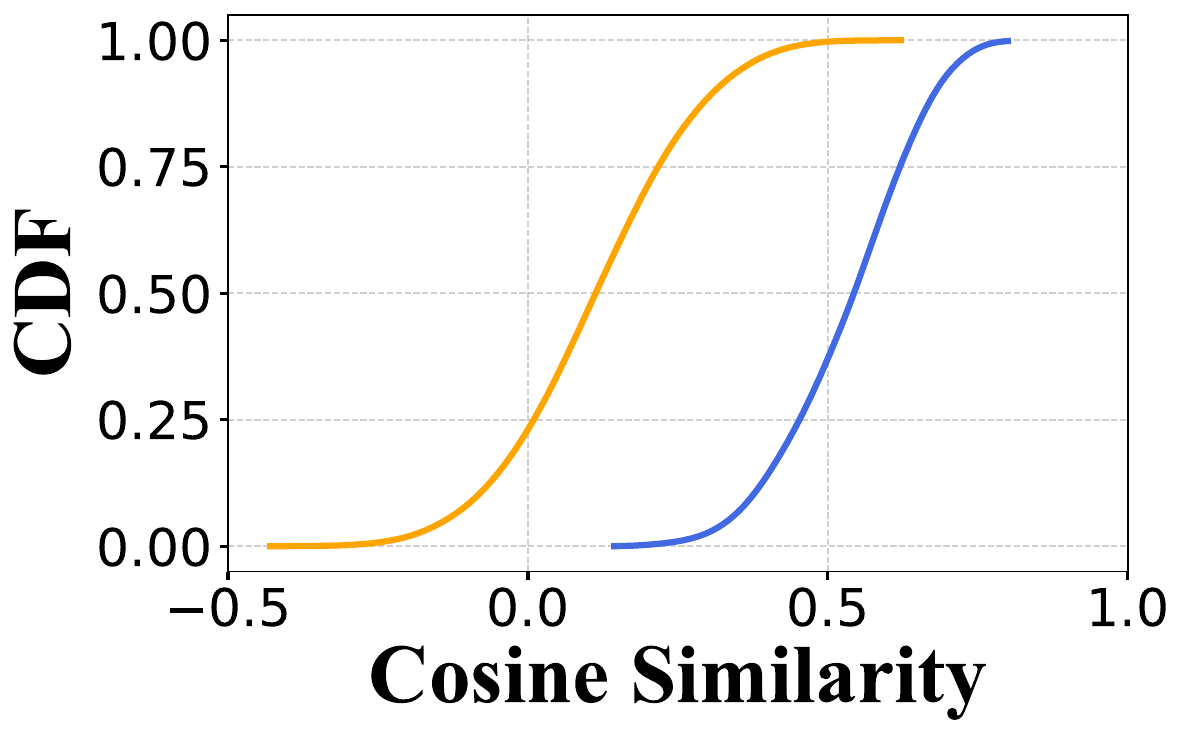}
    	\end{minipage}
    }
    \subfigure[BNE]{
    	\begin{minipage}[b]{0.23\linewidth}
    		\centering
    		\includegraphics[trim=0mm 0mm 0mm 0mm, clip,width=0.95\textwidth]{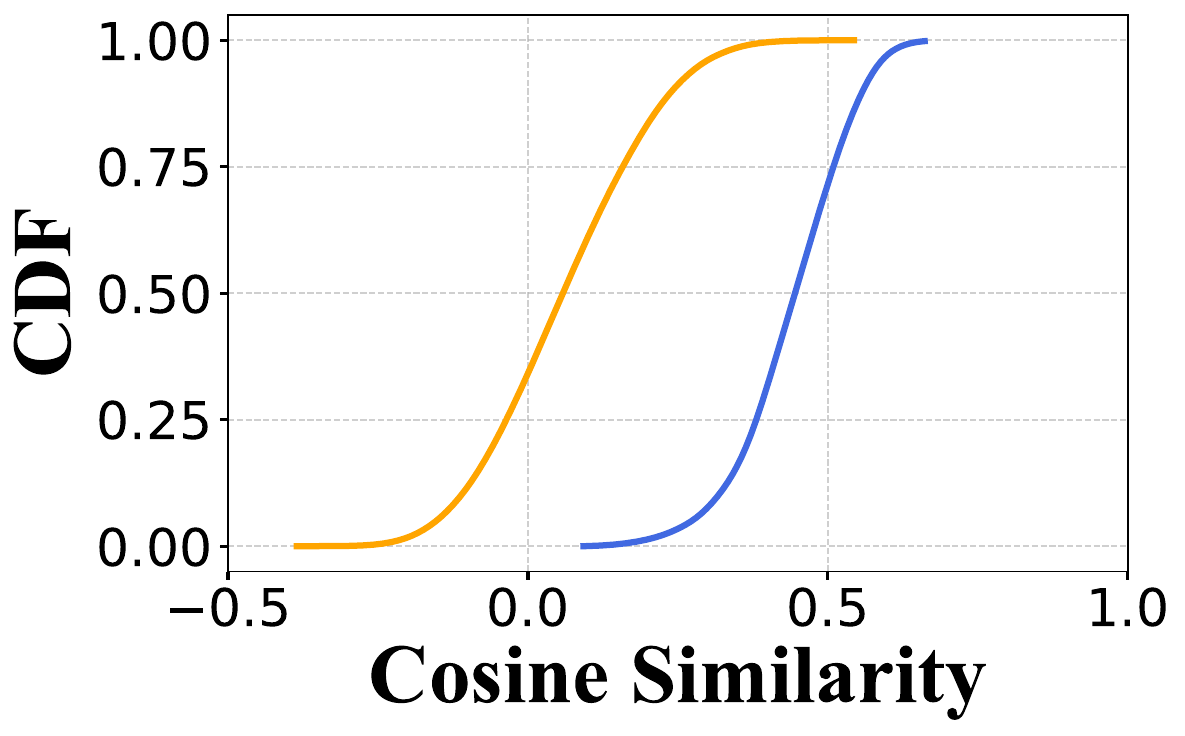}
    	\end{minipage}
    }
    \subfigure[FreeVC]{
    	\begin{minipage}[b]{0.23\linewidth}
    		\centering
    		\includegraphics[trim=0mm 0mm 0mm 0mm, clip,width=0.95\textwidth]{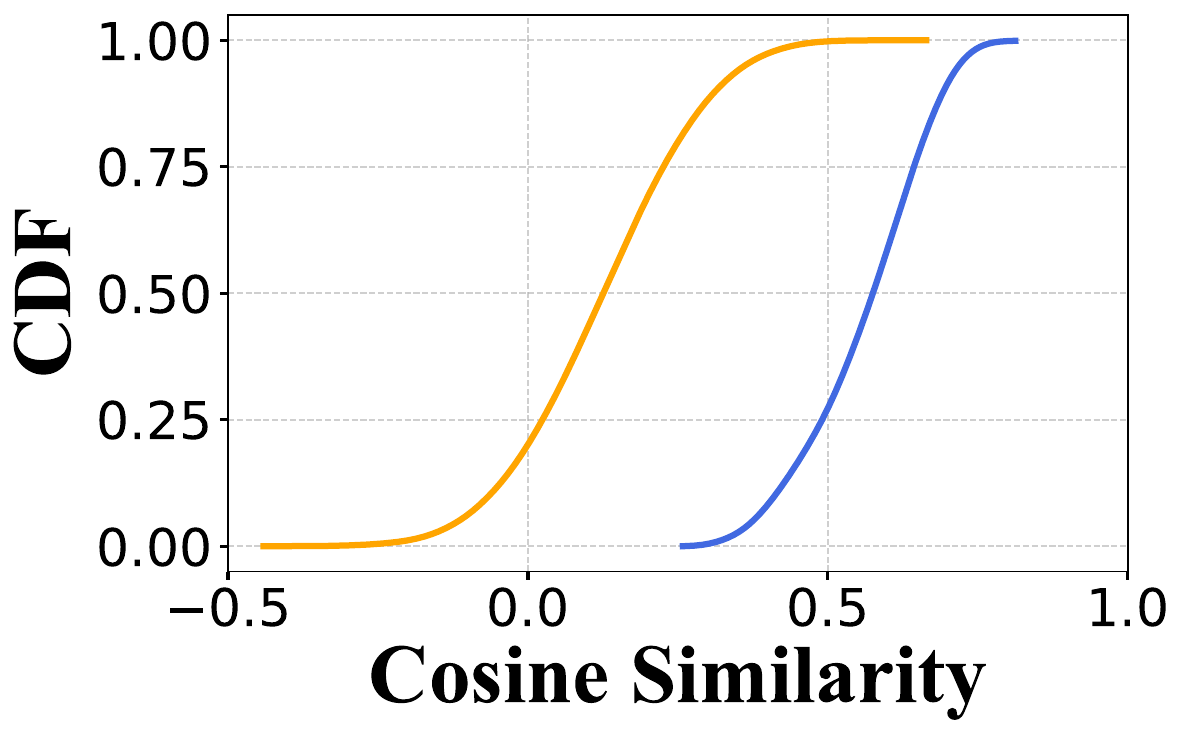}
    	\end{minipage}
    }
    \subfigure[Diff]{
    	\begin{minipage}[b]{0.23\linewidth}
    		\centering
    		\includegraphics[trim=0mm 0mm 0mm 0mm, clip,width=0.95\textwidth]{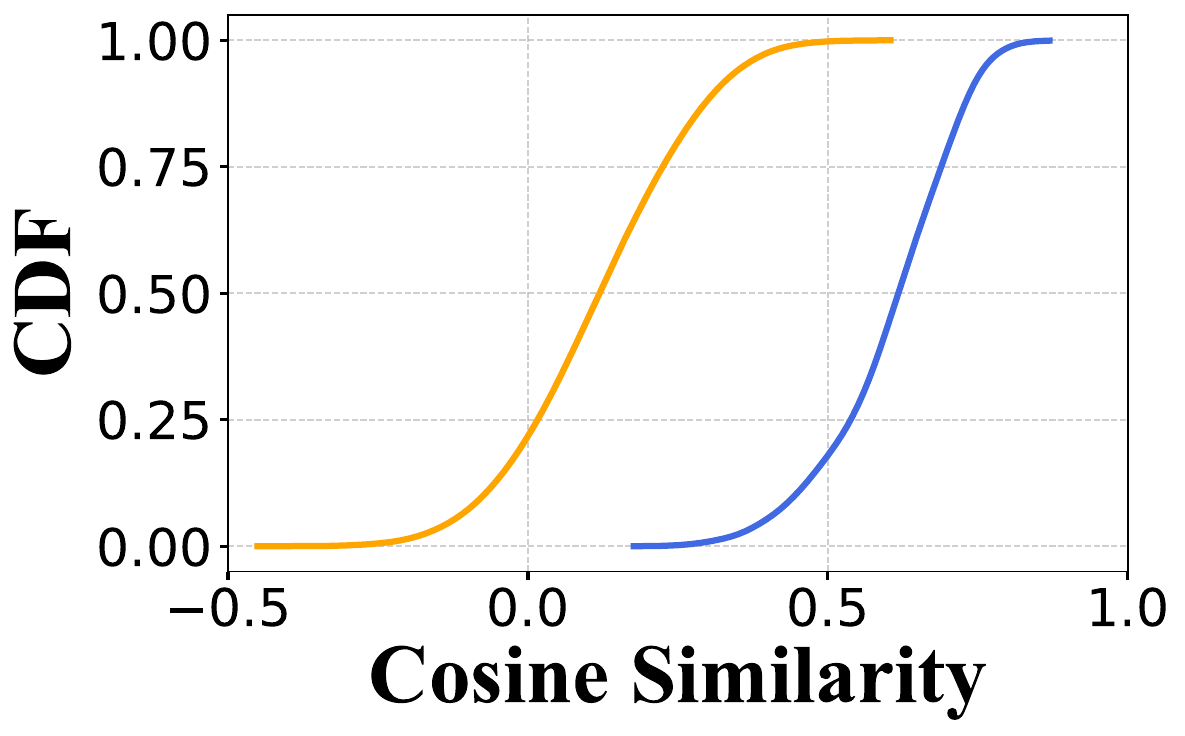}
    	\end{minipage}
    }
    \subfigure[DDDM]{
    	\begin{minipage}[b]{0.23\linewidth}
    		\centering
    		\includegraphics[trim=0mm 0mm 0mm 0mm, clip,width=0.95\textwidth]{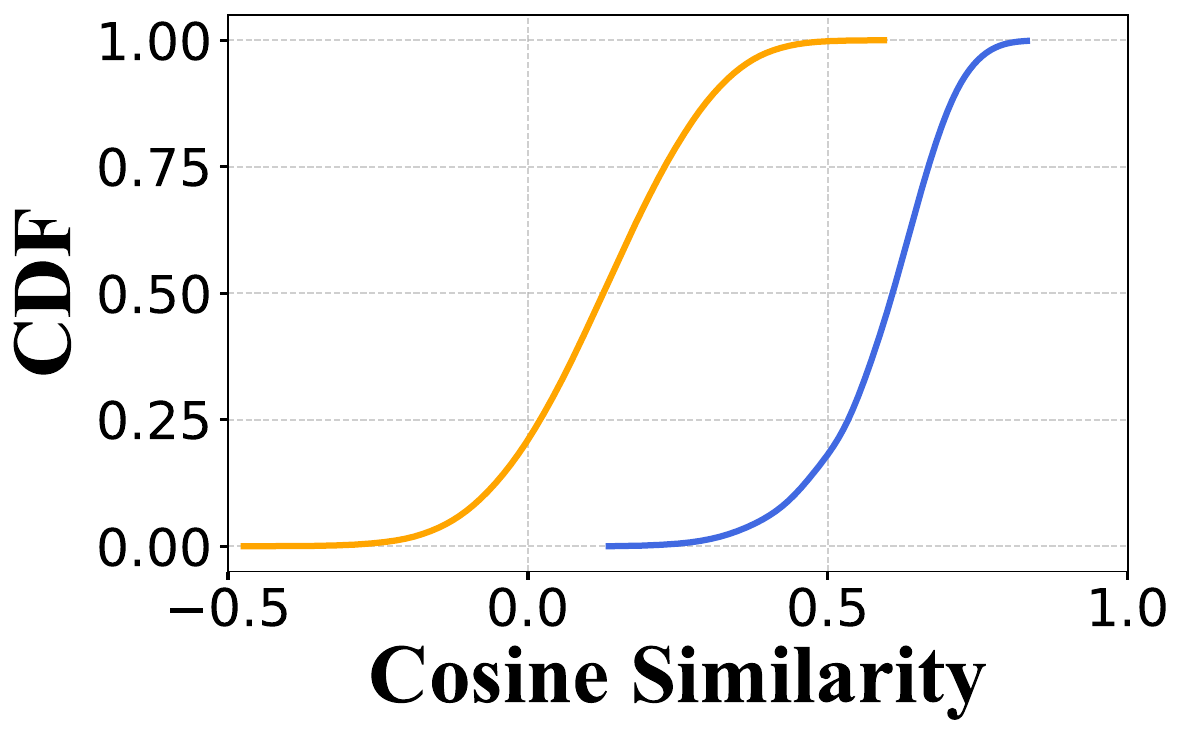}
    	\end{minipage}
    }
 \vspace{-0.4cm}
	\caption{Cumulative distribution functions of \sys against various voice conversion methods.}
	\label{fig:cdf_n}
\end{figure*}

\begin{figure*}[h]
    \centering
    \begin{minipage}[b]{0.85\linewidth}
        \centering
        \includegraphics[trim=0mm 0mm 0mm 0mm, clip, width=\textwidth]{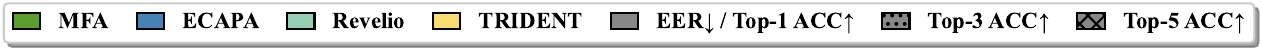}
    \end{minipage} \\
    \vspace{0.1cm}

    \hspace{-0.2cm}
    \begin{minipage}[b]{0.255\textwidth}
        \centering
        \includegraphics[trim=0mm 0mm 0mm 0mm, clip, width=0.95\linewidth]{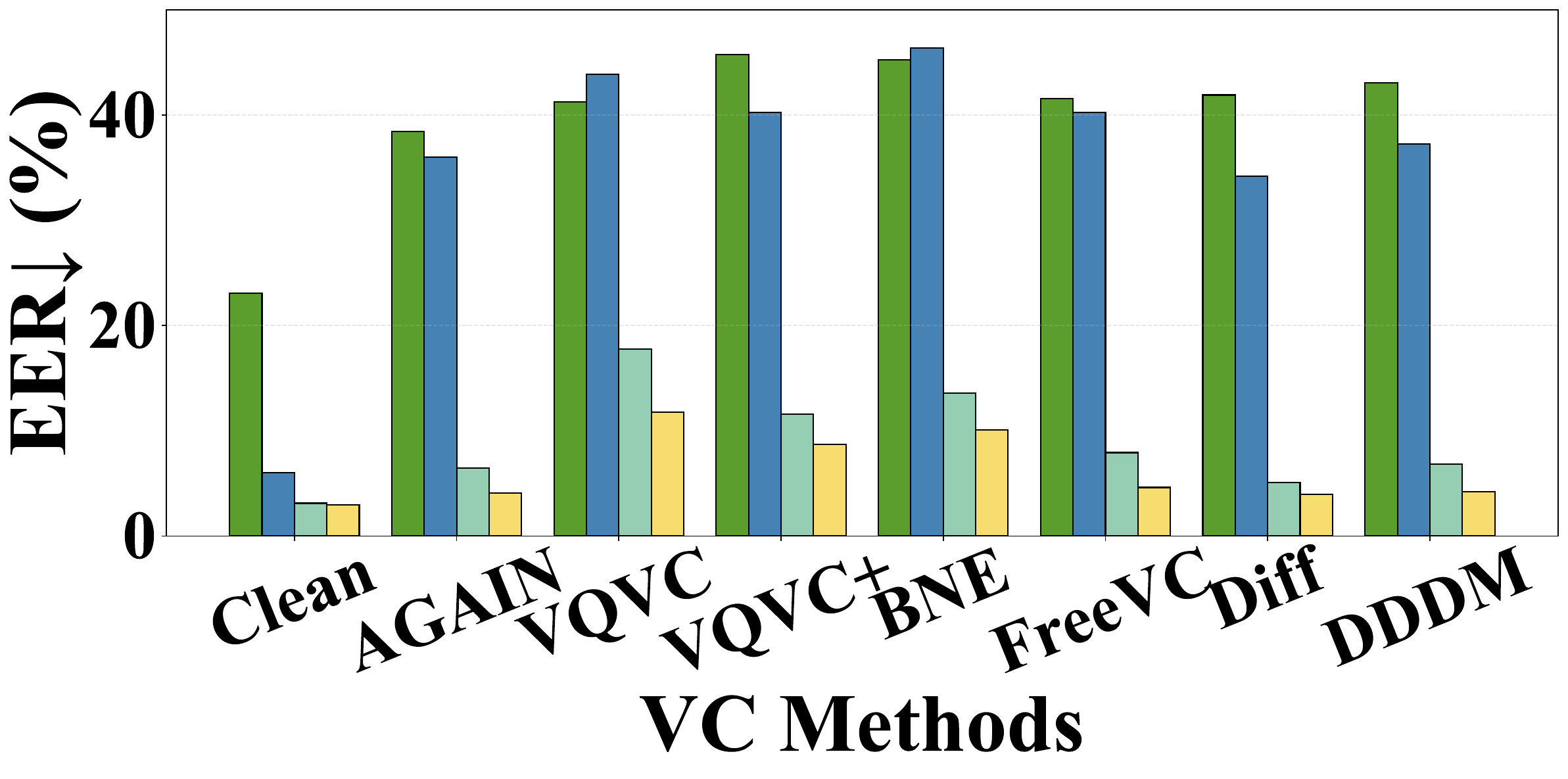}
    \end{minipage}
    \hfill
    \hspace{-0.6cm}
    \begin{minipage}[b]{0.255\textwidth}
        \centering
        \includegraphics[trim=0mm 0mm 0mm 0mm, clip, width=0.95\linewidth]{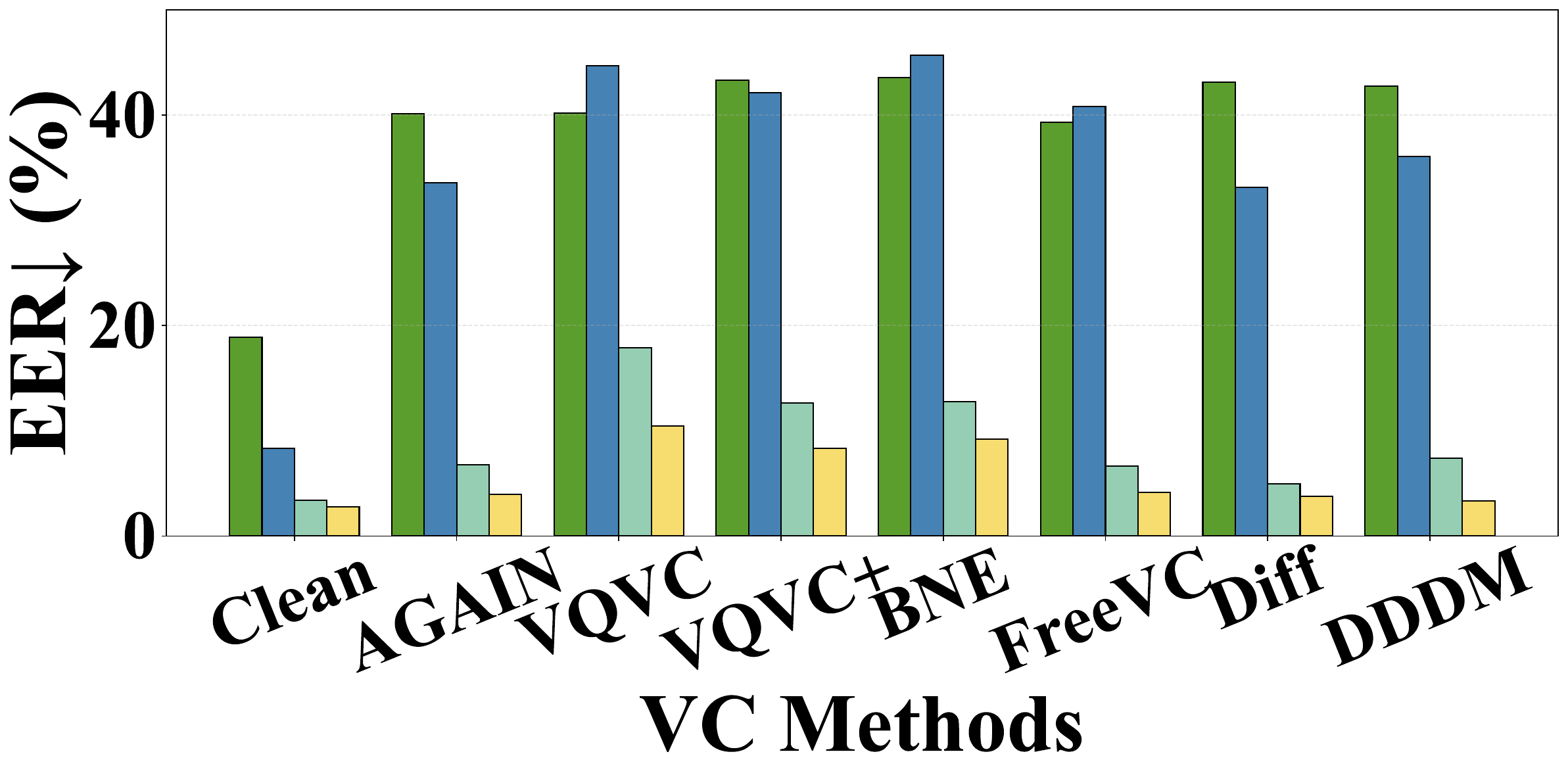}
    \end{minipage}
    \hfill
    \hspace{-0.6cm}
    \begin{minipage}[b]{0.255\textwidth}
        \centering
        \includegraphics[trim=0mm 0mm 0mm 0mm, clip, width=0.95\linewidth]{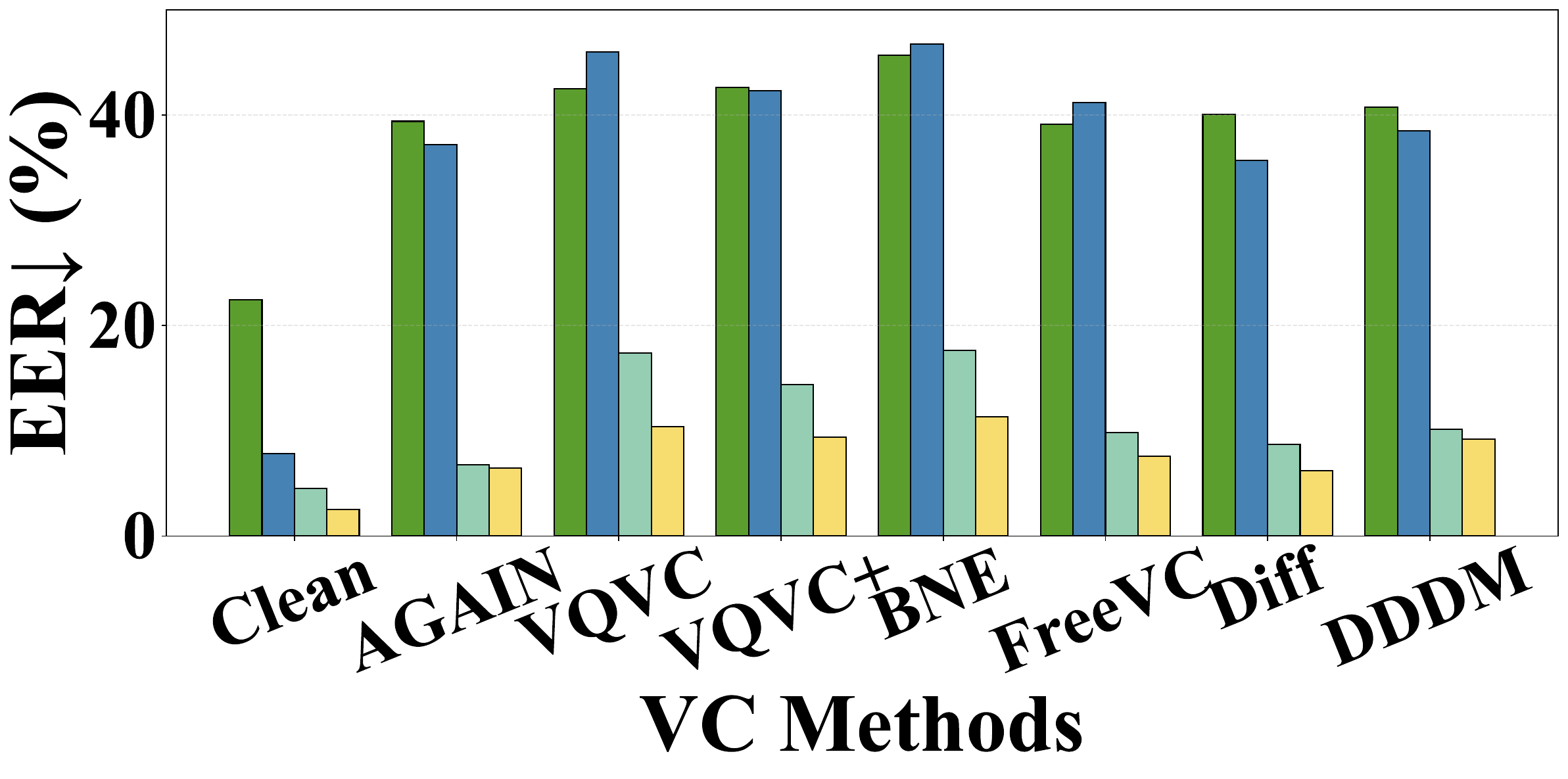}
    \end{minipage}
    \hfill
    \hspace{-0.6cm}
    \begin{minipage}[b]{0.255\textwidth}
        \centering
        \includegraphics[trim=0mm 0mm 0mm 0mm, clip, width=0.95\linewidth]{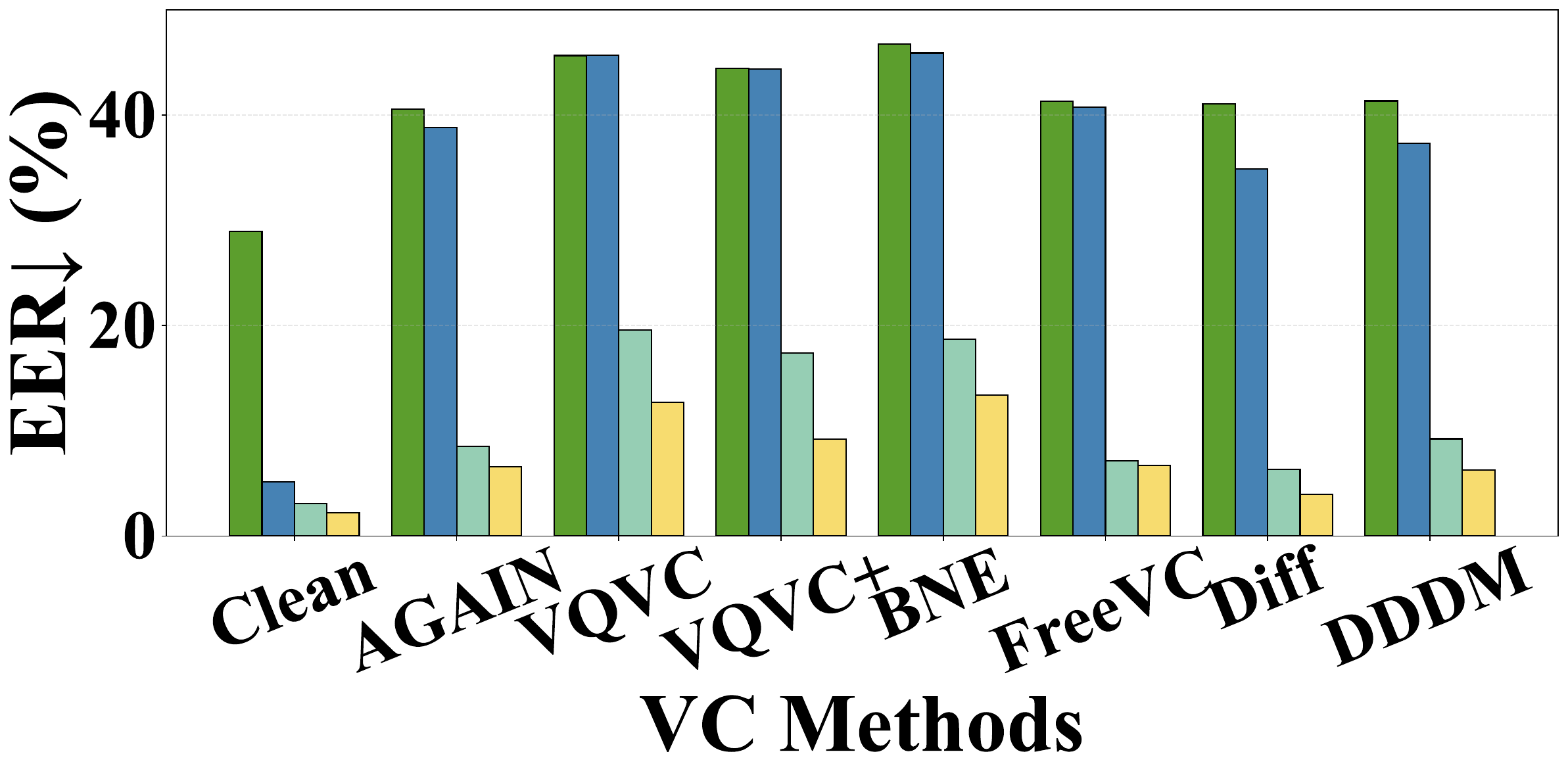}
    \end{minipage}

    \hspace{-0.4cm}
    \subfigure[$\mu$-law]{
        \begin{minipage}[b]{0.261\textwidth}
            \centering
            \includegraphics[trim=0mm 0mm 0mm 0mm, clip, width=0.95\linewidth]{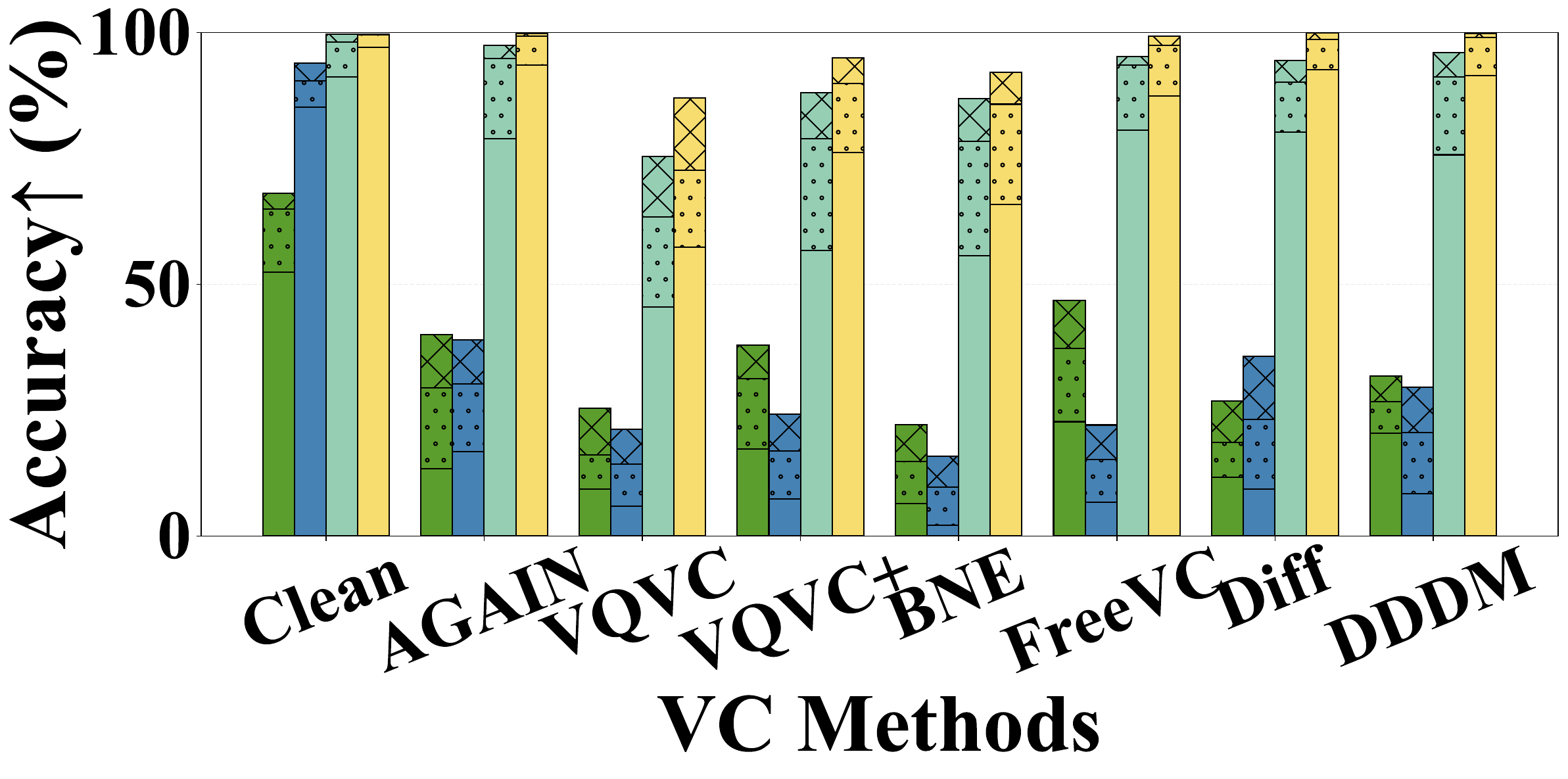}
        \end{minipage}
    }
    \hfill
    \hspace{-0.6cm}
    \subfigure[A-law]{
        \begin{minipage}[b]{0.261\textwidth}
            \centering
            \includegraphics[trim=0mm 0mm 0mm 0mm, clip, width=0.95\linewidth]{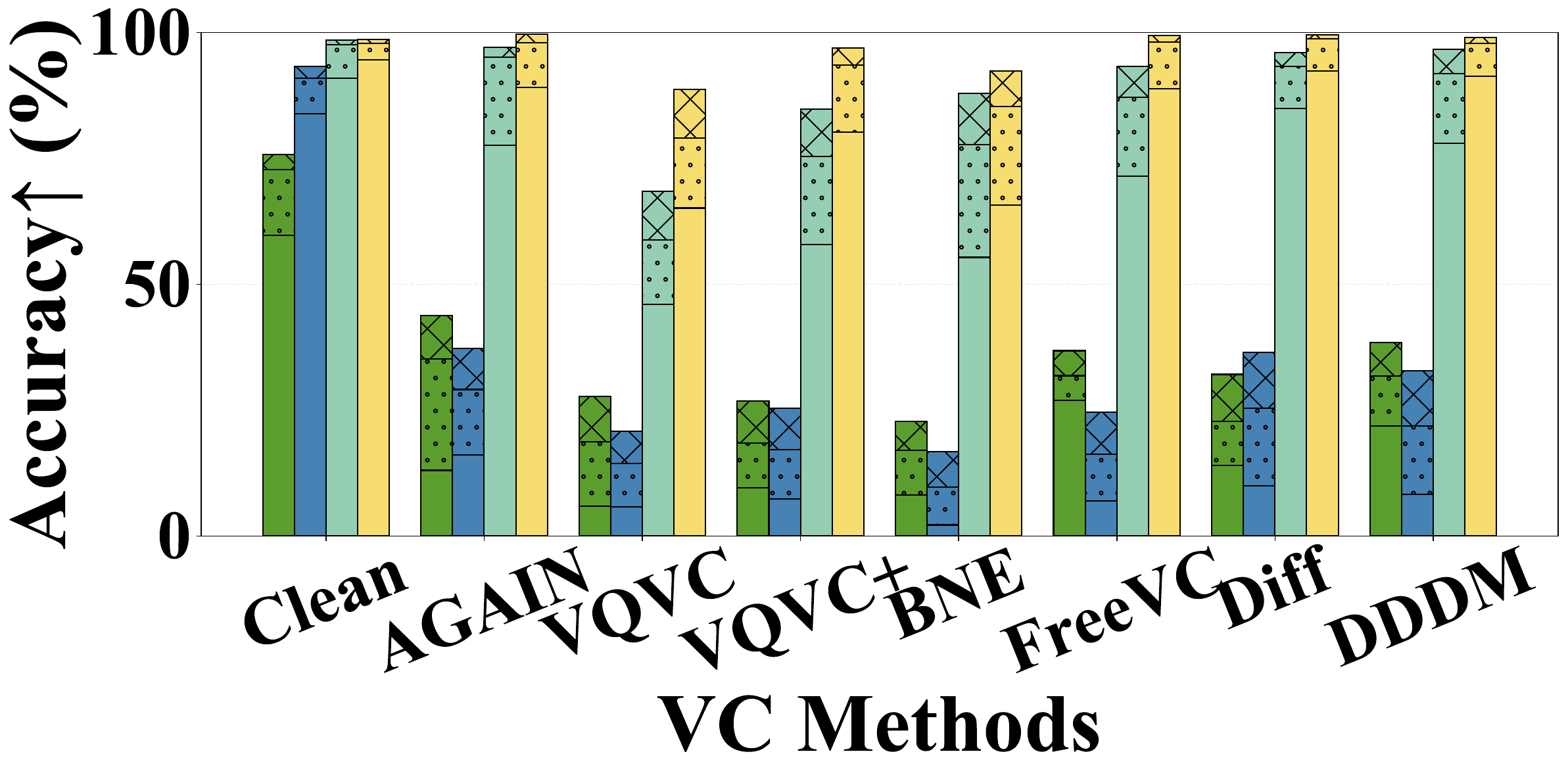}
        \end{minipage}
    }
    \hfill
    \hspace{-0.6cm}
    \subfigure[GSM-FR]{
        \begin{minipage}[b]{0.261\textwidth}
            \centering
            \includegraphics[trim=0mm 0mm 0mm 0mm, clip, width=0.95\linewidth]{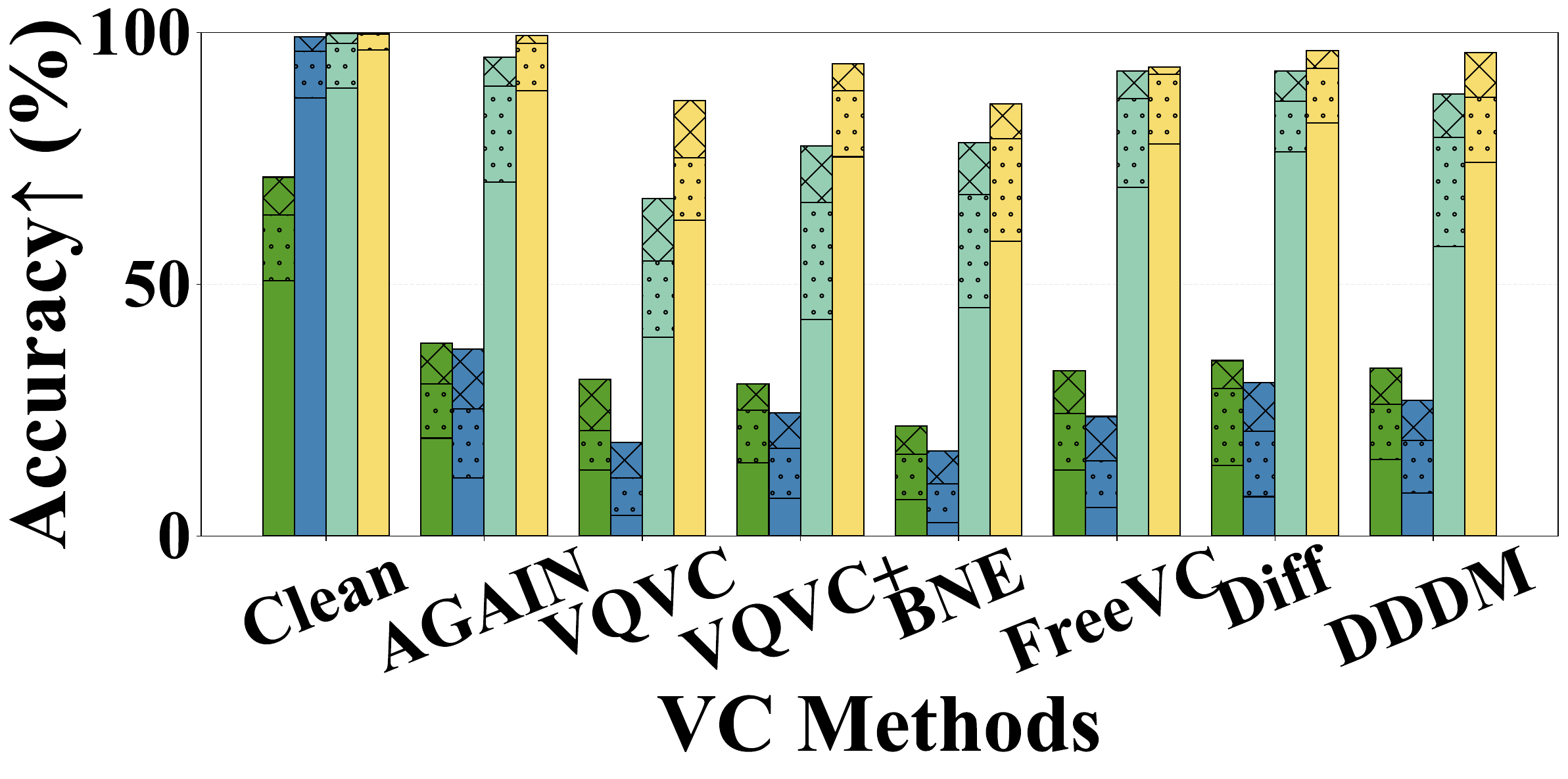}
        \end{minipage}
    }
    \hfill
    \hspace{-0.6cm}
    \subfigure[AMR-NB]{
        \begin{minipage}[b]{0.261\textwidth}
            \centering
            \includegraphics[trim=0mm 0mm 0mm 0mm, clip, width=0.95\linewidth]{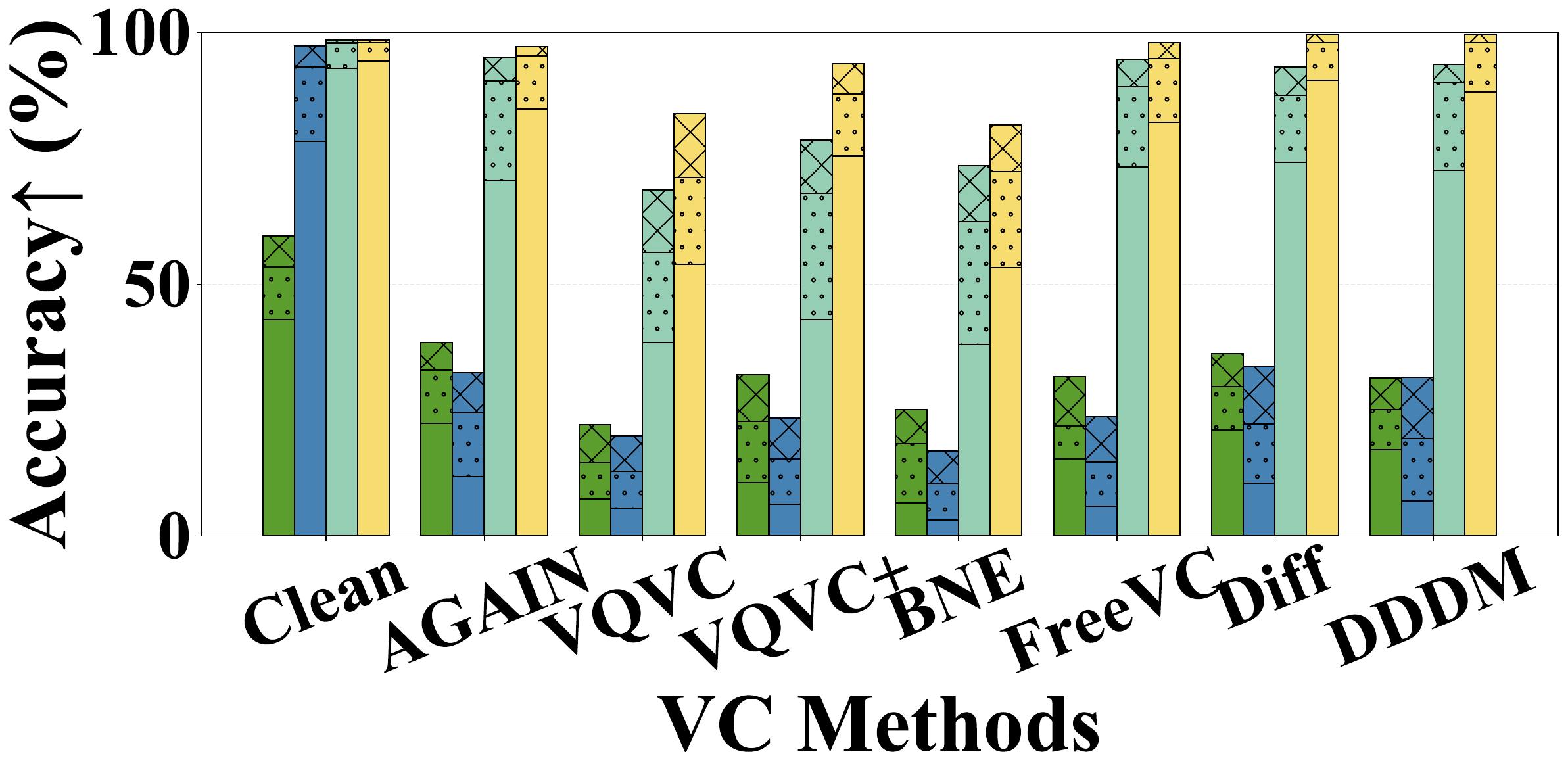}
        \end{minipage}
    }
    \caption{Performance of voiceprint recovery methods over telephony.}
    \label{fig:tel}
\end{figure*}

\begin{figure*}[h]
    \centering
    	\begin{minipage}[b]{0.48\linewidth}
    		\centering
    		\includegraphics[trim=0mm 0mm 0mm 0mm, clip, width=\textwidth]{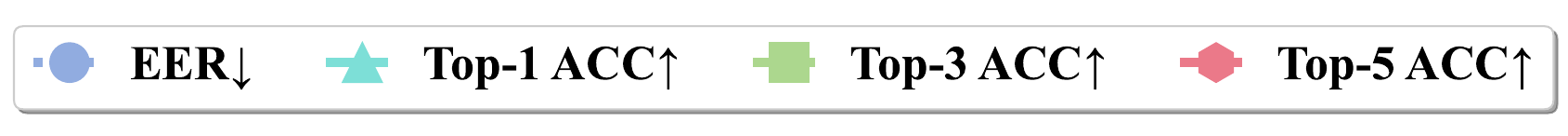}
    	\end{minipage} \\
    \vspace{-0.05cm}
    \subfigure[Clean]{
    	\begin{minipage}[b]{0.23\linewidth}
    		\centering
    		\includegraphics[trim=0mm 0mm 0mm 0mm, clip, width=0.95\textwidth]{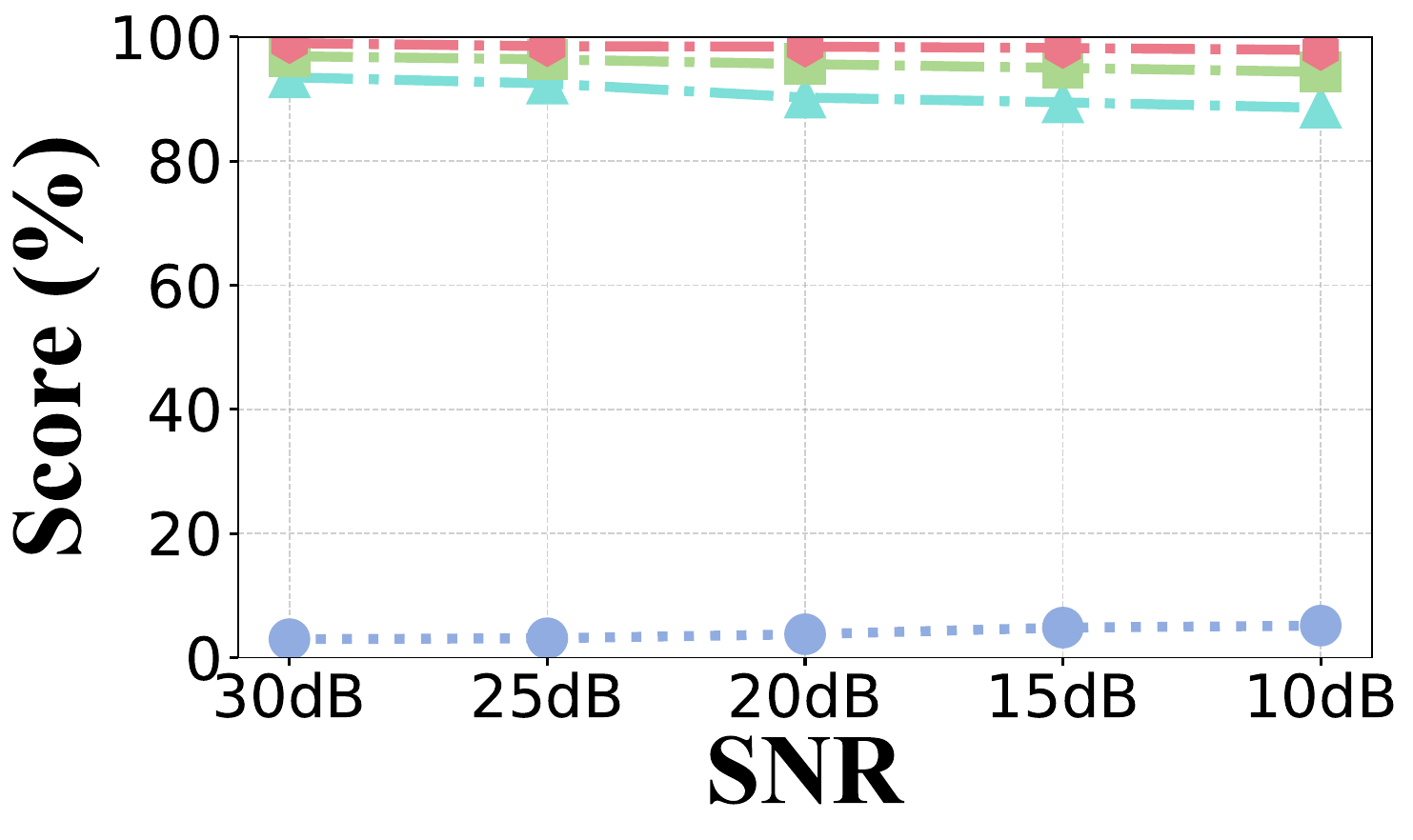}
    		\vspace{-0.1cm}
    	\end{minipage}
    }
    \subfigure[AGAIN]{
    	\begin{minipage}[b]{0.23\linewidth}
    		\centering
    		\includegraphics[trim=0mm 0mm 0mm 0mm, clip, width=0.95\textwidth]{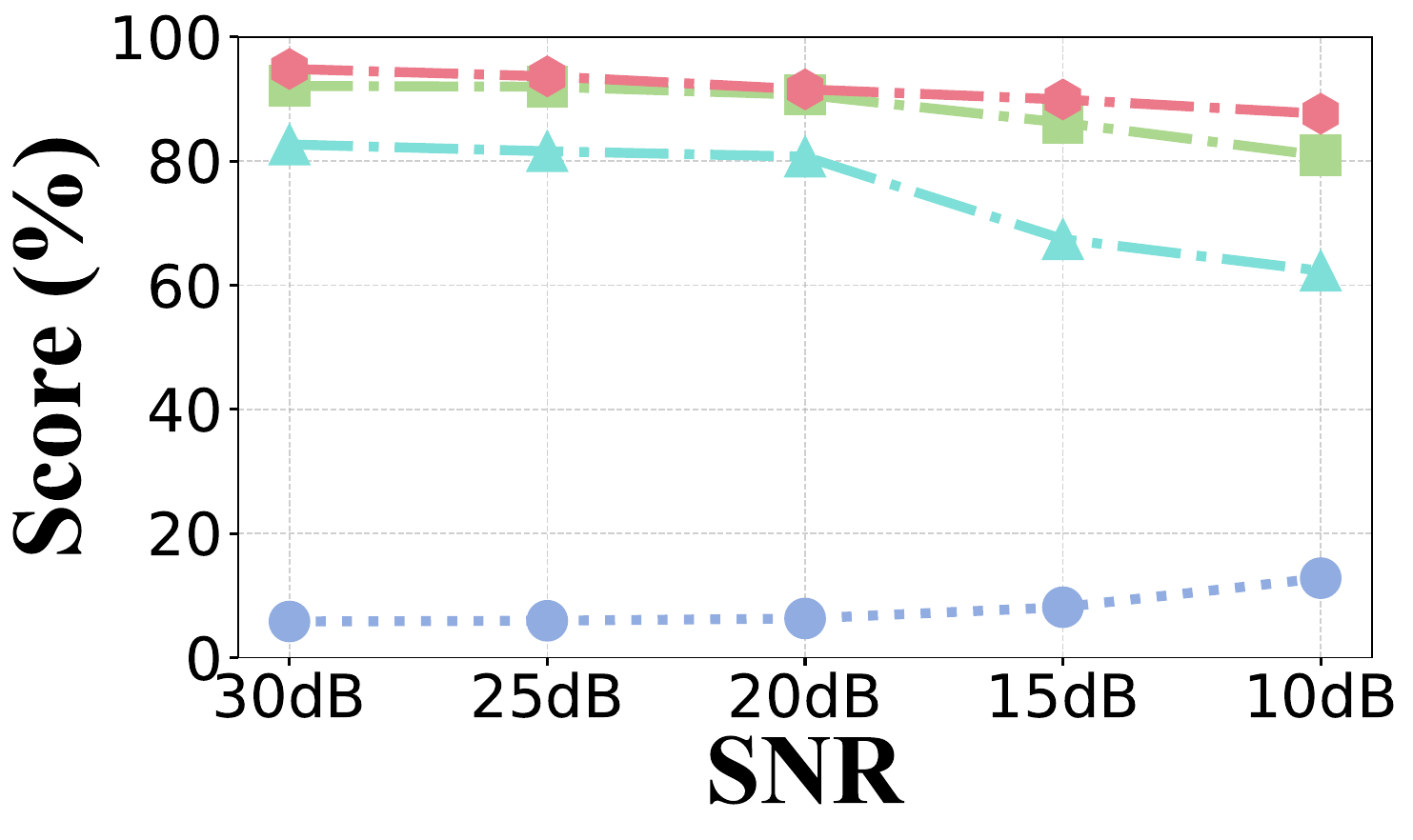}
    		\vspace{-0.1cm}
    	\end{minipage}
    }
    \subfigure[VQVC]{
    	\begin{minipage}[b]{0.23\linewidth}
    		\centering
    		\includegraphics[trim=0mm 0mm 0mm 0mm, clip, width=0.95\textwidth]{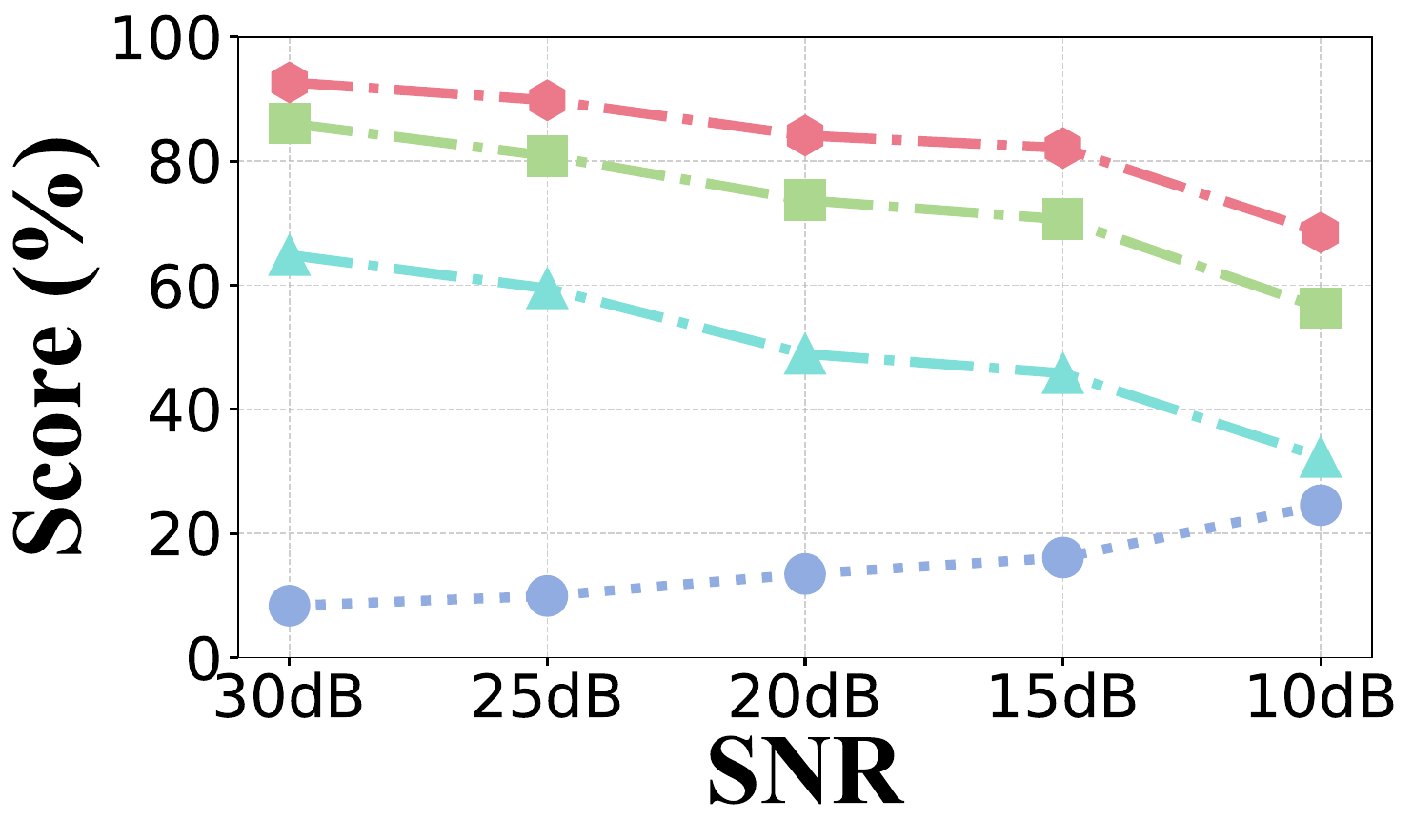}
    		\vspace{-0.1cm}
    	\end{minipage}
    }
  \vspace{-0.2cm}
    \subfigure[VQVC+]{
    	\begin{minipage}[b]{0.23\linewidth}
    		\centering
    		\includegraphics[trim=0mm 0mm 0mm 0mm, clip,width=0.95\textwidth]{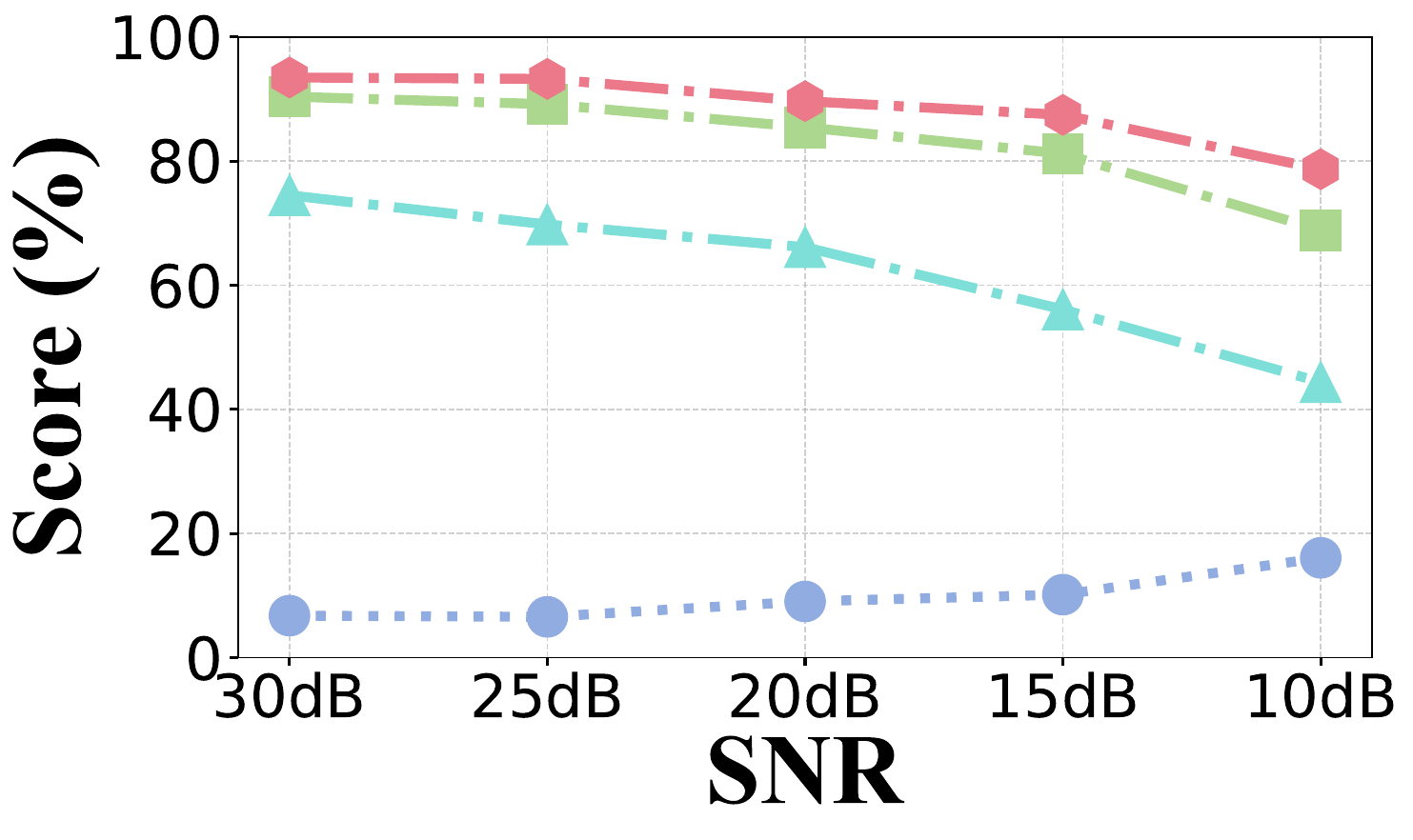}
    		\vspace{-0.1cm}
    	\end{minipage}
    }
    \subfigure[BNE]{
    	\begin{minipage}[b]{0.23\linewidth}
    		\centering
    		\includegraphics[trim=0mm 0mm 0mm 0mm, clip,width=0.95\textwidth]{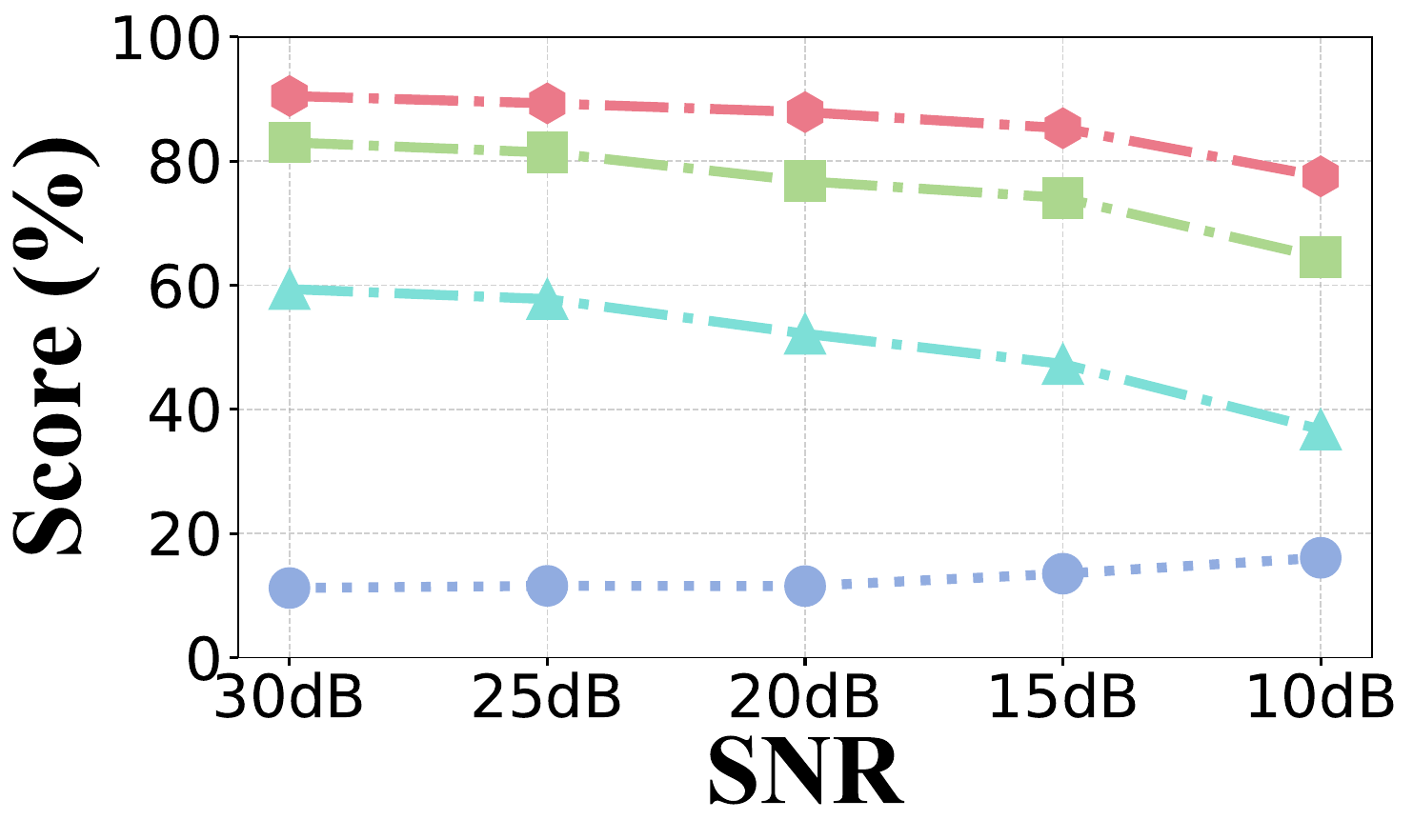}
    		\vspace{-0.1cm}
    	\end{minipage}
    }
    \subfigure[FreeVC]{
    	\begin{minipage}[b]{0.23\linewidth}
    		\centering
    		\includegraphics[trim=0mm 0mm 0mm 0mm, clip,width=0.95\textwidth]{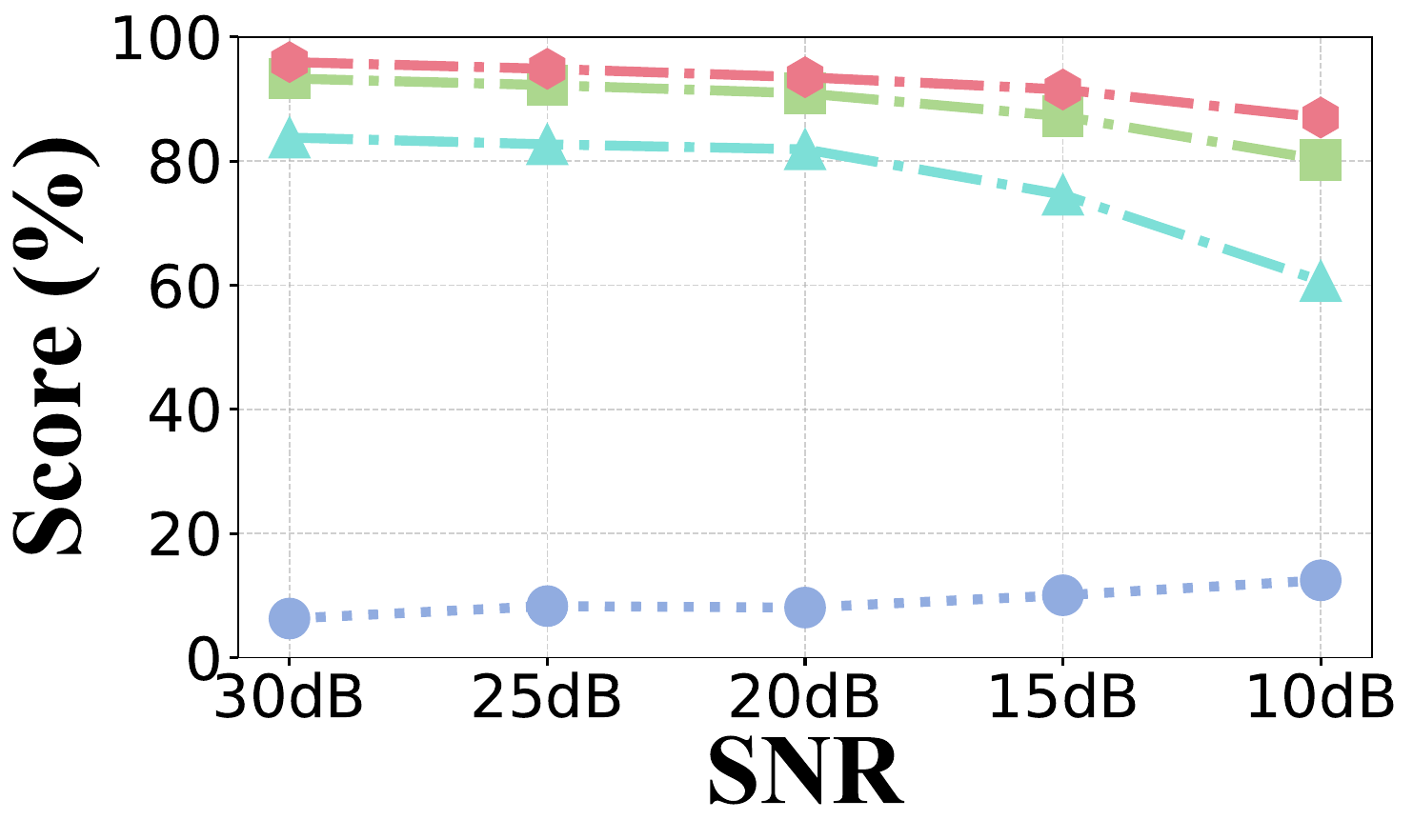}
    		\vspace{-0.1cm}
    	\end{minipage}
    }
    \subfigure[Diff]{
    	\begin{minipage}[b]{0.23\linewidth}
    		\centering
    		\includegraphics[trim=0mm 0mm 0mm 0mm, clip,width=0.95\textwidth]{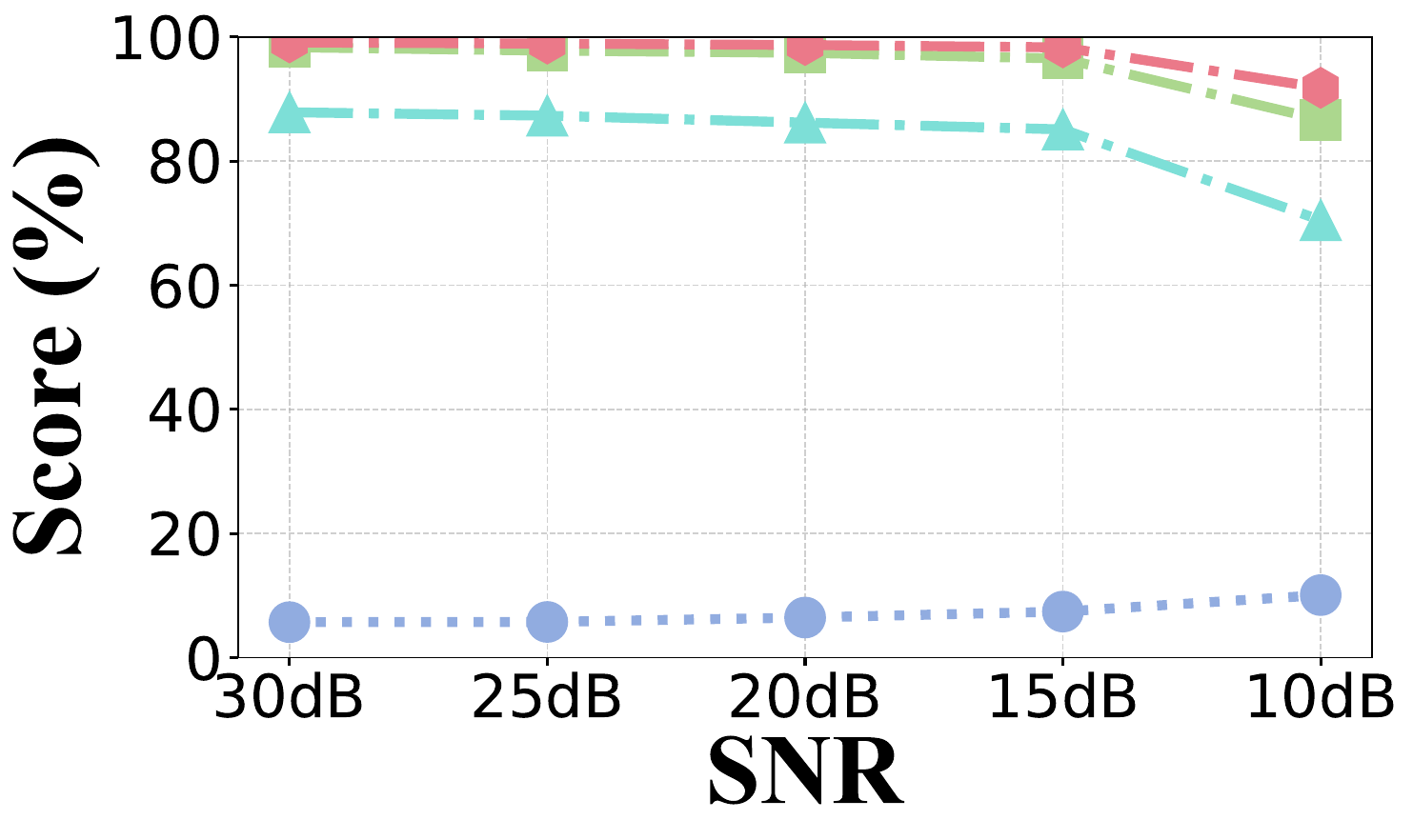}
    		\vspace{-0.1cm}
    	\end{minipage}
    }
    \subfigure[DDDM]{
    	\begin{minipage}[b]{0.23\linewidth}
    		\centering
    		\includegraphics[trim=0mm 0mm 0mm 0mm, clip,width=0.95\textwidth]{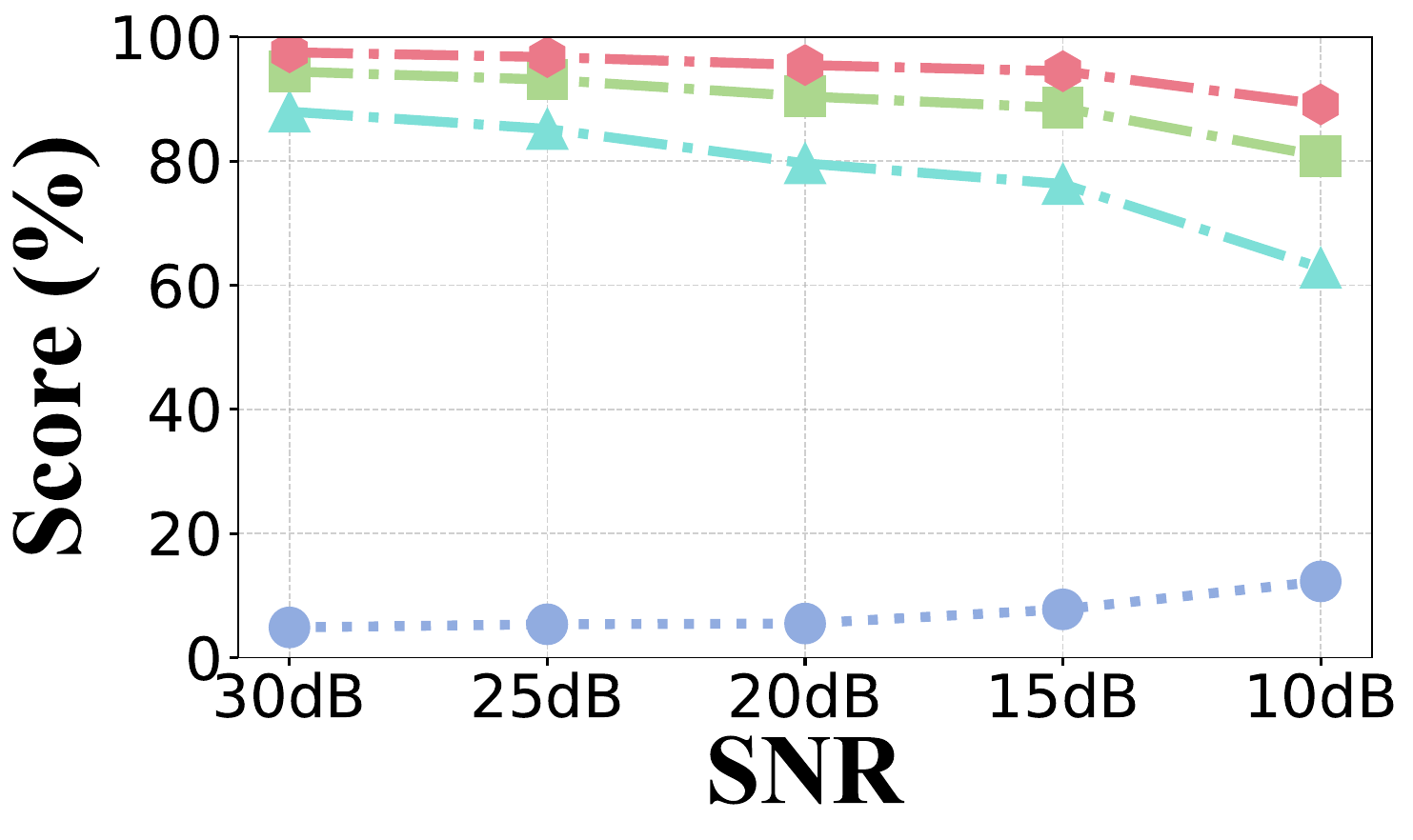}
    		\vspace{-0.1cm}
    	\end{minipage}
    }
 \vspace{-0.4cm}
	\caption{Performance of Revelio under additive white gaussian noise.}
	\label{fig:line_noise_r}
\end{figure*}

\begin{figure*}[h]
    \centering
    	\begin{minipage}[b]{0.48\linewidth}
    		\centering
    		\includegraphics[trim=0mm 0mm 0mm 0mm, clip, width=\textwidth]{Section/Pictures/Draw/Line/legend.pdf}
    	\end{minipage} \\
    \vspace{-0.05cm}
    \subfigure[Clean]{
    	\begin{minipage}[b]{0.23\linewidth}
    		\centering
    		\includegraphics[trim=0mm 0mm 0mm 0mm, clip, width=0.95\textwidth]{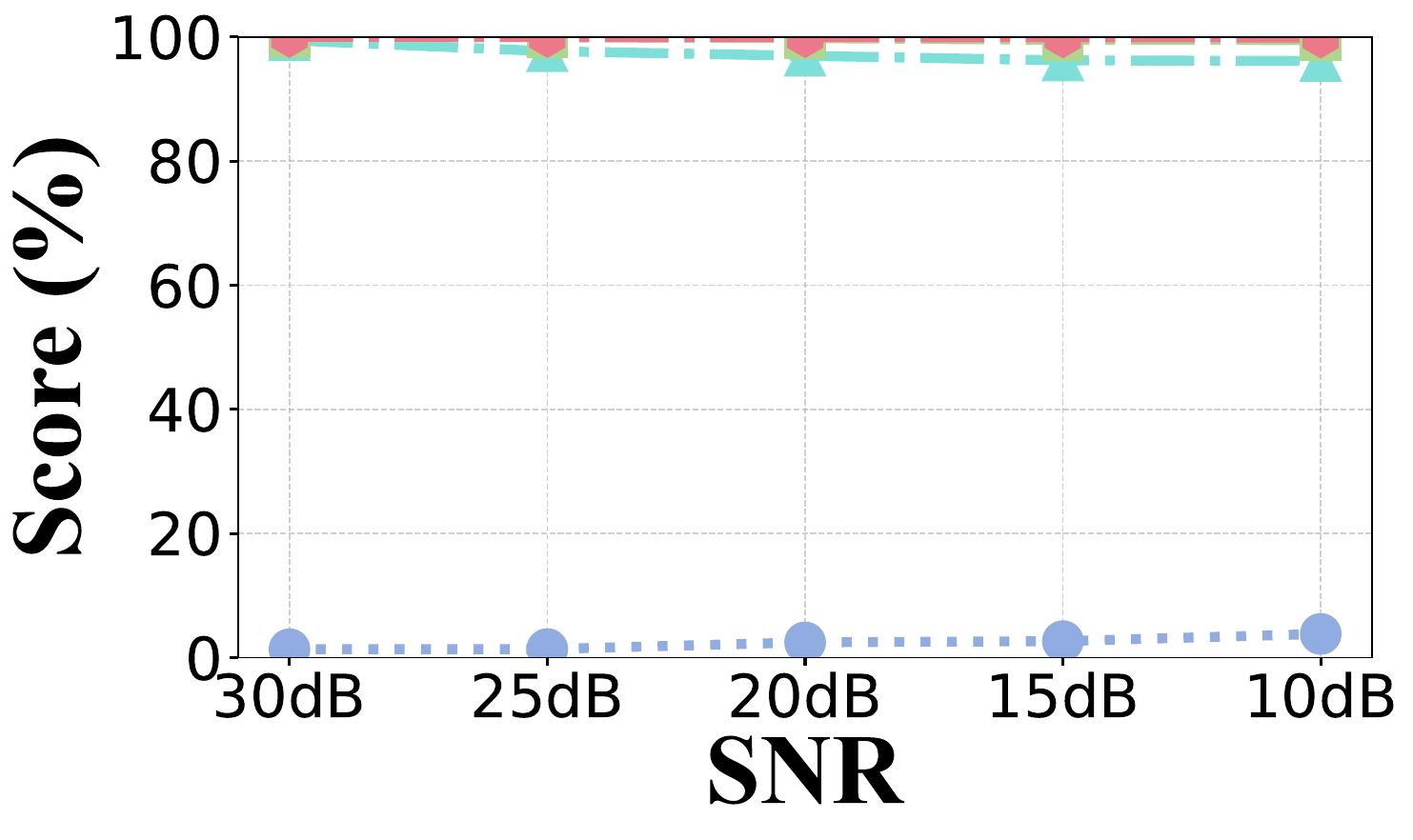}
    		\vspace{-0.1cm}
    	\end{minipage}
    }
    \subfigure[AGAIN]{
    	\begin{minipage}[b]{0.23\linewidth}
    		\centering
    		\includegraphics[trim=0mm 0mm 0mm 0mm, clip, width=0.95\textwidth]{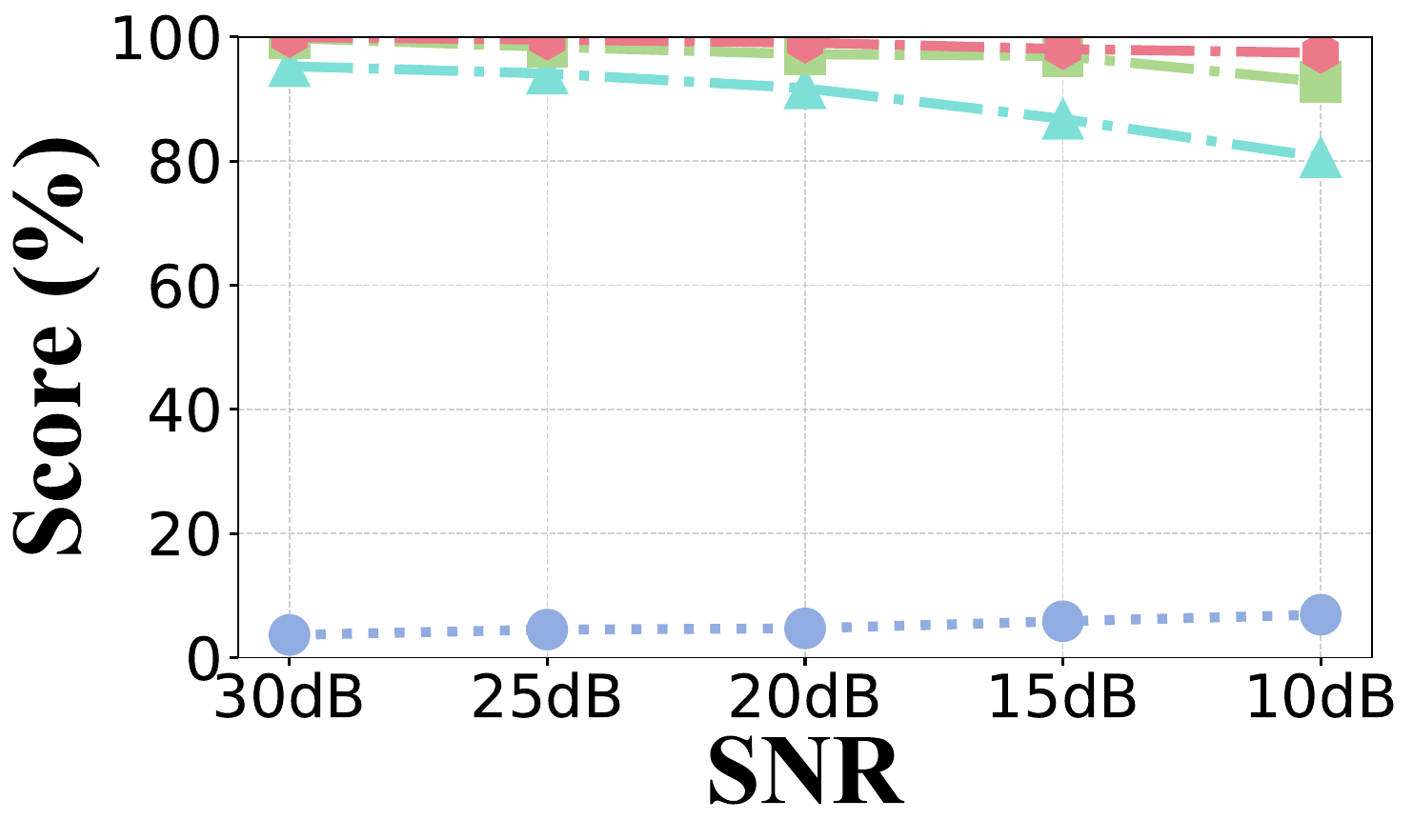}
    		\vspace{-0.1cm}
    	\end{minipage}
    }
    \subfigure[VQVC]{
    	\begin{minipage}[b]{0.23\linewidth}
    		\centering
    		\includegraphics[trim=0mm 0mm 0mm 0mm, clip, width=0.95\textwidth]{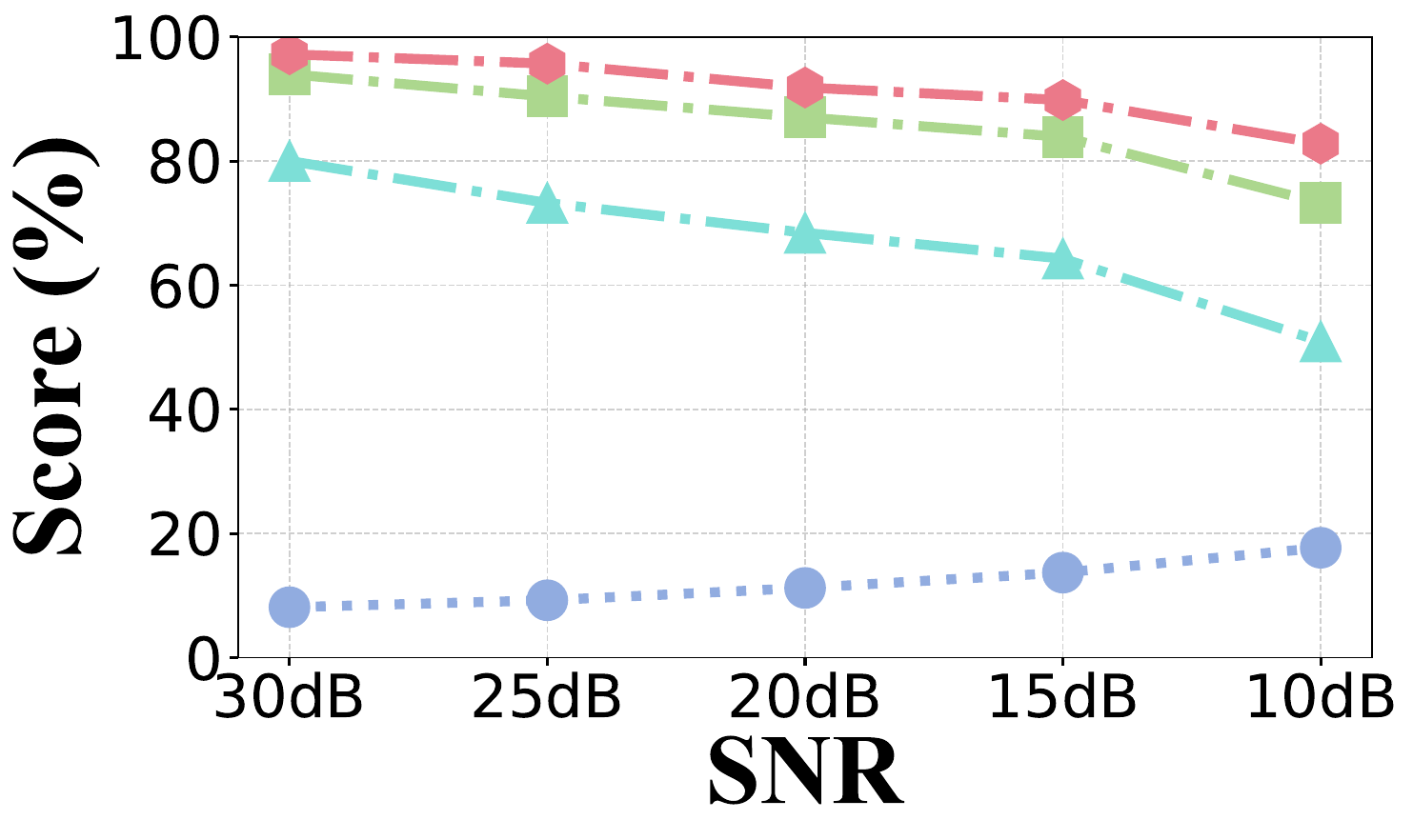}
    		\vspace{-0.1cm}
    	\end{minipage}
    }
  \vspace{-0.2cm}
    \subfigure[VQVC+]{
    	\begin{minipage}[b]{0.23\linewidth}
    		\centering
    		\includegraphics[trim=0mm 0mm 0mm 0mm, clip,width=0.95\textwidth]{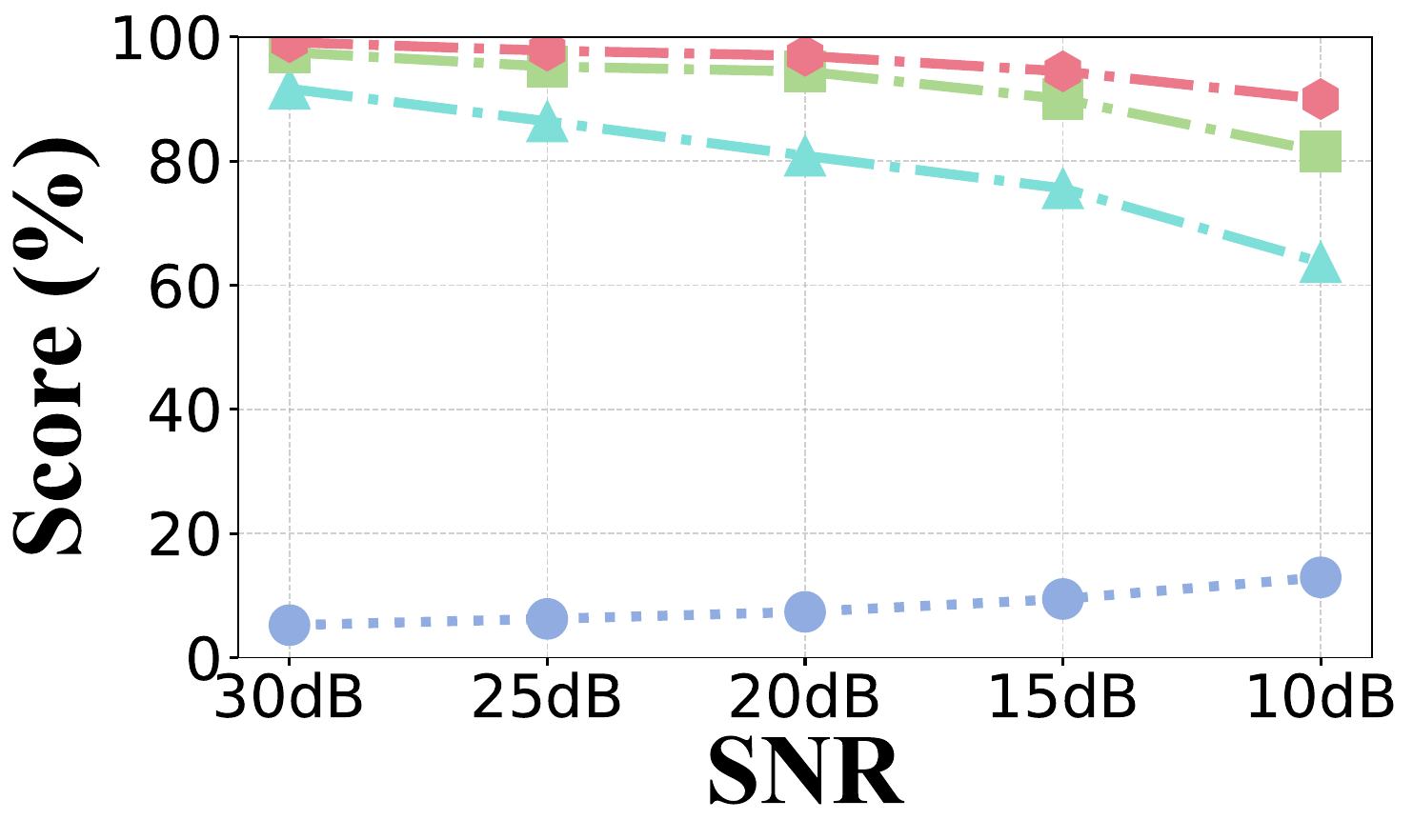}
    		\vspace{-0.1cm}
    	\end{minipage}
    }
    \subfigure[BNE]{
    	\begin{minipage}[b]{0.23\linewidth}
    		\centering
    		\includegraphics[trim=0mm 0mm 0mm 0mm, clip,width=0.95\textwidth]{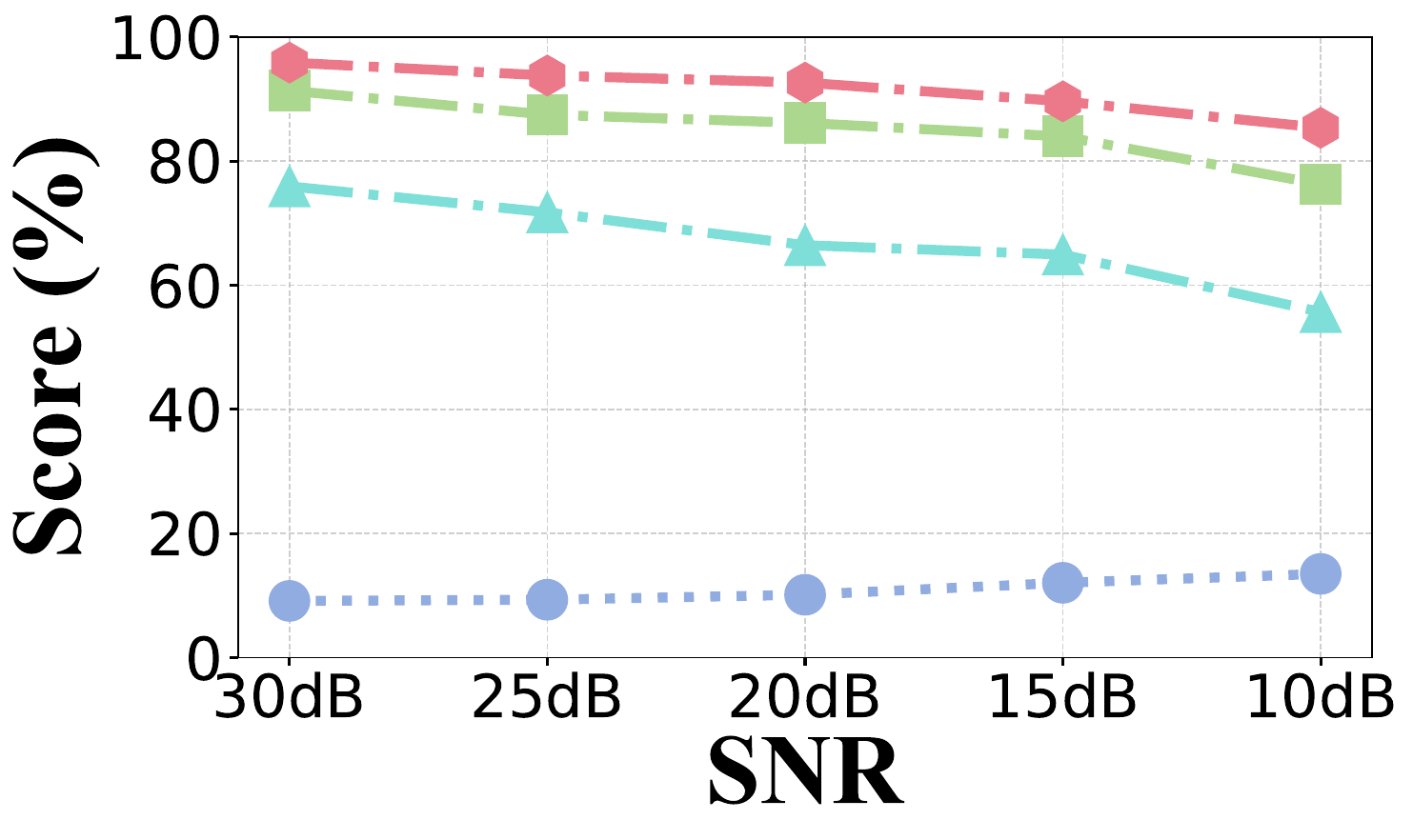}
    		\vspace{-0.1cm}
    	\end{minipage}
    }
    \subfigure[FreeVC]{
    	\begin{minipage}[b]{0.23\linewidth}
    		\centering
    		\includegraphics[trim=0mm 0mm 0mm 0mm, clip,width=0.95\textwidth]{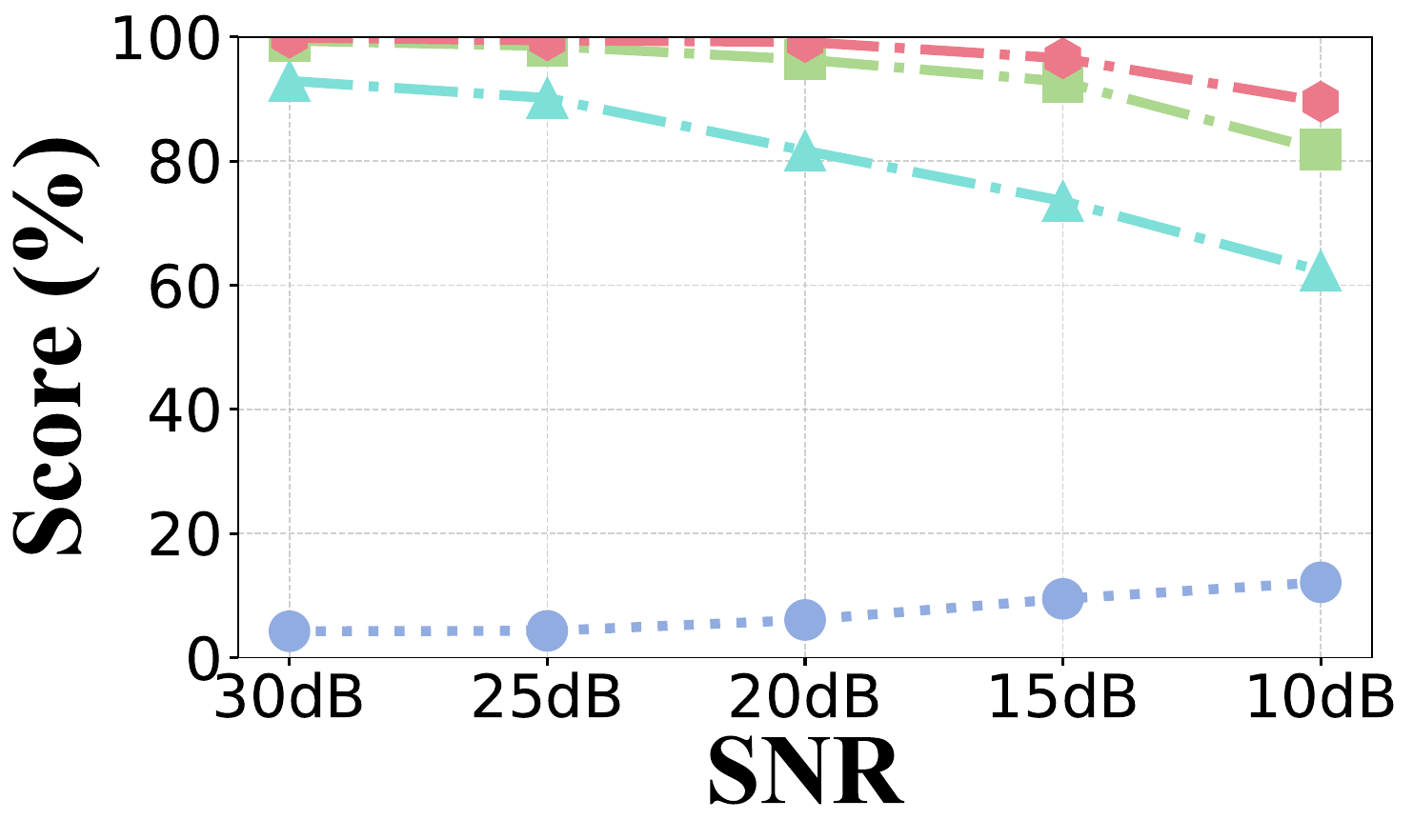}
    		\vspace{-0.1cm}
    	\end{minipage}
    }
    \subfigure[Diff]{
    	\begin{minipage}[b]{0.23\linewidth}
    		\centering
    		\includegraphics[trim=0mm 0mm 0mm 0mm, clip,width=0.95\textwidth]{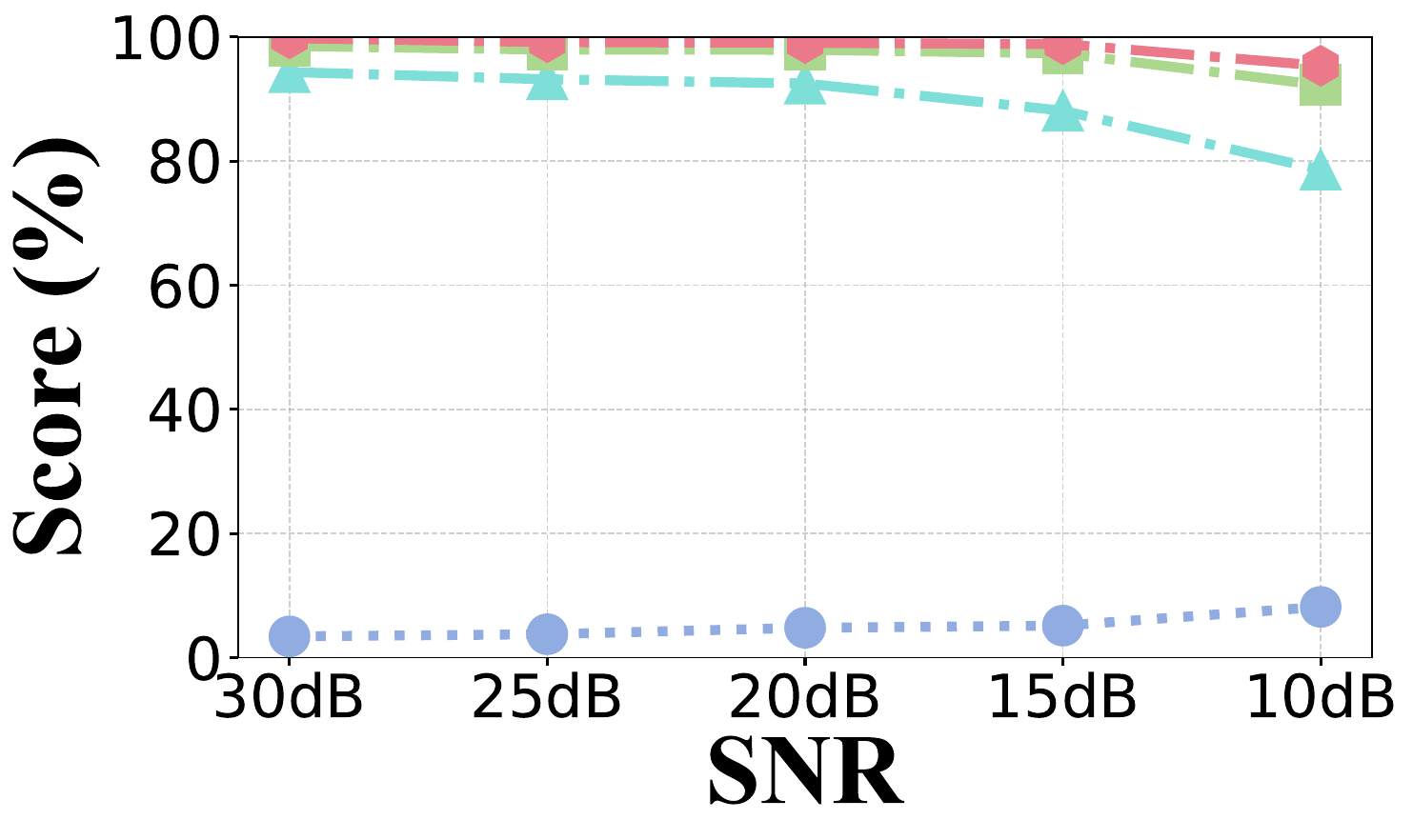}
    		\vspace{-0.1cm}
    	\end{minipage}
    }
    \subfigure[DDDM]{
    	\begin{minipage}[b]{0.23\linewidth}
    		\centering
    		\includegraphics[trim=0mm 0mm 0mm 0mm, clip,width=0.95\textwidth]{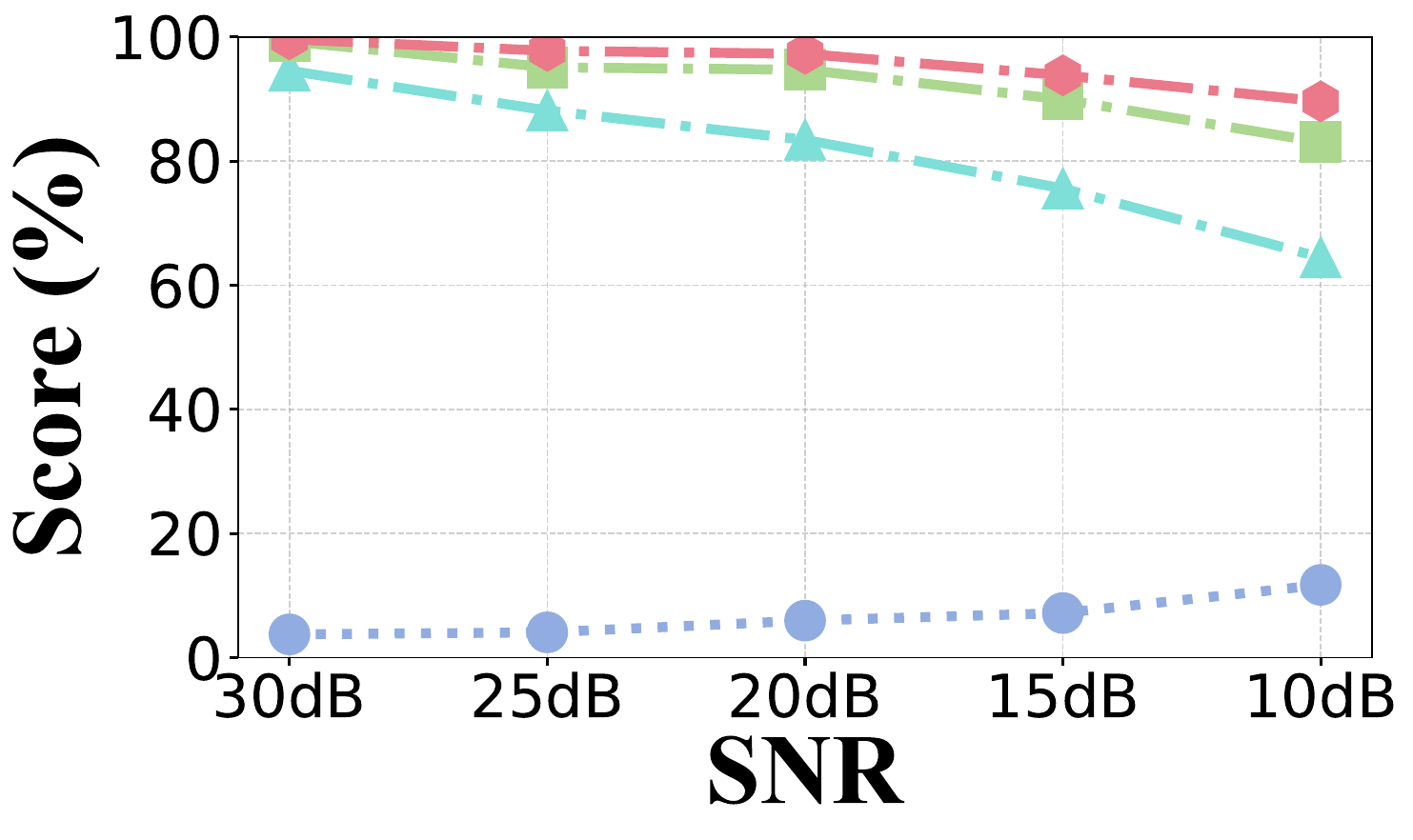}
    		\vspace{-0.1cm}
    	\end{minipage}
    }
 \vspace{-0.4cm}
	\caption{Performance of \sys under additive white gaussian noise.}
	\label{fig:line_noise}
\end{figure*}

\begin{figure*}[h]
    \centering
    	\begin{minipage}[b]{0.48\linewidth}
    		\centering
    		\includegraphics[trim=0mm 0mm 0mm 0mm, clip, width=\textwidth]{Section/Pictures/Draw/Line/legend.pdf}
    	\end{minipage} \\
    \vspace{-0.05cm}
    \subfigure[Clean]{
    	\begin{minipage}[b]{0.23\linewidth}
    		\centering
    		\includegraphics[trim=0mm 0mm 0mm 0mm, clip, width=0.95\textwidth]{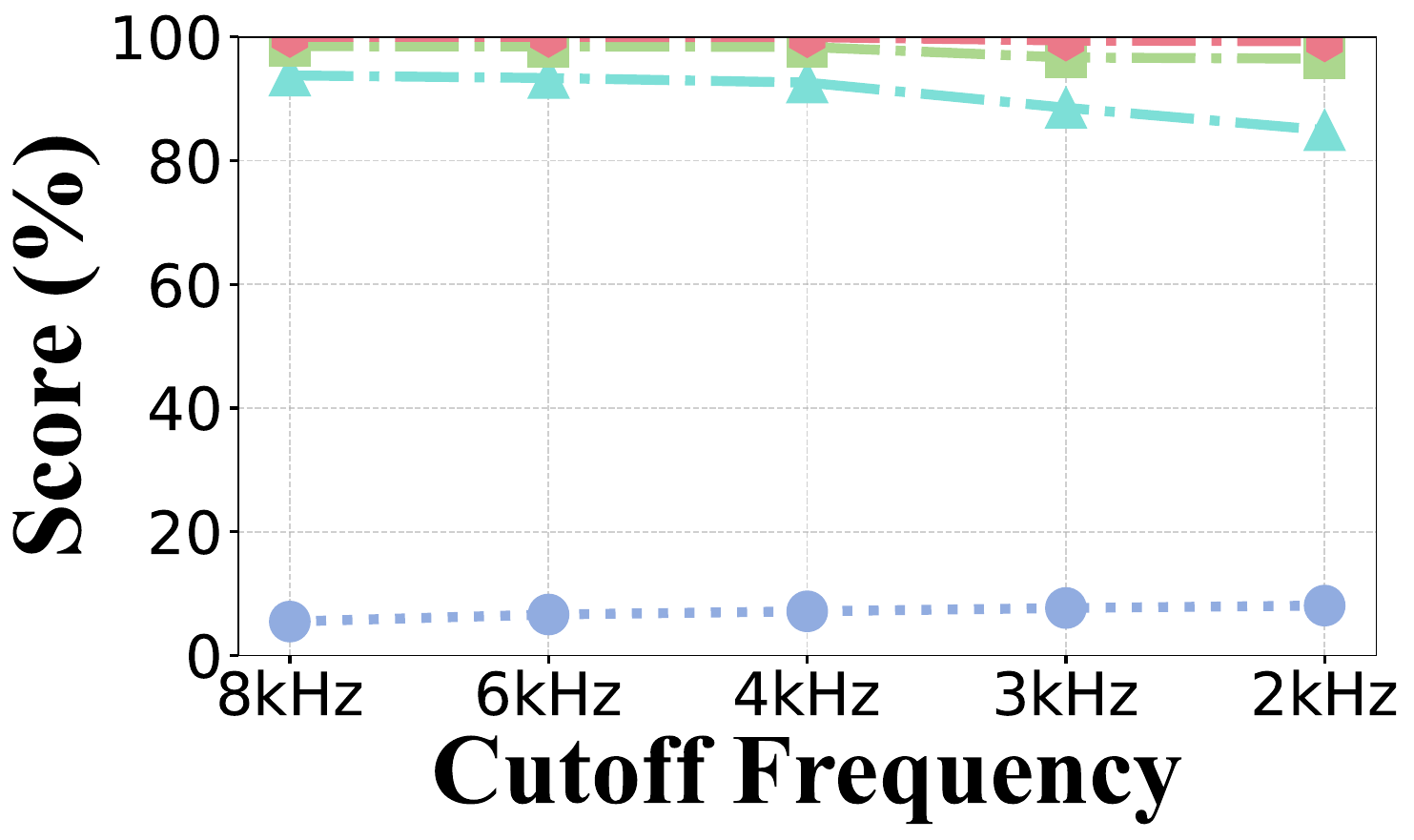}
    		\vspace{-0.1cm}
    	\end{minipage}
    }
    \subfigure[AGAIN]{
    	\begin{minipage}[b]{0.23\linewidth}
    		\centering
    		\includegraphics[trim=0mm 0mm 0mm 0mm, clip, width=0.95\textwidth]{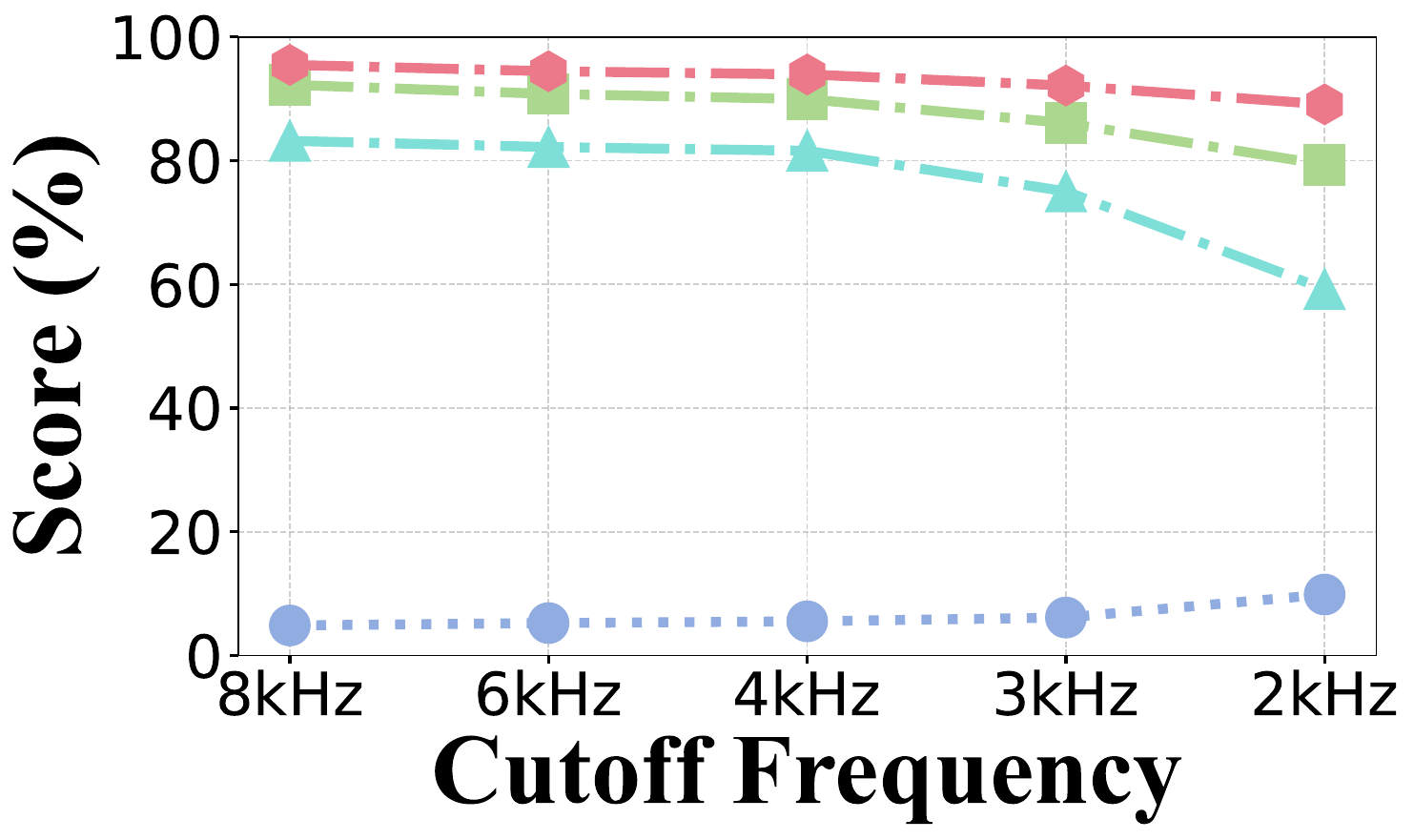}
    		\vspace{-0.1cm}
    	\end{minipage}
    }
    \subfigure[VQVC]{
    	\begin{minipage}[b]{0.23\linewidth}
    		\centering
    		\includegraphics[trim=0mm 0mm 0mm 0mm, clip, width=0.95\textwidth]{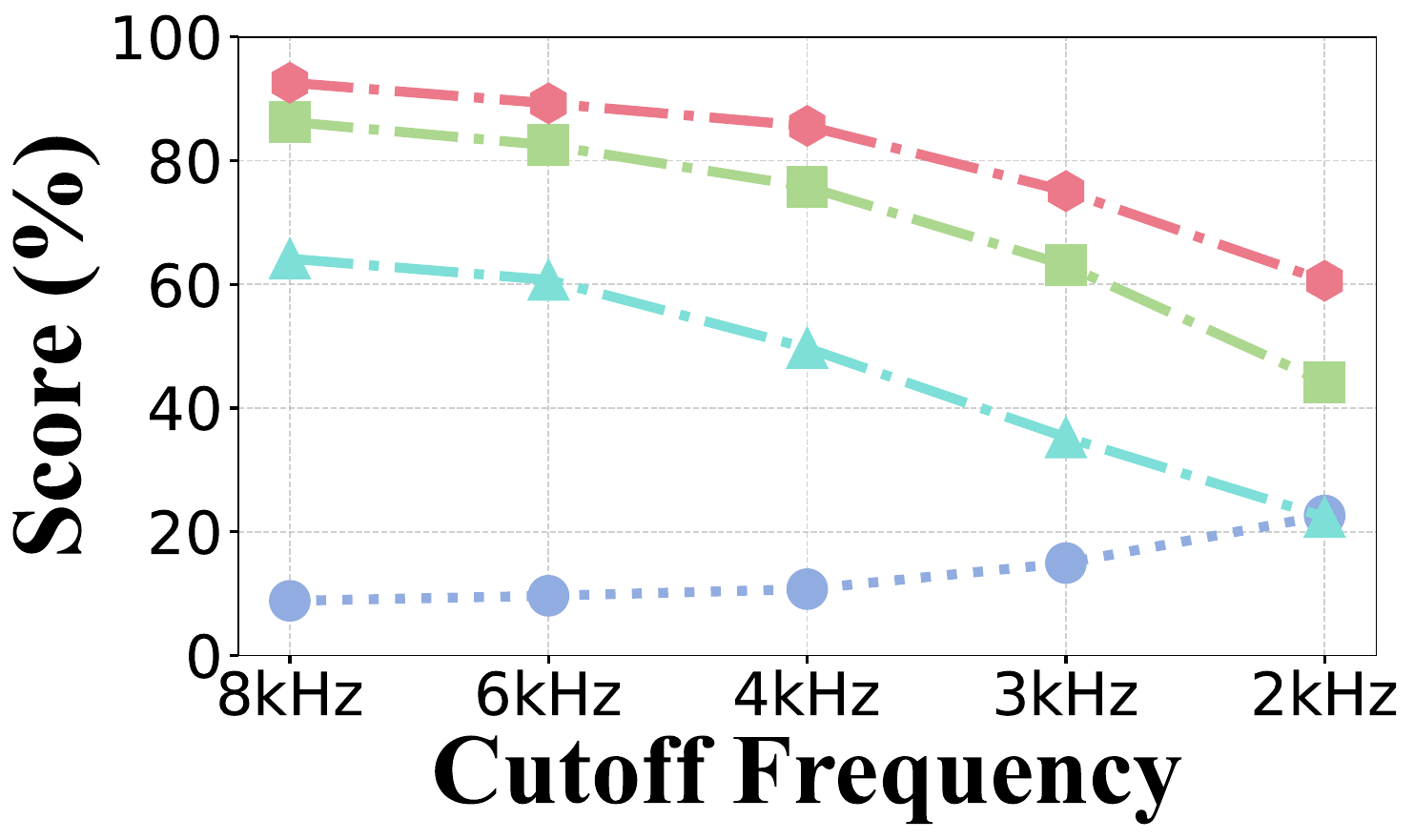}
    		\vspace{-0.1cm}
    	\end{minipage}
    }
  \vspace{-0.2cm}
    \subfigure[VQVC+]{
    	\begin{minipage}[b]{0.23\linewidth}
    		\centering
    		\includegraphics[trim=0mm 0mm 0mm 0mm, clip,width=0.95\textwidth]{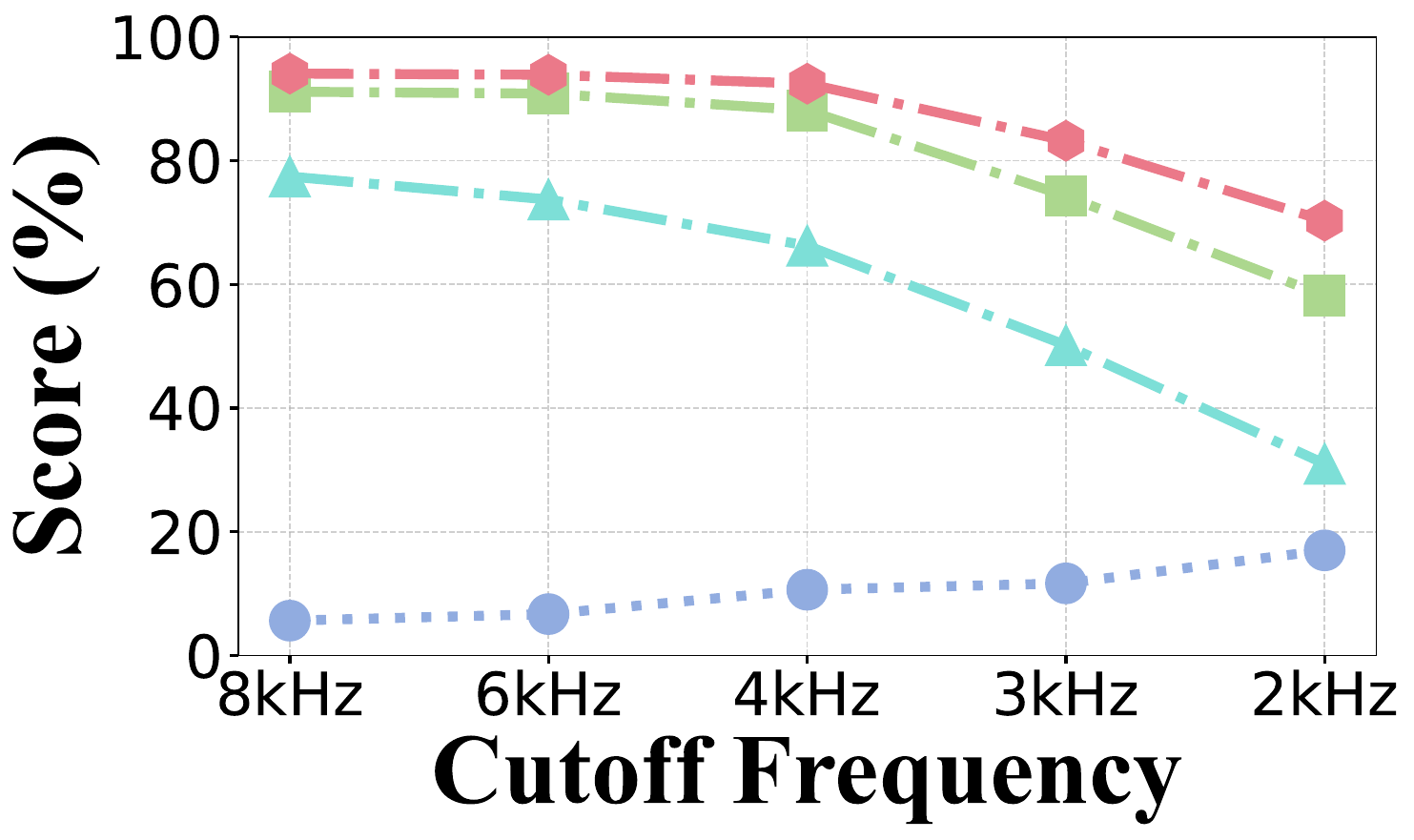}
    		\vspace{-0.1cm}
    	\end{minipage}
    }
    \subfigure[BNE]{
    	\begin{minipage}[b]{0.23\linewidth}
    		\centering
    		\includegraphics[trim=0mm 0mm 0mm 0mm, clip,width=0.95\textwidth]{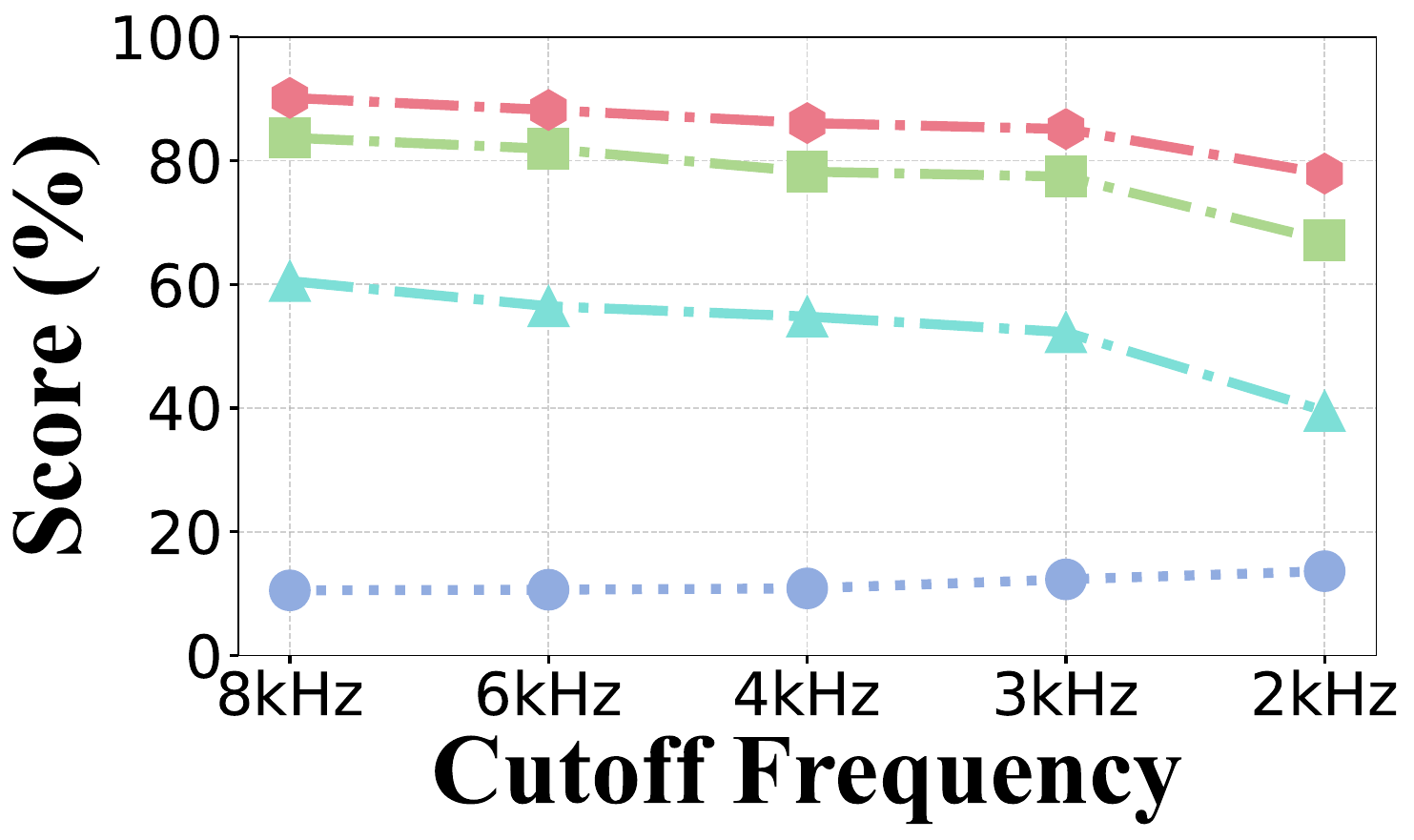}
    		\vspace{-0.1cm}
    	\end{minipage}
    }
    \subfigure[FreeVC]{
    	\begin{minipage}[b]{0.23\linewidth}
    		\centering
    		\includegraphics[trim=0mm 0mm 0mm 0mm, clip,width=0.95\textwidth]{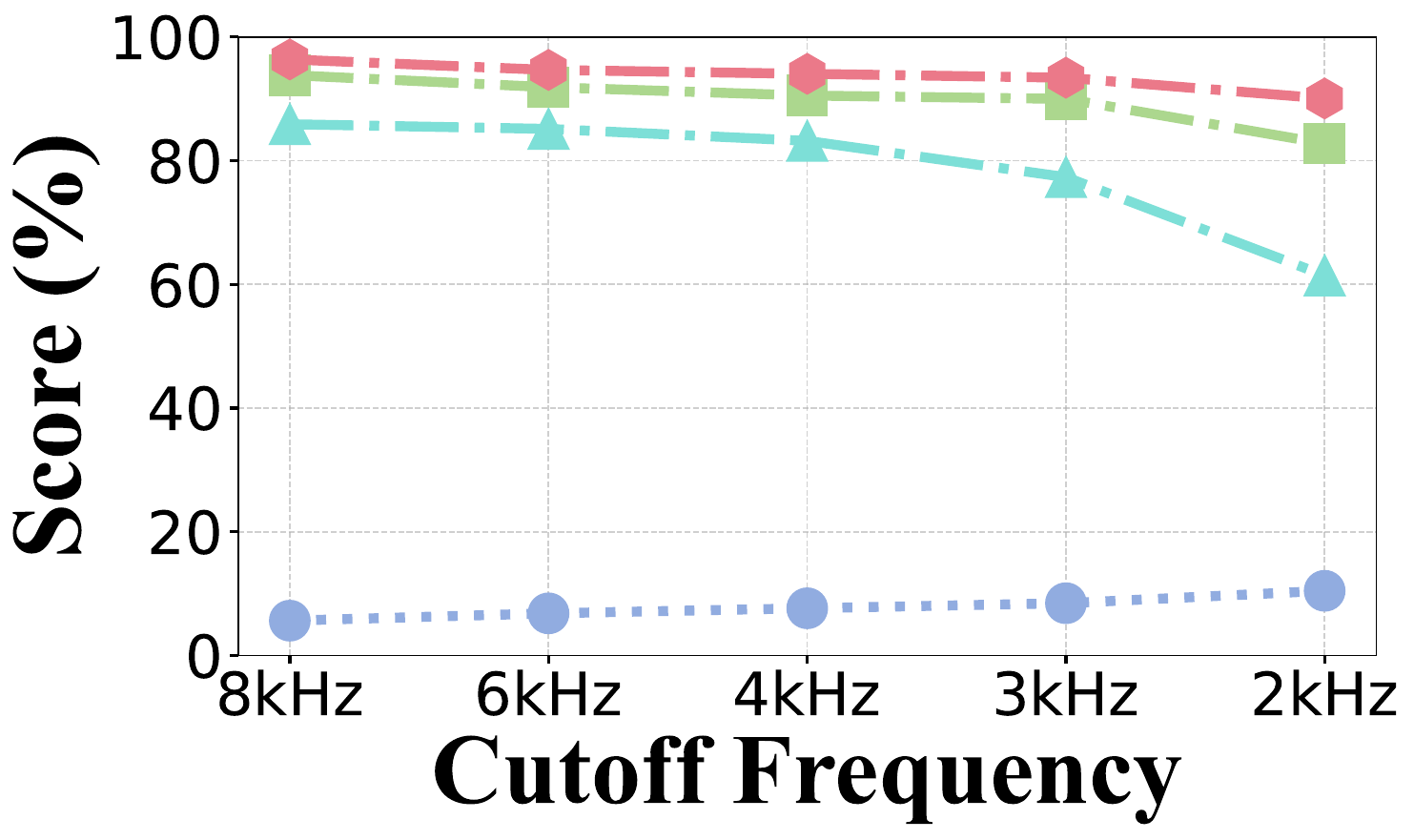}
    		\vspace{-0.1cm}
    	\end{minipage}
    }
    \subfigure[Diff]{
    	\begin{minipage}[b]{0.23\linewidth}
    		\centering
    		\includegraphics[trim=0mm 0mm 0mm 0mm, clip,width=0.95\textwidth]{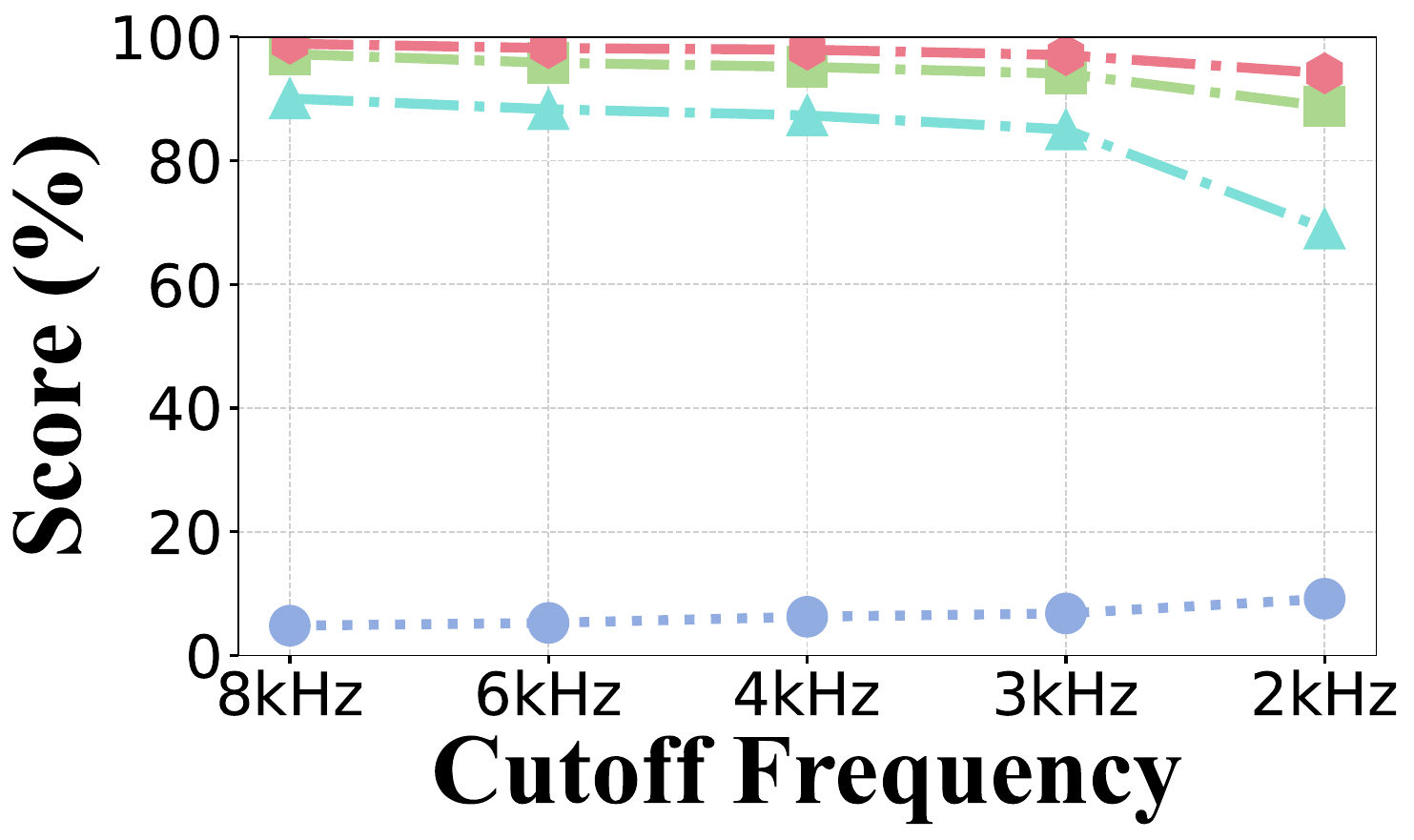}
    		\vspace{-0.1cm}
    	\end{minipage}
    }
    \subfigure[DDDM]{
    	\begin{minipage}[b]{0.23\linewidth}
    		\centering
    		\includegraphics[trim=0mm 0mm 0mm 0mm, clip,width=0.95\textwidth]{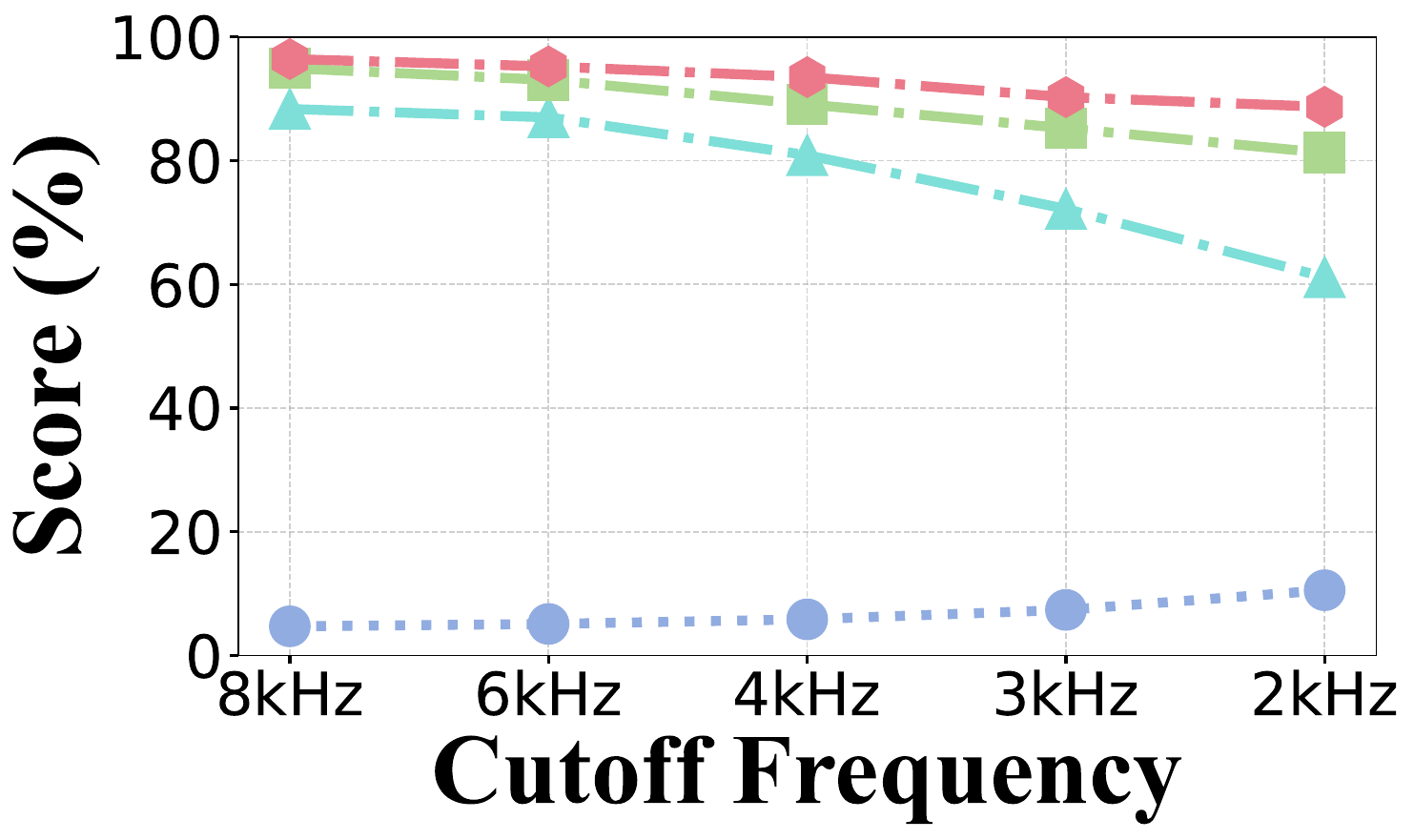}
    		\vspace{-0.1cm}
    	\end{minipage}
    }
 \vspace{-0.3cm}
	\caption{Performance of Revelio under low-pass filtering.}
	\label{fig:line_filter_r}
\end{figure*}

\begin{figure*}[h]
    \centering
    	\begin{minipage}[b]{0.48\linewidth}
    		\centering
    		\includegraphics[trim=0mm 0mm 0mm 0mm, clip, width=\textwidth]{Section/Pictures/Draw/Line/legend.pdf}
    	\end{minipage} \\
    \vspace{-0.05cm}
    \subfigure[Clean]{
    	\begin{minipage}[b]{0.23\linewidth}
    		\centering
    		\includegraphics[trim=0mm 0mm 0mm 0mm, clip, width=0.95\textwidth]{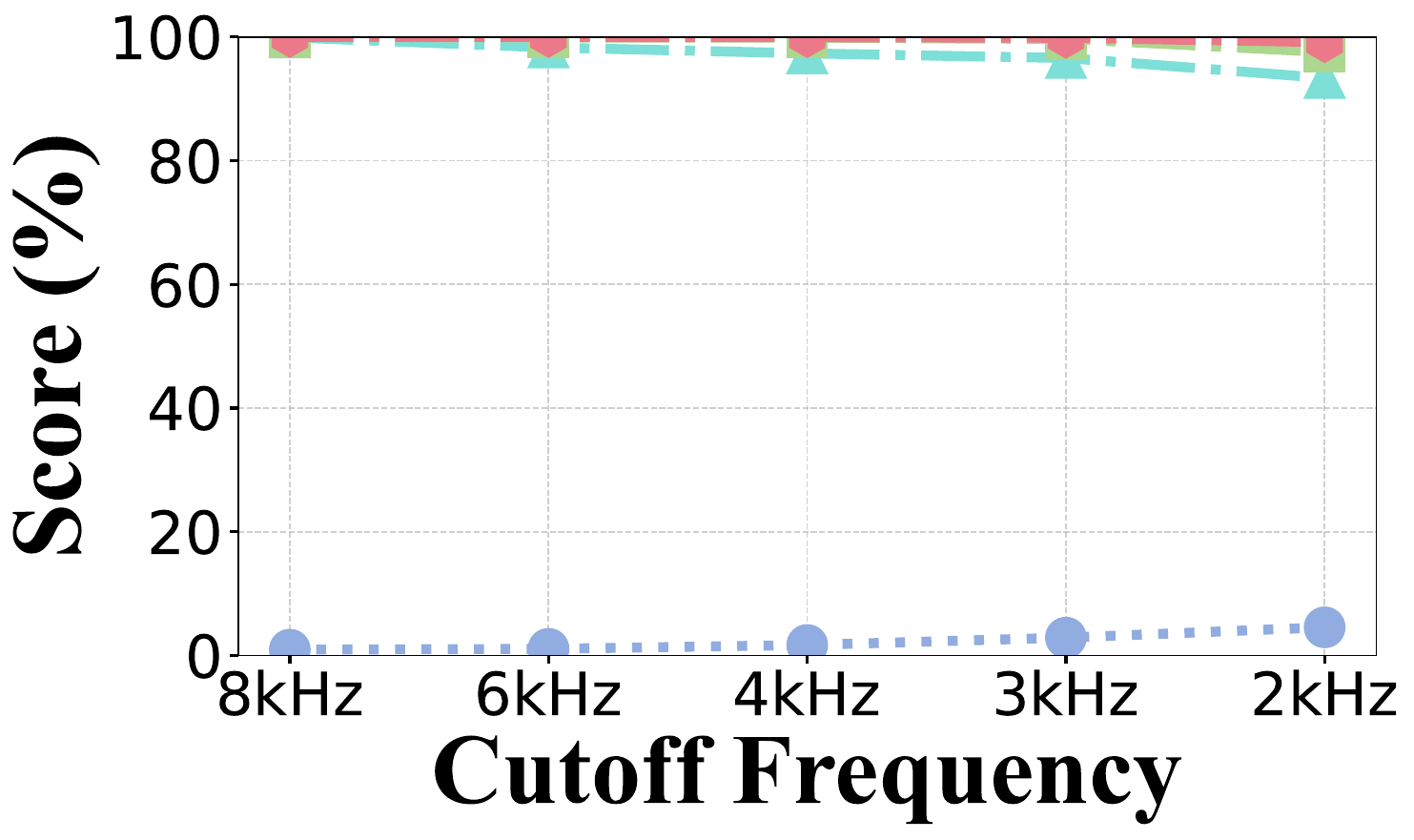}
    		\vspace{-0.1cm}
    	\end{minipage}
    }
    \subfigure[AGAIN]{
    	\begin{minipage}[b]{0.23\linewidth}
    		\centering
    		\includegraphics[trim=0mm 0mm 0mm 0mm, clip, width=0.95\textwidth]{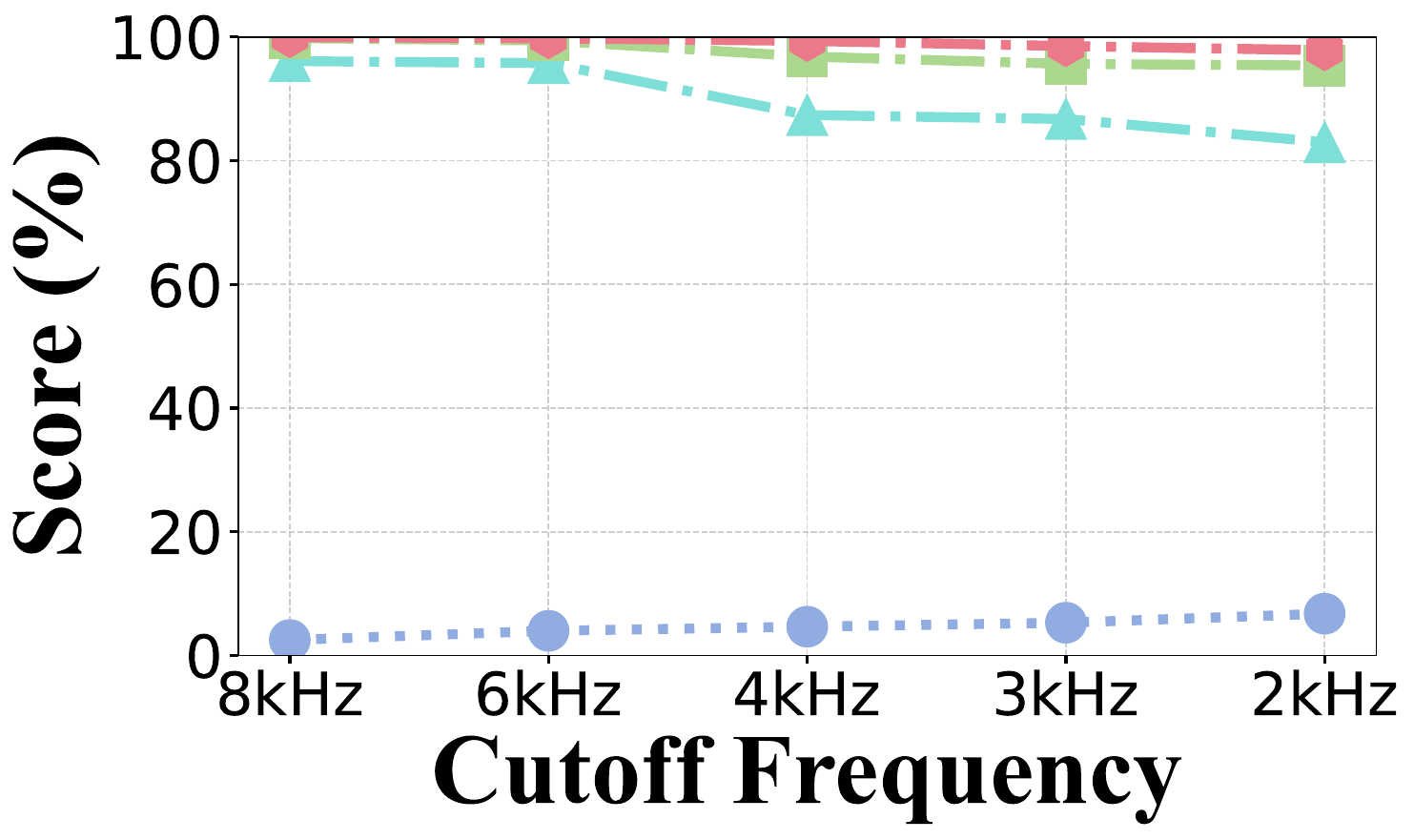}
    		\vspace{-0.1cm}
    	\end{minipage}
    }
    \subfigure[VQVC]{
    	\begin{minipage}[b]{0.23\linewidth}
    		\centering
    		\includegraphics[trim=0mm 0mm 0mm 0mm, clip, width=0.95\textwidth]{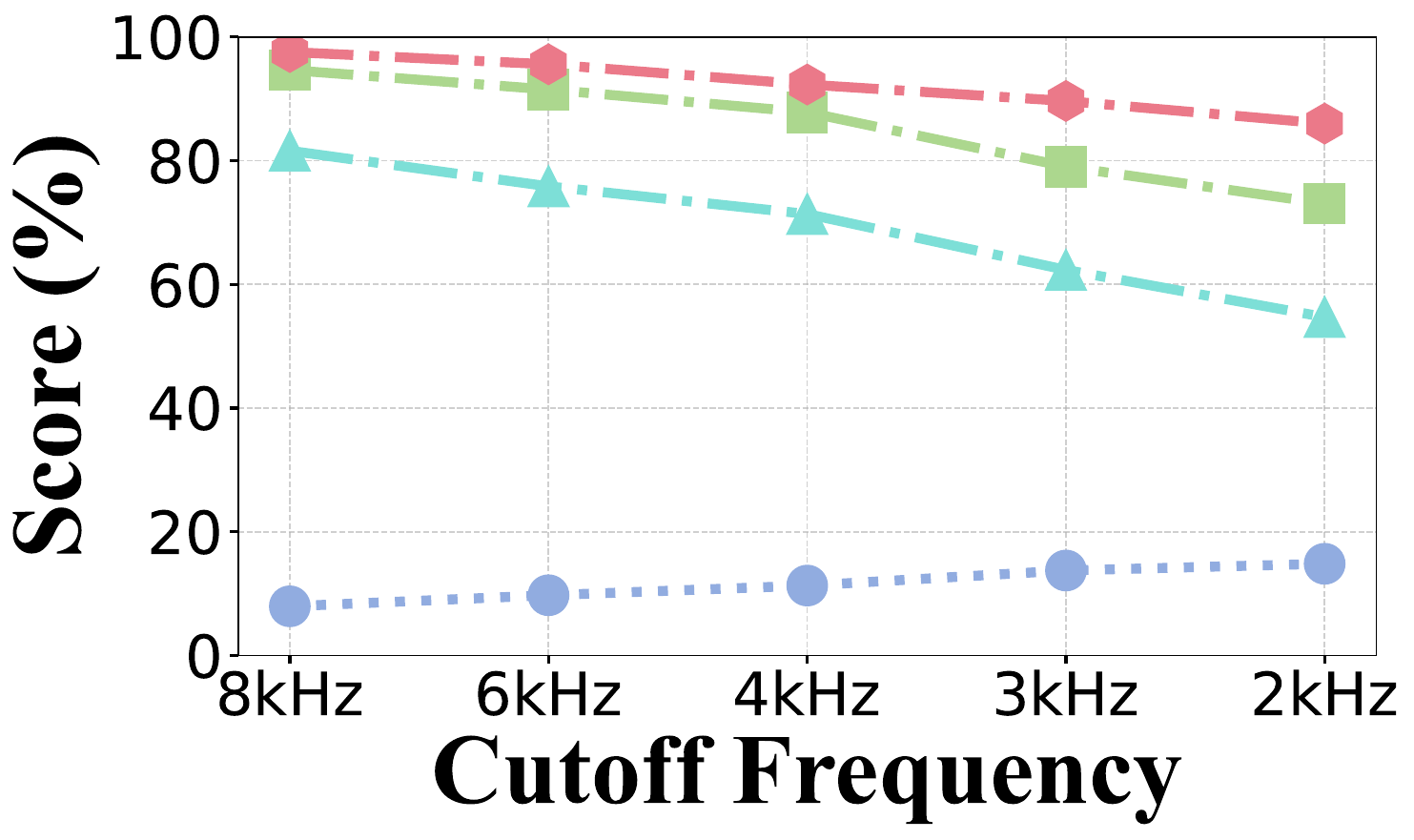}
    		\vspace{-0.1cm}
    	\end{minipage}
    }
  \vspace{-0.2cm}
    \subfigure[VQVC+]{
    	\begin{minipage}[b]{0.23\linewidth}
    		\centering
    		\includegraphics[trim=0mm 0mm 0mm 0mm, clip,width=0.95\textwidth]{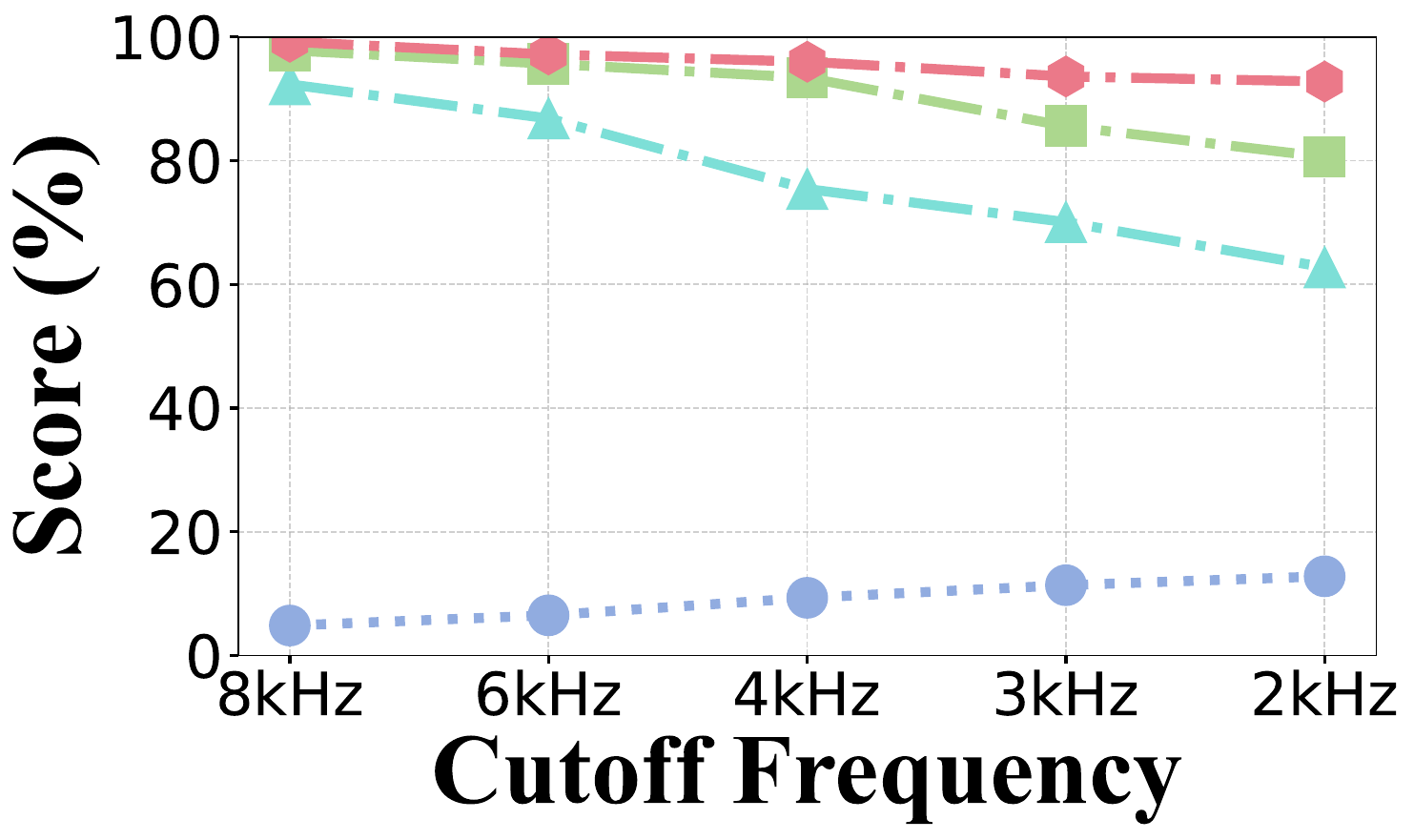}
    		\vspace{-0.1cm}
    	\end{minipage}
    }
    \subfigure[BNE]{
    	\begin{minipage}[b]{0.23\linewidth}
    		\centering
    		\includegraphics[trim=0mm 0mm 0mm 0mm, clip,width=0.95\textwidth]{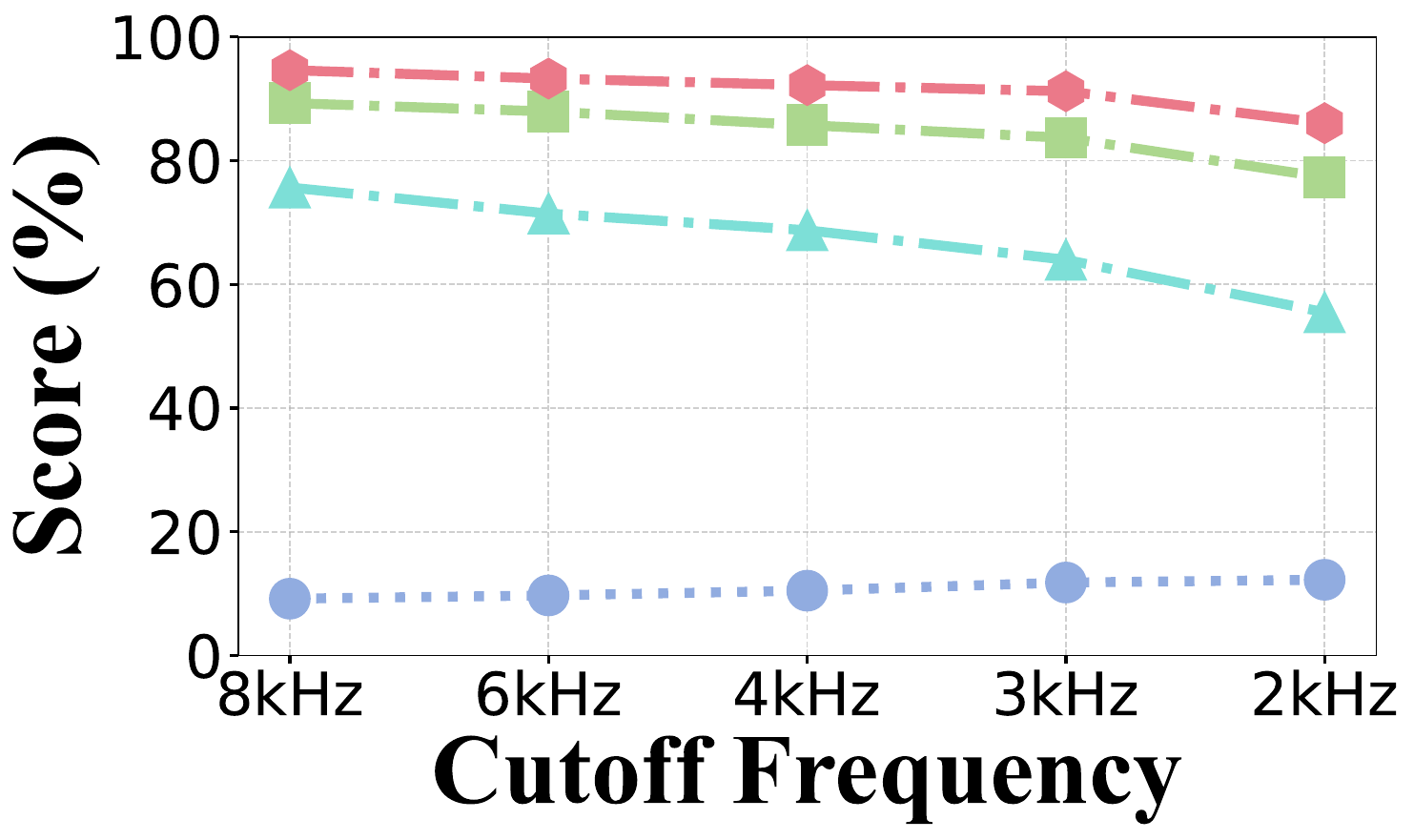}
    		\vspace{-0.1cm}
    	\end{minipage}
    }
    \subfigure[FreeVC]{
    	\begin{minipage}[b]{0.23\linewidth}
    		\centering
    		\includegraphics[trim=0mm 0mm 0mm 0mm, clip,width=0.95\textwidth]{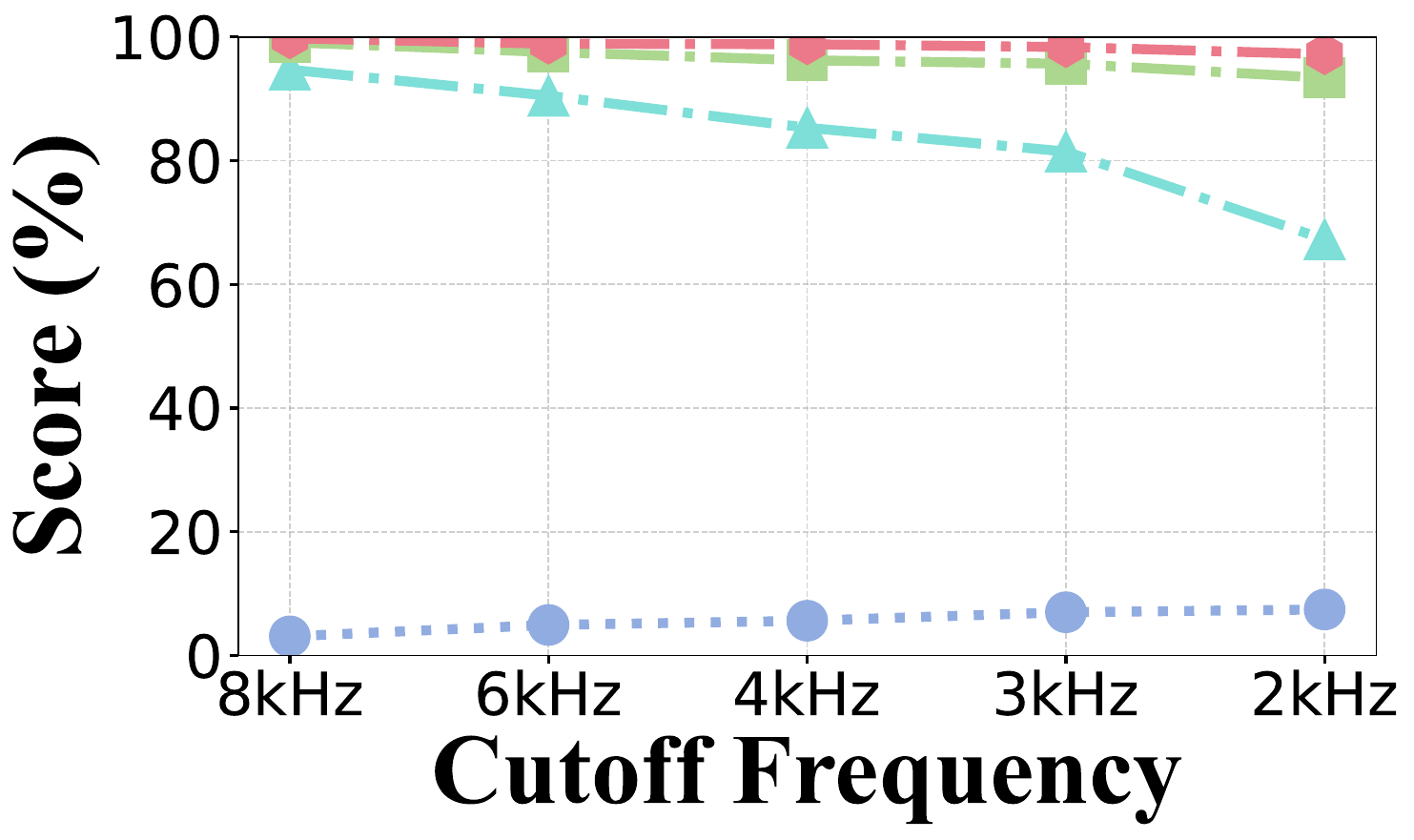}
    		\vspace{-0.1cm}
    	\end{minipage}
    }
    \subfigure[Diff]{
    	\begin{minipage}[b]{0.23\linewidth}
    		\centering
    		\includegraphics[trim=0mm 0mm 0mm 0mm, clip,width=0.95\textwidth]{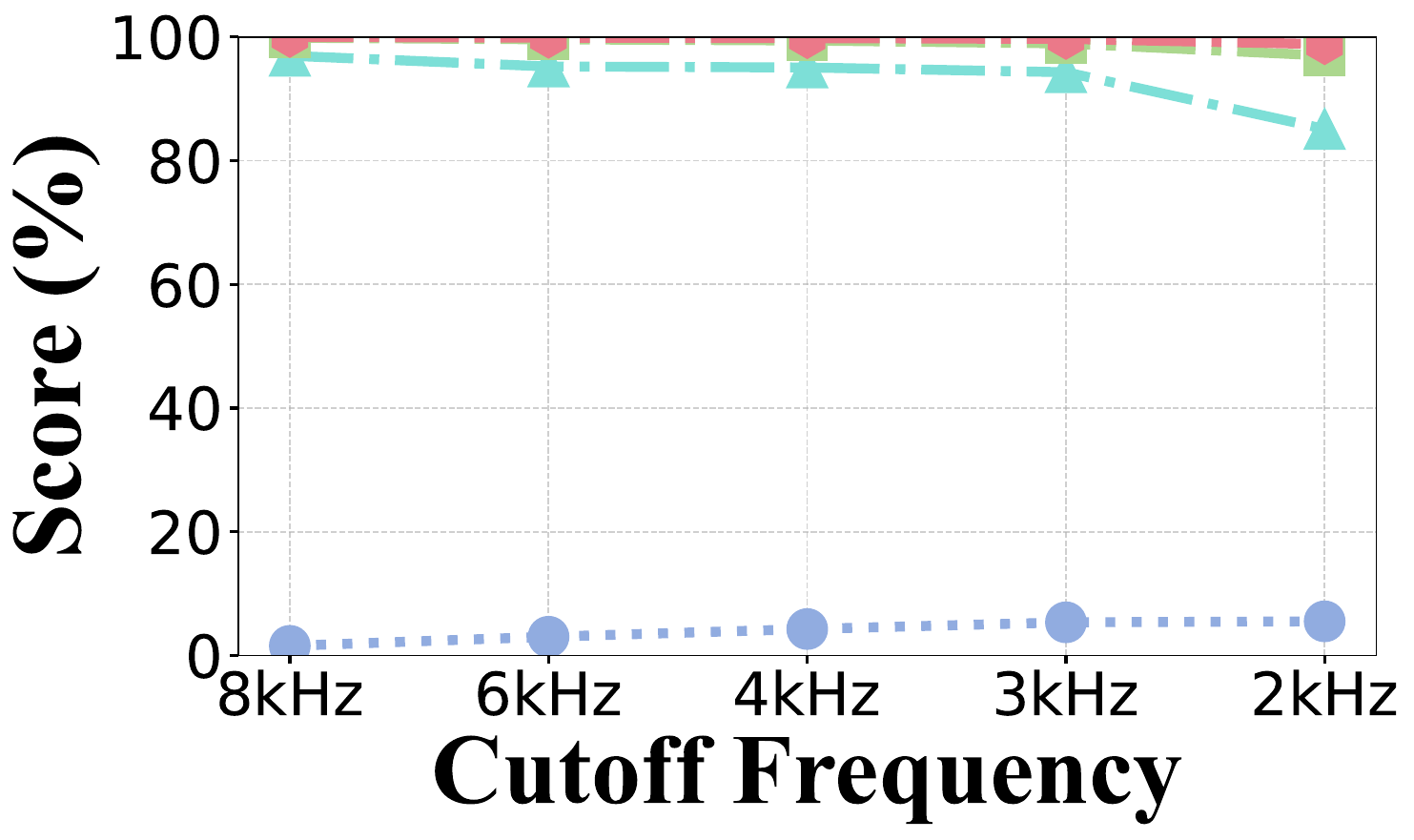}
    		\vspace{-0.1cm}
    	\end{minipage}
    }
    \subfigure[DDDM]{
    	\begin{minipage}[b]{0.23\linewidth}
    		\centering
    		\includegraphics[trim=0mm 0mm 0mm 0mm, clip,width=0.95\textwidth]{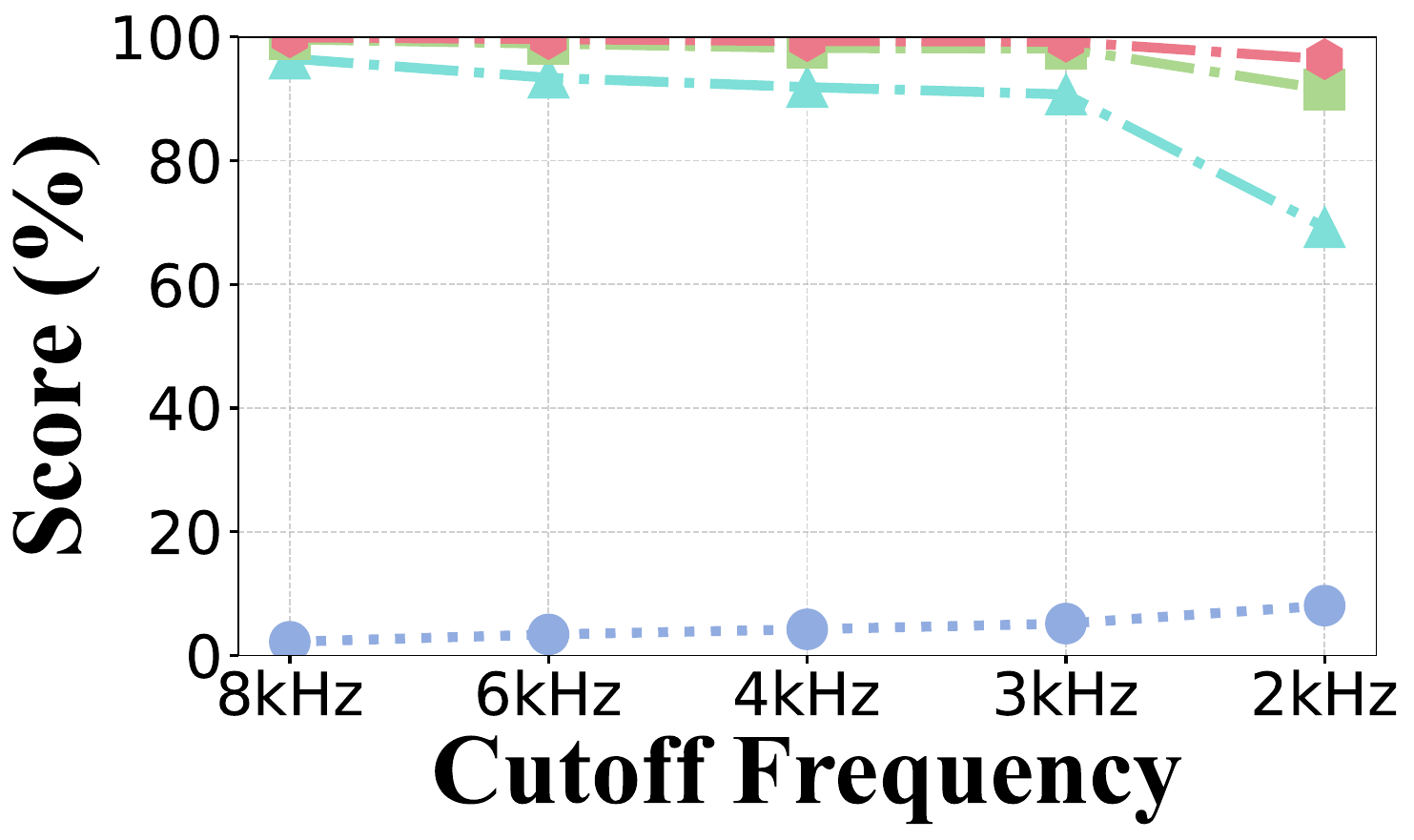}
    		\vspace{-0.1cm}
    	\end{minipage}
    }
 \vspace{-0.3cm}
	\caption{Performance of \sys under low-pass filtering.}
	\label{fig:line_filter}
\end{figure*}

\begin{figure*}[h]
    \centering
    	\begin{minipage}[b]{0.48\linewidth}
    		\centering
    		\includegraphics[trim=0mm 0mm 0mm 0mm, clip, width=\textwidth]{Section/Pictures/Draw/Line/legend.pdf}
    	\end{minipage} \\
    \vspace{-0.05cm}
    \subfigure[Clean]{
    	\begin{minipage}[b]{0.23\linewidth}
    		\centering
    		\includegraphics[trim=0mm 0mm 0mm 0mm, clip, width=0.95\textwidth]{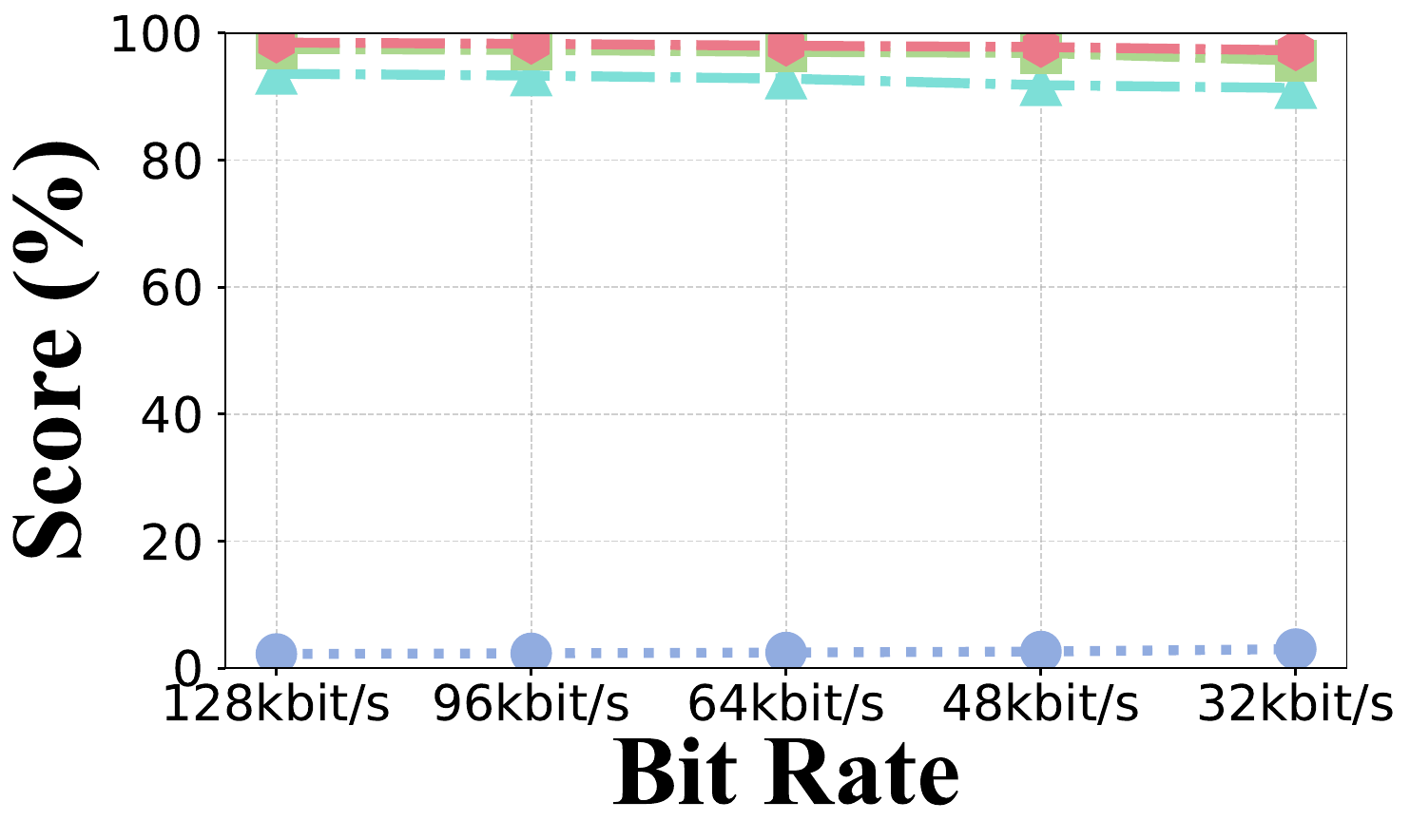}
    		\vspace{-0.1cm}
    	\end{minipage}
    }
    \subfigure[AGAIN]{
    	\begin{minipage}[b]{0.23\linewidth}
    		\centering
    		\includegraphics[trim=0mm 0mm 0mm 0mm, clip, width=0.95\textwidth]{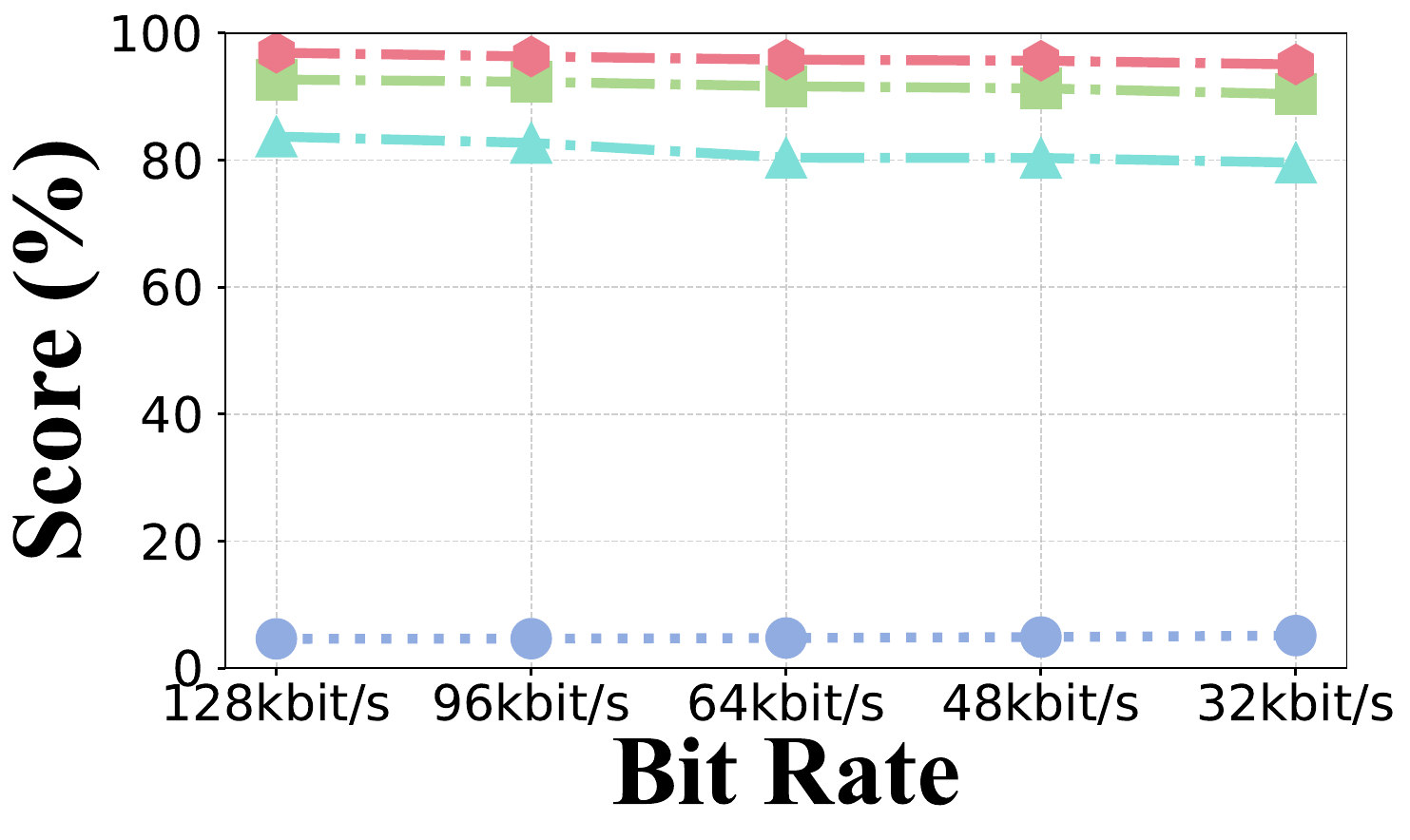}
    		\vspace{-0.1cm}
    	\end{minipage}
    }
    \subfigure[VQVC]{
    	\begin{minipage}[b]{0.23\linewidth}
    		\centering
    		\includegraphics[trim=0mm 0mm 0mm 0mm, clip, width=0.95\textwidth]{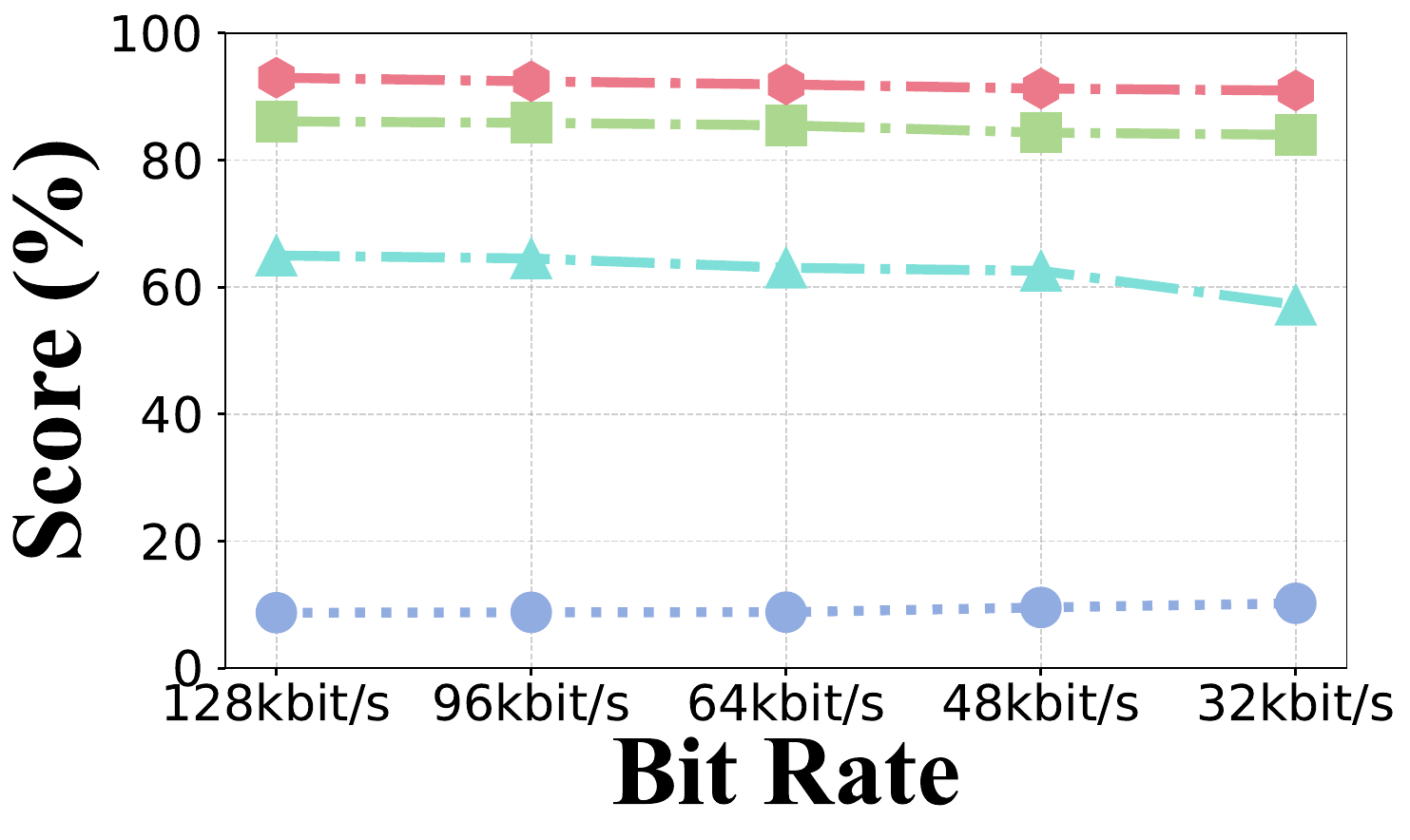}
    		\vspace{-0.1cm}
    	\end{minipage}
    }
  \vspace{-0.2cm}
    \subfigure[VQVC+]{
    	\begin{minipage}[b]{0.23\linewidth}
    		\centering
    		\includegraphics[trim=0mm 0mm 0mm 0mm, clip,width=0.95\textwidth]{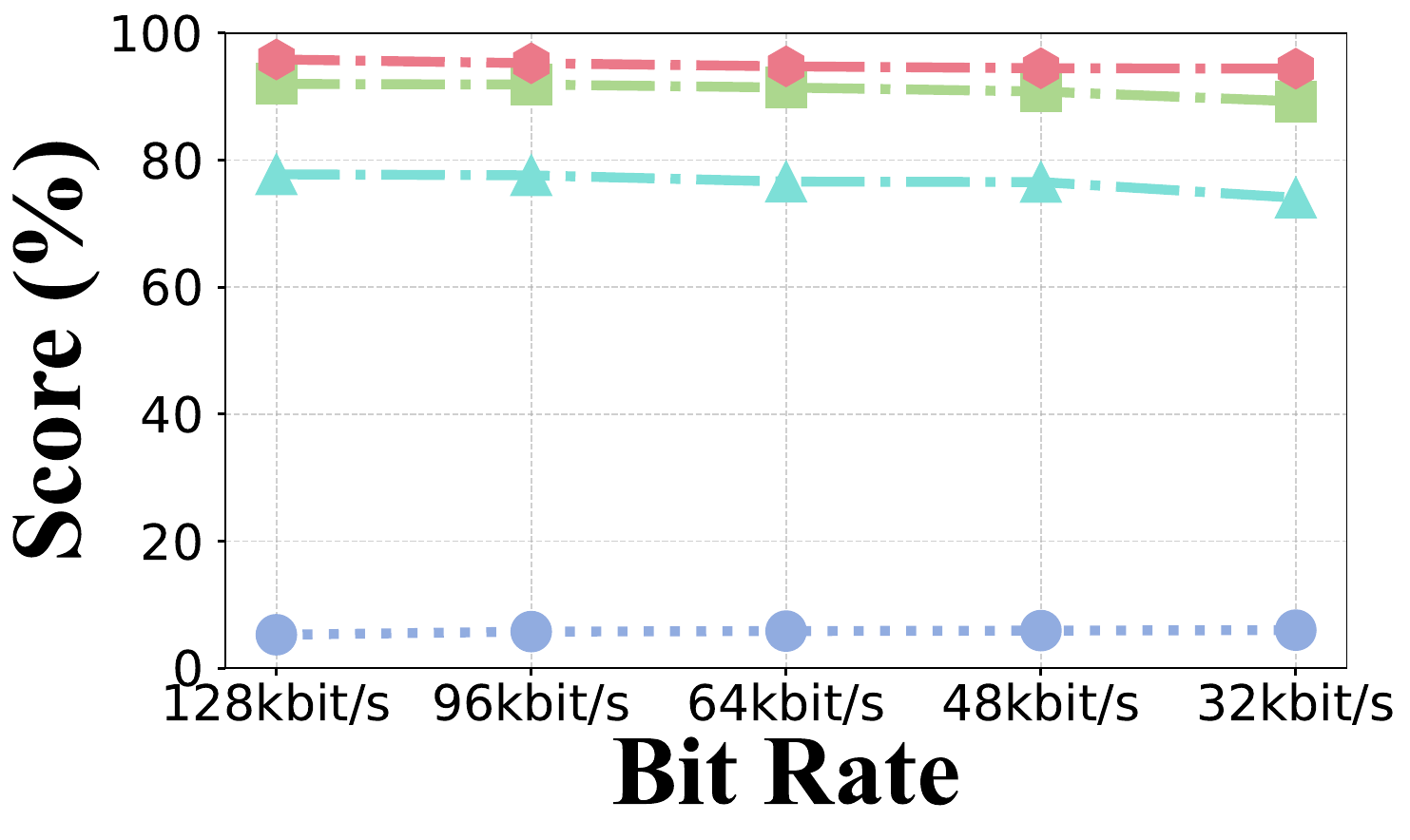}
    		\vspace{-0.1cm}
    	\end{minipage}
    }
    \subfigure[BNE]{
    	\begin{minipage}[b]{0.23\linewidth}
    		\centering
    		\includegraphics[trim=0mm 0mm 0mm 0mm, clip,width=0.95\textwidth]{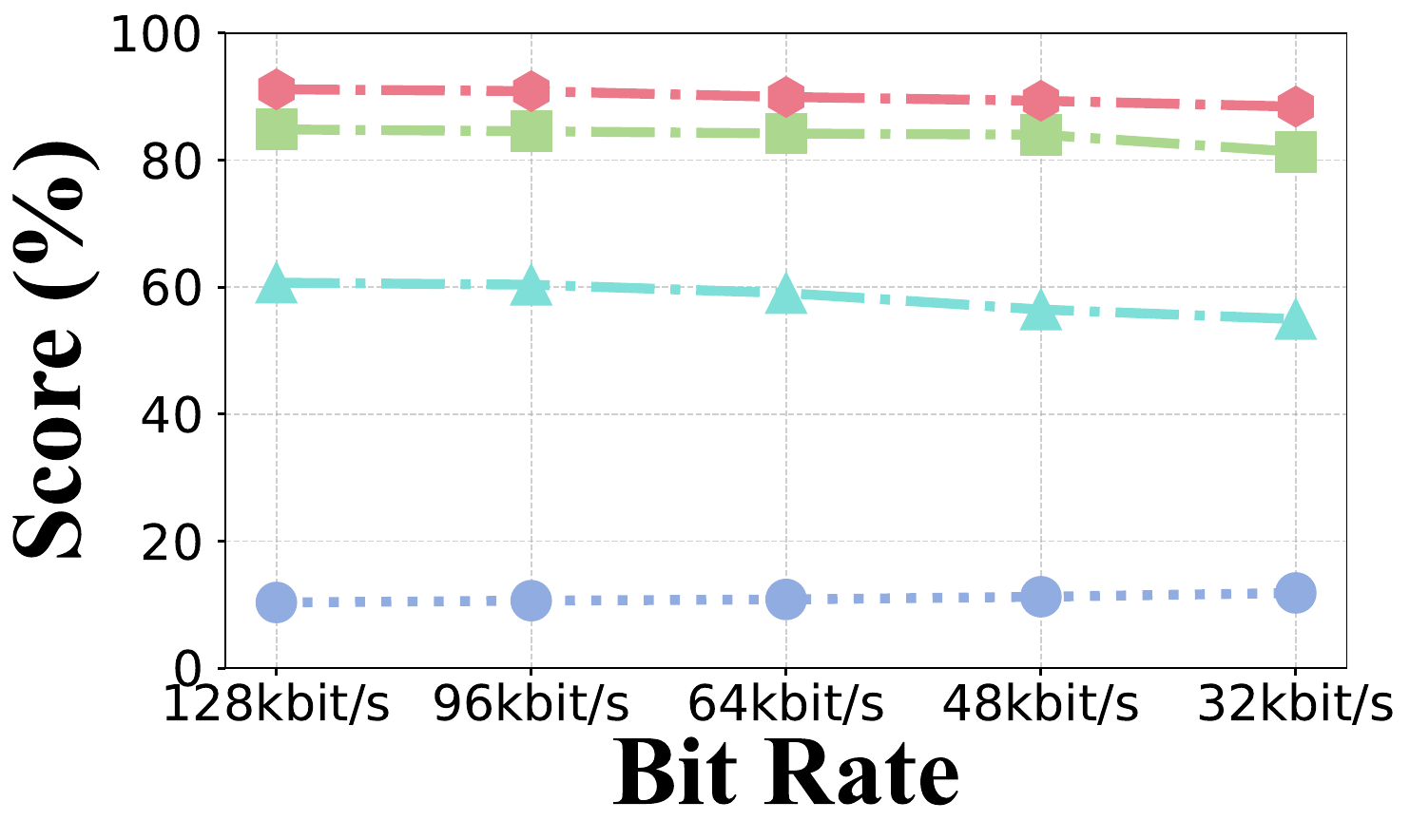}
    		\vspace{-0.1cm}
    	\end{minipage}
    }
    \subfigure[FreeVC]{
    	\begin{minipage}[b]{0.23\linewidth}
    		\centering
    		\includegraphics[trim=0mm 0mm 0mm 0mm, clip,width=0.95\textwidth]{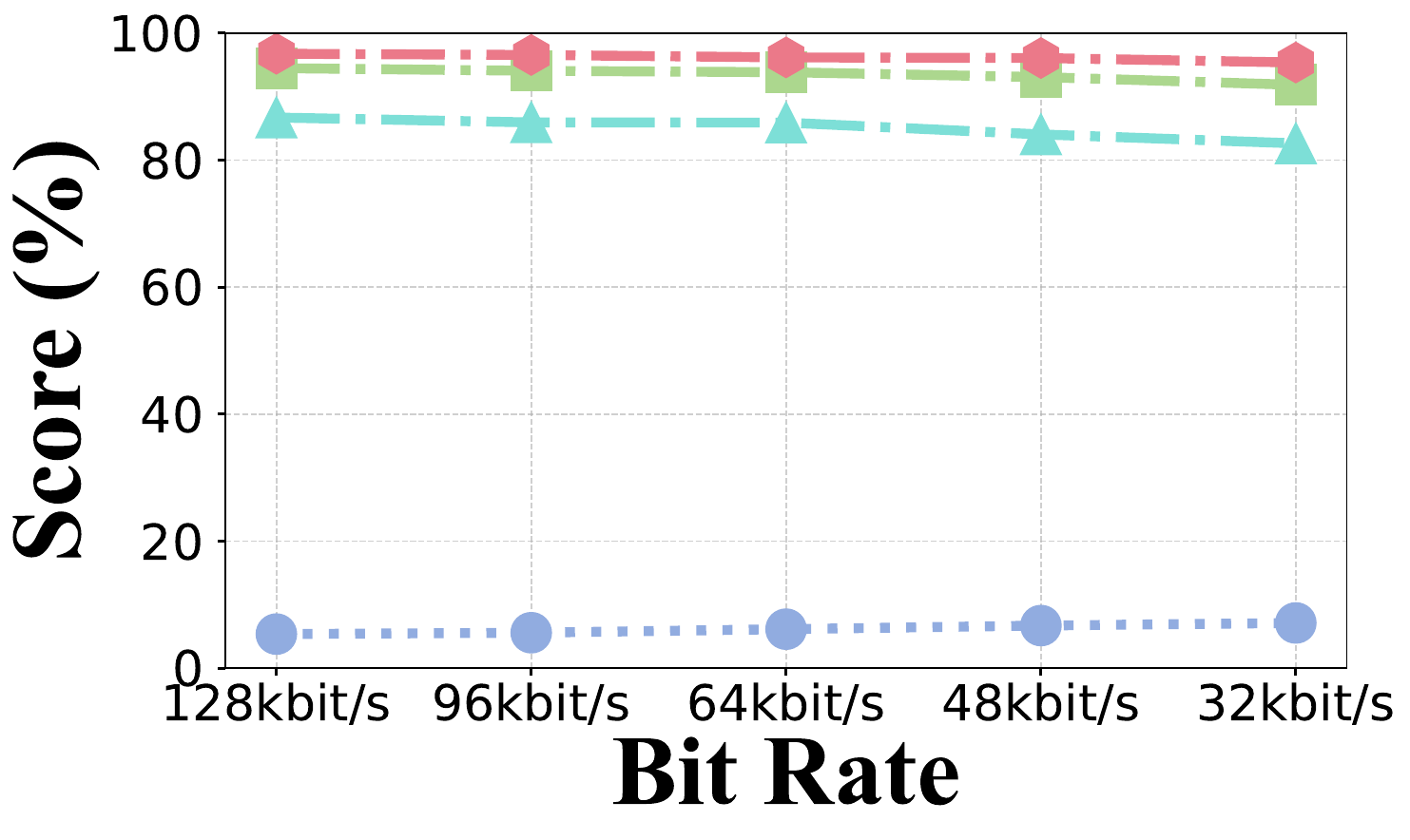}
    		\vspace{-0.1cm}
    	\end{minipage}
    }
    \subfigure[Diff]{
    	\begin{minipage}[b]{0.23\linewidth}
    		\centering
    		\includegraphics[trim=0mm 0mm 0mm 0mm, clip,width=0.95\textwidth]{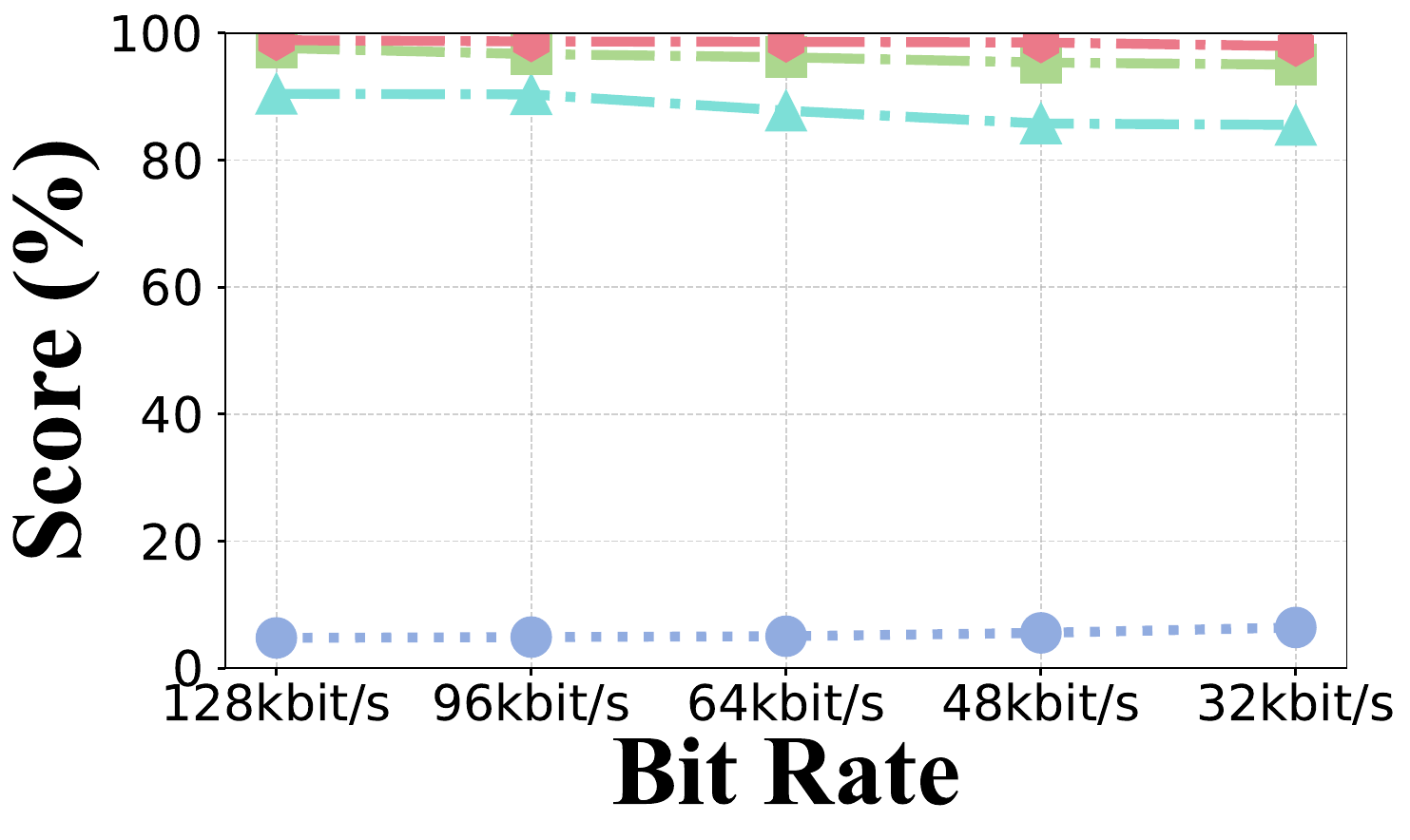}
    		\vspace{-0.1cm}
    	\end{minipage}
    }
    \subfigure[DDDM]{
    	\begin{minipage}[b]{0.23\linewidth}
    		\centering
    		\includegraphics[trim=0mm 0mm 0mm 0mm, clip,width=0.95\textwidth]{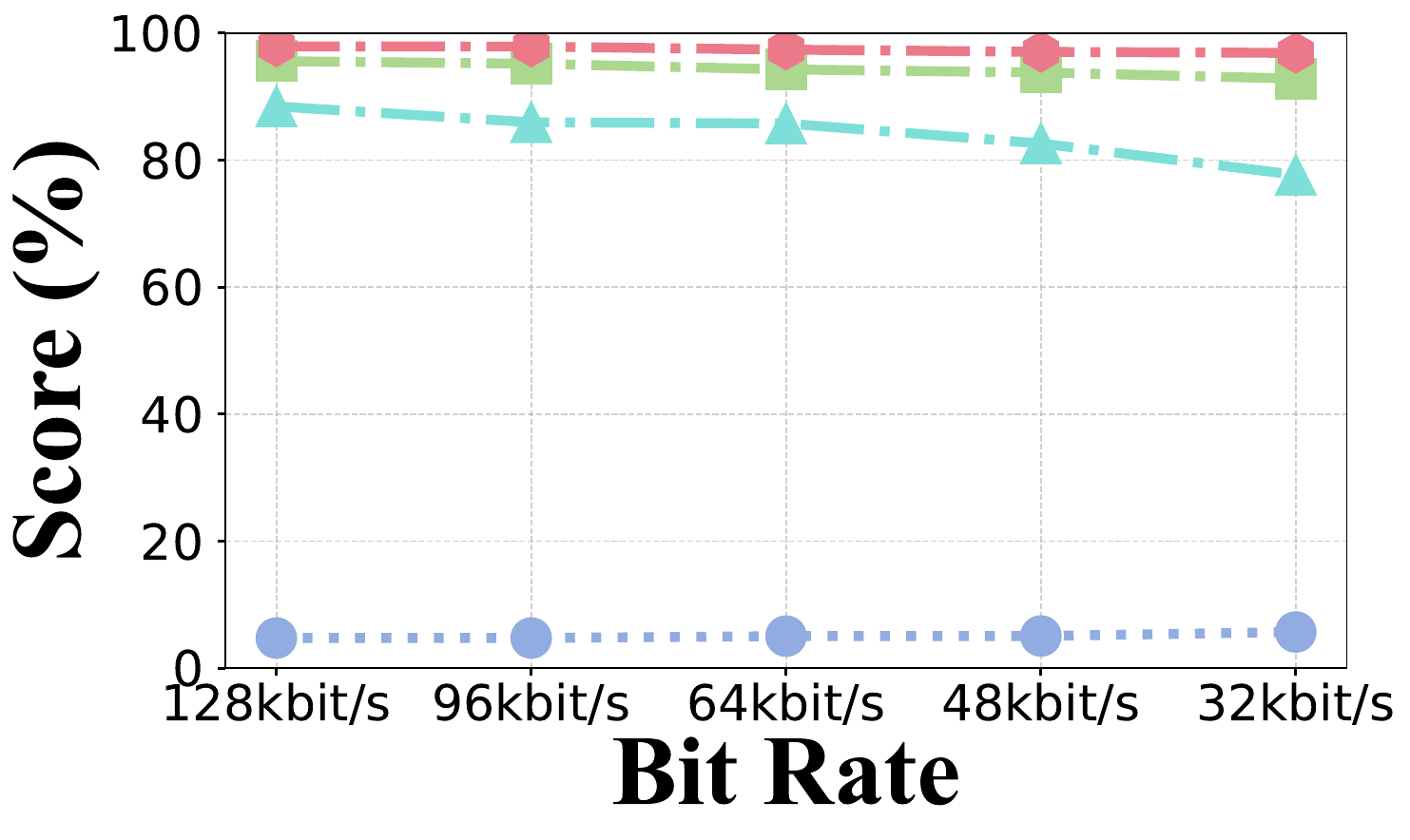}
    		\vspace{-0.1cm}
    	\end{minipage}
    }
 \vspace{-0.3cm}
	\caption{Performance of Revelio under MP3 compression.}
	\label{fig:line_compression_r}
\end{figure*}

\begin{figure*}[h]
    \centering
    	\begin{minipage}[b]{0.48\linewidth}
    		\centering
    		\includegraphics[trim=0mm 0mm 0mm 0mm, clip, width=\textwidth]{Section/Pictures/Draw/Line/legend.pdf}
    	\end{minipage} \\
    \vspace{-0.05cm}
    \subfigure[Clean]{
    	\begin{minipage}[b]{0.23\linewidth}
    		\centering
    		\includegraphics[trim=0mm 0mm 0mm 0mm, clip, width=0.95\textwidth]{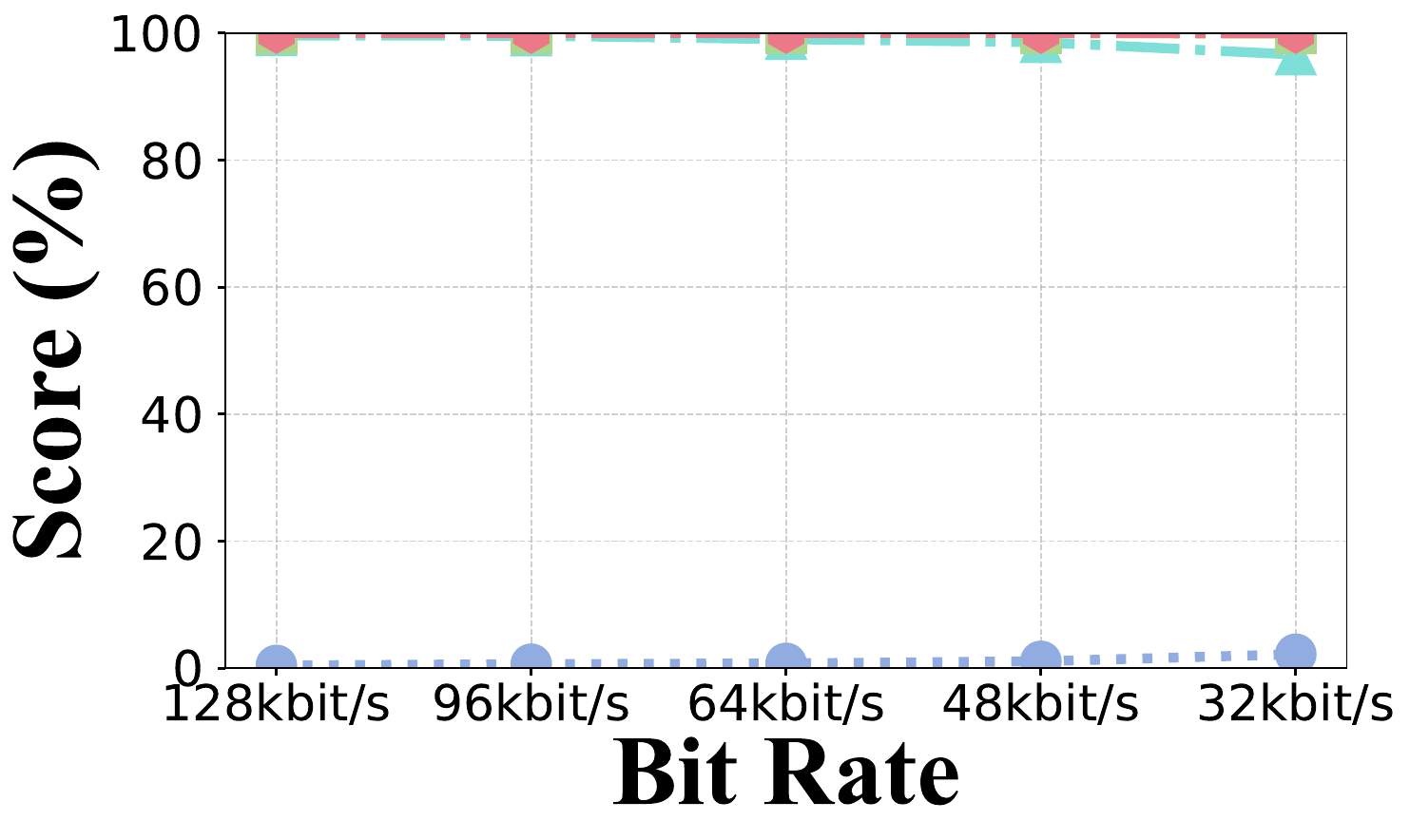}
    		\vspace{-0.1cm}
    	\end{minipage}
    }
    \subfigure[AGAIN]{
    	\begin{minipage}[b]{0.23\linewidth}
    		\centering
    		\includegraphics[trim=0mm 0mm 0mm 0mm, clip, width=0.95\textwidth]{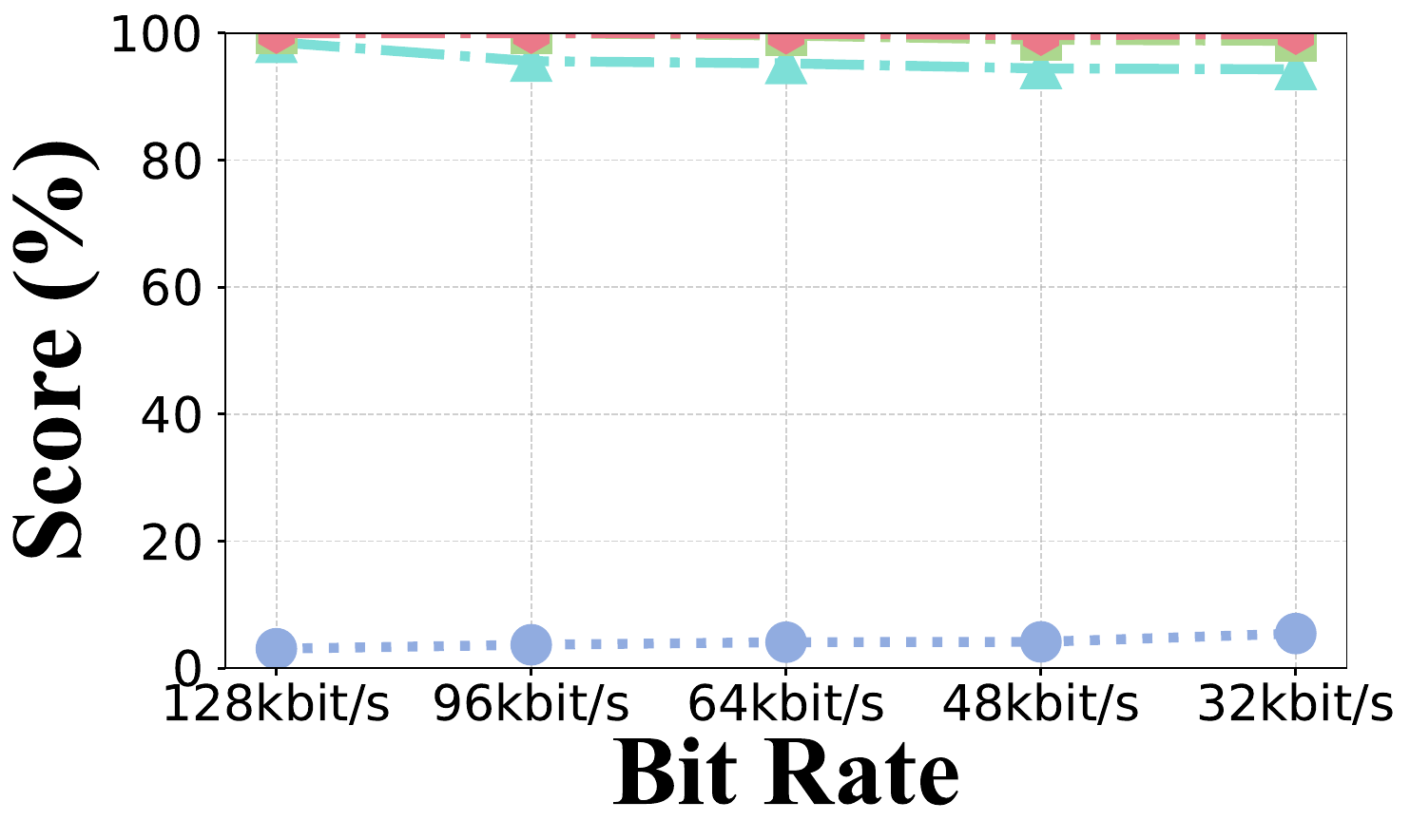}
    		\vspace{-0.1cm}
    	\end{minipage}
    }
    \subfigure[VQVC]{
    	\begin{minipage}[b]{0.23\linewidth}
    		\centering
    		\includegraphics[trim=0mm 0mm 0mm 0mm, clip, width=0.95\textwidth]{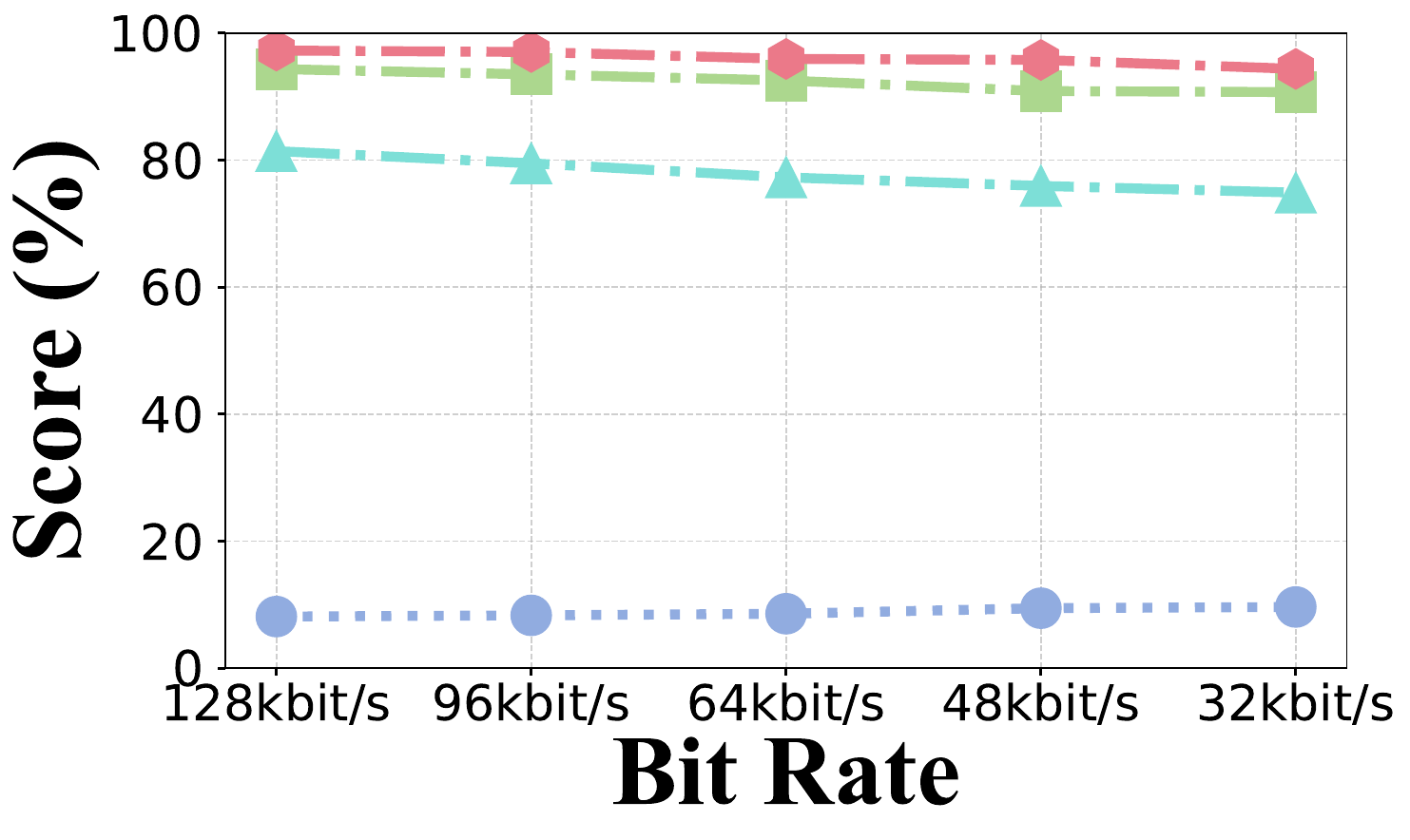}
    		\vspace{-0.1cm}
    	\end{minipage}
    }
  \vspace{-0.2cm}
    \subfigure[VQVC+]{
    	\begin{minipage}[b]{0.23\linewidth}
    		\centering
    		\includegraphics[trim=0mm 0mm 0mm 0mm, clip,width=0.95\textwidth]{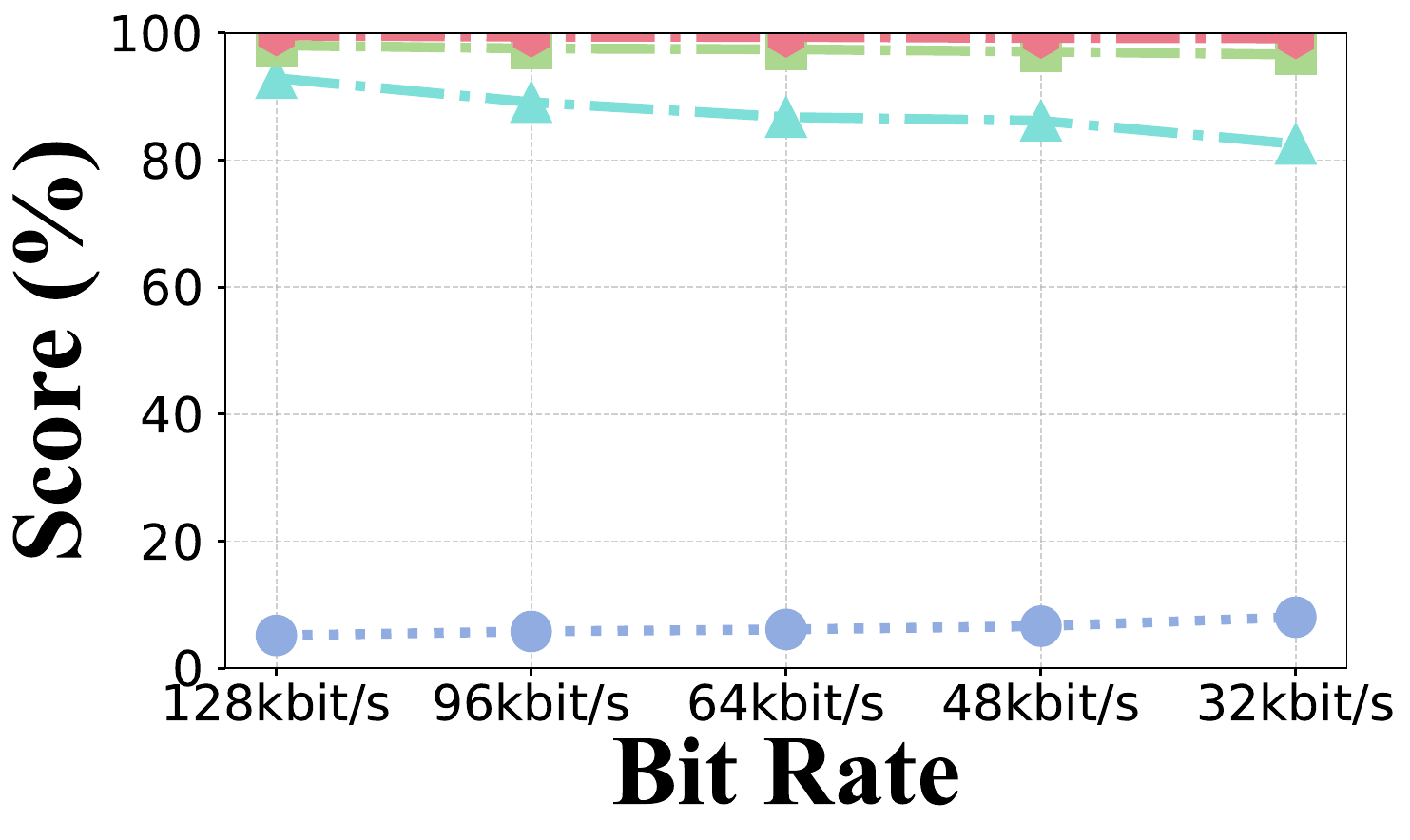}
    		\vspace{-0.1cm}
    	\end{minipage}
    }
    \subfigure[BNE]{
    	\begin{minipage}[b]{0.23\linewidth}
    		\centering
    		\includegraphics[trim=0mm 0mm 0mm 0mm, clip,width=0.95\textwidth]{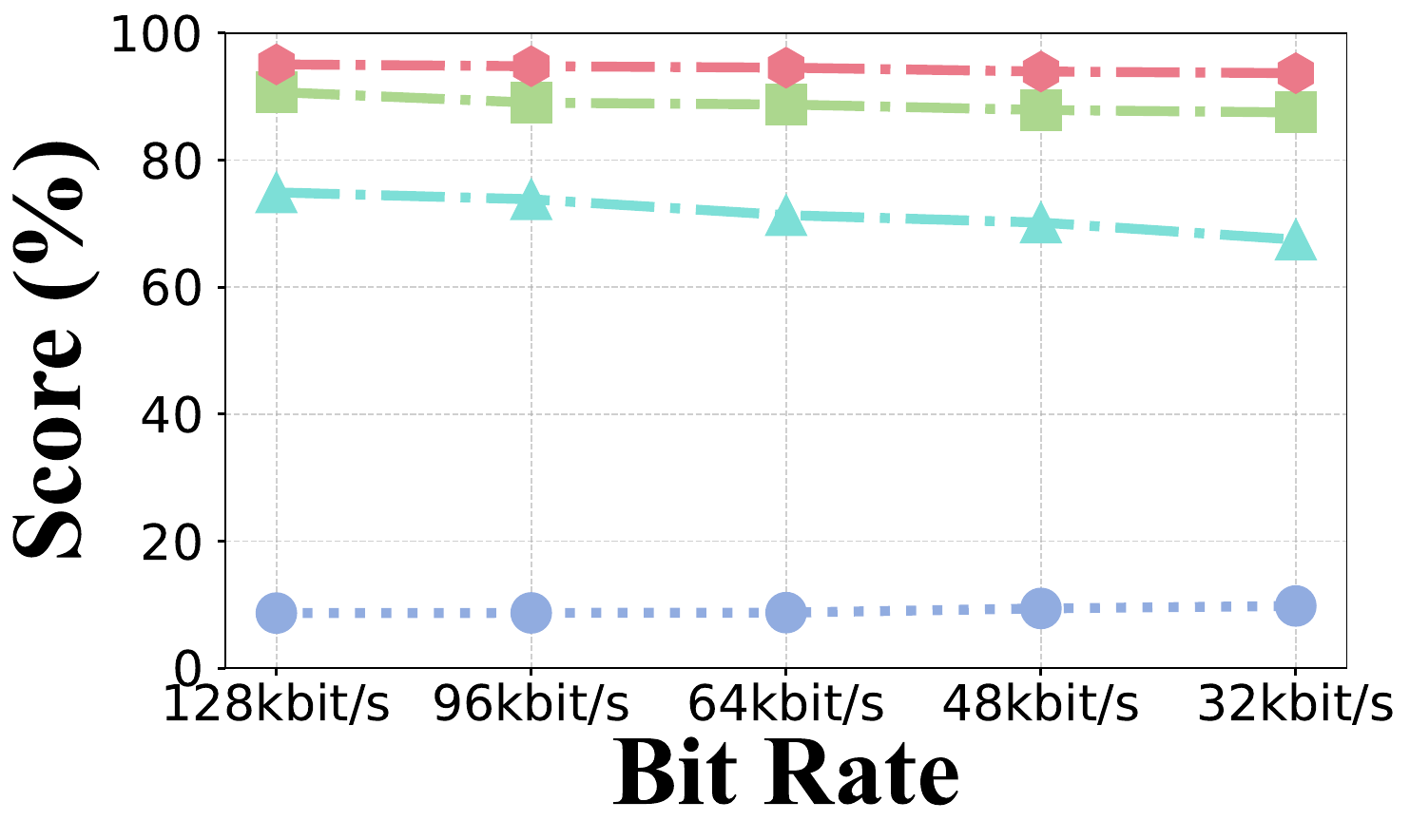}
    		\vspace{-0.1cm}
    	\end{minipage}
    }
    \subfigure[FreeVC]{
    	\begin{minipage}[b]{0.23\linewidth}
    		\centering
    		\includegraphics[trim=0mm 0mm 0mm 0mm, clip,width=0.95\textwidth]{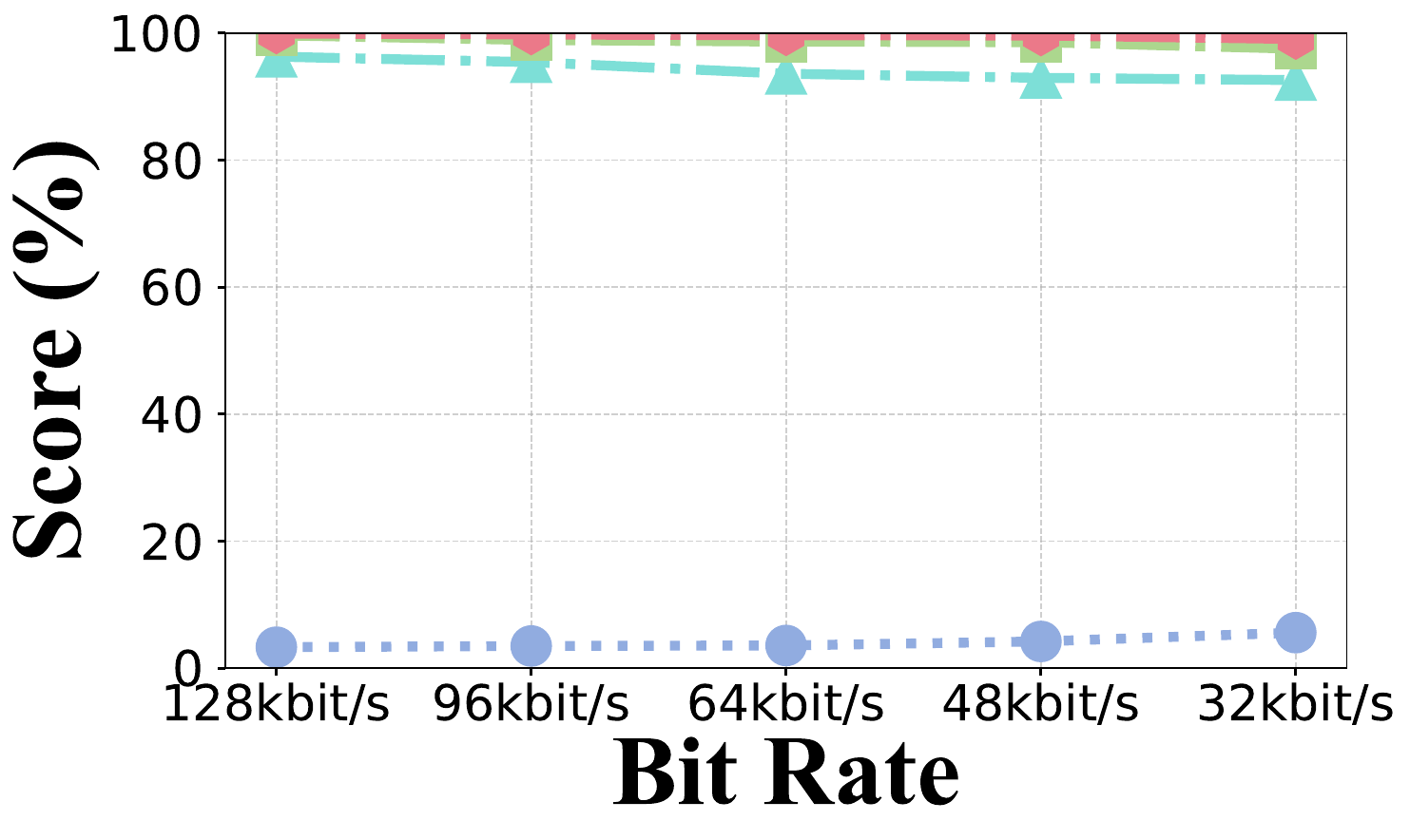}
    		\vspace{-0.1cm}
    	\end{minipage}
    }
    \subfigure[Diff]{
    	\begin{minipage}[b]{0.23\linewidth}
    		\centering
    		\includegraphics[trim=0mm 0mm 0mm 0mm, clip,width=0.95\textwidth]{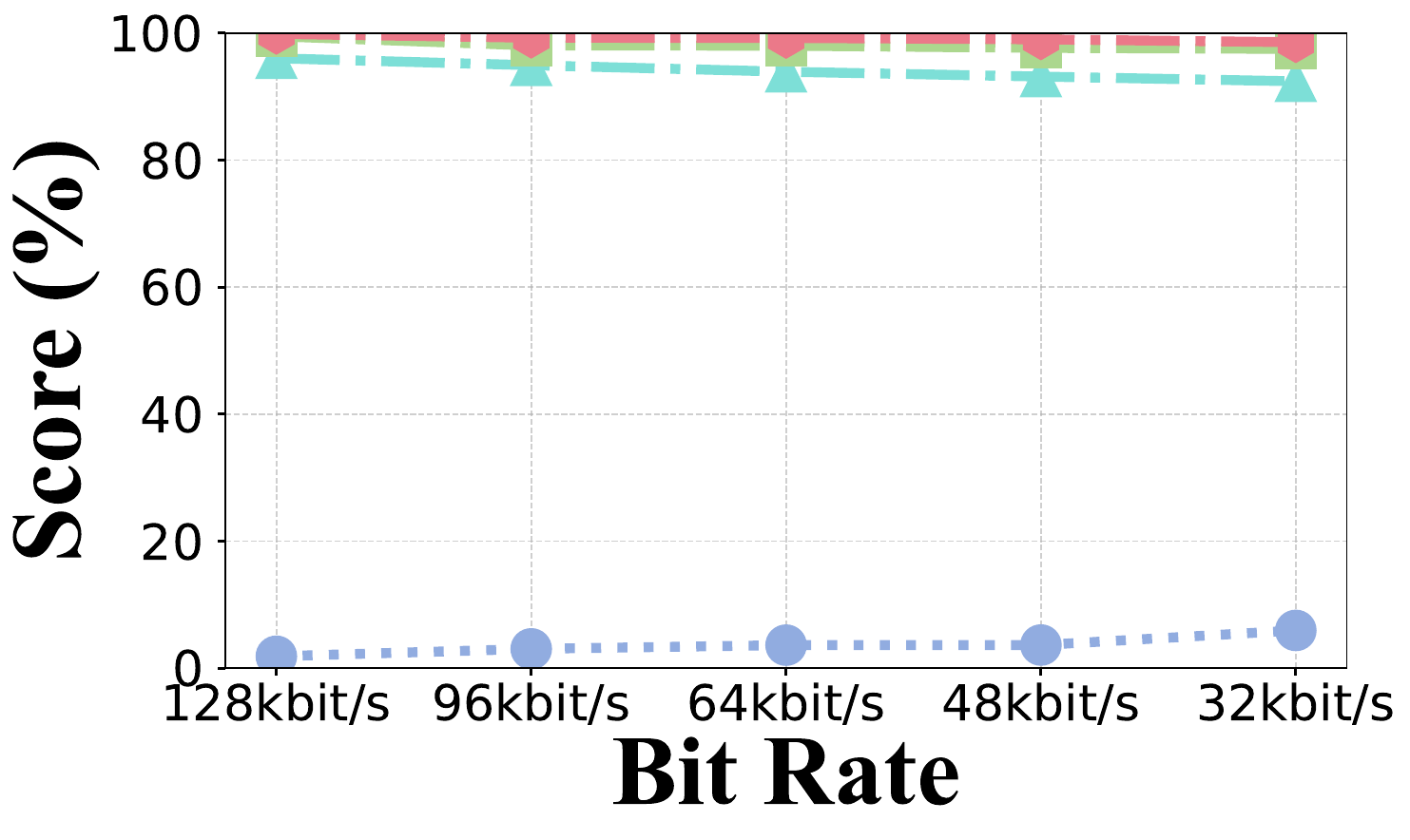}
    		\vspace{-0.1cm}
    	\end{minipage}
    }
    \subfigure[DDDM]{
    	\begin{minipage}[b]{0.23\linewidth}
    		\centering
    		\includegraphics[trim=0mm 0mm 0mm 0mm, clip,width=0.95\textwidth]{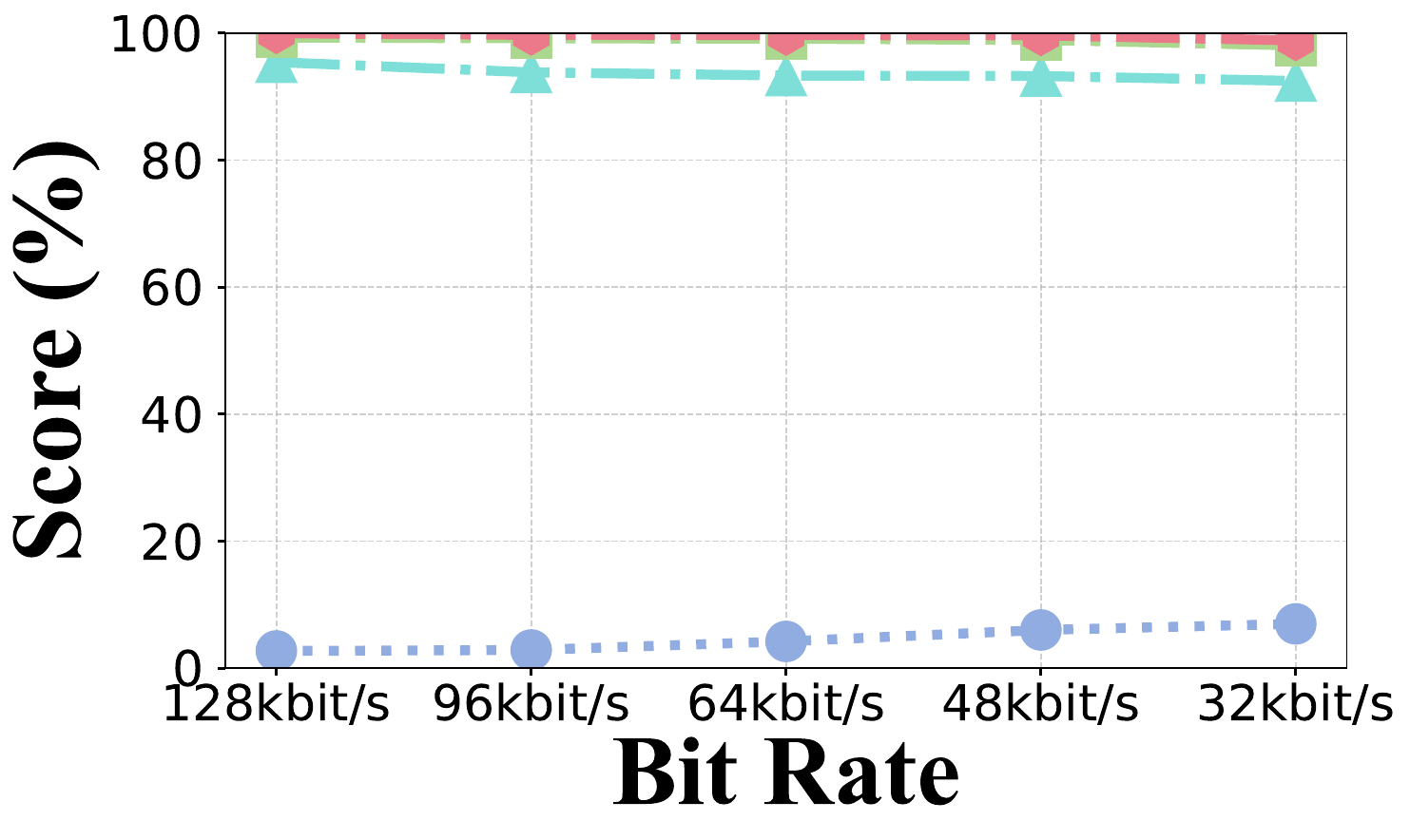}
    		\vspace{-0.1cm}
    	\end{minipage}
    }
 \vspace{-0.3cm}
	\caption{Performance of \sys under MP3 compression.}
	\label{fig:line_compression}
\end{figure*}

\begin{figure*}[h]
    \centering
    \begin{minipage}[b]{0.85\linewidth}
        \centering
        \includegraphics[trim=0mm 0mm 0mm 0mm, clip, width=\textwidth]{Section/Pictures/Draw/tel/legend.pdf}
    \end{minipage} \\
    \vspace{0.1cm}
    
    \hspace{-0.2cm}
    \begin{minipage}[b]{0.255\textwidth}
        \centering
        \includegraphics[trim=0mm 0mm 0mm 0mm, clip, width=0.95\linewidth]{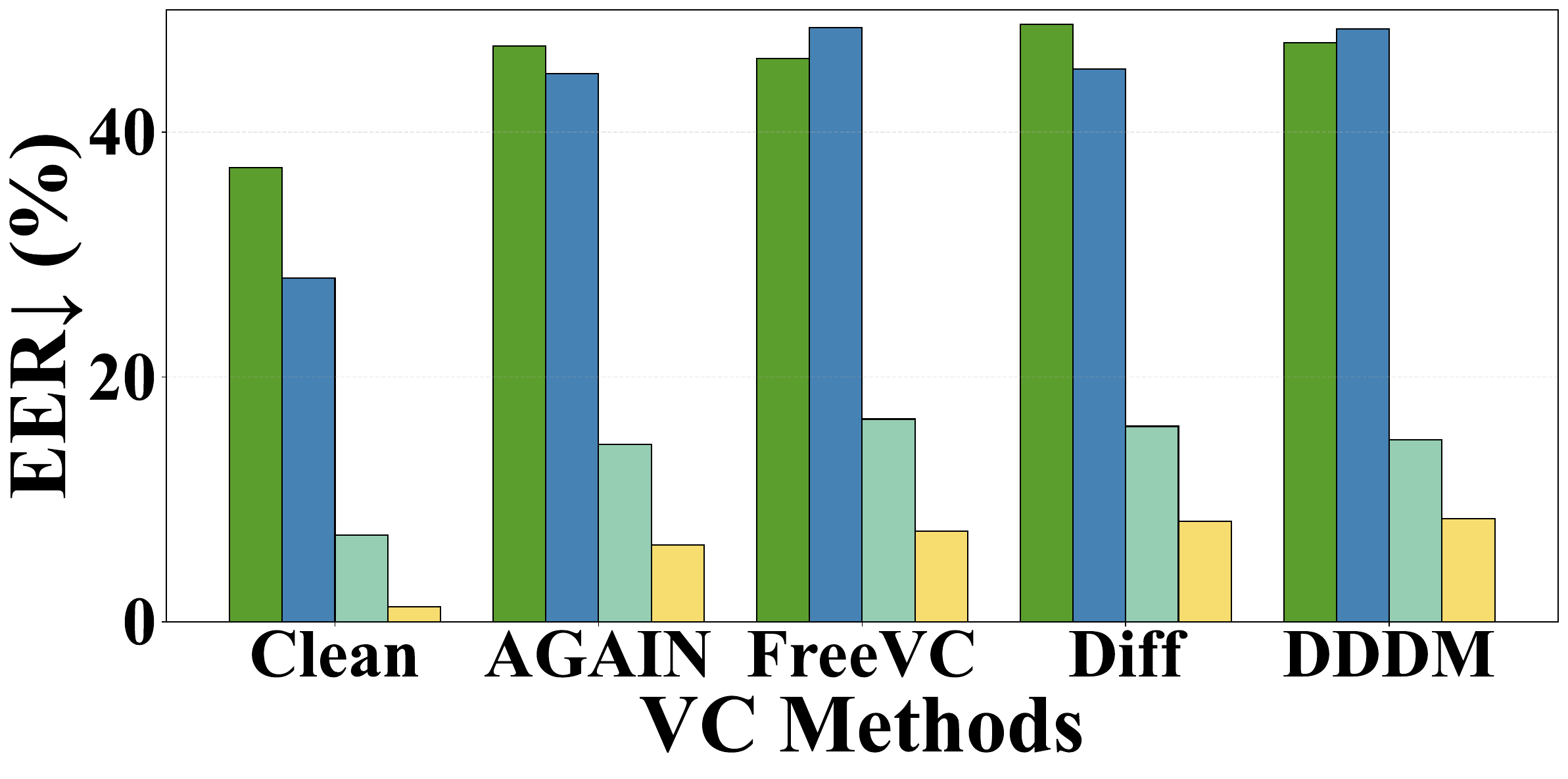}
    \end{minipage}
    \hfill
    \hspace{-0.6cm}
    \begin{minipage}[b]{0.255\textwidth}
        \centering
        \includegraphics[trim=0mm 0mm 0mm 0mm, clip, width=0.95\linewidth]{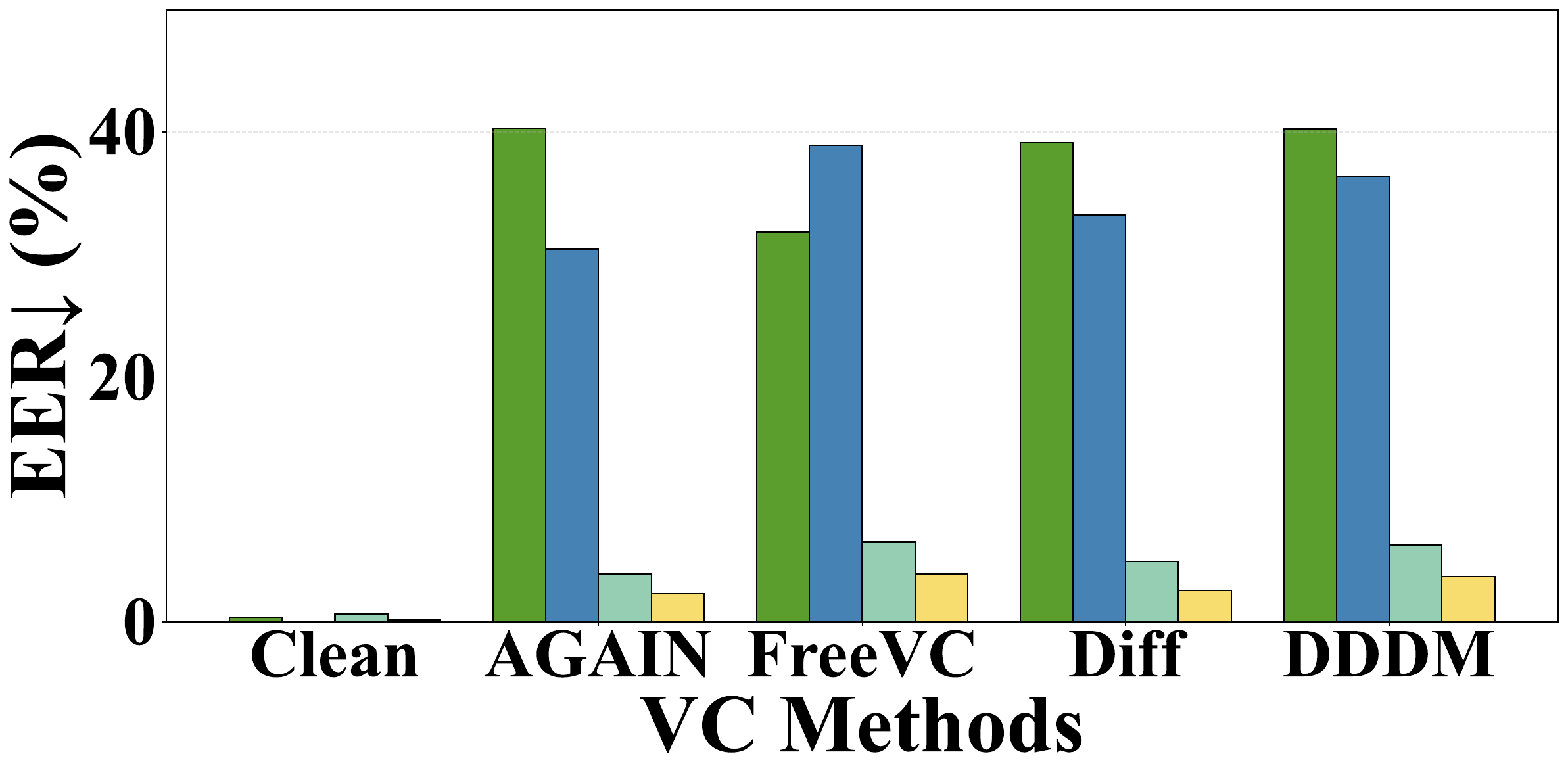}
    \end{minipage}
    \hfill
    \hspace{-0.6cm}
    \begin{minipage}[b]{0.255\textwidth}
        \centering
        \includegraphics[trim=0mm 0mm 0mm 0mm, clip, width=0.95\linewidth]{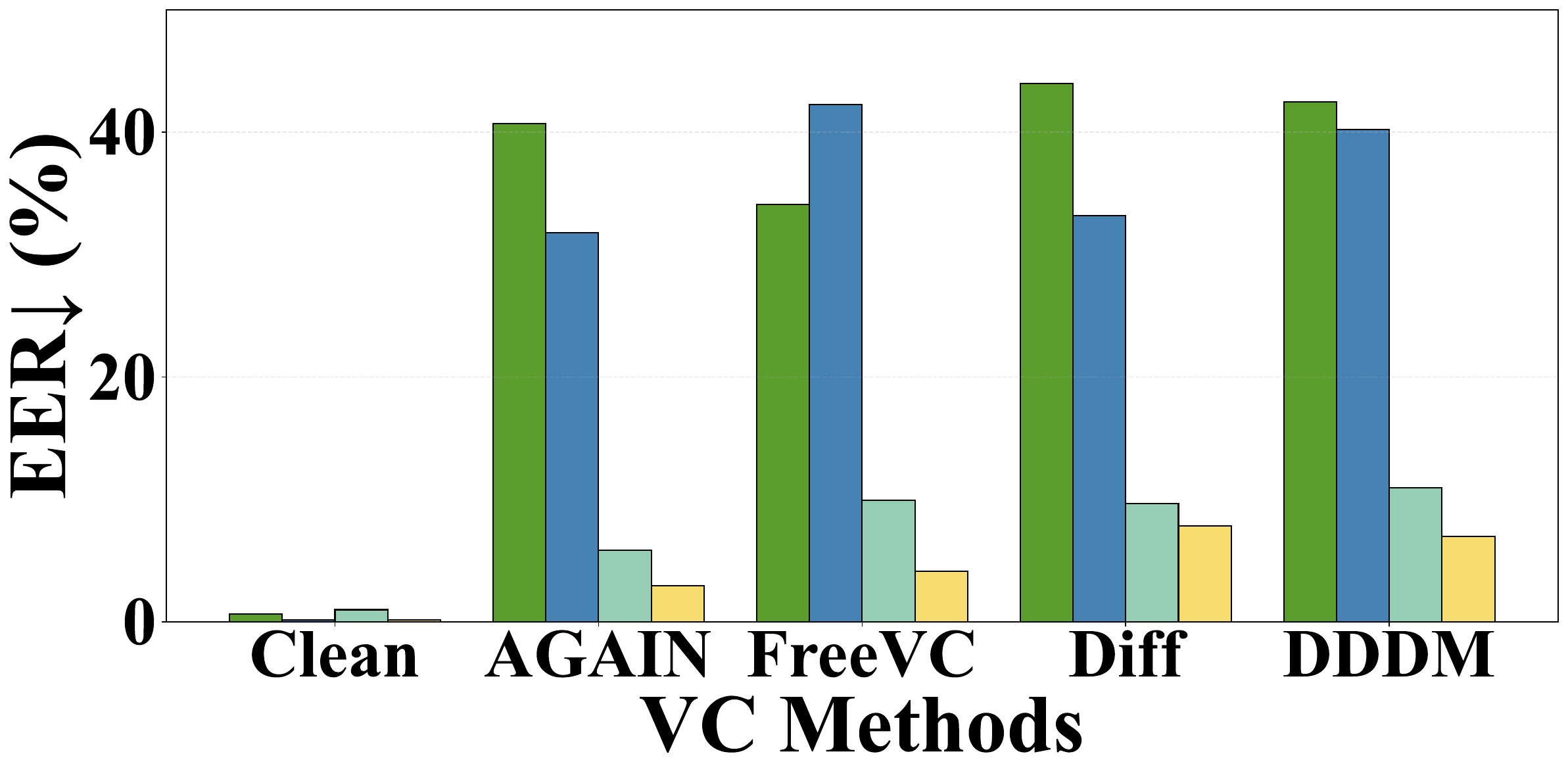}
    \end{minipage}
    \hfill
    \hspace{-0.6cm}
    \begin{minipage}[b]{0.255\textwidth}
        \centering
        \includegraphics[trim=0mm 0mm 0mm 0mm, clip, width=0.95\linewidth]{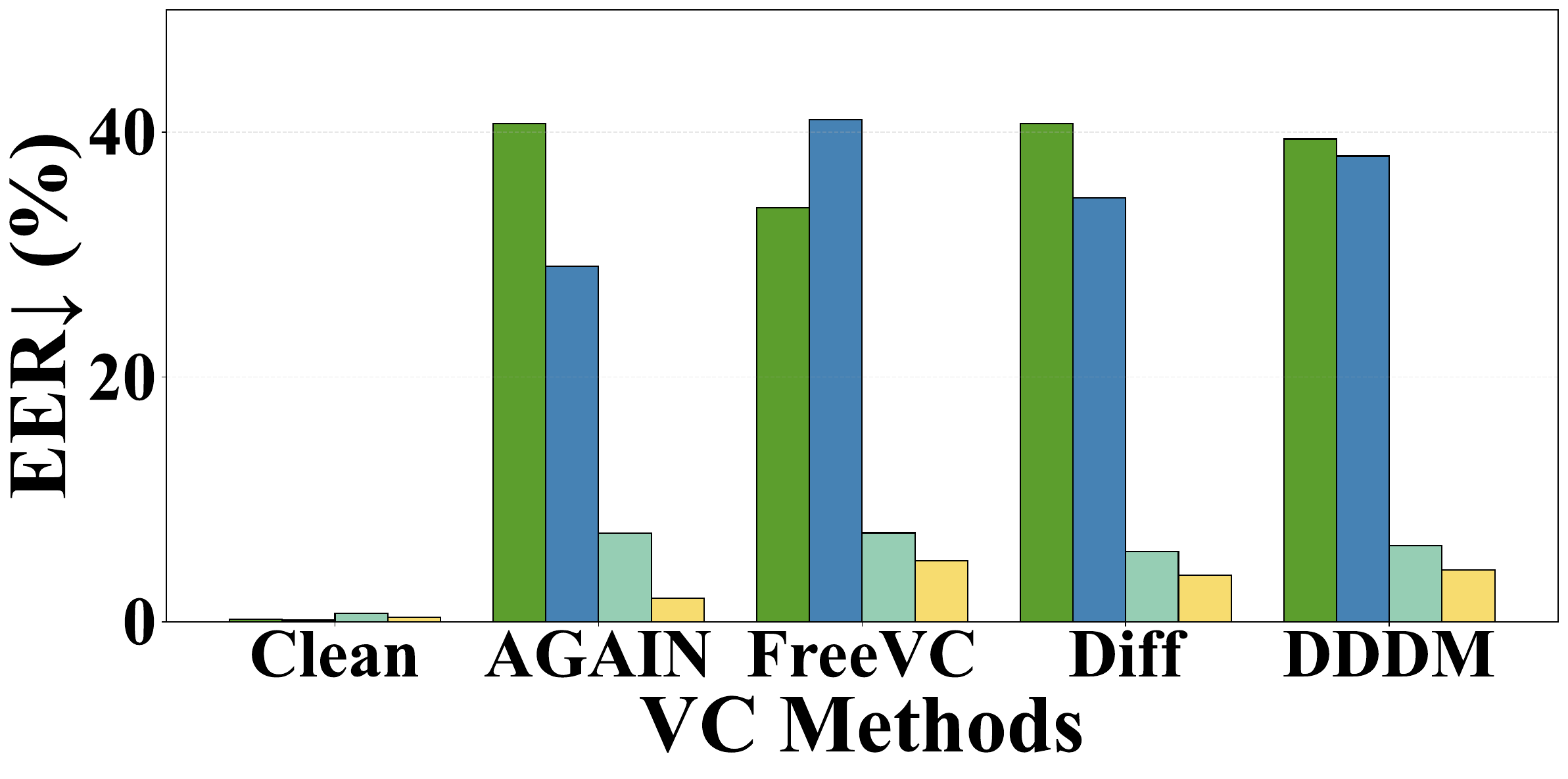}
    \end{minipage}

    \vspace{-0.1cm}
    \hspace{-0.4cm}
    \subfigure[English (VCTK)]{
    \begin{minipage}[b]{0.261\textwidth}
        \centering
        \includegraphics[trim=0mm 0mm 0mm 0mm, clip, width=0.95\linewidth]{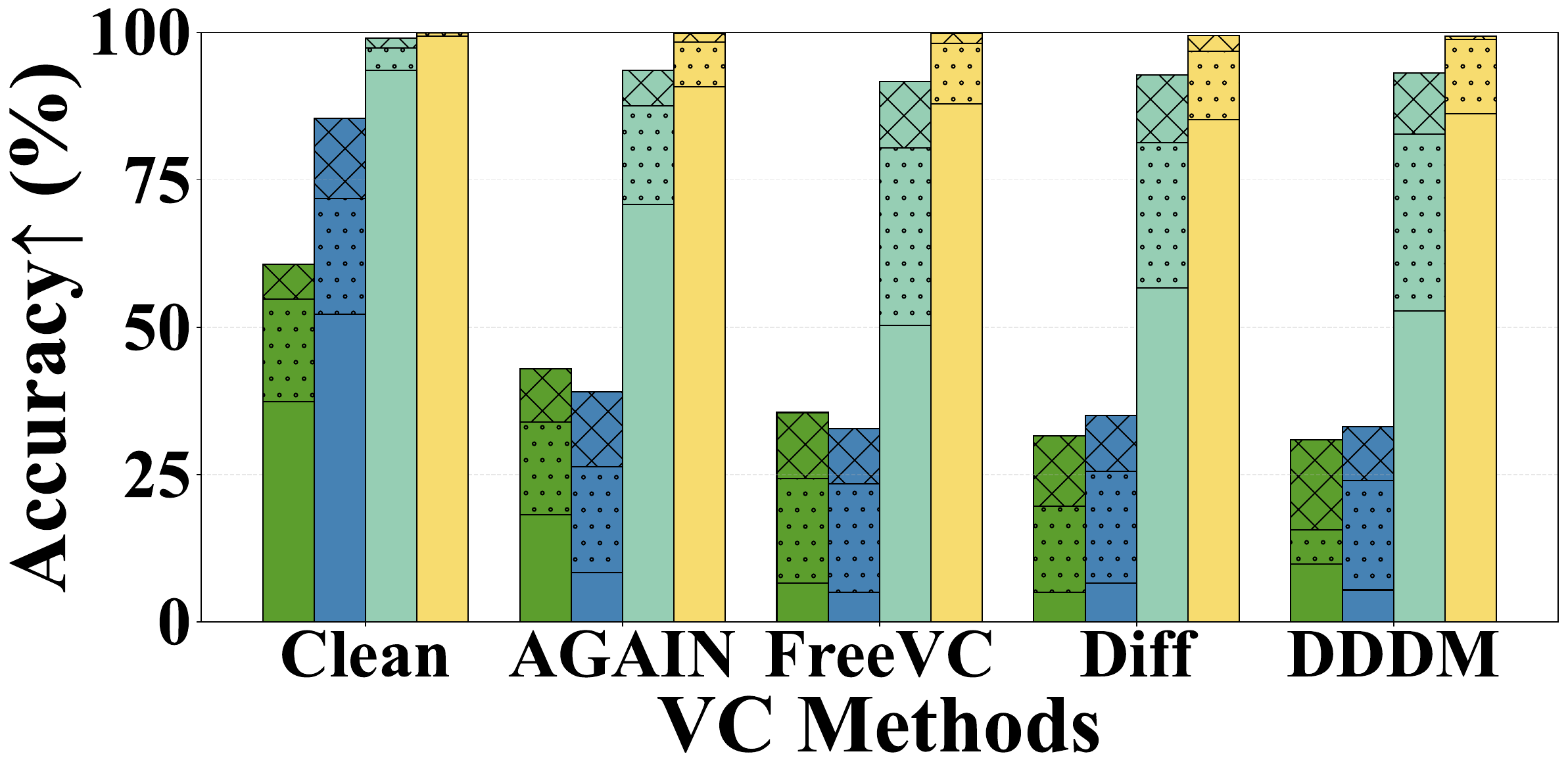}
    \end{minipage}
    }
    \hfill
    \hspace{-0.6cm}
    \subfigure[Spanish]{
        \begin{minipage}[b]{0.261\textwidth}
            \centering
            \includegraphics[trim=0mm 0mm 0mm 0mm, clip, width=0.95\linewidth]{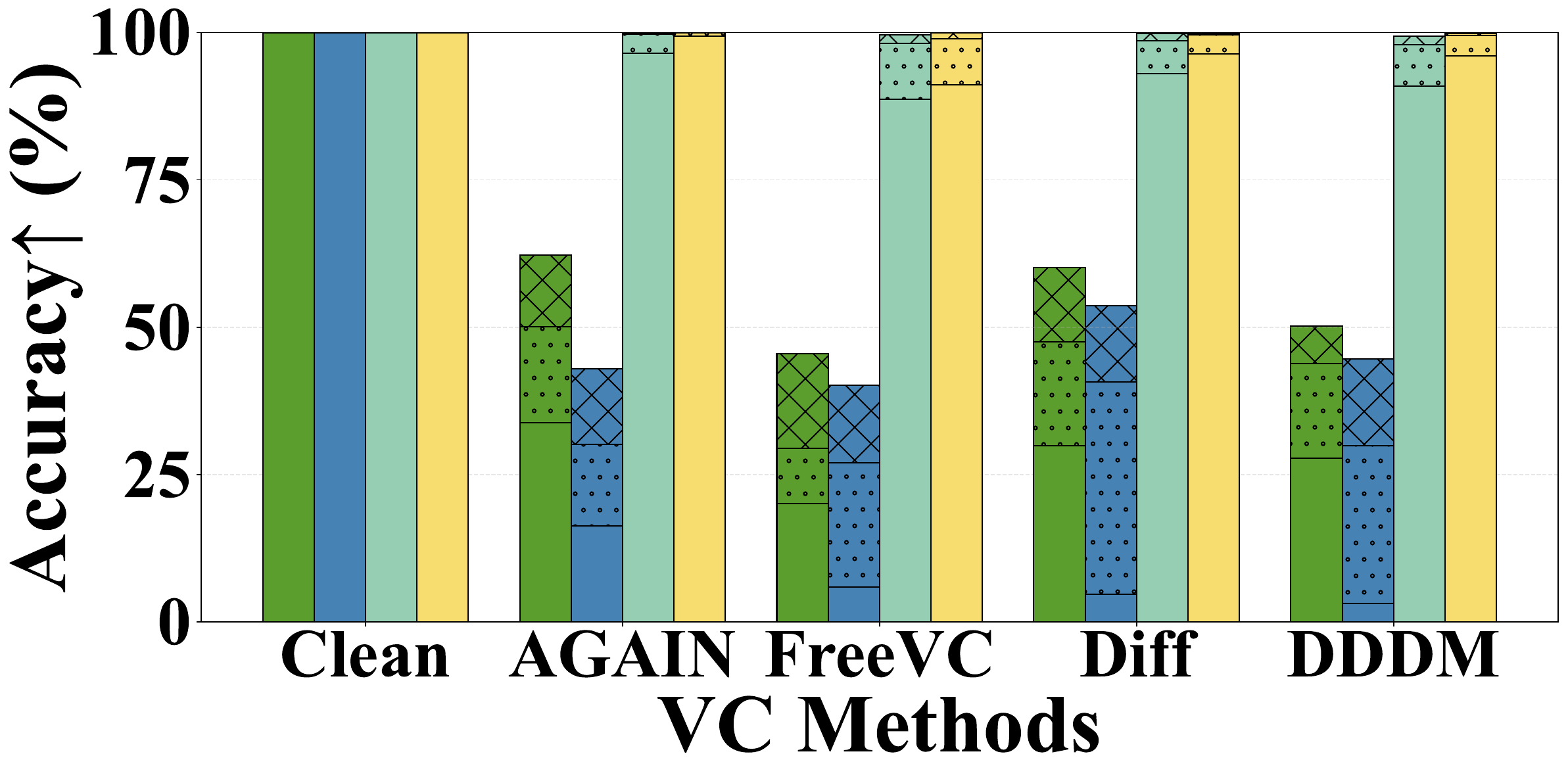}
        \end{minipage}
    }
    \hfill
    \hspace{-0.6cm}
    \subfigure[French]{
        \begin{minipage}[b]{0.261\textwidth}
            \centering
            \includegraphics[trim=0mm 0mm 0mm 0mm, clip, width=0.95\linewidth]{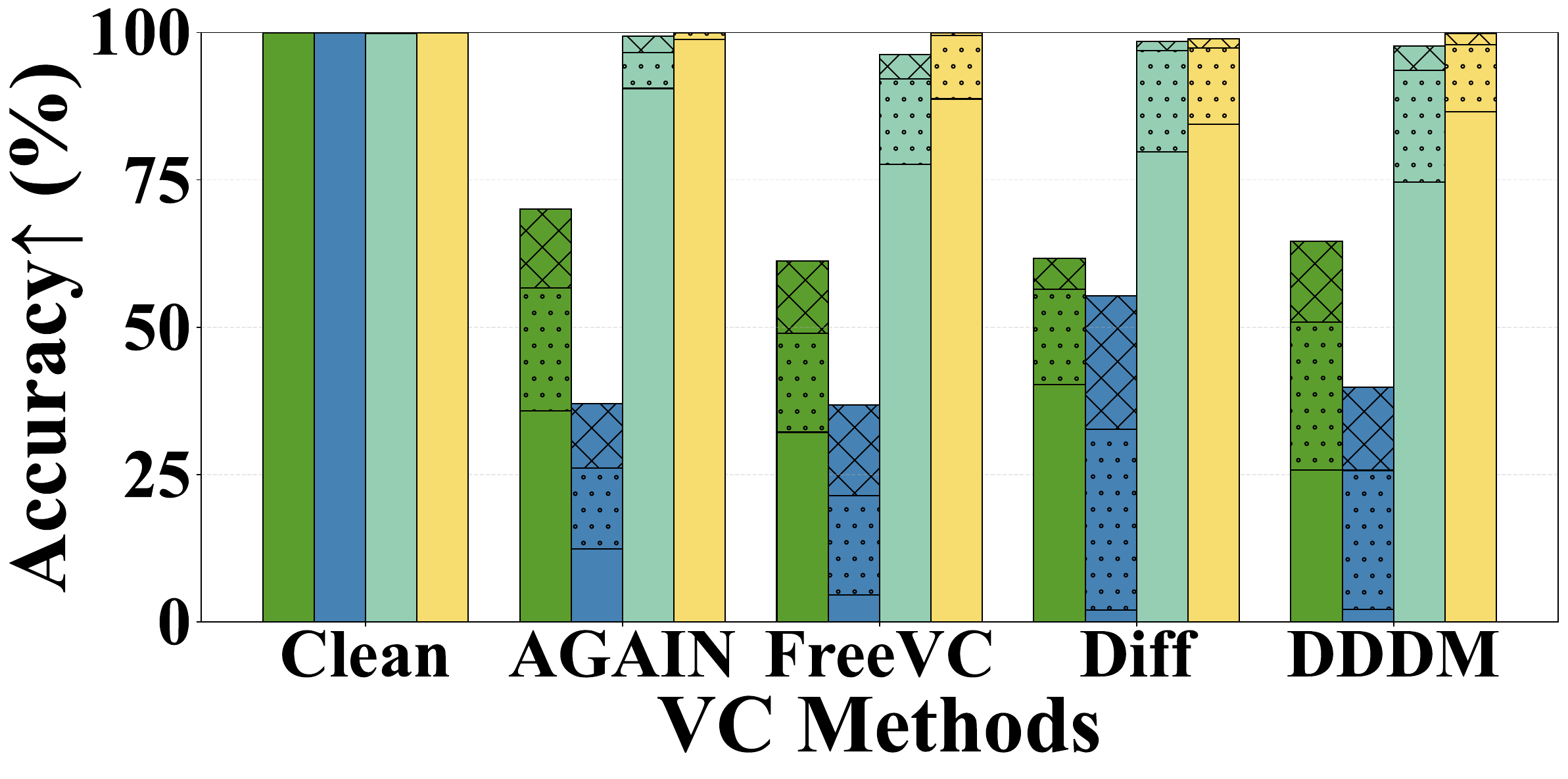}
        \end{minipage}
    }
    \hfill
    \hspace{-0.6cm}
    \subfigure[German]{
        \begin{minipage}[b]{0.261\textwidth}
            \centering
            \includegraphics[trim=0mm 0mm 0mm 0mm, clip, width=0.95\linewidth]{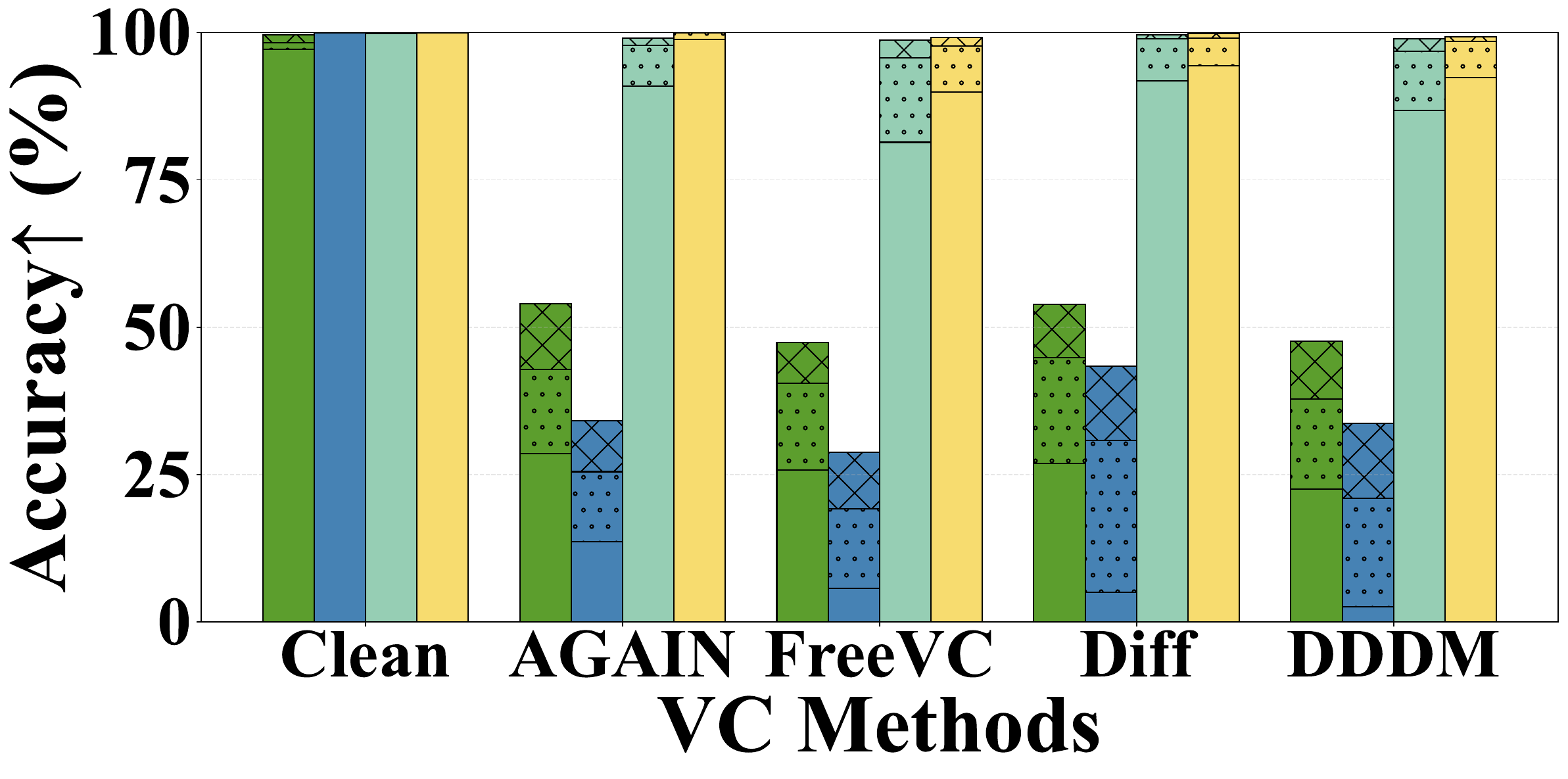}
        \end{minipage}
    }

    \begin{minipage}[b]{0.255\textwidth}
        \centering
        \includegraphics[trim=0mm 0mm 0mm 0mm, clip, width=0.95\linewidth]{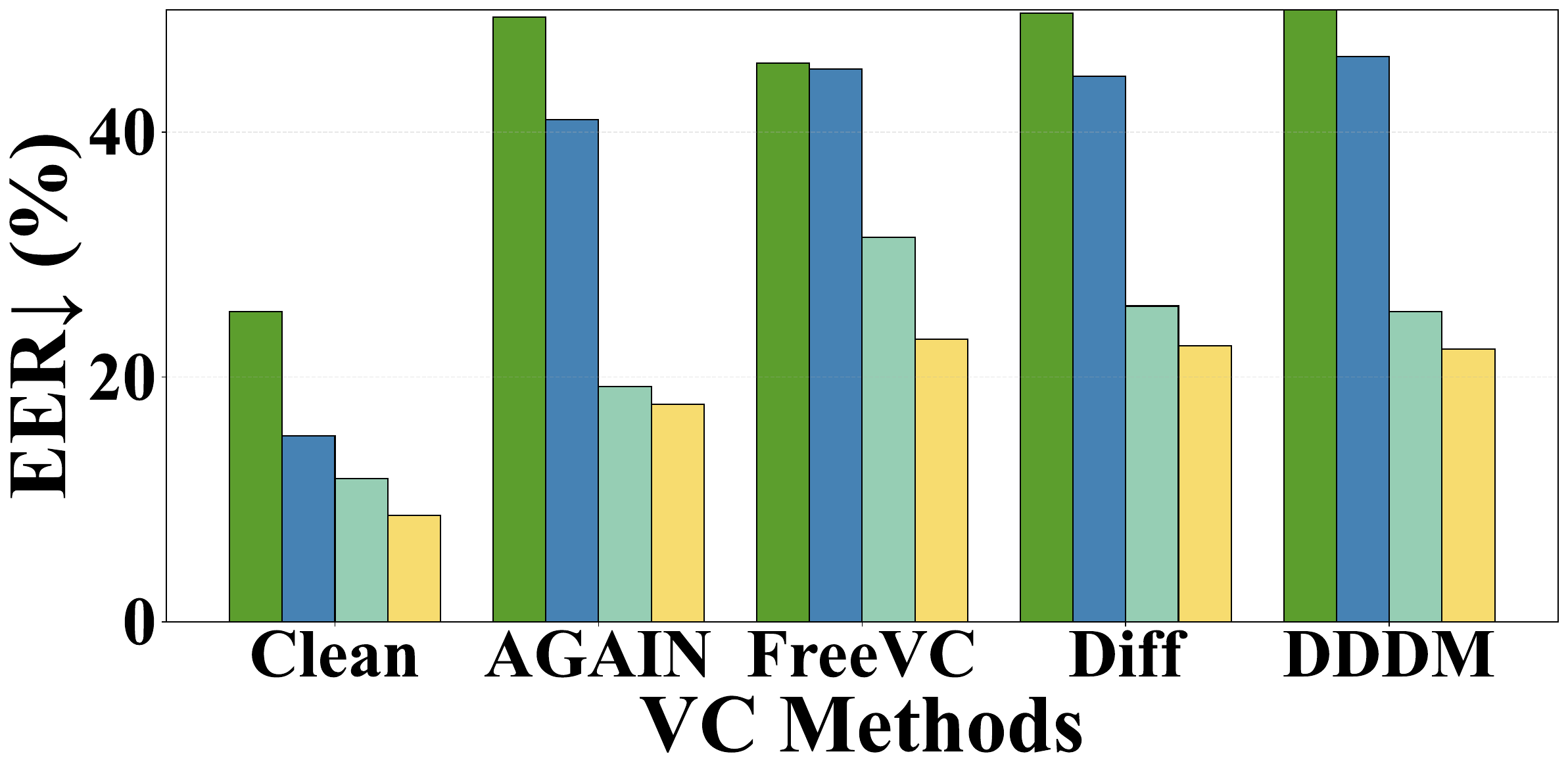}
    \end{minipage}
    \begin{minipage}[b]{0.255\textwidth}
        \centering
        \includegraphics[trim=0mm 0mm 0mm 0mm, clip, width=0.95\linewidth]{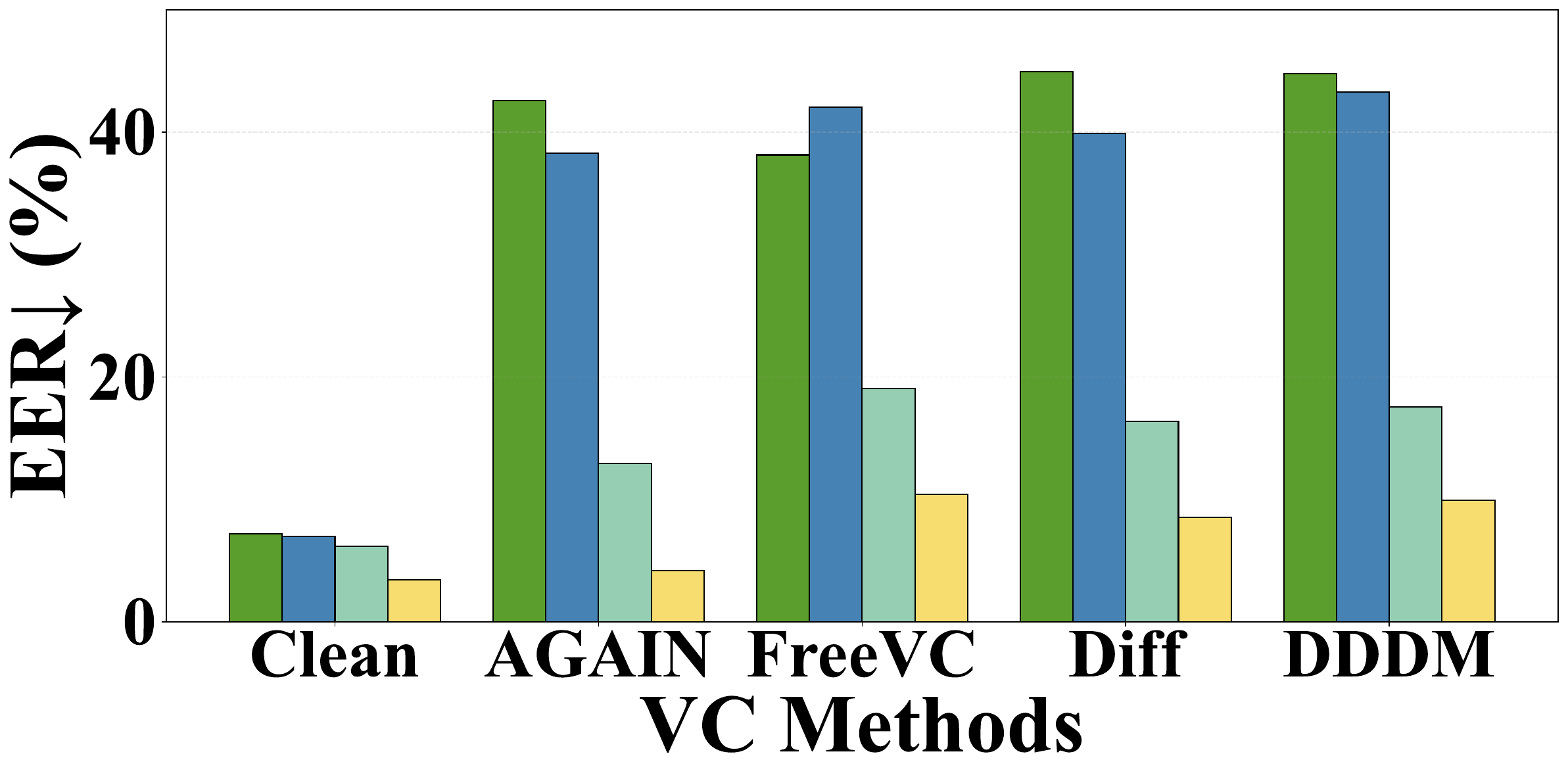}
    \end{minipage}
    \begin{minipage}[b]{0.255\textwidth}
        \centering
        \includegraphics[trim=0mm 0mm 0mm 0mm, clip, width=0.95\linewidth]{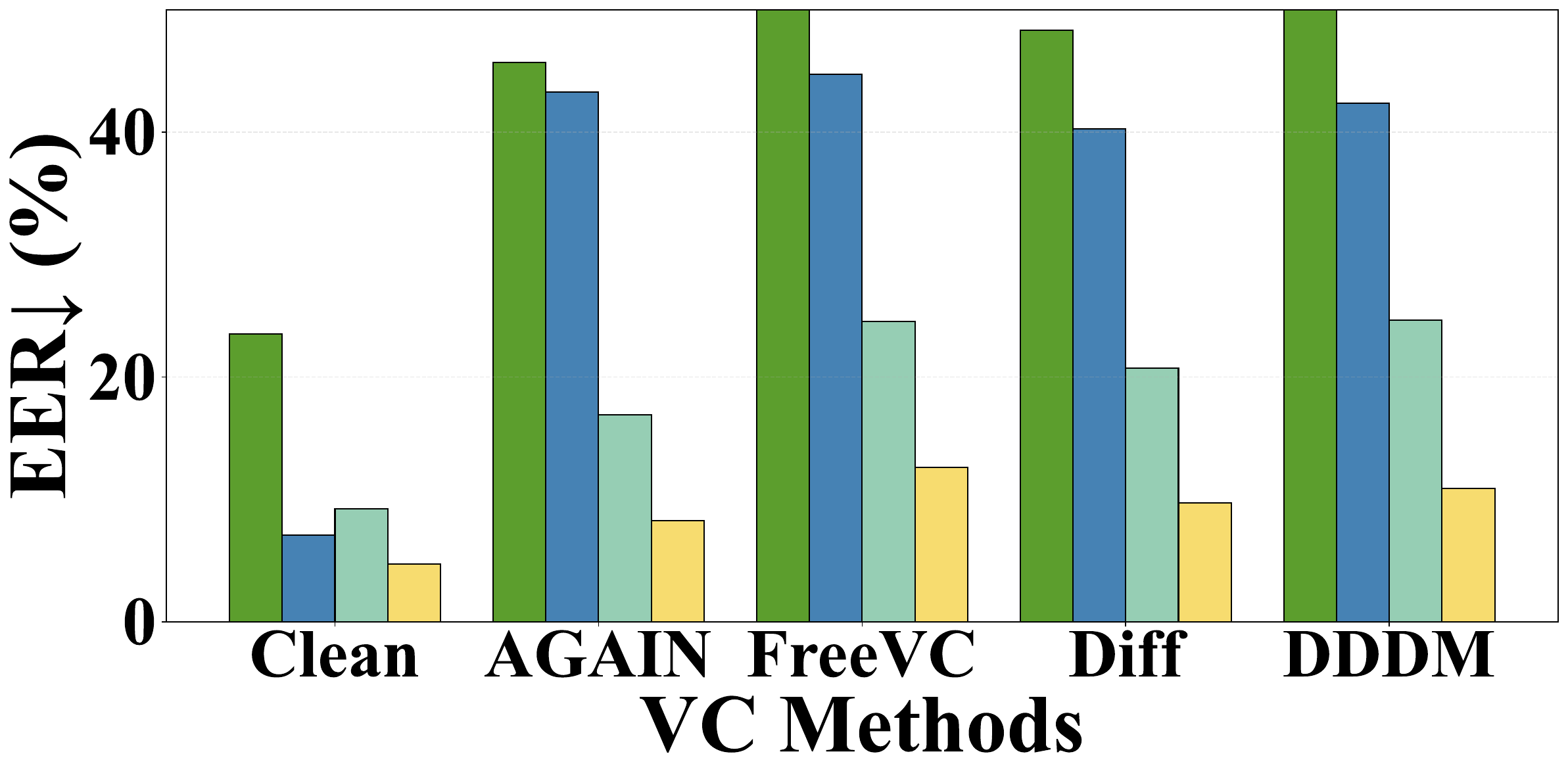}
    \end{minipage}

    \vspace{-0.1cm}
    \subfigure[Chinese]{
        \begin{minipage}[b]{0.261\textwidth}
            \centering
            \includegraphics[trim=0mm 0mm 0mm 0mm, clip, width=0.95\linewidth]{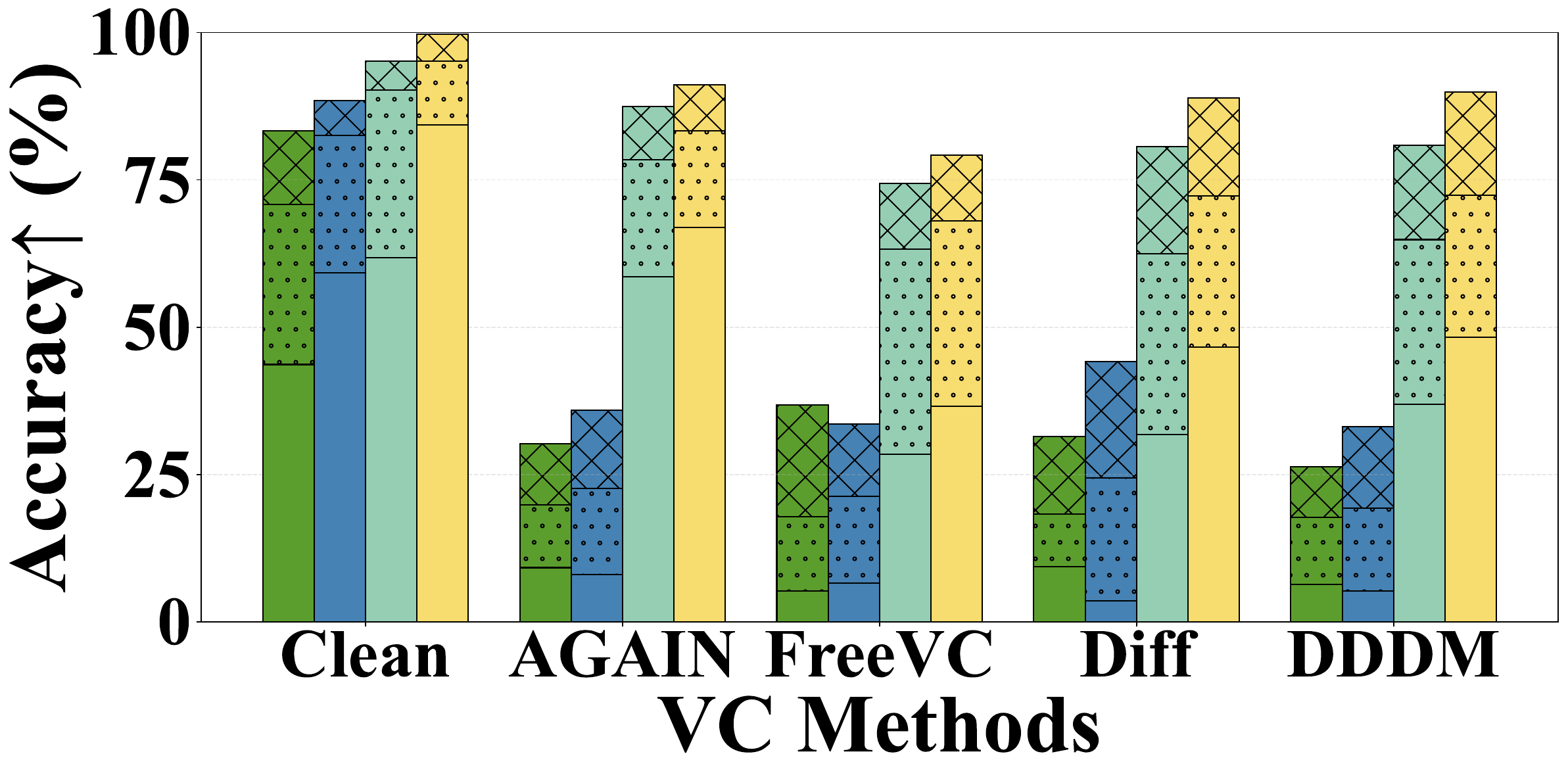}
        \end{minipage}
    }
    \hspace{-0.34cm}
    \subfigure[Amharic]{
        \begin{minipage}[b]{0.261\textwidth}
            \centering
            \includegraphics[trim=0mm 0mm 0mm 0mm, clip, width=0.95\linewidth]{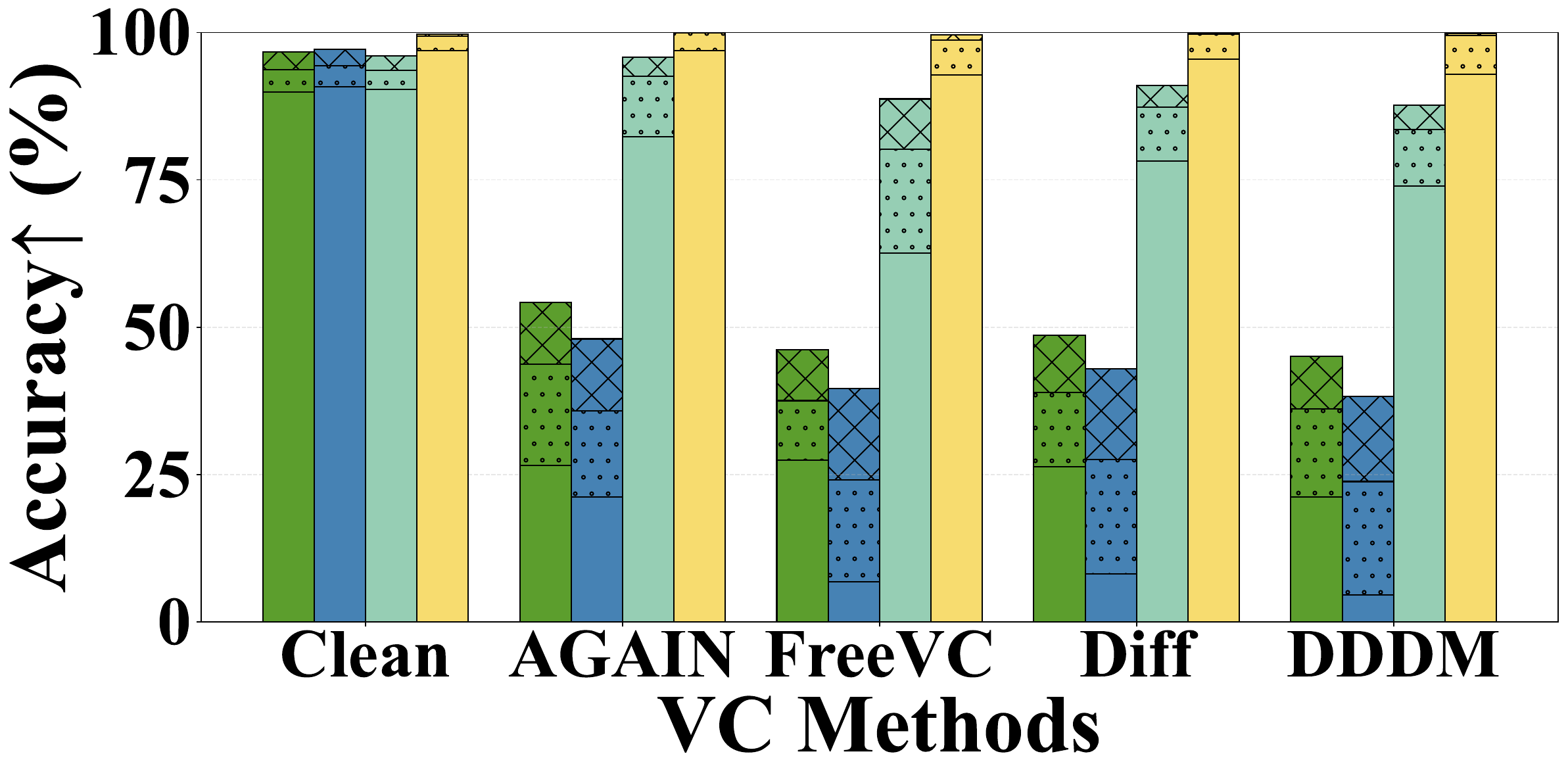}
        \end{minipage}
    }
    \hspace{-0.34cm}
    \subfigure[Finnish]{
        \begin{minipage}[b]{0.261\textwidth}
            \centering
            \includegraphics[trim=0mm 0mm 0mm 0mm, clip, width=0.95\linewidth]{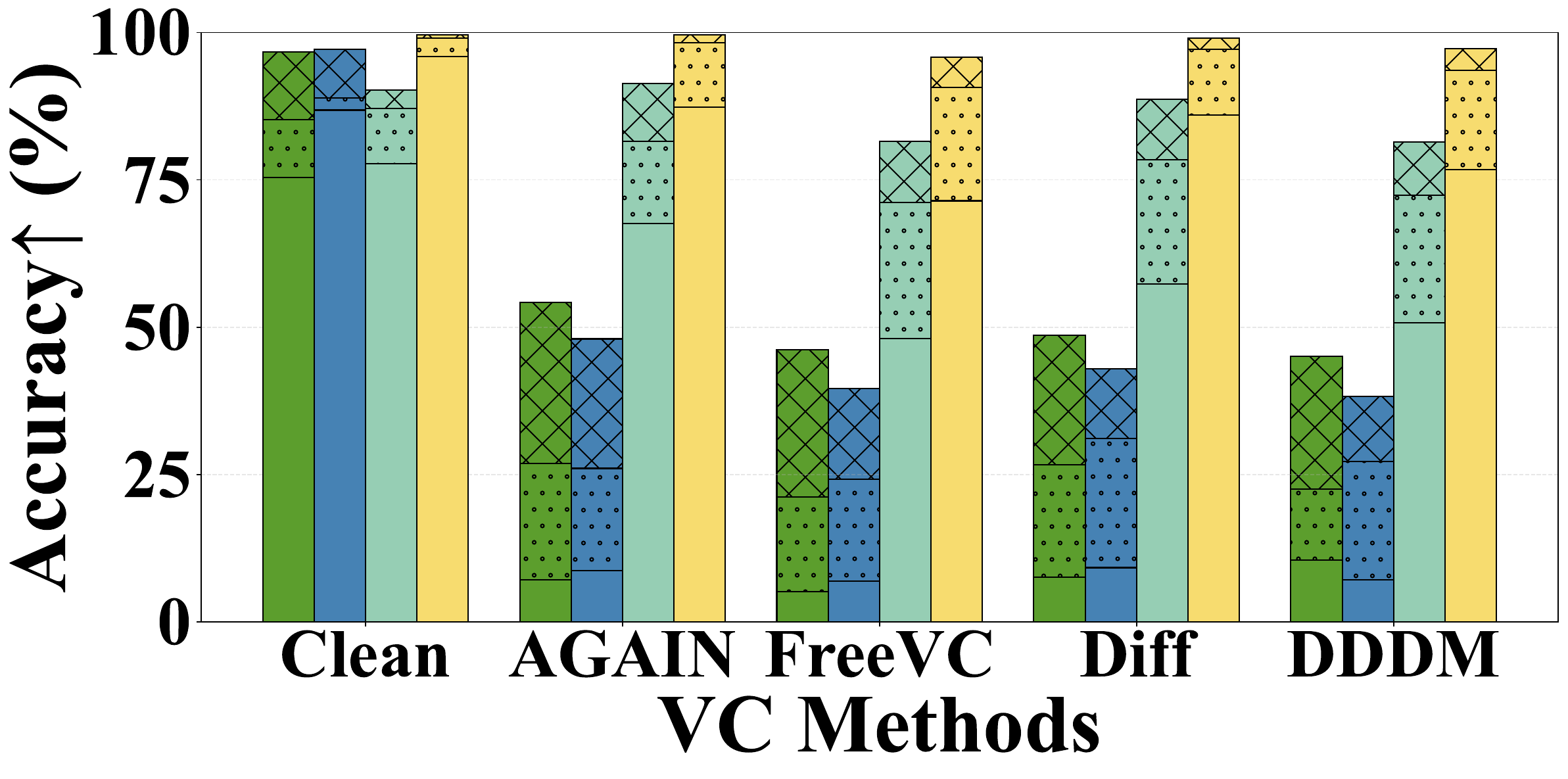}
        \end{minipage}
    }
    
    \vspace{-0.3cm}
    \caption{Performance of voiceprint recovery methods over unseen language.}
    \label{fig:lang}
\end{figure*}

\begin{figure*}[h]
    \centering
    	\begin{minipage}[b]{0.48\linewidth}
    		\centering
    		\includegraphics[trim=0mm 0mm 0mm 0mm, clip, width=\textwidth]{Section/Pictures/Draw/Line/legend.pdf}
    	\end{minipage} \\
    \vspace{-0.05cm}
    \subfigure[Clean]{
    	\begin{minipage}[b]{0.23\linewidth}
    		\centering
    		\includegraphics[trim=0mm 0mm 0mm 0mm, clip, width=0.95\textwidth]{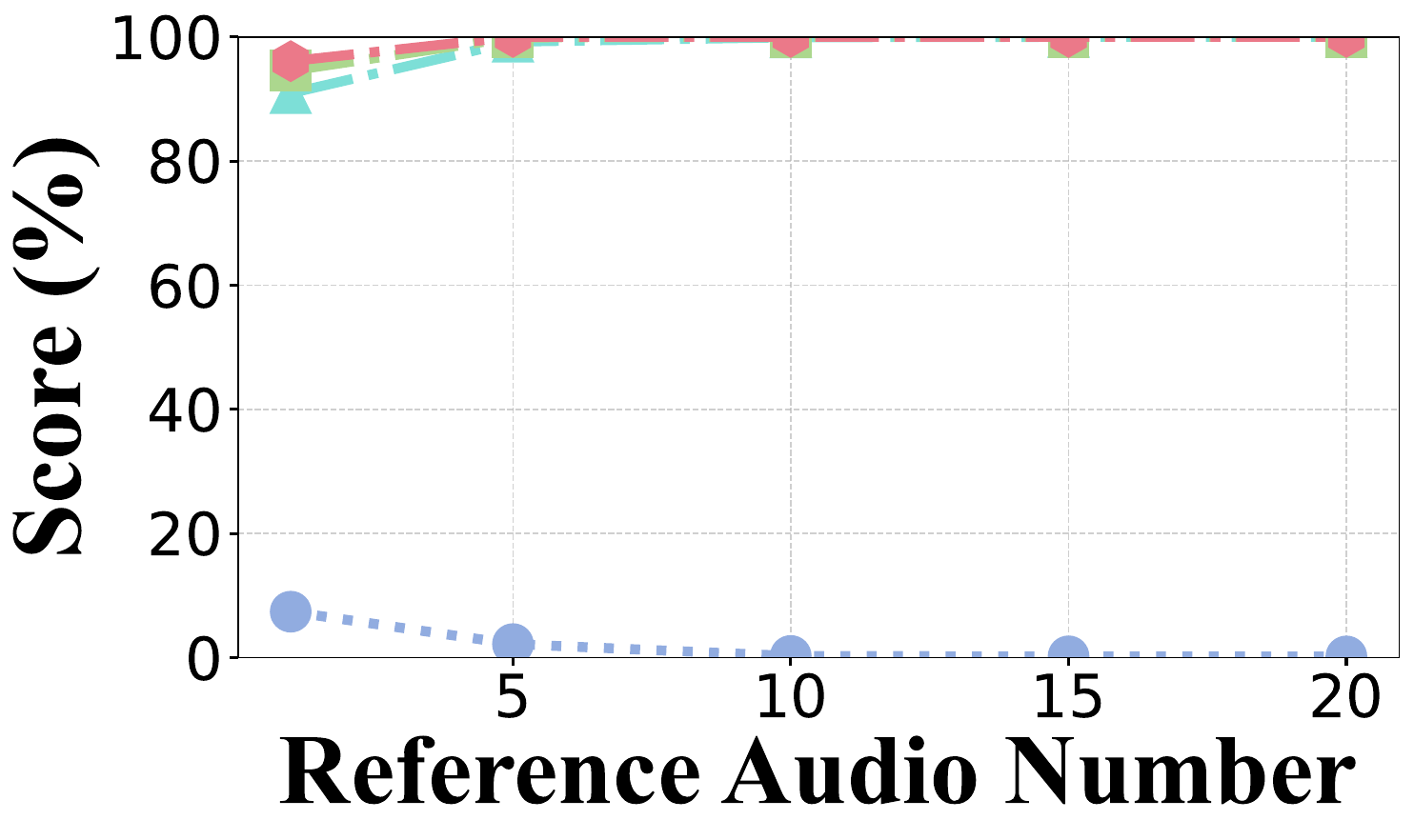}
    		\vspace{-0.1cm}
    	\end{minipage}
    }
    \subfigure[AGAIN]{
    	\begin{minipage}[b]{0.23\linewidth}
    		\centering
    		\includegraphics[trim=0mm 0mm 0mm 0mm, clip, width=0.95\textwidth]{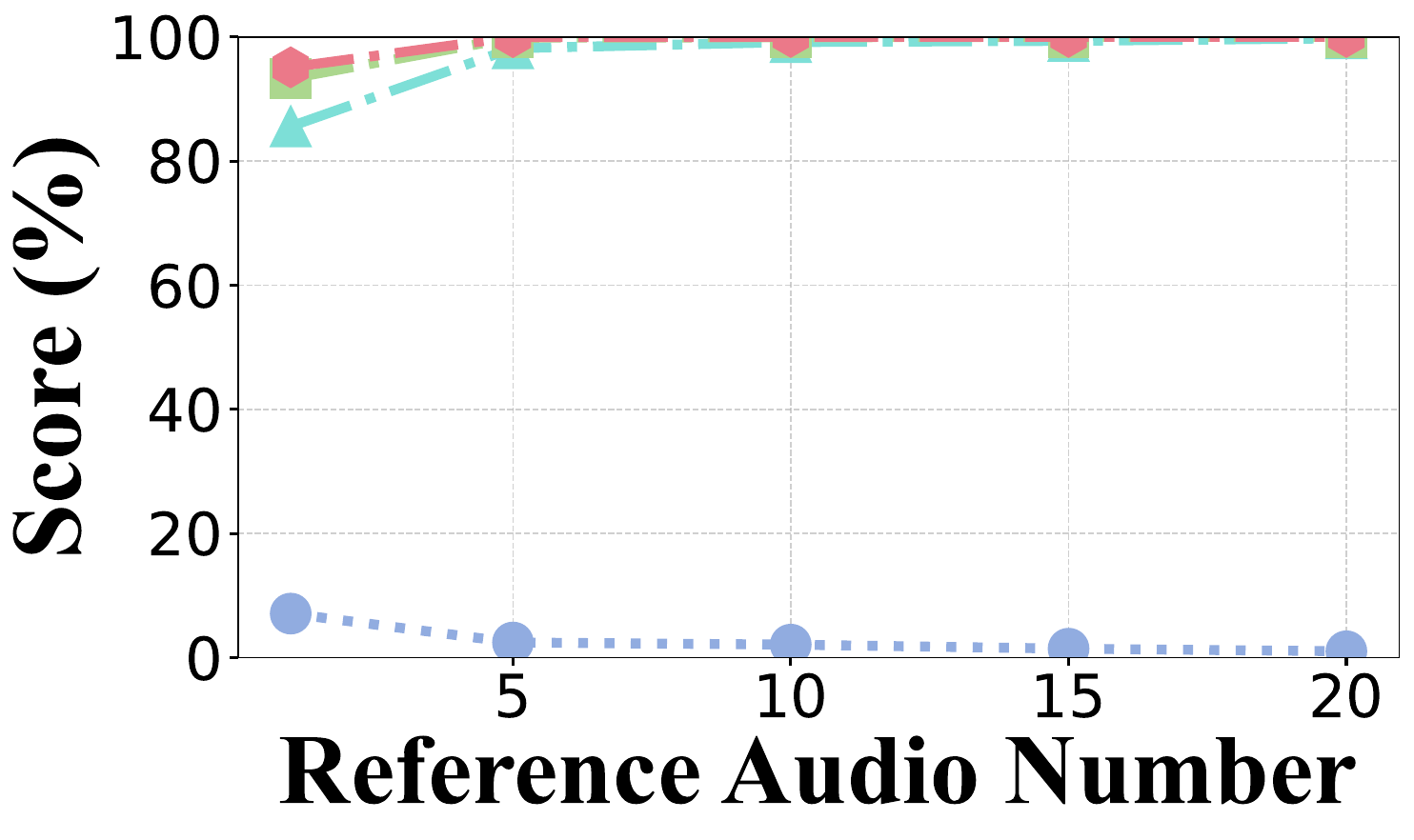}
    		\vspace{-0.1cm}
    	\end{minipage}
    }
    \subfigure[VQVC]{
    	\begin{minipage}[b]{0.23\linewidth}
    		\centering
    		\includegraphics[trim=0mm 0mm 0mm 0mm, clip, width=0.95\textwidth]{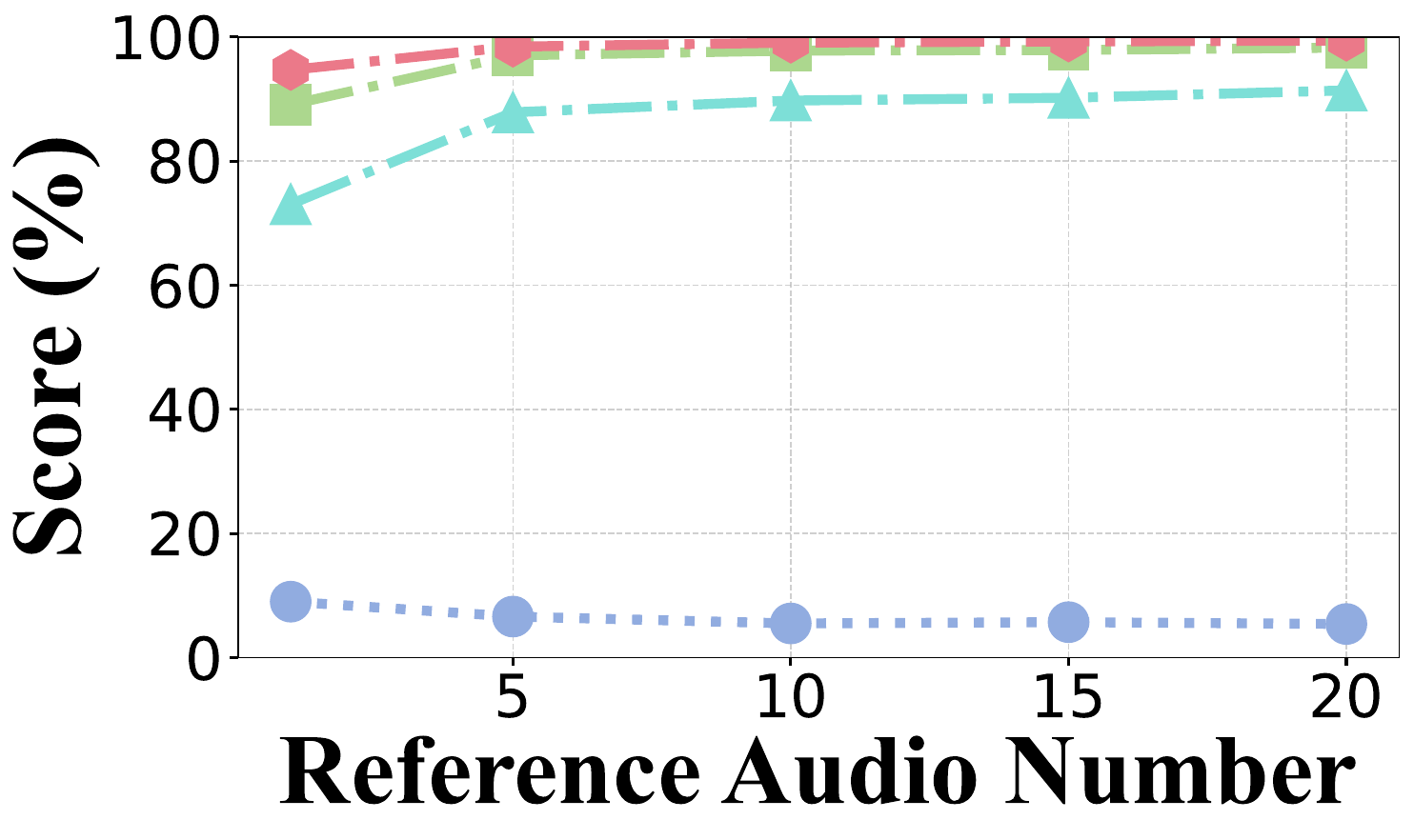}
    		\vspace{-0.1cm}
    	\end{minipage}
    }
  \vspace{-0.2cm}
    \subfigure[VQVC+]{
    	\begin{minipage}[b]{0.23\linewidth}
    		\centering
    		\includegraphics[trim=0mm 0mm 0mm 0mm, clip,width=0.95\textwidth]{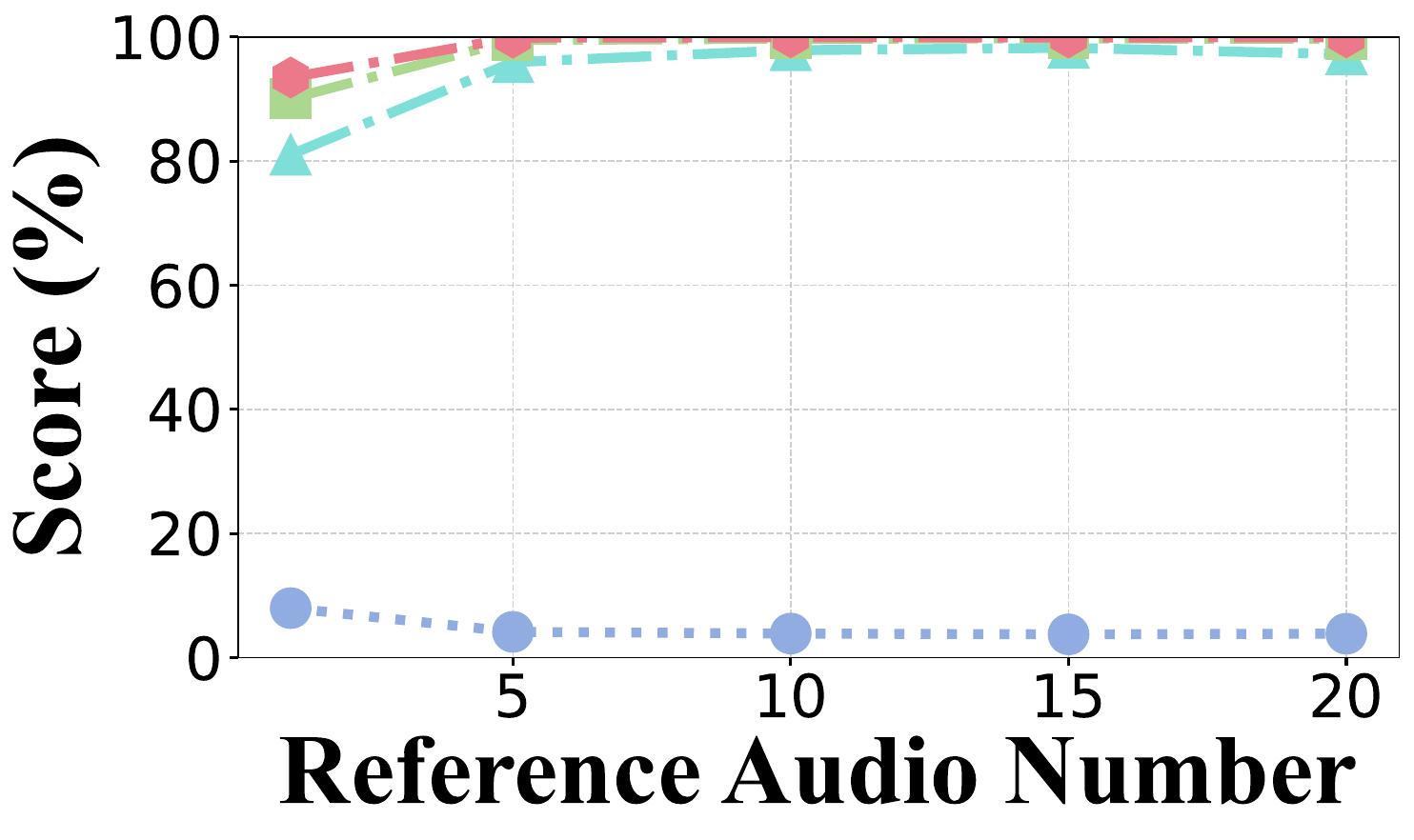}
    		\vspace{-0.1cm}
    	\end{minipage}
    }
    \subfigure[BNE]{
    	\begin{minipage}[b]{0.23\linewidth}
    		\centering
    		\includegraphics[trim=0mm 0mm 0mm 0mm, clip,width=0.95\textwidth]{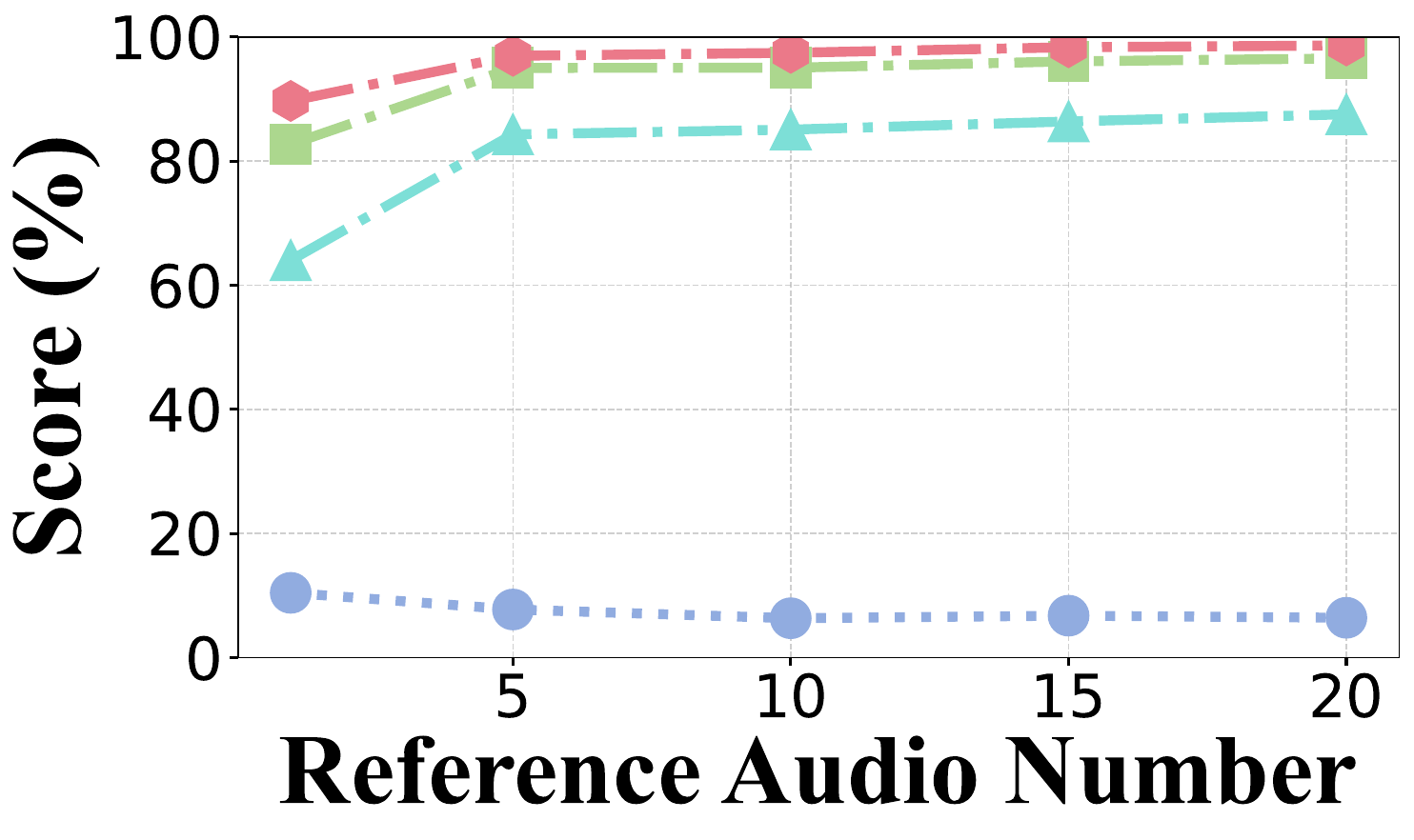}
    		\vspace{-0.1cm}
    	\end{minipage}
    }
    \subfigure[FreeVC]{
    	\begin{minipage}[b]{0.23\linewidth}
    		\centering
    		\includegraphics[trim=0mm 0mm 0mm 0mm, clip,width=0.95\textwidth]{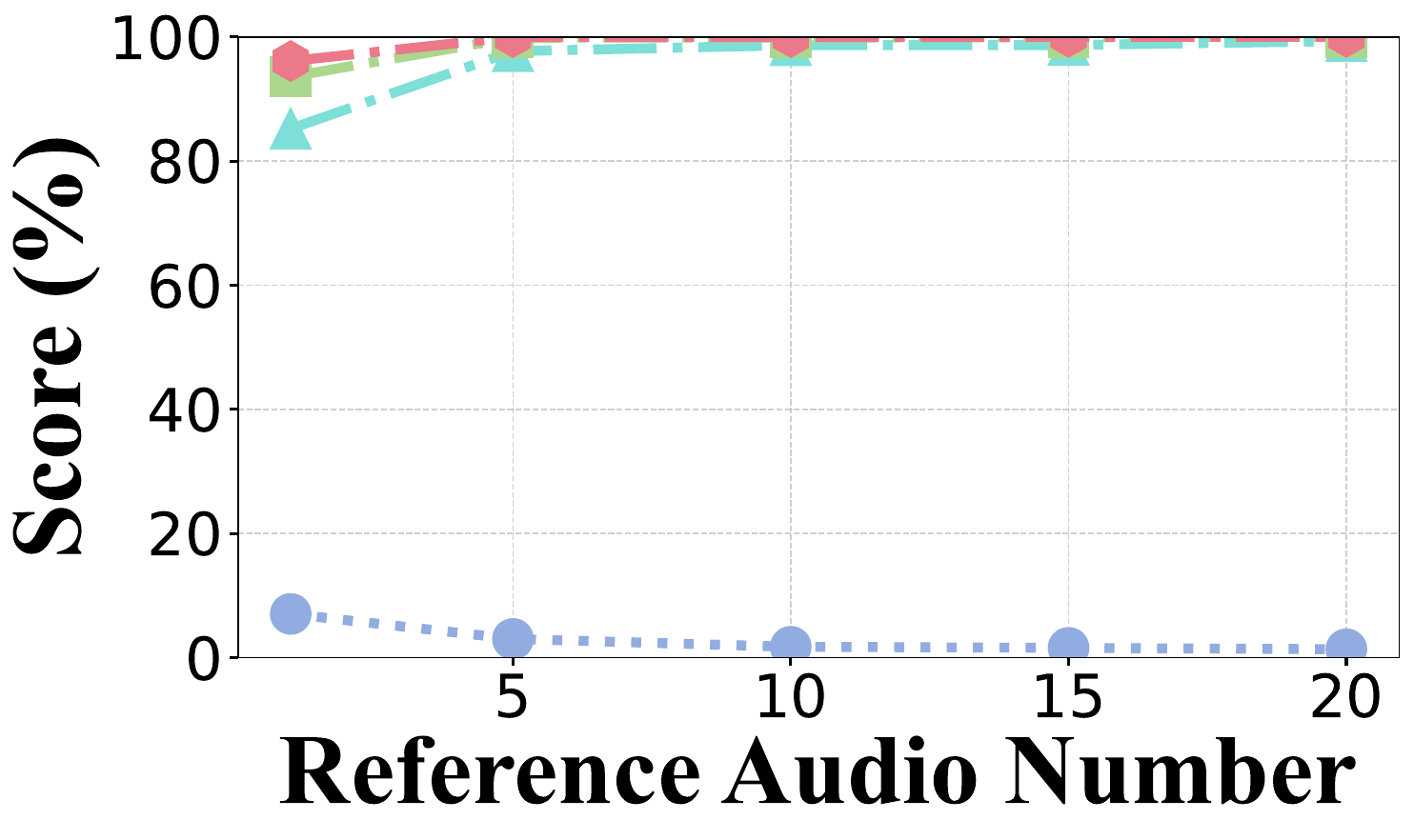}
    		\vspace{-0.1cm}
    	\end{minipage}
    }
    \subfigure[Diff]{
    	\begin{minipage}[b]{0.23\linewidth}
    		\centering
    		\includegraphics[trim=0mm 0mm 0mm 0mm, clip,width=0.95\textwidth]{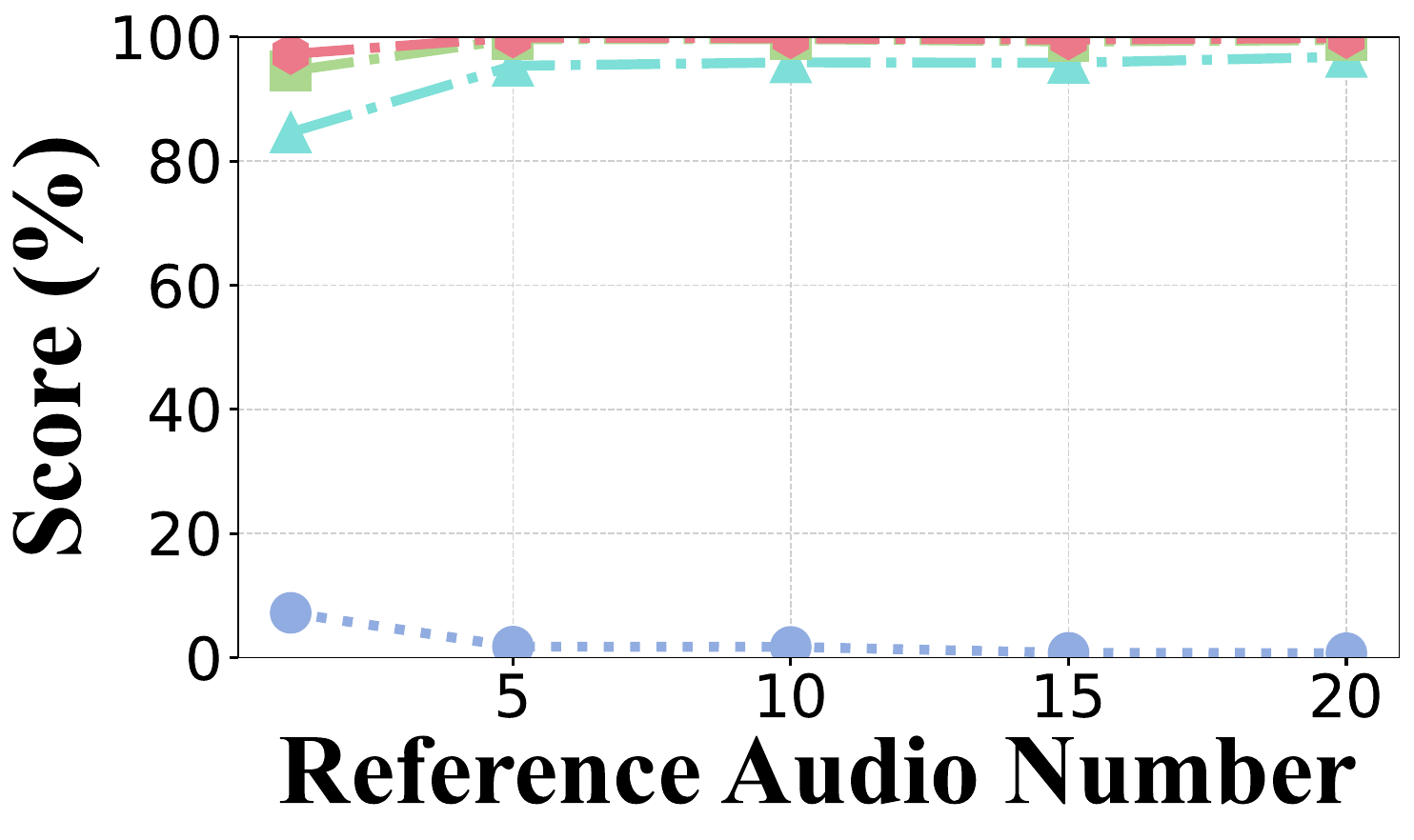}
    		\vspace{-0.1cm}
    	\end{minipage}
    }
    \subfigure[DDDM]{
    	\begin{minipage}[b]{0.23\linewidth}
    		\centering
    		\includegraphics[trim=0mm 0mm 0mm 0mm, clip,width=0.95\textwidth]{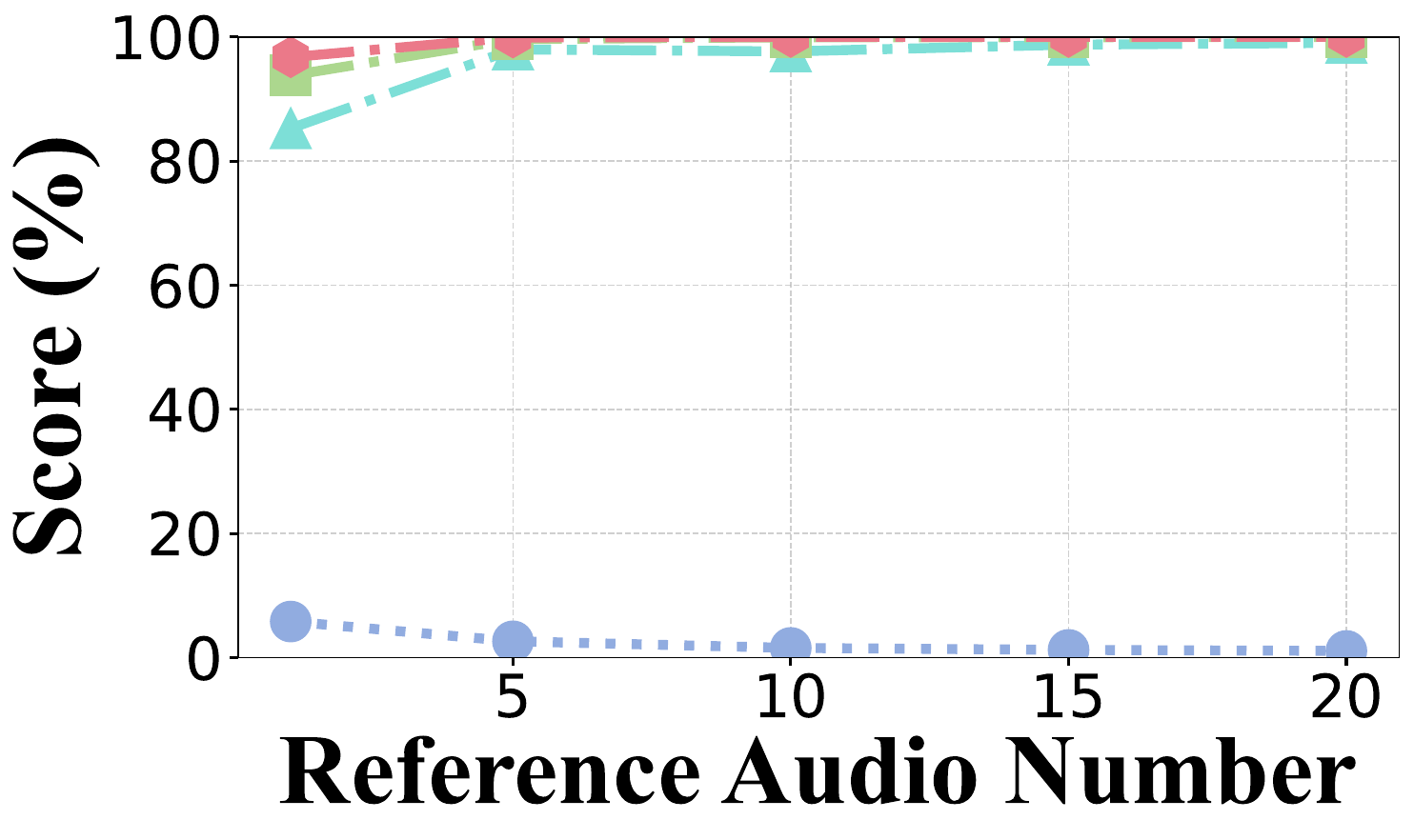}
    		\vspace{-0.1cm}
    	\end{minipage}
    }
 \vspace{-0.3cm}
	\caption{Impact of reference audio number.}
	\label{fig:line_ref}
\end{figure*}

\begin{figure*}[h]
    \centering
    	\begin{minipage}[b]{0.48\linewidth}
    		\centering
    		\includegraphics[trim=0mm 0mm 0mm 0mm, clip, width=\textwidth]{Section/Pictures/Draw/Line/legend.pdf}
    	\end{minipage} \\
    \vspace{-0.05cm}
    \subfigure[Clean]{
    	\begin{minipage}[b]{0.23\linewidth}
    		\centering
    		\includegraphics[trim=0mm 0mm 0mm 0mm, clip, width=0.95\textwidth]{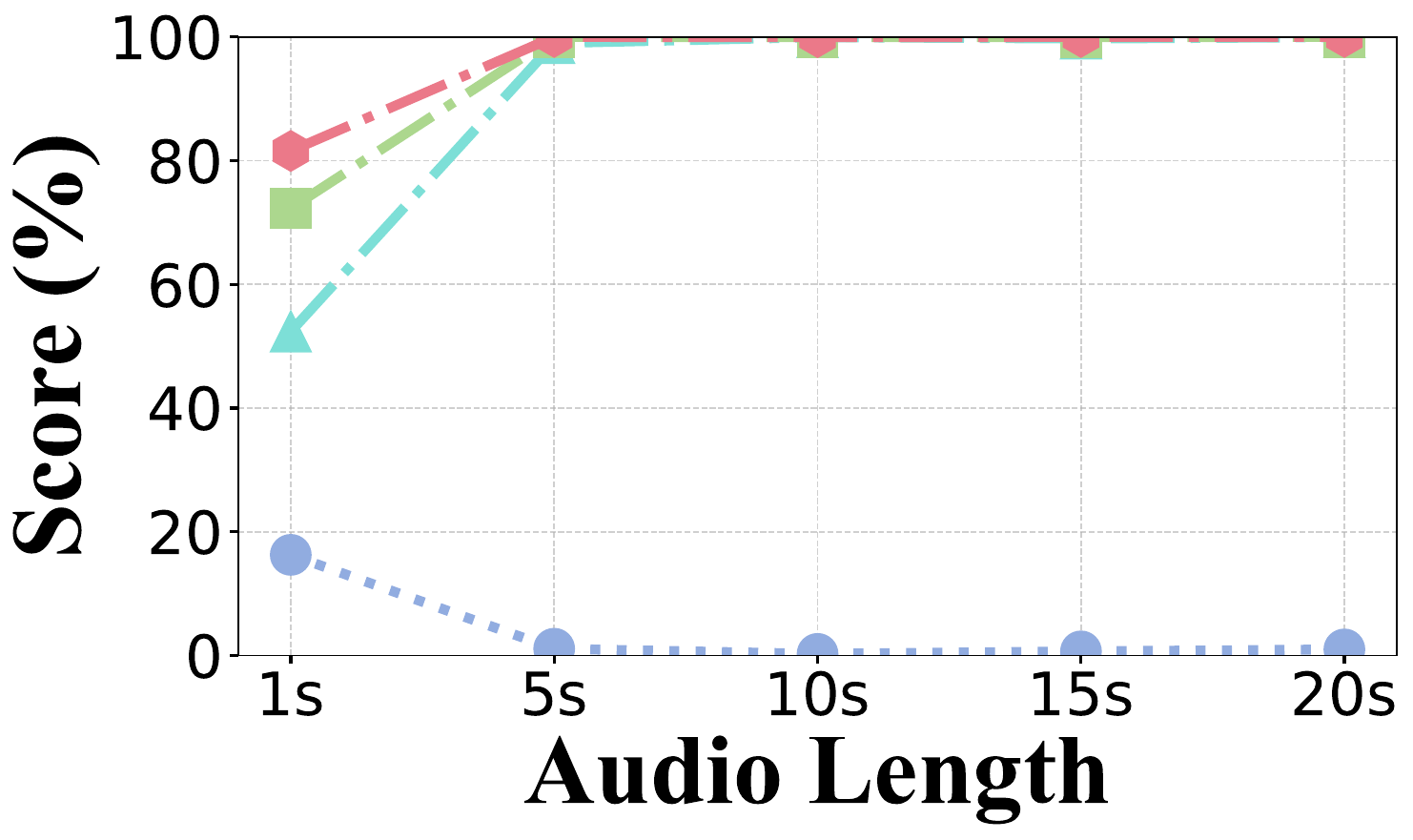}
    		\vspace{-0.1cm}
    	\end{minipage}
    }
    \subfigure[AGAIN]{
    	\begin{minipage}[b]{0.23\linewidth}
    		\centering
    		\includegraphics[trim=0mm 0mm 0mm 0mm, clip, width=0.95\textwidth]{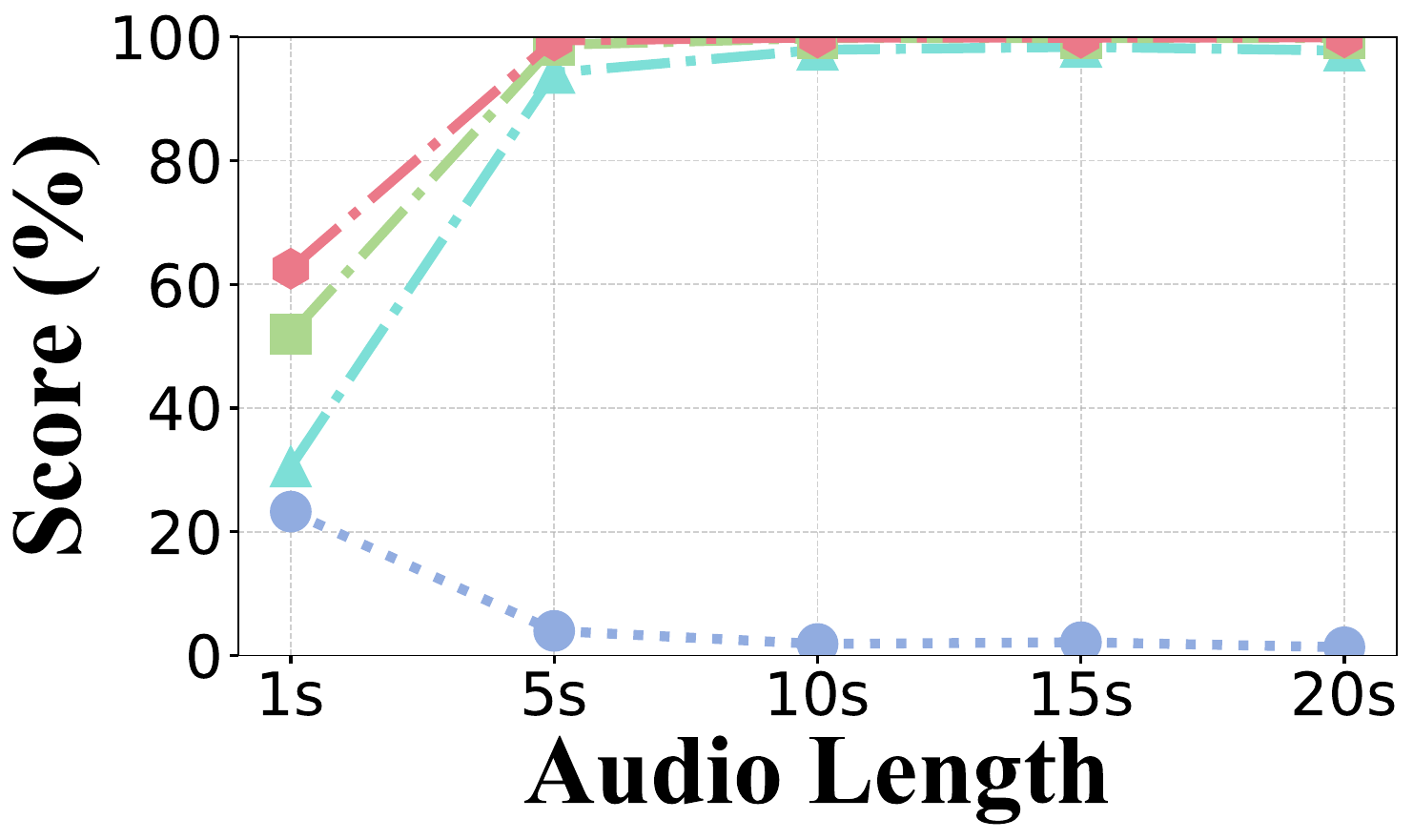}
    		\vspace{-0.1cm}
    	\end{minipage}
    }
    \subfigure[VQVC]{
    	\begin{minipage}[b]{0.23\linewidth}
    		\centering
    		\includegraphics[trim=0mm 0mm 0mm 0mm, clip, width=0.95\textwidth]{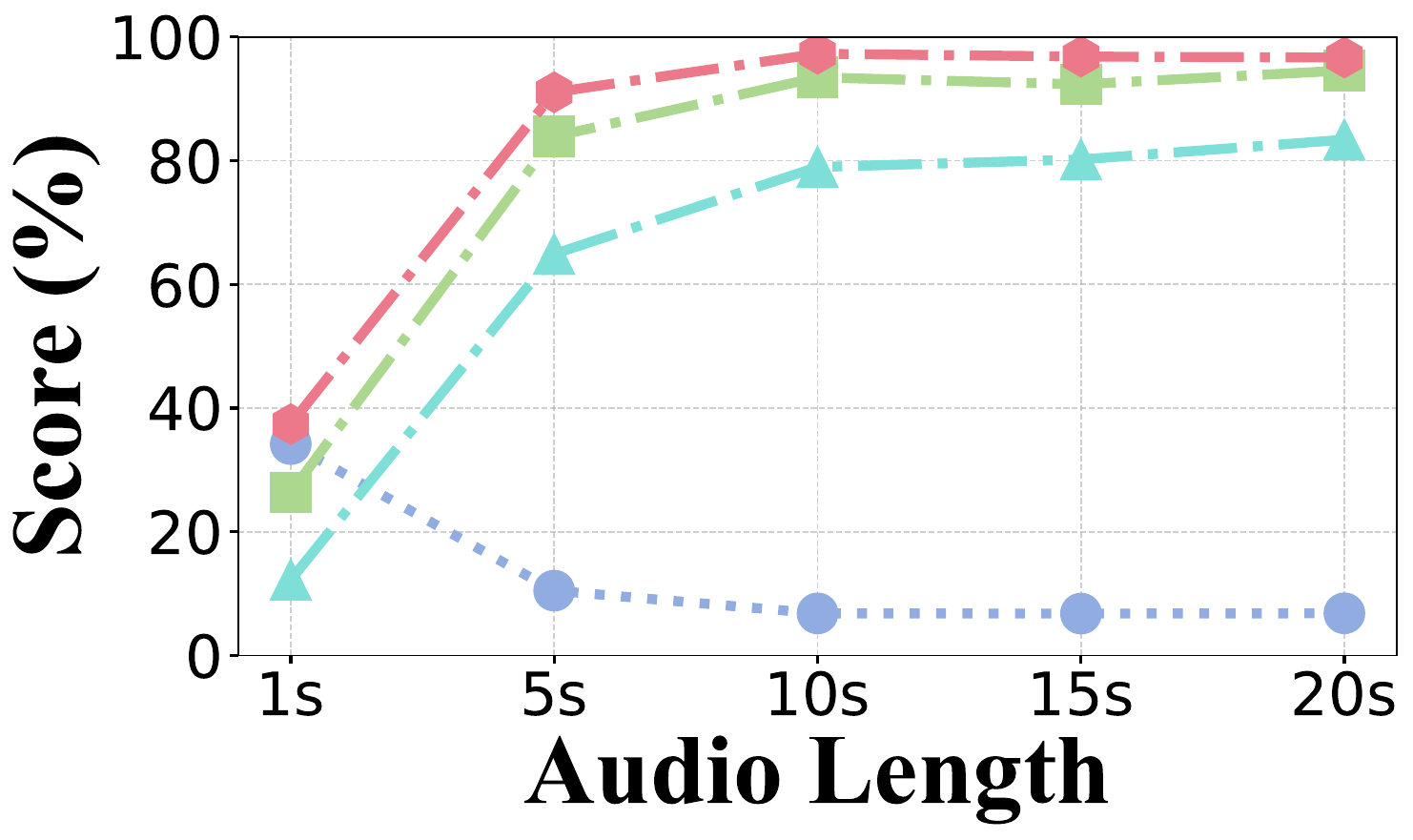}
    		\vspace{-0.1cm}
    	\end{minipage}
    }
  \vspace{-0.2cm}
    \subfigure[VQVC+]{
    	\begin{minipage}[b]{0.23\linewidth}
    		\centering
    		\includegraphics[trim=0mm 0mm 0mm 0mm, clip,width=0.95\textwidth]{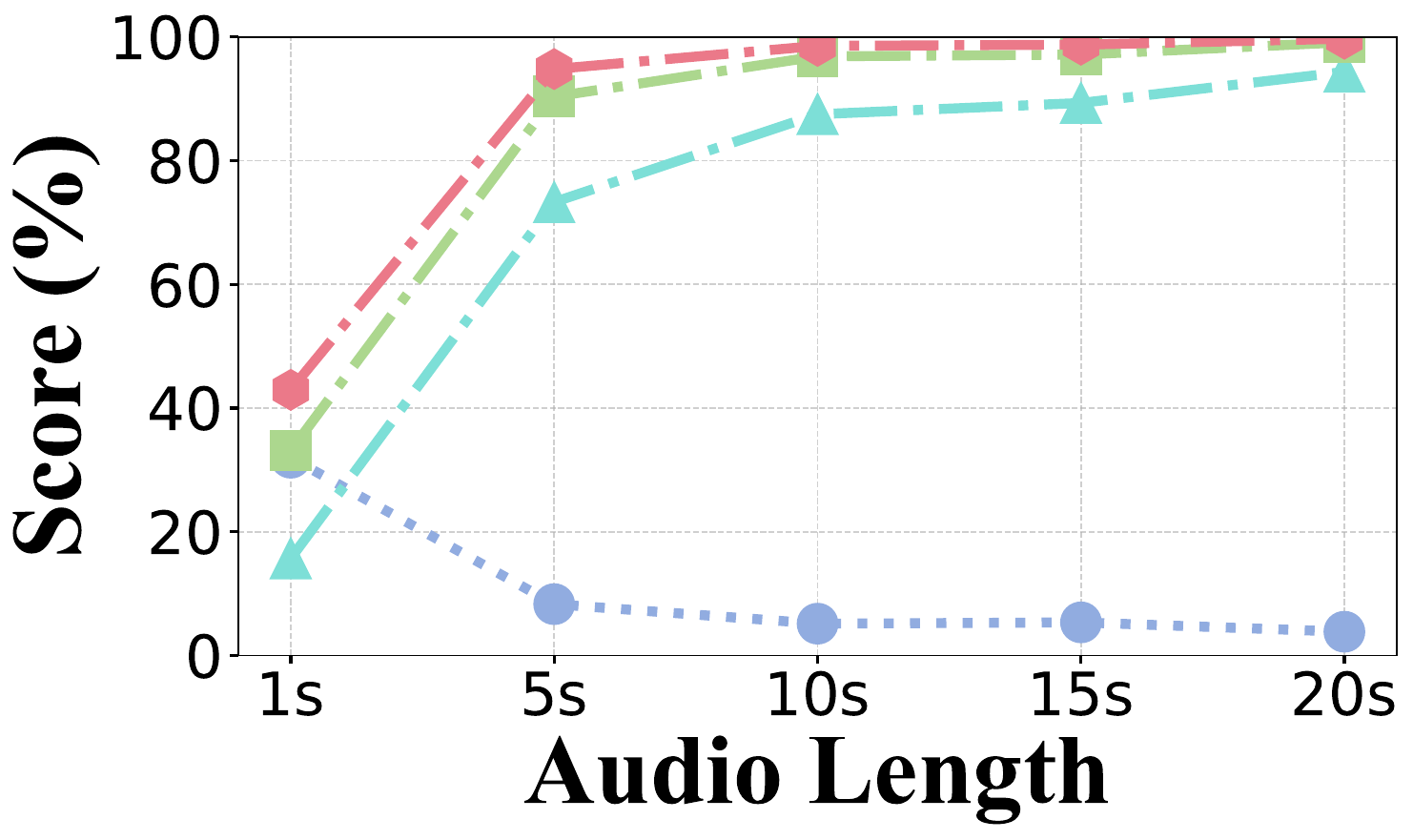}
    		\vspace{-0.1cm}
    	\end{minipage}
    }
    \subfigure[BNE]{
    	\begin{minipage}[b]{0.23\linewidth}
    		\centering
    		\includegraphics[trim=0mm 0mm 0mm 0mm, clip,width=0.95\textwidth]{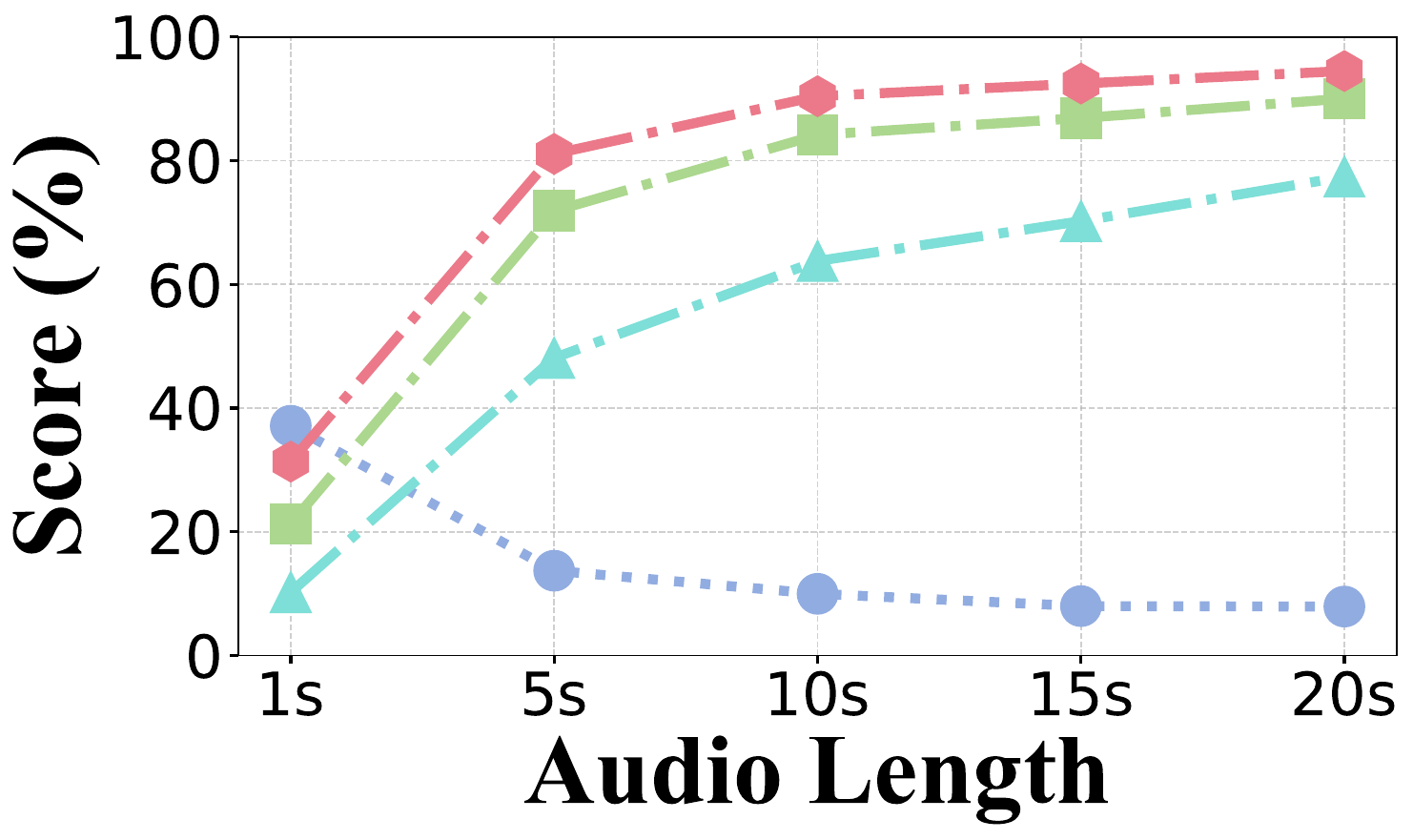}
    		\vspace{-0.1cm}
    	\end{minipage}
    }
    \subfigure[FreeVC]{
    	\begin{minipage}[b]{0.23\linewidth}
    		\centering
    		\includegraphics[trim=0mm 0mm 0mm 0mm, clip,width=0.95\textwidth]{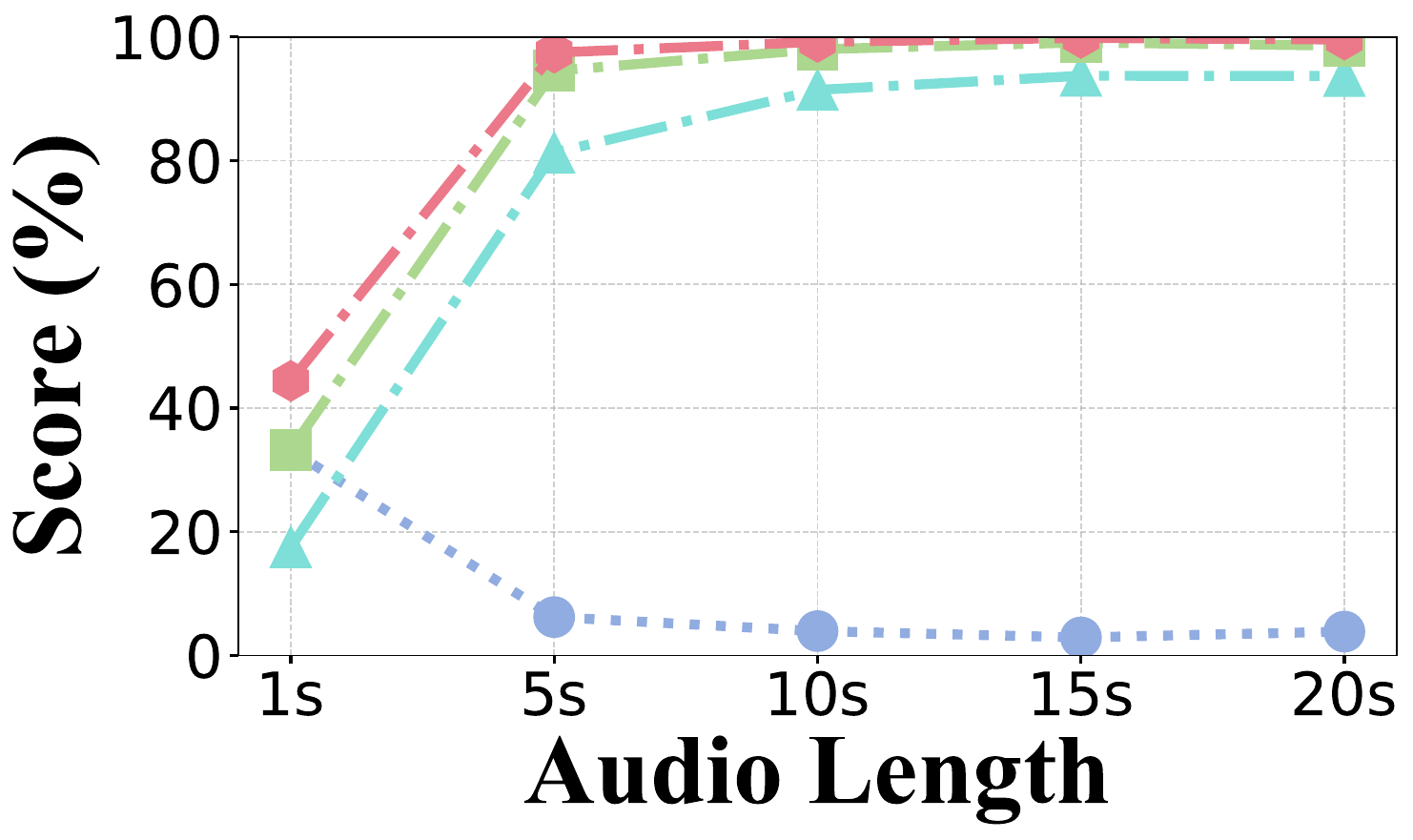}
    		\vspace{-0.1cm}
    	\end{minipage}
    }
    \subfigure[Diff]{
    	\begin{minipage}[b]{0.23\linewidth}
    		\centering
    		\includegraphics[trim=0mm 0mm 0mm 0mm, clip,width=0.95\textwidth]{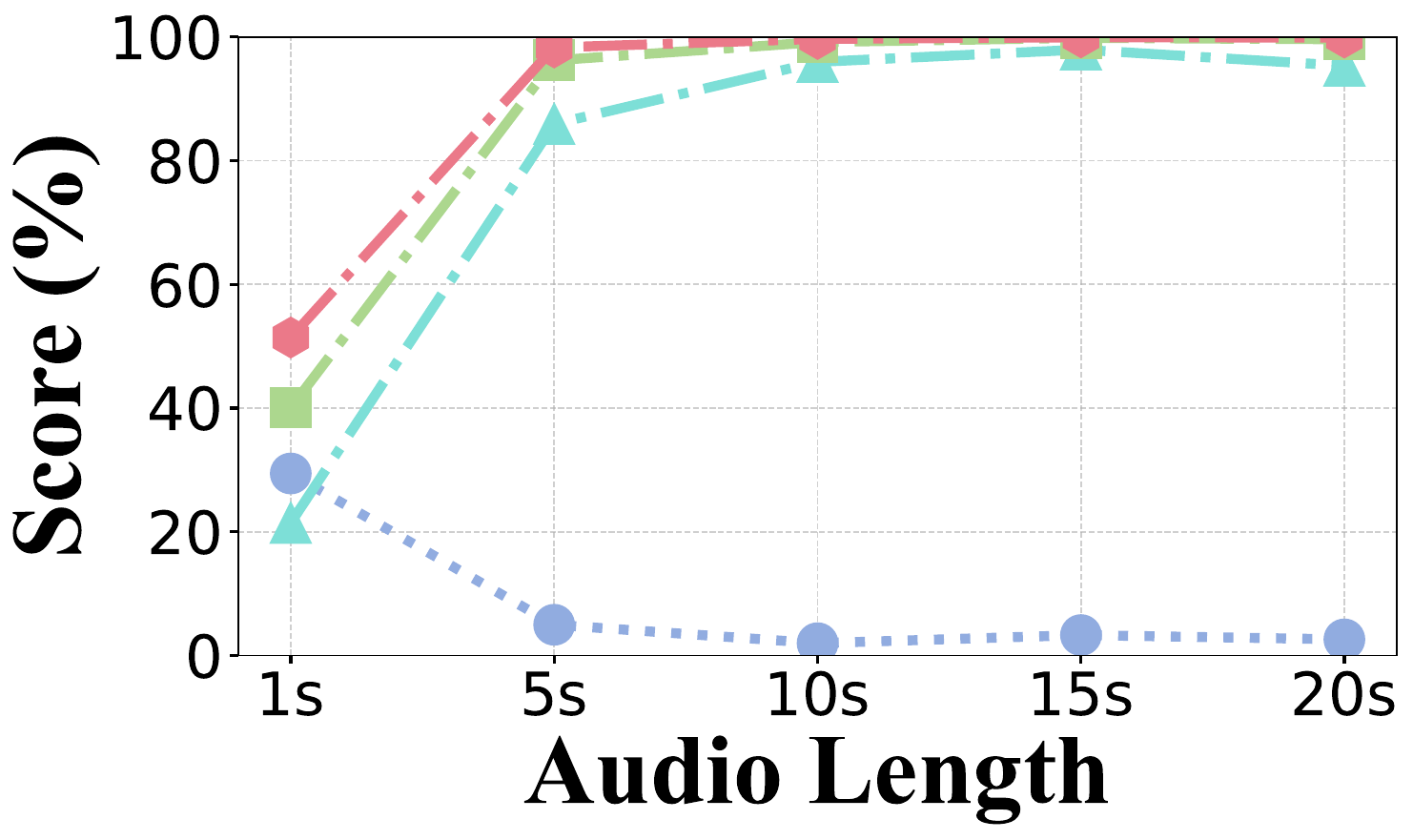}
    		\vspace{-0.1cm}
    	\end{minipage}
    }
    \subfigure[DDDM]{
    	\begin{minipage}[b]{0.23\linewidth}
    		\centering
    		\includegraphics[trim=0mm 0mm 0mm 0mm, clip,width=0.95\textwidth]{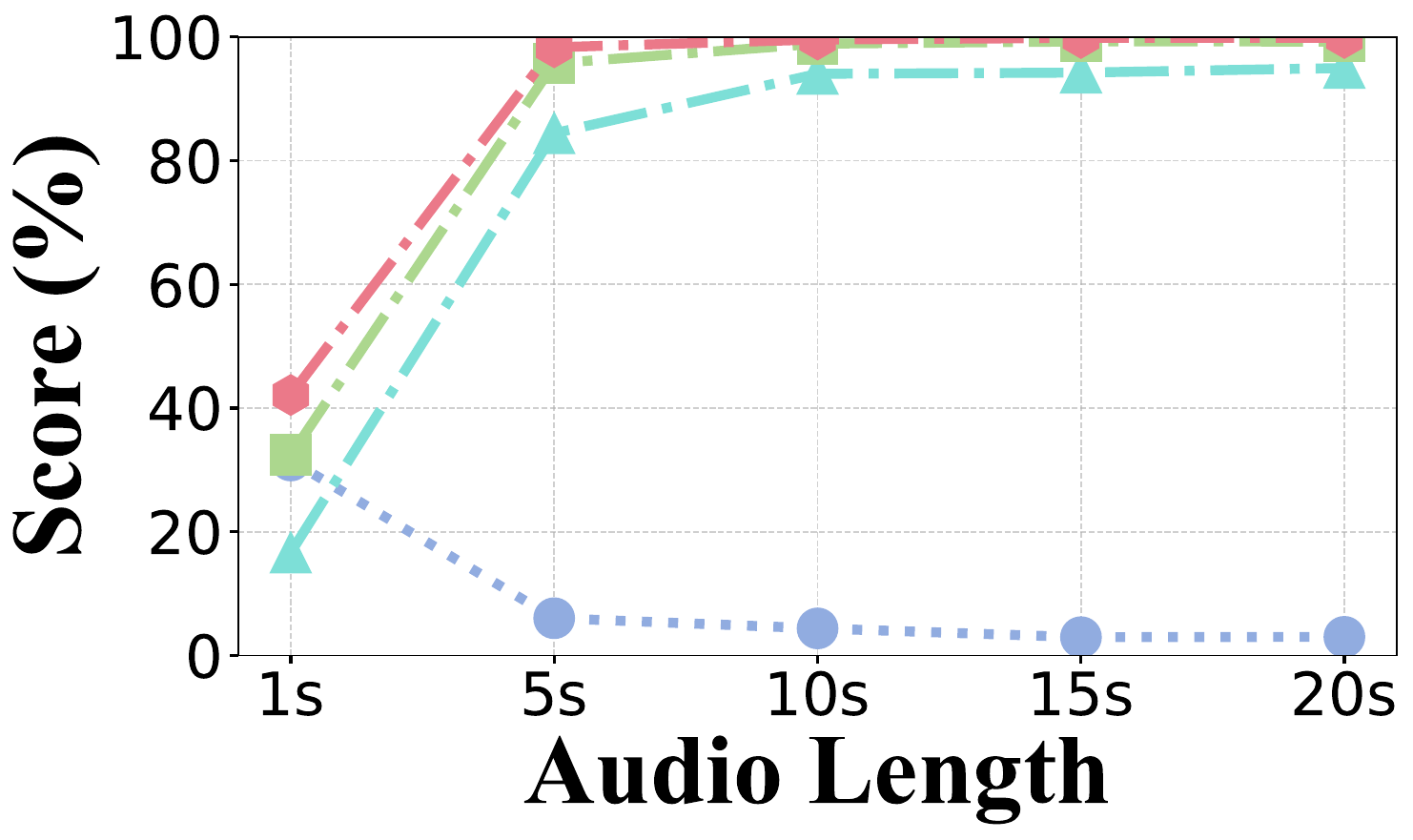}
    		\vspace{-0.1cm}
    	\end{minipage}
    }
 \vspace{-0.3cm}
	\caption{Impact of audio length.}
	\label{fig:line_len}
\end{figure*}

\begin{figure*}[h]
    \centering
    	\begin{minipage}[b]{0.48\linewidth}
    		\centering
    		\includegraphics[trim=0mm 0mm 0mm 0mm, clip, width=\textwidth]{Section/Pictures/Draw/Line/legend.pdf}
    	\end{minipage} \\
    \vspace{-0.05cm}
    \subfigure[Clean]{
    	\begin{minipage}[b]{0.23\linewidth}
    		\centering
    		\includegraphics[trim=0mm 0mm 0mm 0mm, clip, width=0.95\textwidth]{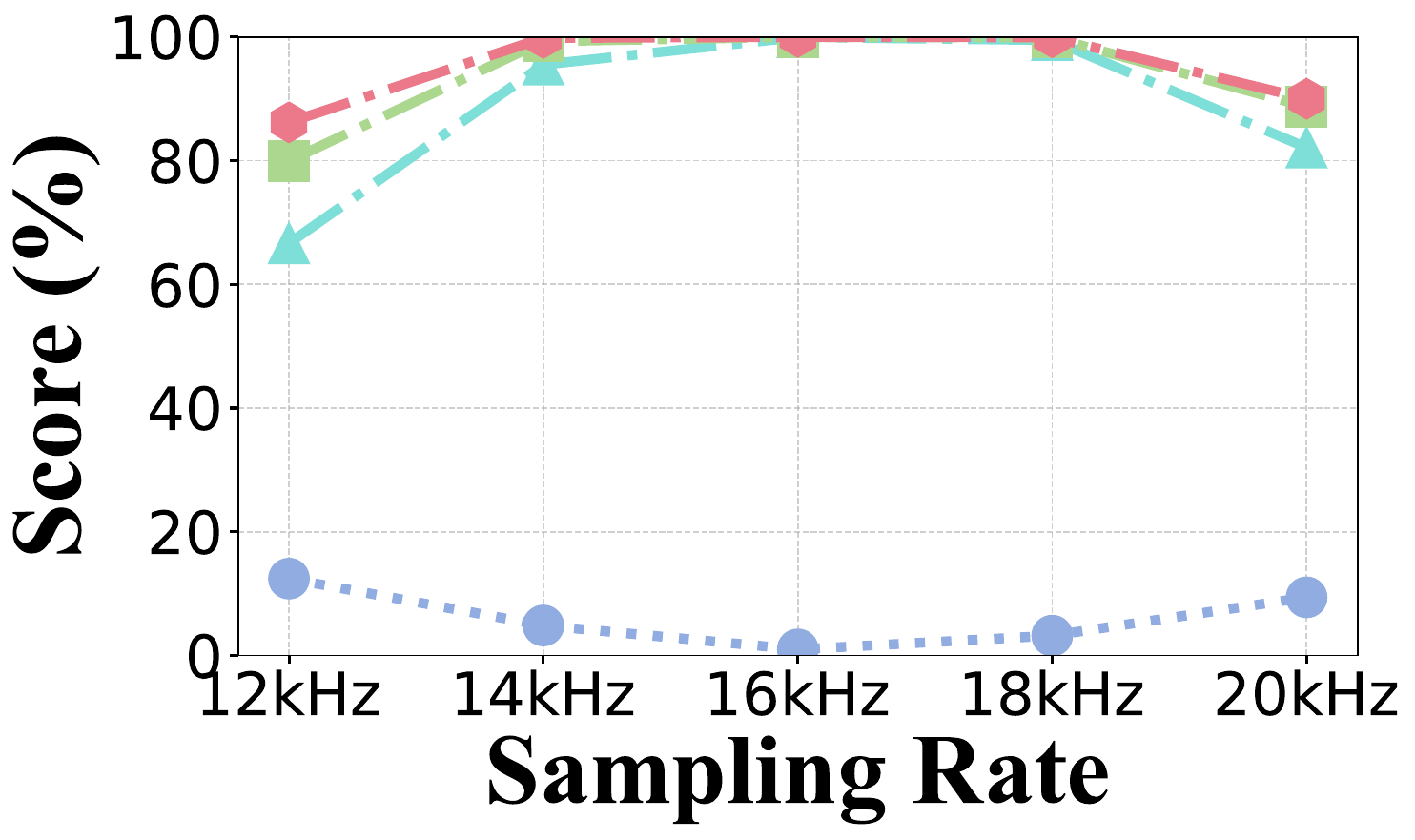}
    		\vspace{-0.1cm}
    	\end{minipage}
    }
    \subfigure[AGAIN]{
    	\begin{minipage}[b]{0.23\linewidth}
    		\centering
    		\includegraphics[trim=0mm 0mm 0mm 0mm, clip, width=0.95\textwidth]{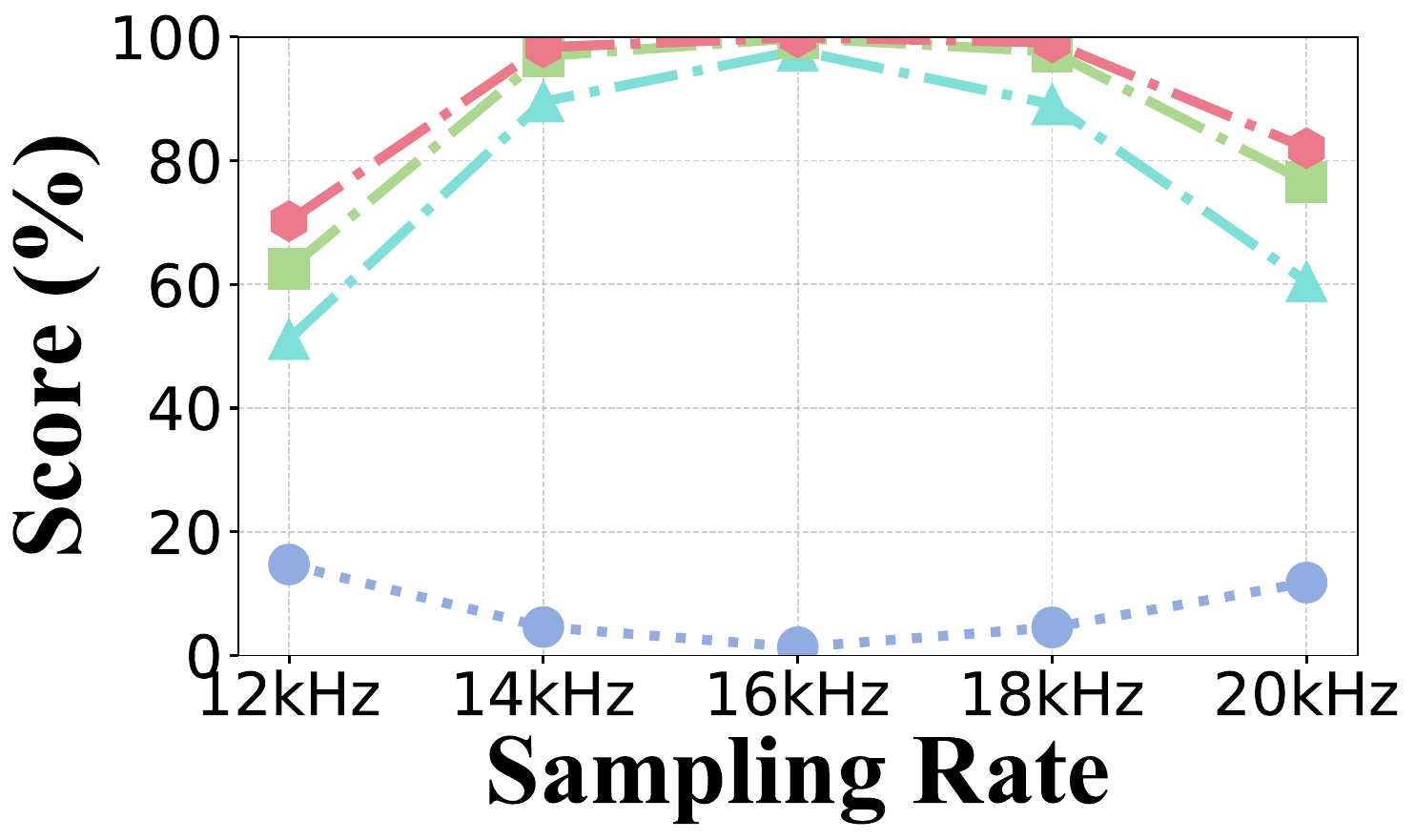}
    		\vspace{-0.1cm}
    	\end{minipage}
    }
    \subfigure[VQVC]{
    	\begin{minipage}[b]{0.23\linewidth}
    		\centering
    		\includegraphics[trim=0mm 0mm 0mm 0mm, clip, width=0.95\textwidth]{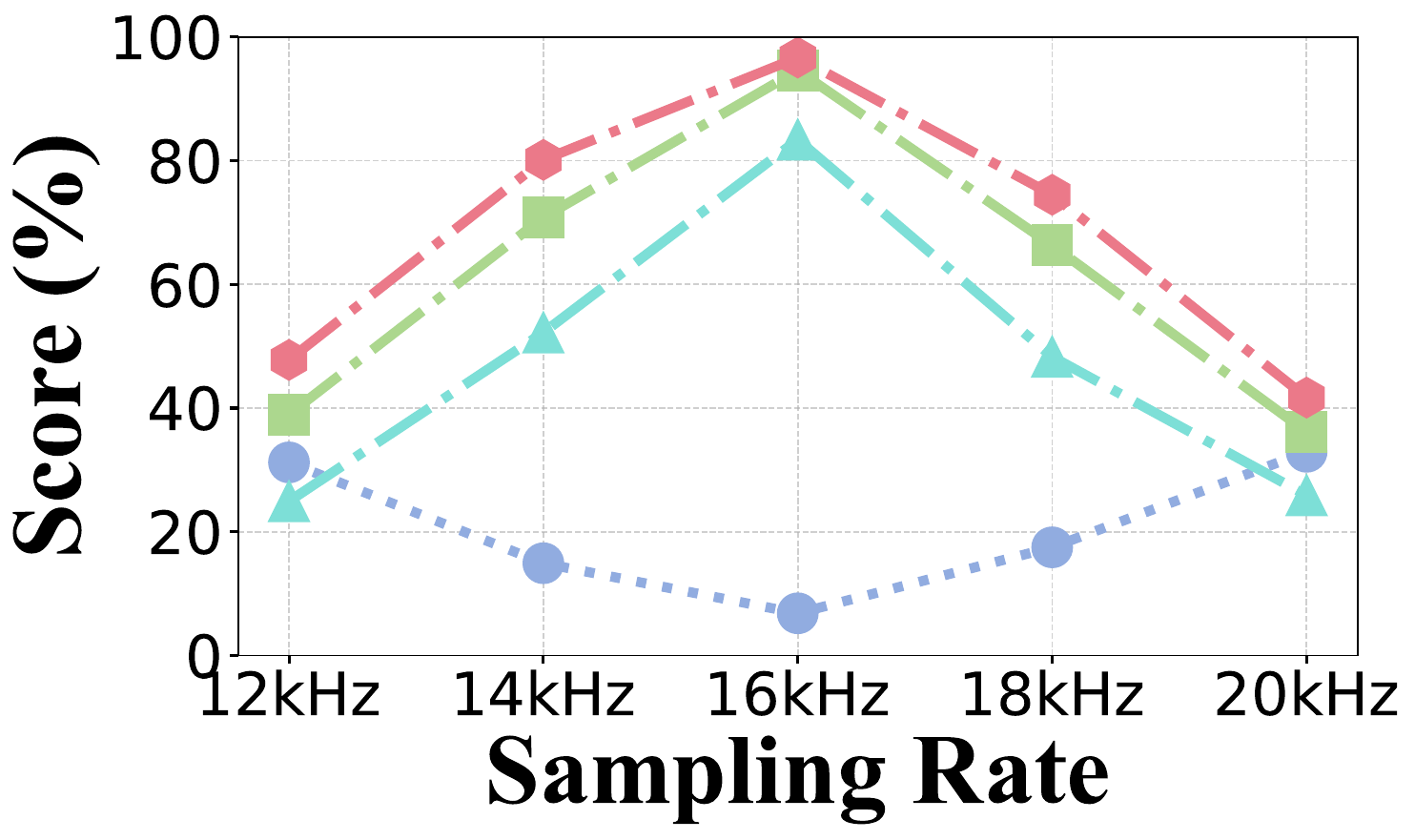}
    		\vspace{-0.1cm}
    	\end{minipage}
    }
  \vspace{-0.2cm}
    \subfigure[VQVC+]{
    	\begin{minipage}[b]{0.23\linewidth}
    		\centering
    		\includegraphics[trim=0mm 0mm 0mm 0mm, clip,width=0.95\textwidth]{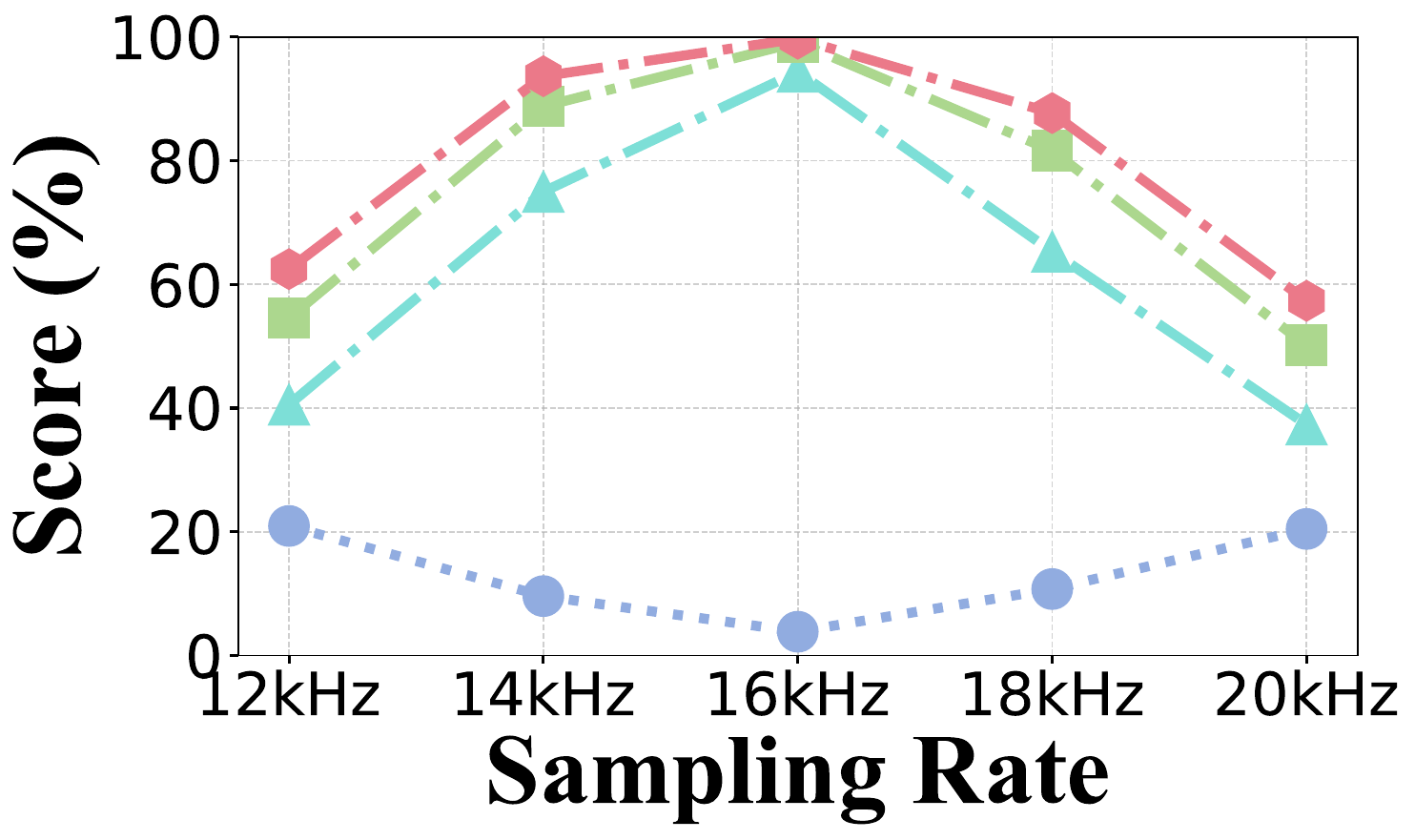}
    		\vspace{-0.1cm}
    	\end{minipage}
    }
    \subfigure[BNE]{
    	\begin{minipage}[b]{0.23\linewidth}
    		\centering
    		\includegraphics[trim=0mm 0mm 0mm 0mm, clip,width=0.95\textwidth]{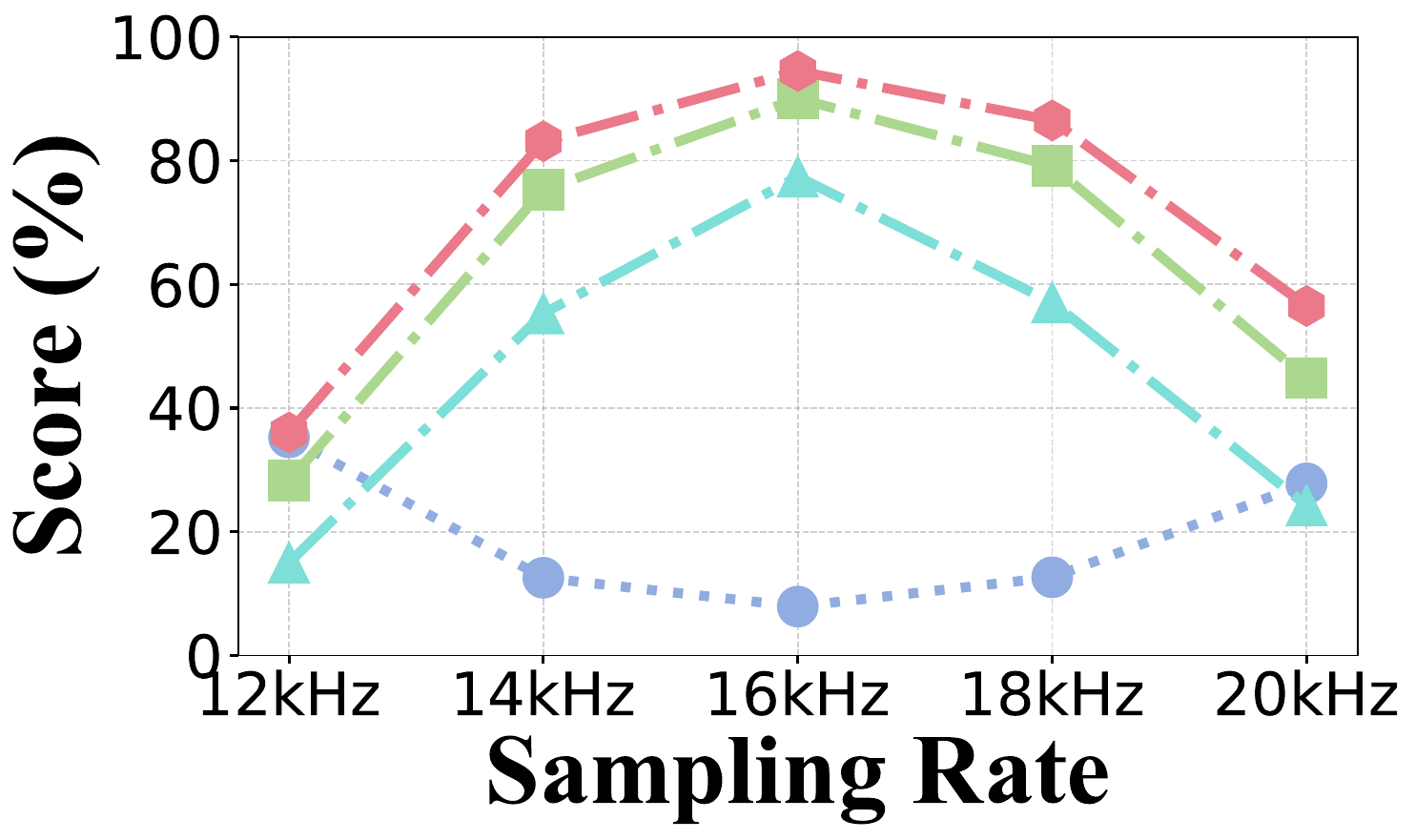}
    		\vspace{-0.1cm}
    	\end{minipage}
    }
    \subfigure[FreeVC]{
    	\begin{minipage}[b]{0.23\linewidth}
    		\centering
    		\includegraphics[trim=0mm 0mm 0mm 0mm, clip,width=0.95\textwidth]{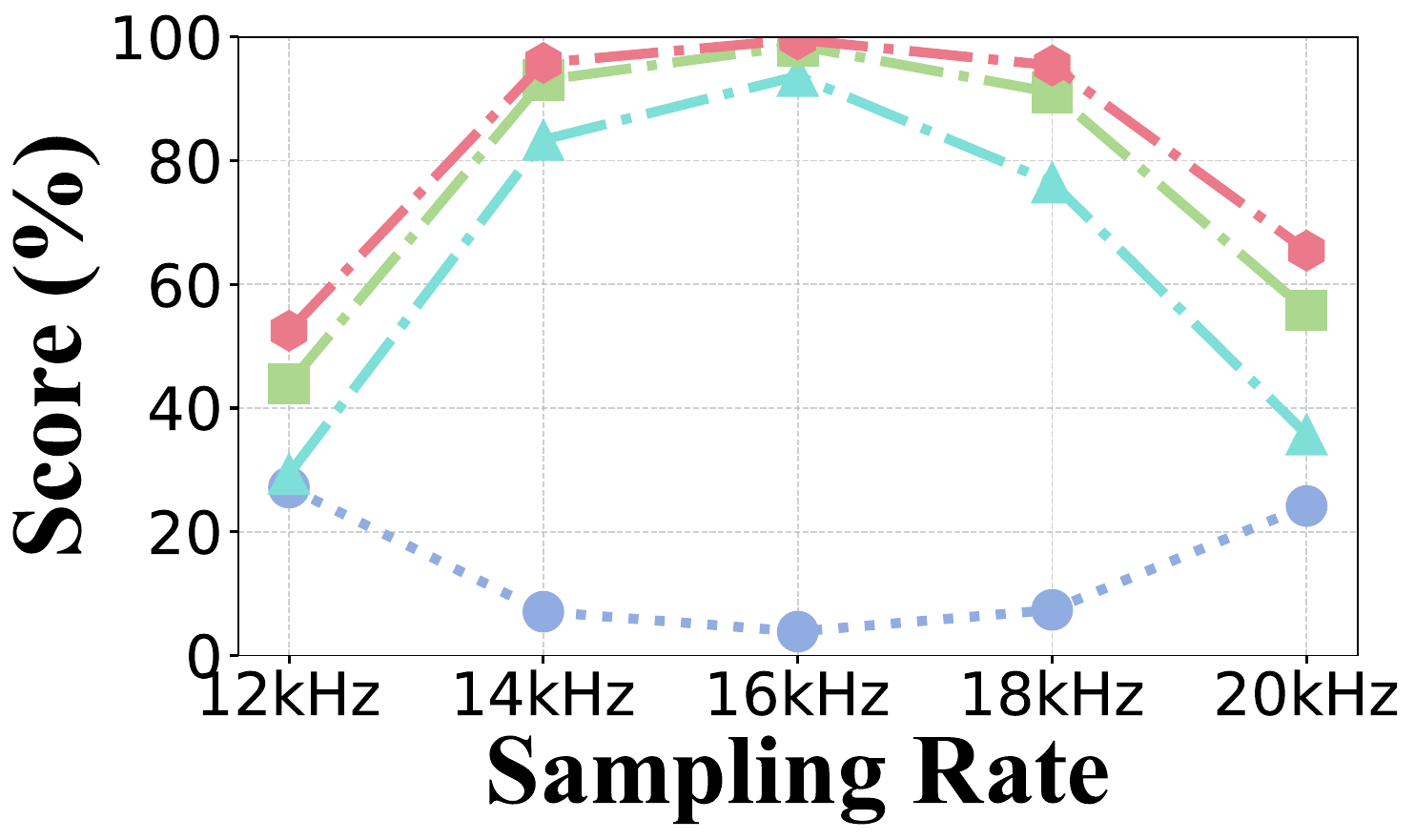}
    		\vspace{-0.1cm}
    	\end{minipage}
    }
    \subfigure[Diff]{
    	\begin{minipage}[b]{0.23\linewidth}
    		\centering
    		\includegraphics[trim=0mm 0mm 0mm 0mm, clip,width=0.95\textwidth]{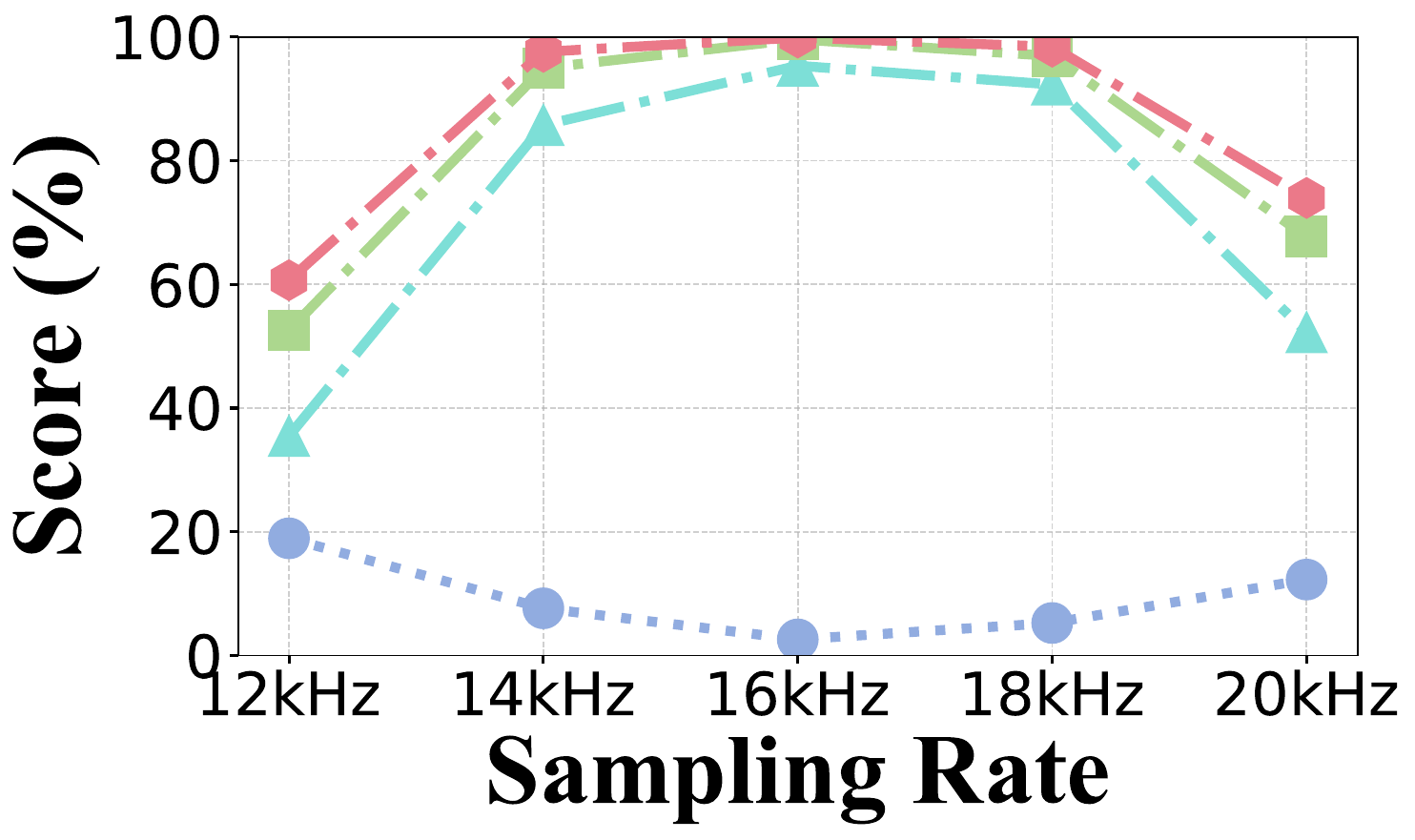}
    		\vspace{-0.1cm}
    	\end{minipage}
    }
    \subfigure[DDDM]{
    	\begin{minipage}[b]{0.23\linewidth}
    		\centering
    		\includegraphics[trim=0mm 0mm 0mm 0mm, clip,width=0.95\textwidth]{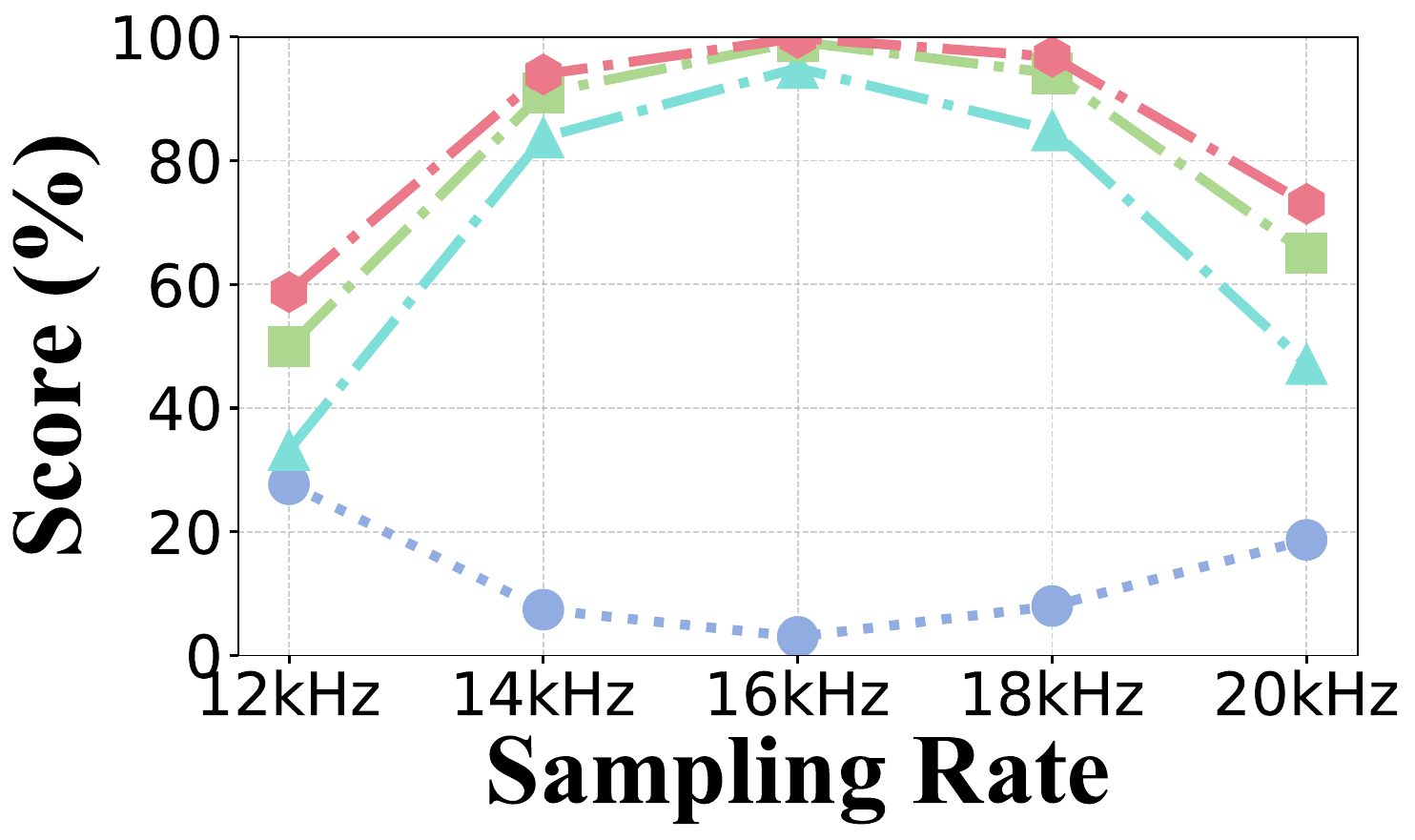}
    		\vspace{-0.1cm}
    	\end{minipage}
    }
 \vspace{-0.3cm}
	\caption{Impact of sampling rate.}
	\label{fig:line_sr}
\end{figure*}

\begin{figure*}[h]
    \centering
    	\begin{minipage}[b]{0.48\linewidth}
    		\centering
    		\includegraphics[trim=0mm 0mm 0mm 0mm, clip, width=\textwidth]{Section/Pictures/Draw/Line/legend.pdf}
    	\end{minipage} \\
    \vspace{-0.05cm}
    \subfigure[Clean]{
    	\begin{minipage}[b]{0.23\linewidth}
    		\centering
    		\includegraphics[trim=0mm 0mm 0mm 0mm, clip, width=0.95\textwidth]{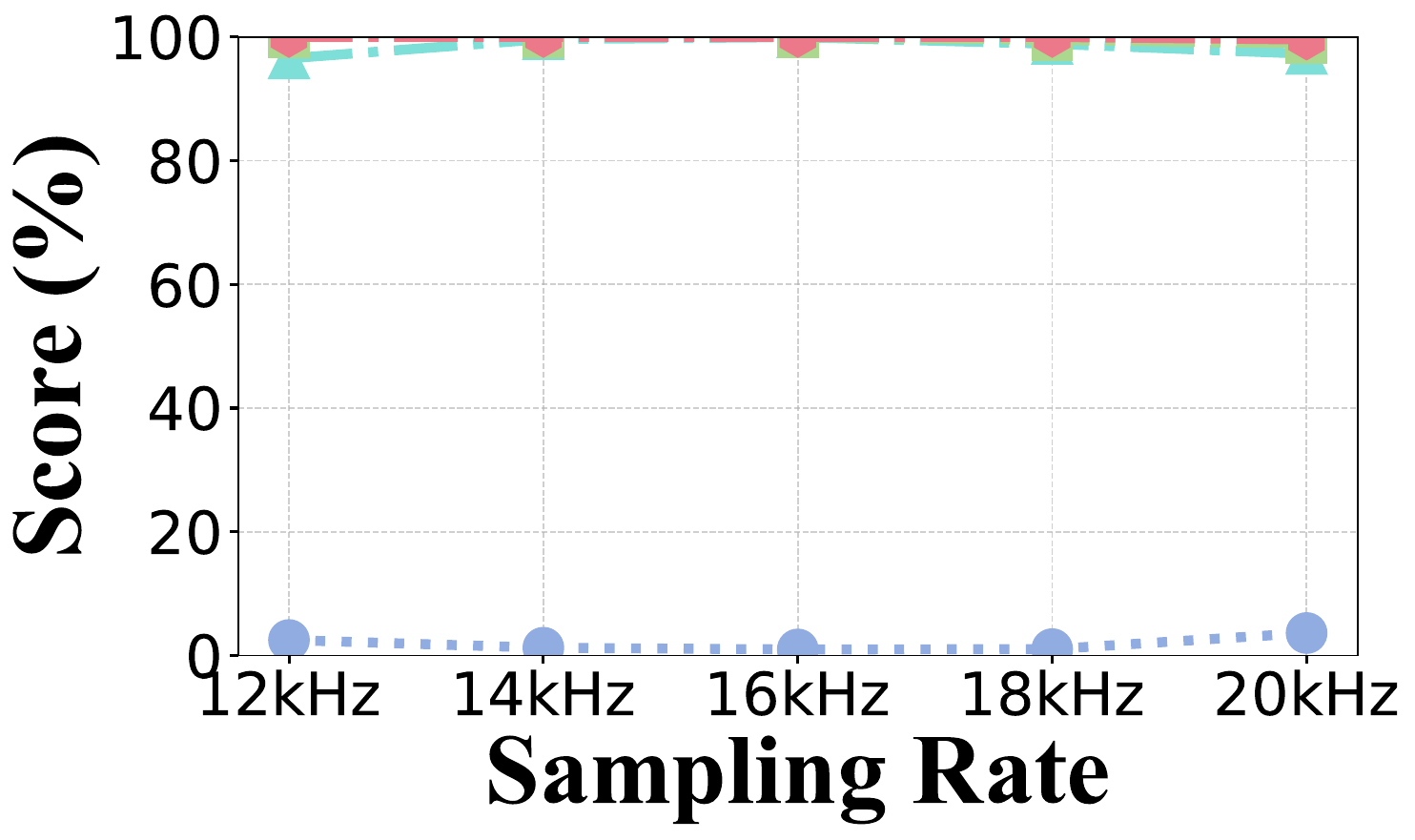}
    		\vspace{-0.1cm}
    	\end{minipage}
    }
    \subfigure[AGAIN]{
    	\begin{minipage}[b]{0.23\linewidth}
    		\centering
    		\includegraphics[trim=0mm 0mm 0mm 0mm, clip, width=0.95\textwidth]{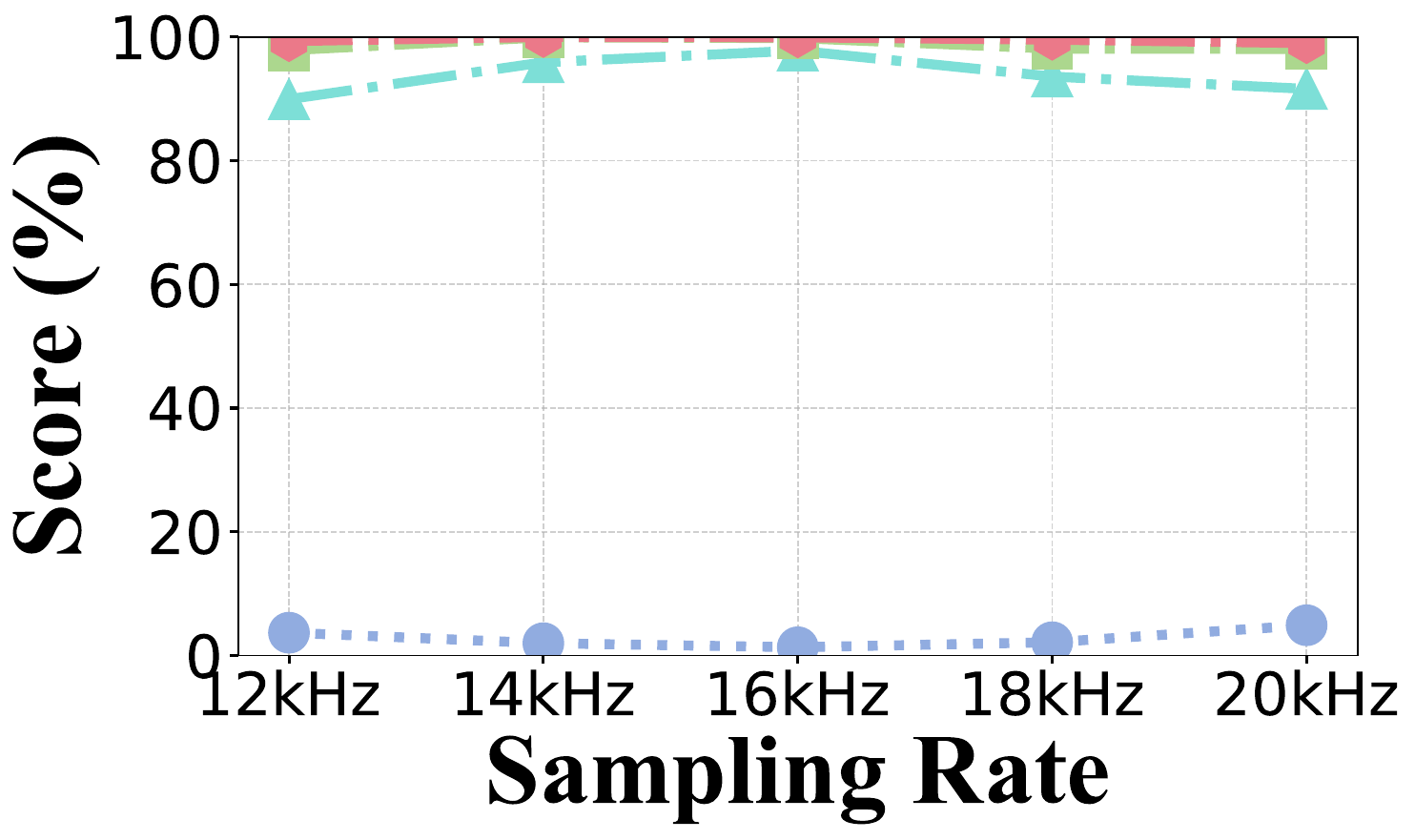}
    		\vspace{-0.1cm}
    	\end{minipage}
    }
    \subfigure[VQVC]{
    	\begin{minipage}[b]{0.23\linewidth}
    		\centering
    		\includegraphics[trim=0mm 0mm 0mm 0mm, clip, width=0.95\textwidth]{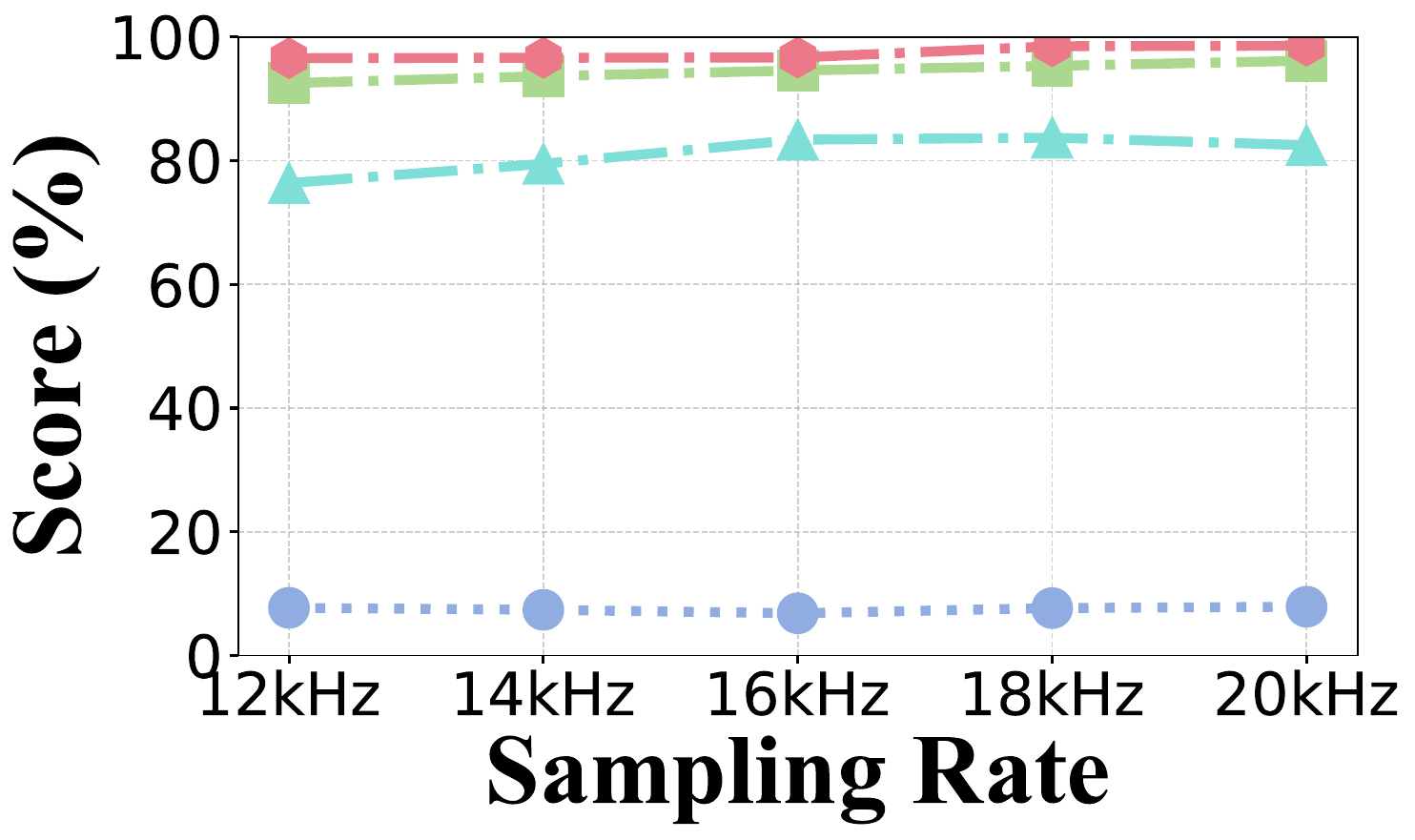}
    		\vspace{-0.1cm}
    	\end{minipage}
    }
  \vspace{-0.2cm}
    \subfigure[VQVC+]{
    	\begin{minipage}[b]{0.23\linewidth}
    		\centering
    		\includegraphics[trim=0mm 0mm 0mm 0mm, clip,width=0.95\textwidth]{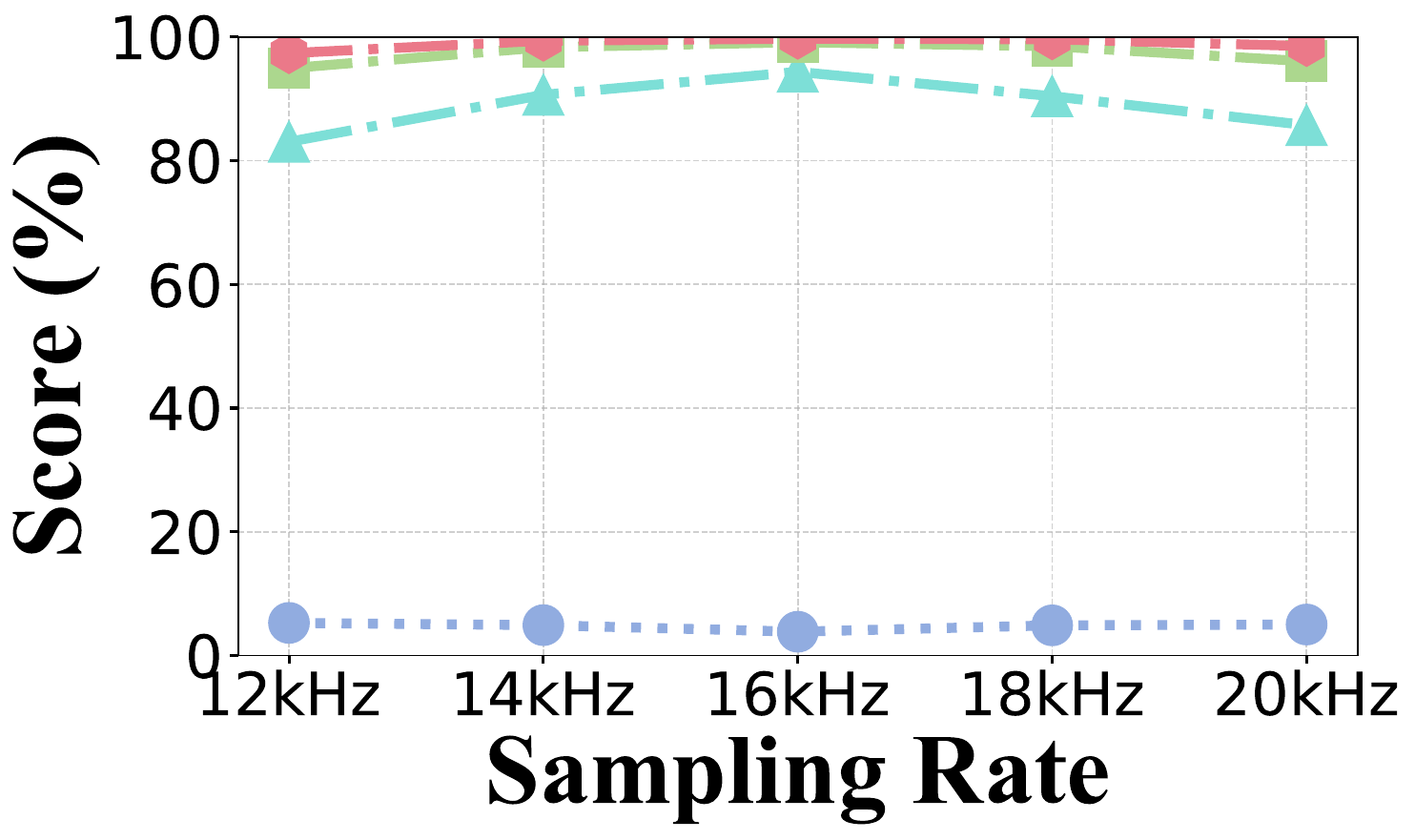}
    		\vspace{-0.1cm}
    	\end{minipage}
    }
    \subfigure[BNE]{
    	\begin{minipage}[b]{0.23\linewidth}
    		\centering
    		\includegraphics[trim=0mm 0mm 0mm 0mm, clip,width=0.95\textwidth]{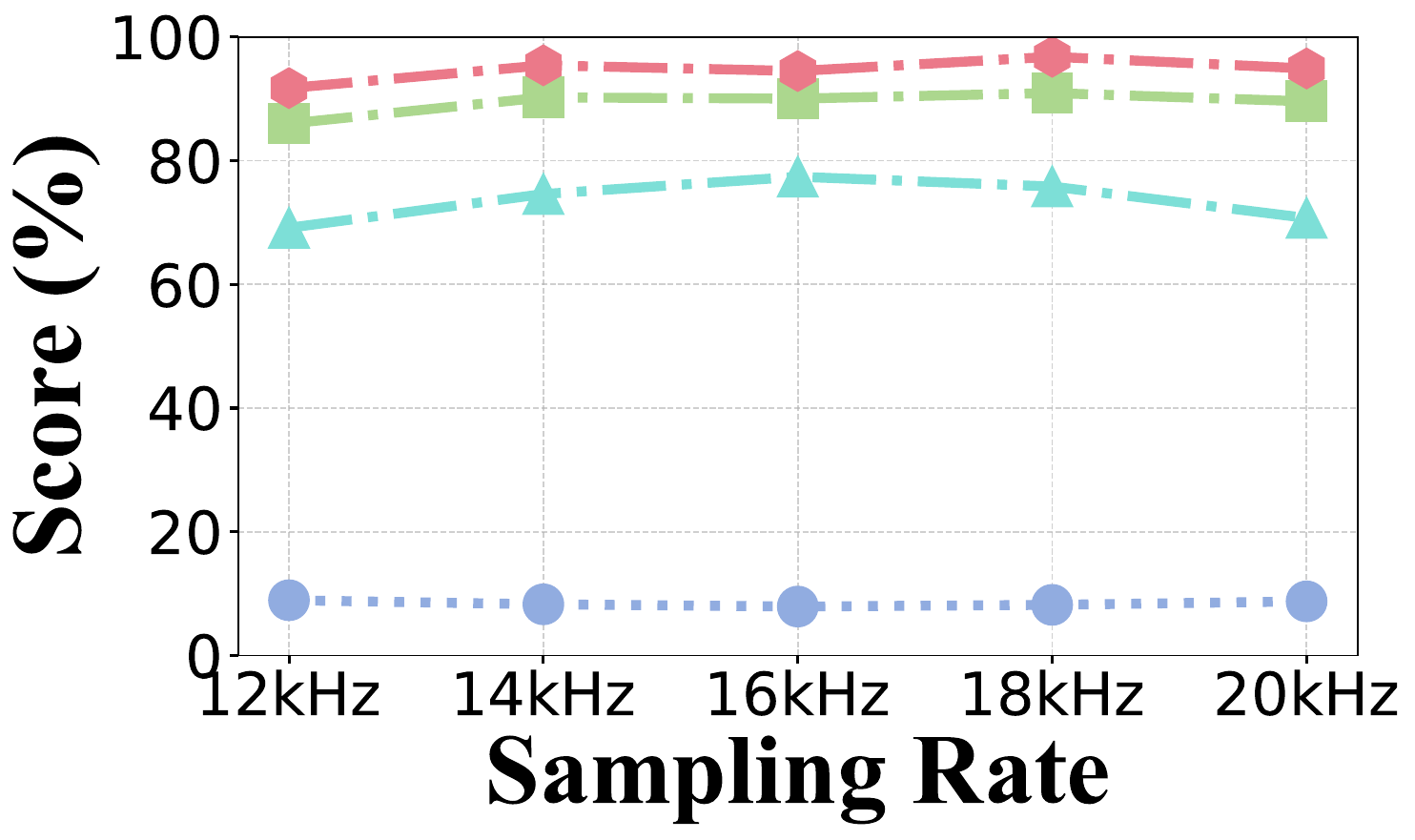}
    		\vspace{-0.1cm}
    	\end{minipage}
    }
    \subfigure[FreeVC]{
    	\begin{minipage}[b]{0.23\linewidth}
    		\centering
    		\includegraphics[trim=0mm 0mm 0mm 0mm, clip,width=0.95\textwidth]{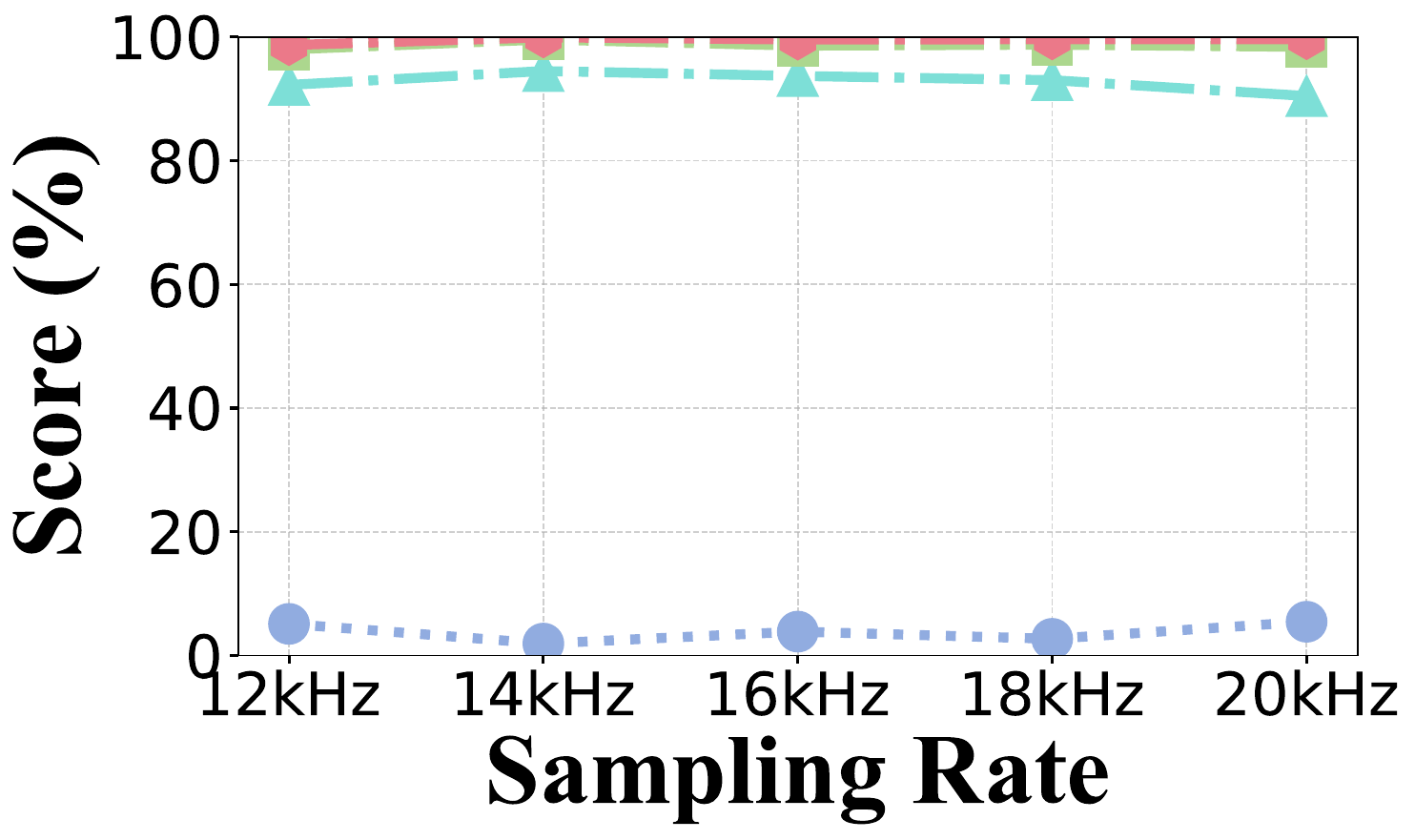}
    		\vspace{-0.1cm}
    	\end{minipage}
    }
    \subfigure[Diff]{
    	\begin{minipage}[b]{0.23\linewidth}
    		\centering
    		\includegraphics[trim=0mm 0mm 0mm 0mm, clip,width=0.95\textwidth]{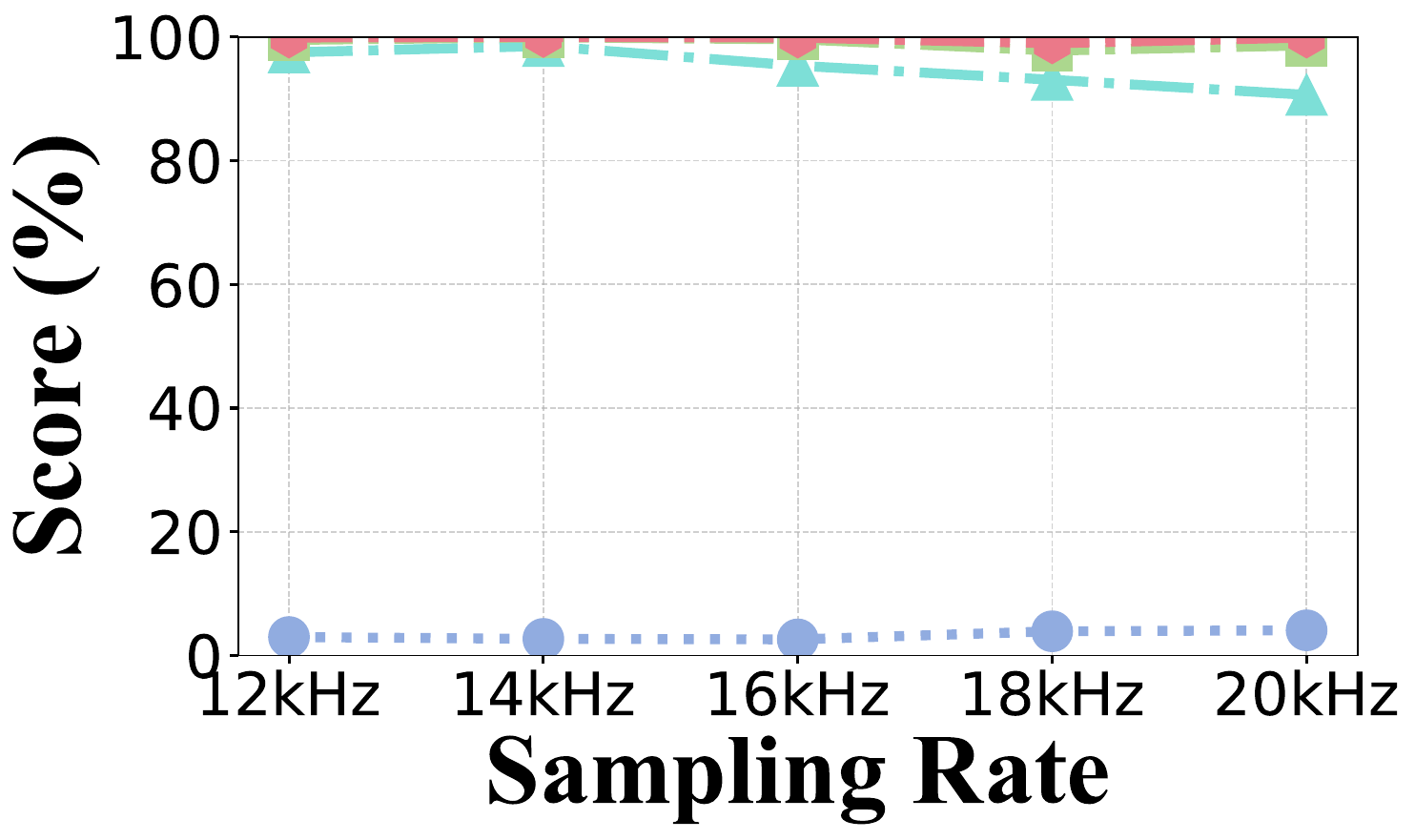}
    		\vspace{-0.1cm}
    	\end{minipage}
    }
    \subfigure[DDDM]{
    	\begin{minipage}[b]{0.23\linewidth}
    		\centering
    		\includegraphics[trim=0mm 0mm 0mm 0mm, clip,width=0.95\textwidth]{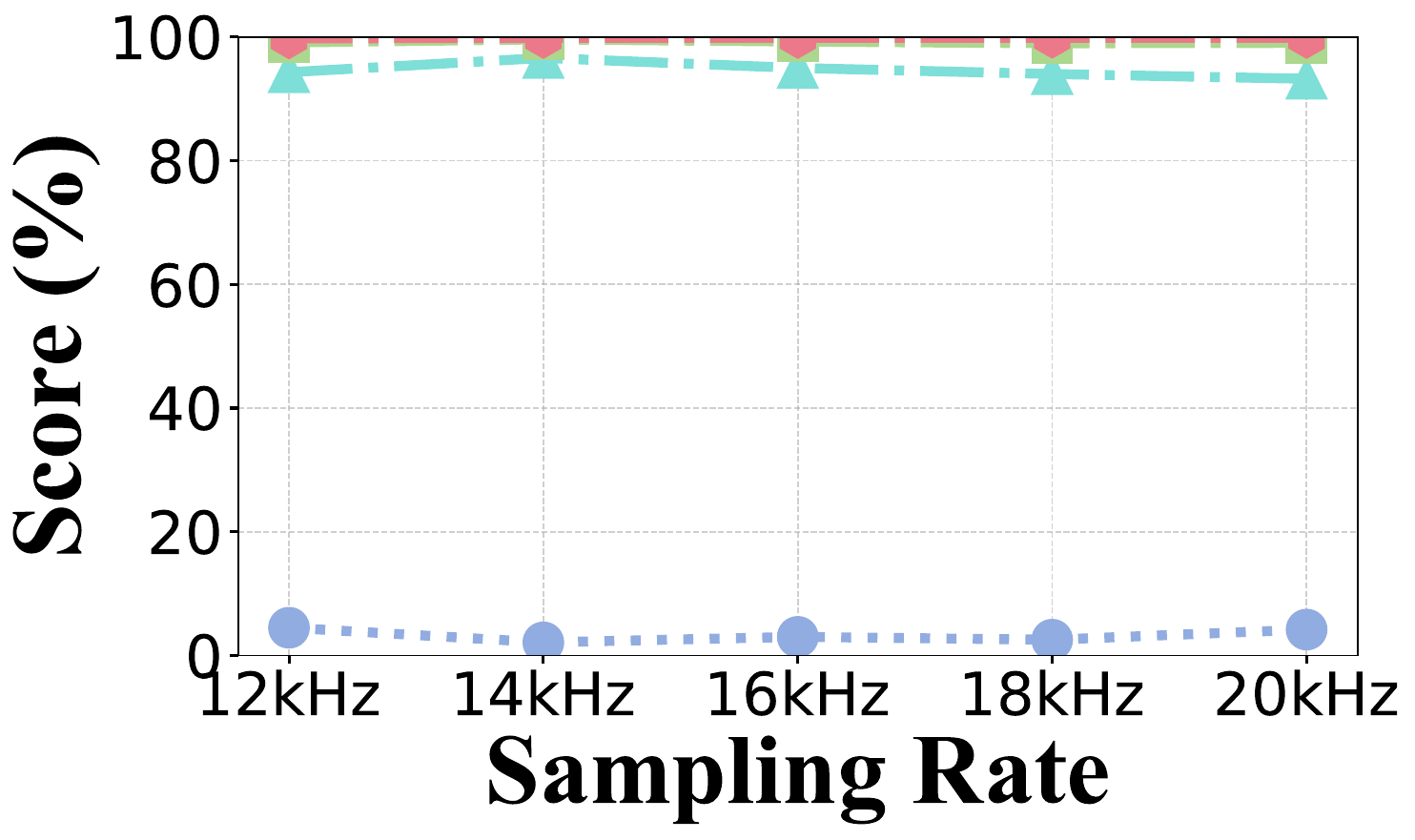}
    		\vspace{-0.1cm}
    	\end{minipage}
    }
 \vspace{-0.3cm}
	\caption{Impact of sampling rate after resampling.}
	\label{fig:line_sr_c}
\end{figure*}

\begin{figure*}[h]
    \centering
    	\begin{minipage}[b]{0.48\linewidth}
    		\centering
    		\includegraphics[trim=0mm 0mm 0mm 0mm, clip, width=\textwidth]{Section/Pictures/Draw/Line/legend.pdf}
    	\end{minipage} \\
    \vspace{-0.05cm}
    \subfigure[Clean]{
    	\begin{minipage}[b]{0.23\linewidth}
    		\centering
    		\includegraphics[trim=0mm 0mm 0mm 0mm, clip, width=0.95\textwidth]{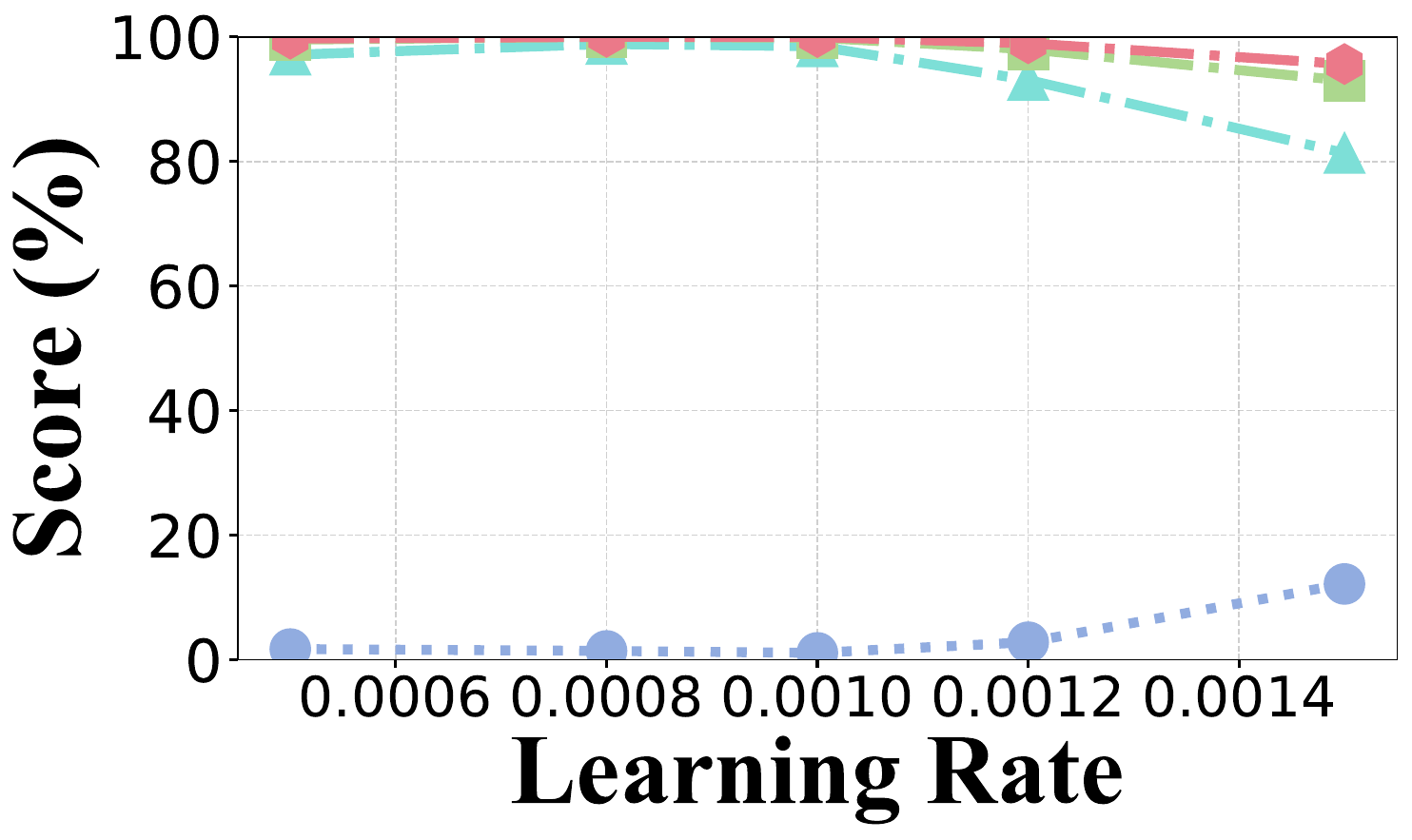}
    		\vspace{-0.1cm}
    	\end{minipage}
    }
    \subfigure[AGAIN]{
    	\begin{minipage}[b]{0.23\linewidth}
    		\centering
    		\includegraphics[trim=0mm 0mm 0mm 0mm, clip, width=0.95\textwidth]{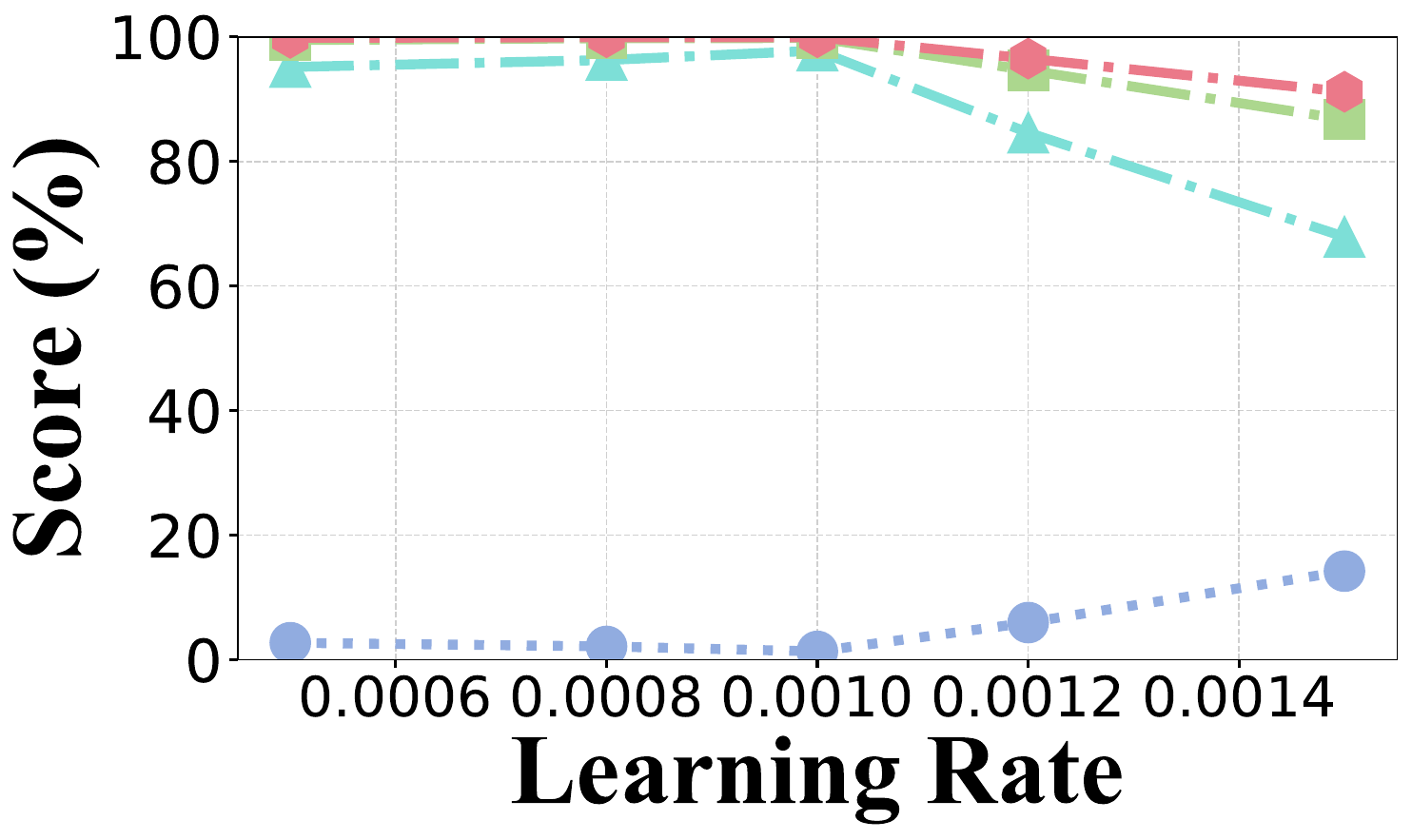}
    		\vspace{-0.1cm}
    	\end{minipage}
    }
    \subfigure[VQVC]{
    	\begin{minipage}[b]{0.23\linewidth}
    		\centering
    		\includegraphics[trim=0mm 0mm 0mm 0mm, clip, width=0.95\textwidth]{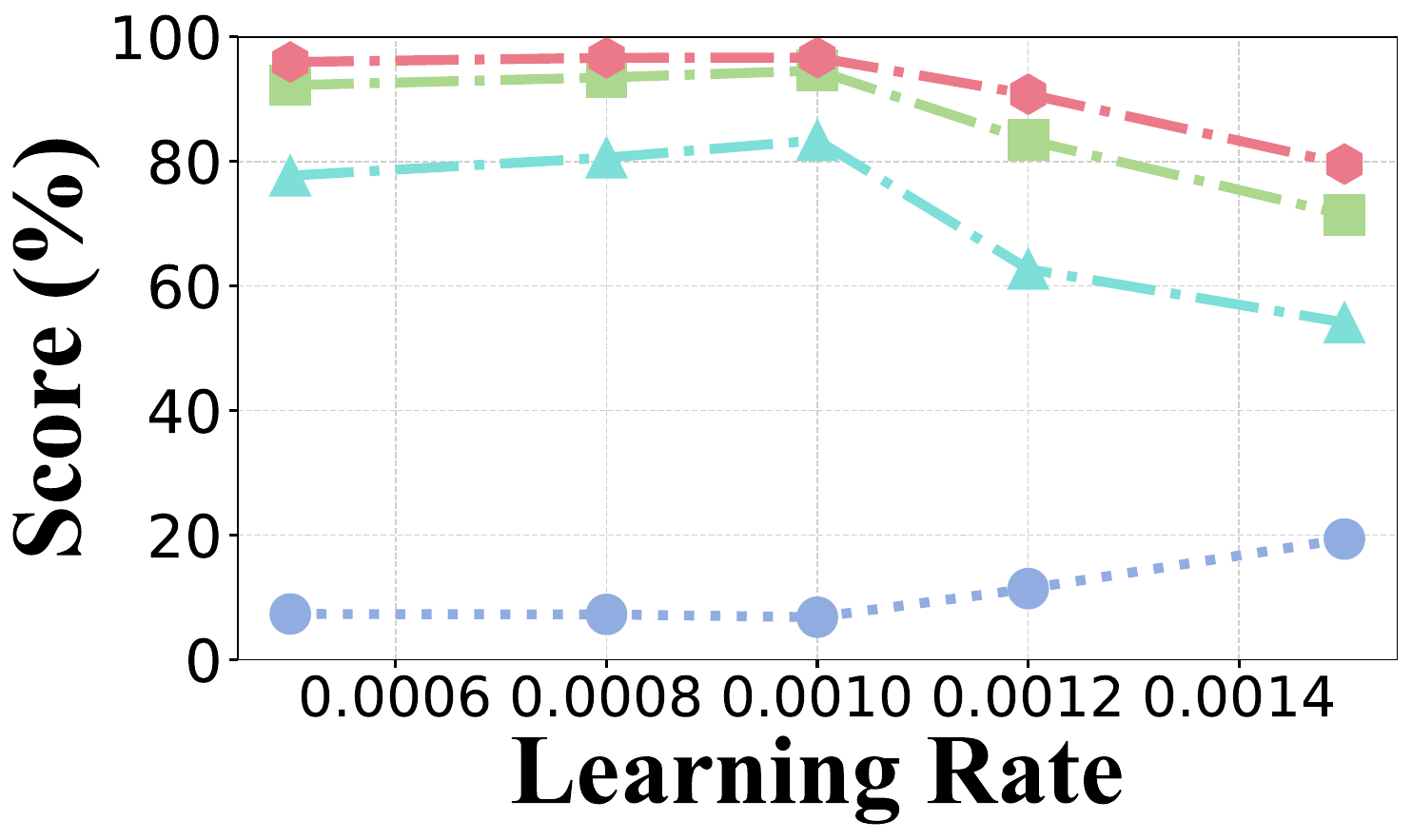}
    		\vspace{-0.1cm}
    	\end{minipage}
    }
  \vspace{-0.2cm}
    \subfigure[VQVC+]{
    	\begin{minipage}[b]{0.23\linewidth}
    		\centering
    		\includegraphics[trim=0mm 0mm 0mm 0mm, clip,width=0.95\textwidth]{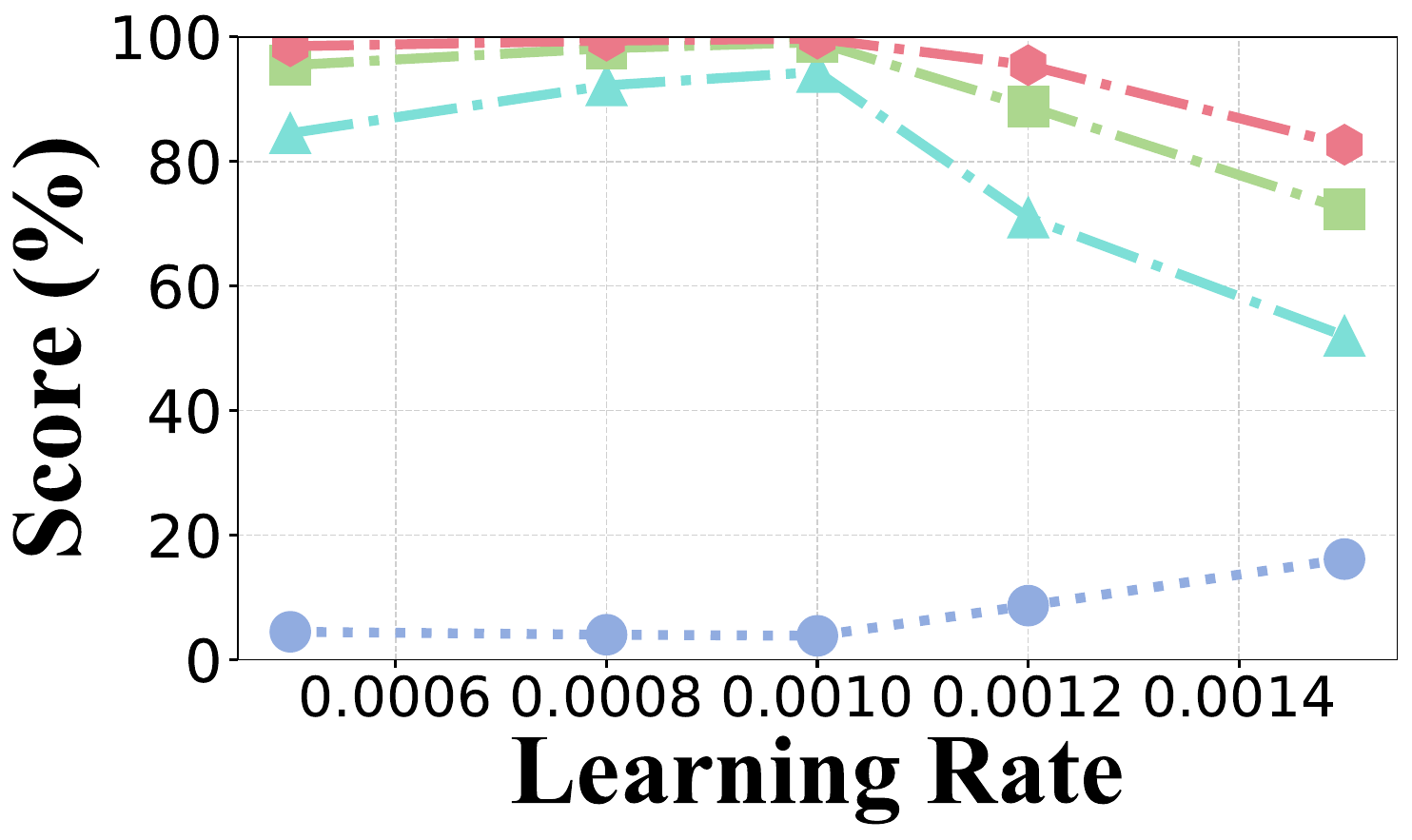}
    		\vspace{-0.1cm}
    	\end{minipage}
    }
    \subfigure[BNE]{
    	\begin{minipage}[b]{0.23\linewidth}
    		\centering
    		\includegraphics[trim=0mm 0mm 0mm 0mm, clip,width=0.95\textwidth]{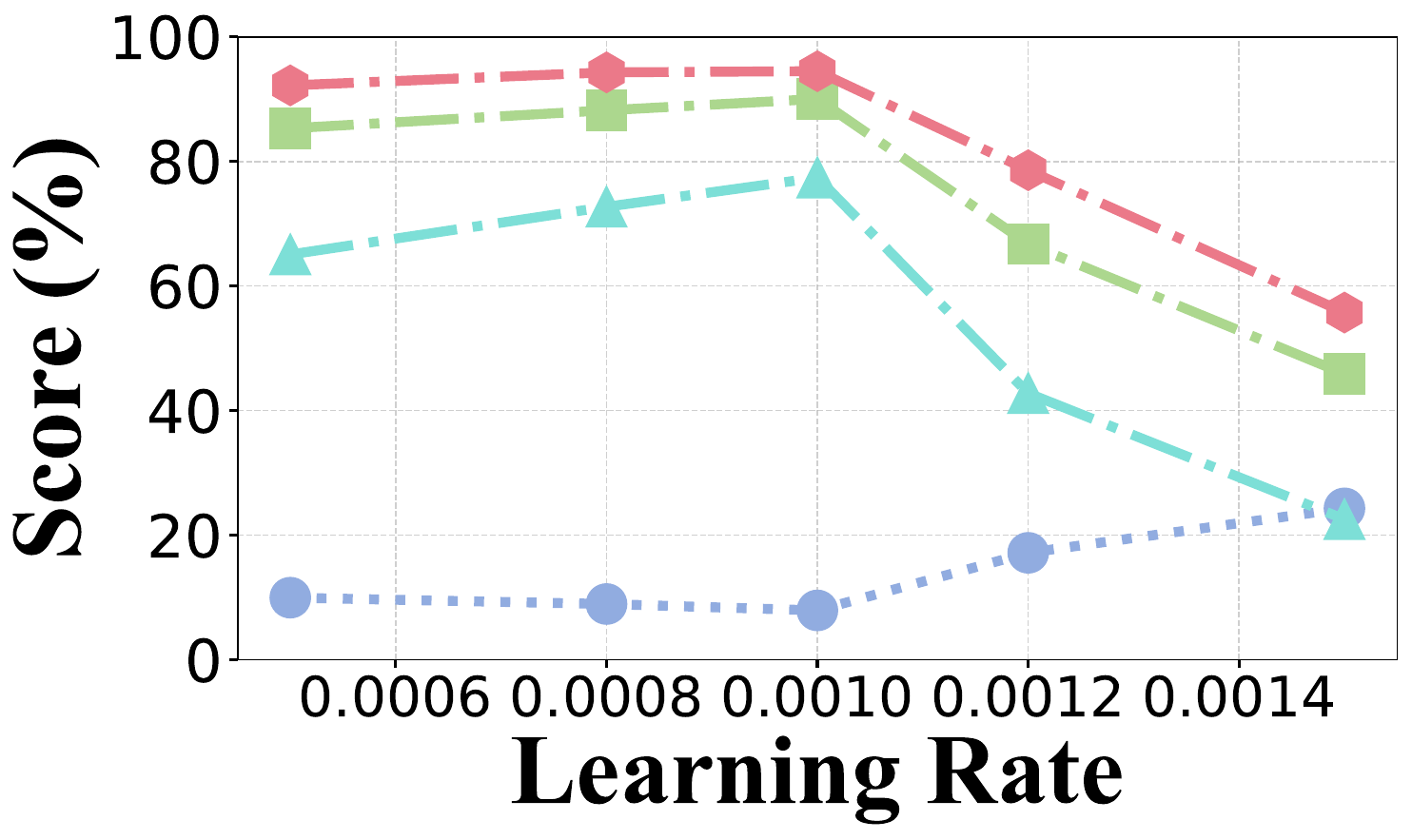}
    		\vspace{-0.1cm}
    	\end{minipage}
    }
    \subfigure[FreeVC]{
    	\begin{minipage}[b]{0.23\linewidth}
    		\centering
    		\includegraphics[trim=0mm 0mm 0mm 0mm, clip,width=0.95\textwidth]{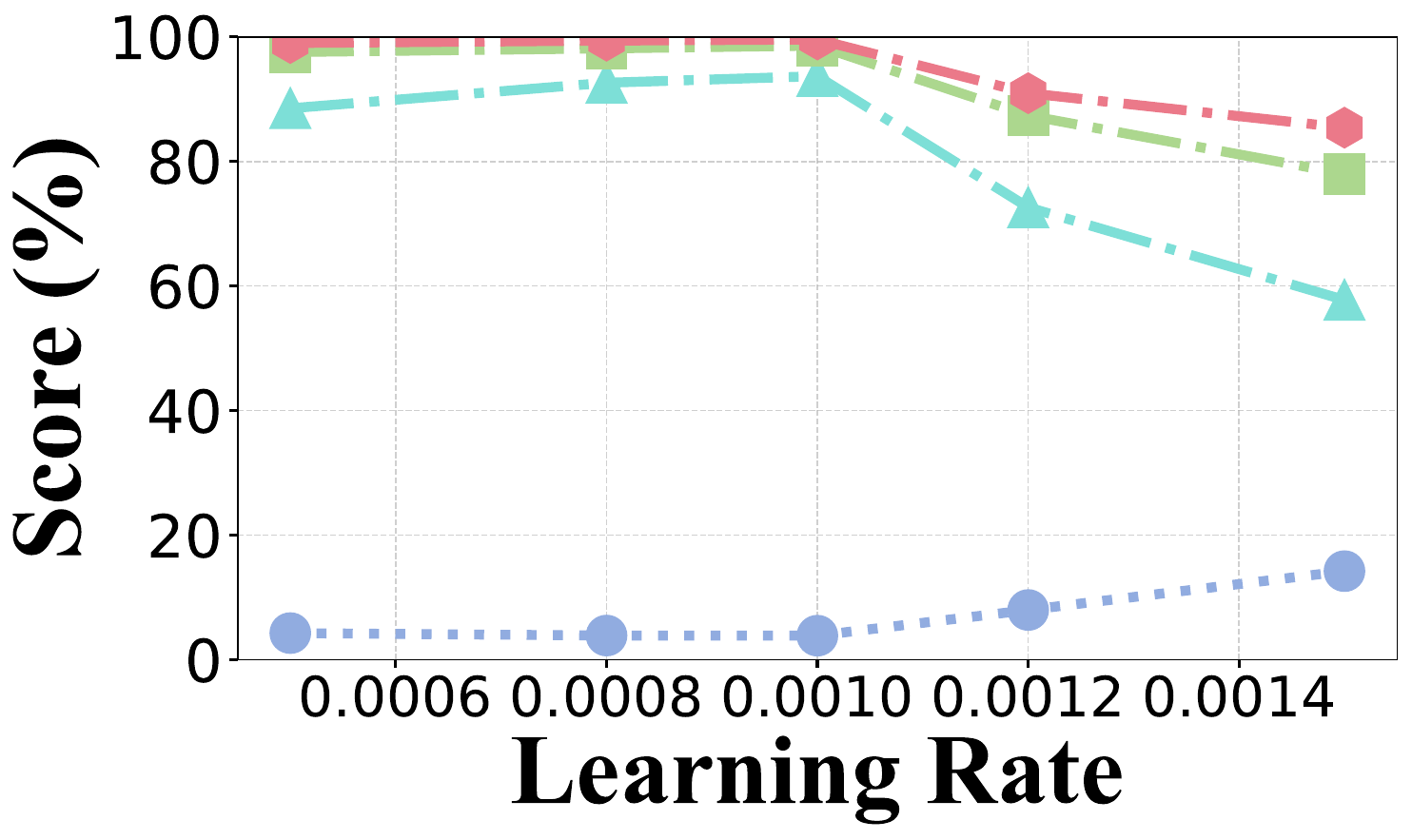}
    		\vspace{-0.1cm}
    	\end{minipage}
    }
    \subfigure[Diff]{
    	\begin{minipage}[b]{0.23\linewidth}
    		\centering
    		\includegraphics[trim=0mm 0mm 0mm 0mm, clip,width=0.95\textwidth]{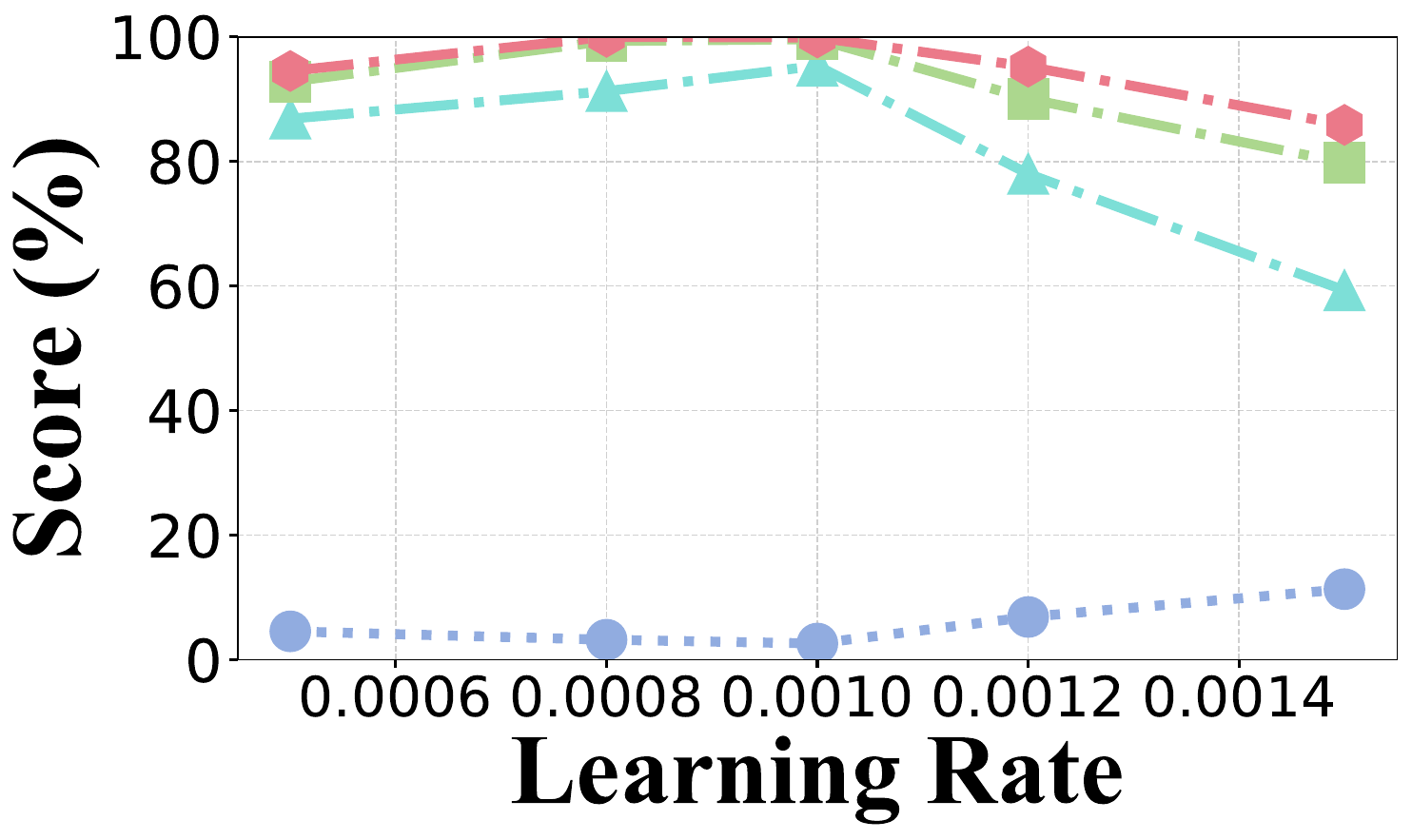}
    		\vspace{-0.1cm}
    	\end{minipage}
    }
    \subfigure[DDDM]{
    	\begin{minipage}[b]{0.23\linewidth}
    		\centering
    		\includegraphics[trim=0mm 0mm 0mm 0mm, clip,width=0.95\textwidth]{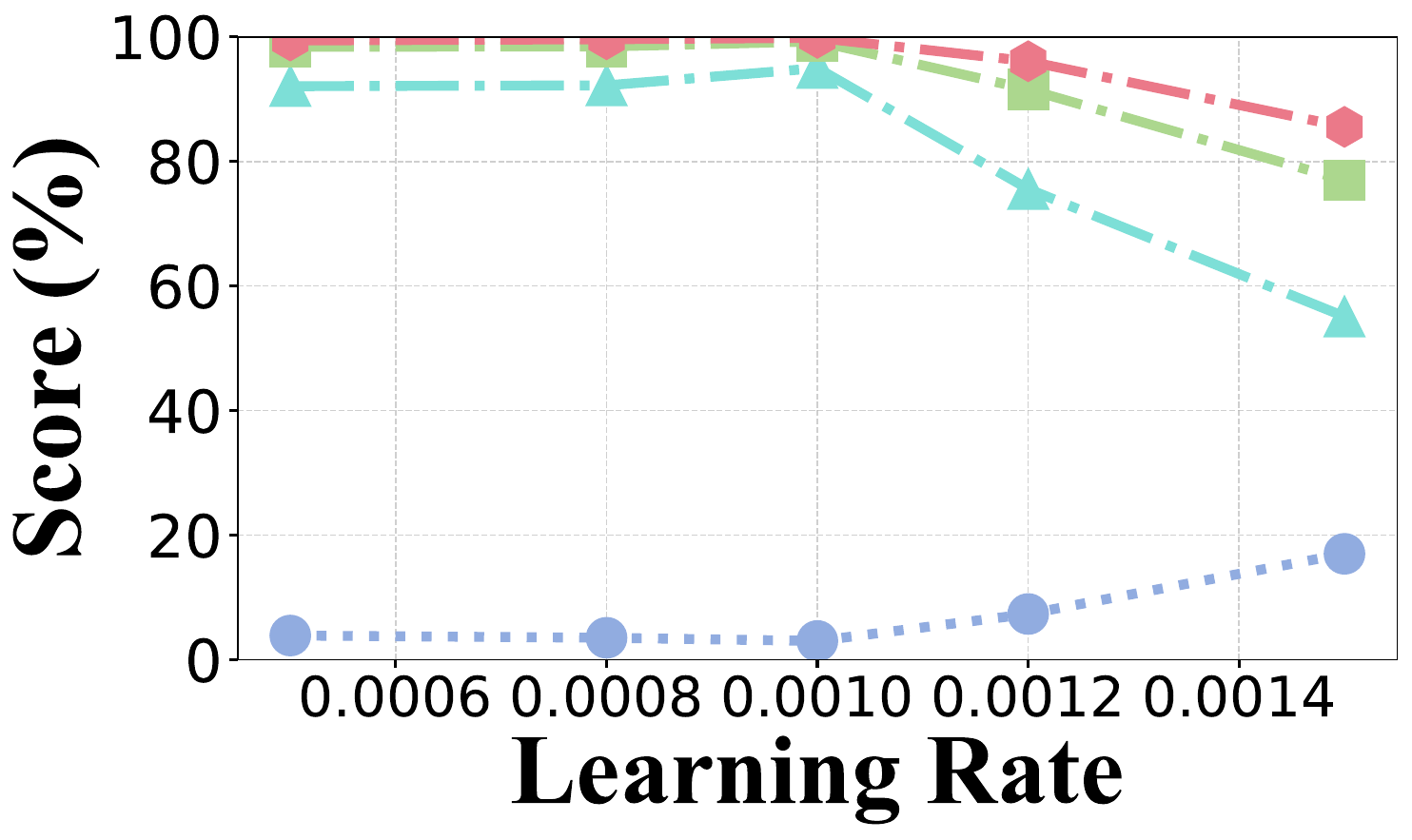}
    		\vspace{-0.1cm}
    	\end{minipage}
    }
 \vspace{-0.3cm}
	\caption{Impact of learning rate.}
	\label{fig:line_lr}
\end{figure*}

\begin{figure*}[h]
    \centering
    	\begin{minipage}[b]{0.48\linewidth}
    		\centering
    		\includegraphics[trim=0mm 0mm 0mm 0mm, clip, width=\textwidth]{Section/Pictures/Draw/Line/legend.pdf}
    	\end{minipage} \\
    \vspace{-0.05cm}
    \subfigure[Clean]{
    	\begin{minipage}[b]{0.23\linewidth}
    		\centering
    		\includegraphics[trim=0mm 0mm 0mm 0mm, clip, width=0.95\textwidth]{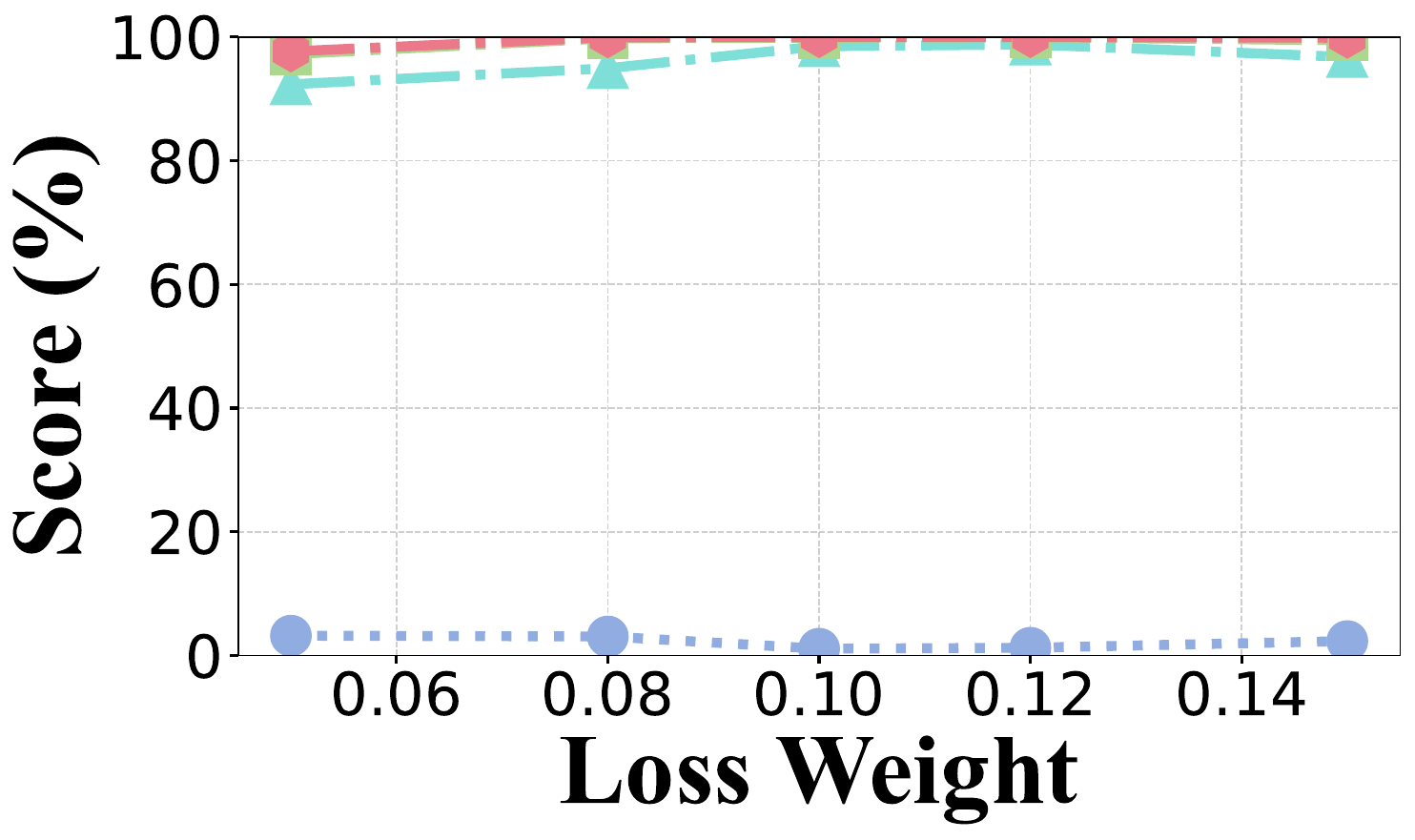}
    		\vspace{-0.1cm}
    	\end{minipage}
    }
    \subfigure[AGAIN]{
    	\begin{minipage}[b]{0.23\linewidth}
    		\centering
    		\includegraphics[trim=0mm 0mm 0mm 0mm, clip, width=0.95\textwidth]{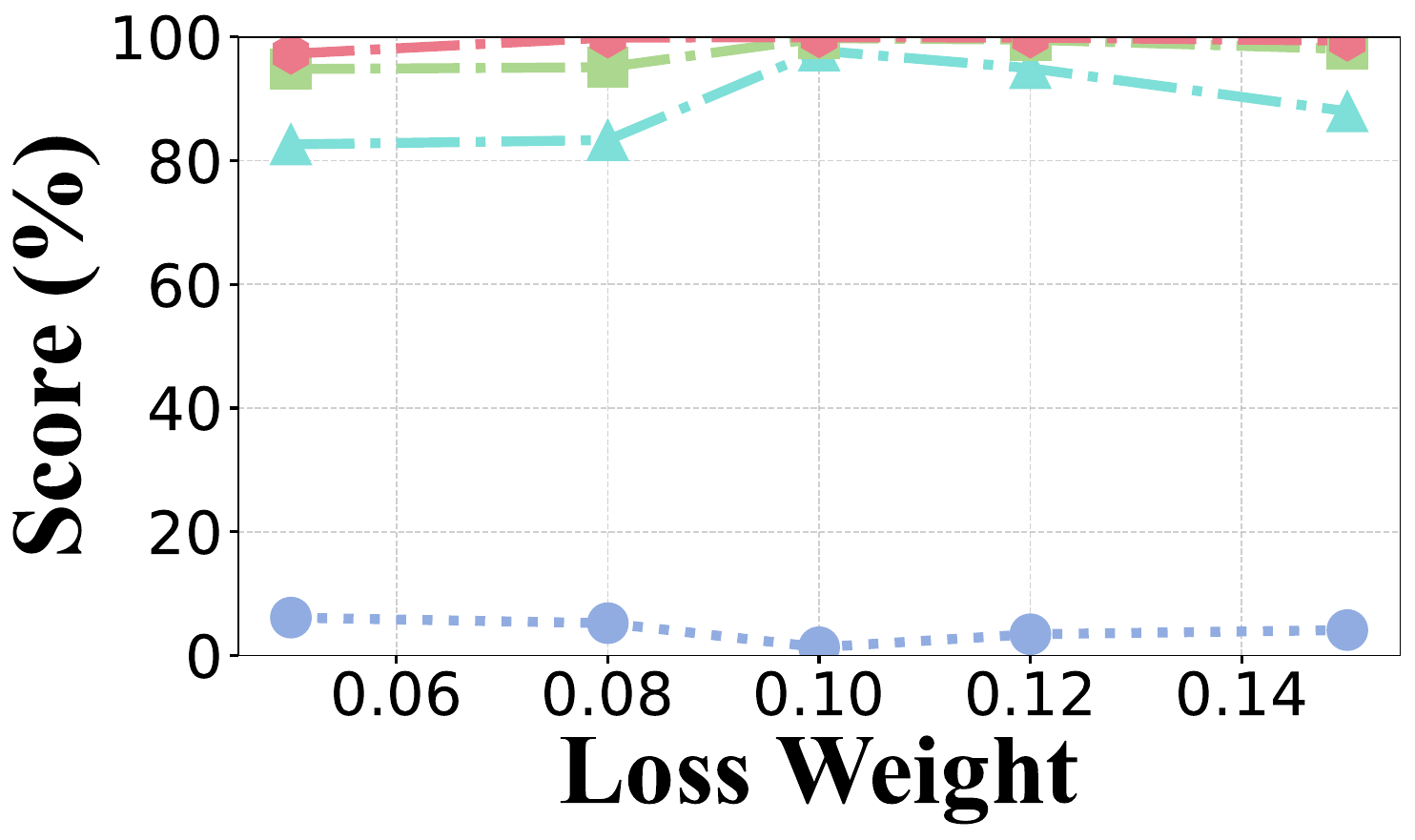}
    		\vspace{-0.1cm}
    	\end{minipage}
    }
    \subfigure[VQVC]{
    	\begin{minipage}[b]{0.23\linewidth}
    		\centering
    		\includegraphics[trim=0mm 0mm 0mm 0mm, clip, width=0.95\textwidth]{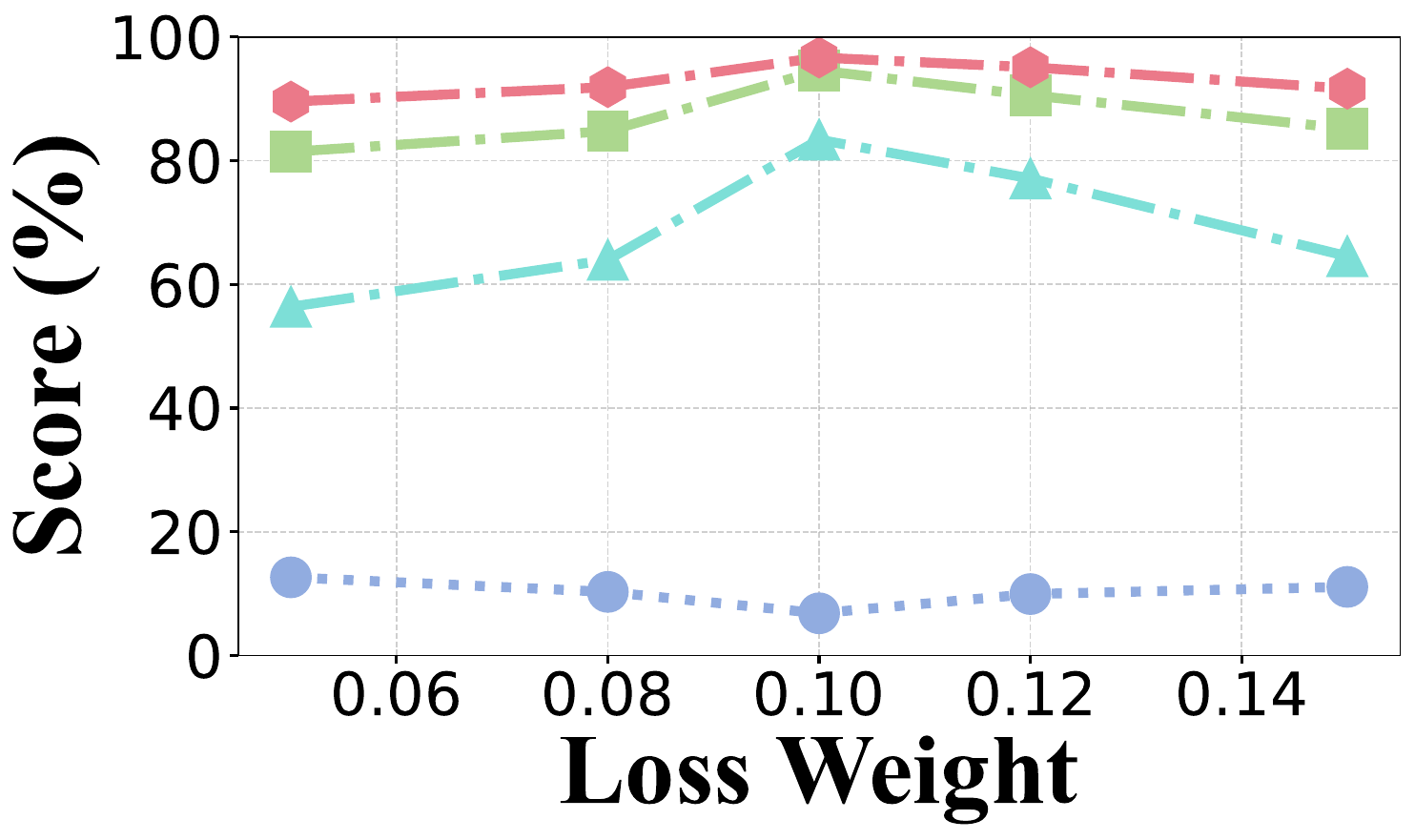}
    		\vspace{-0.1cm}
    	\end{minipage}
    }
  \vspace{-0.2cm}
    \subfigure[VQVC+]{
    	\begin{minipage}[b]{0.23\linewidth}
    		\centering
    		\includegraphics[trim=0mm 0mm 0mm 0mm, clip,width=0.95\textwidth]{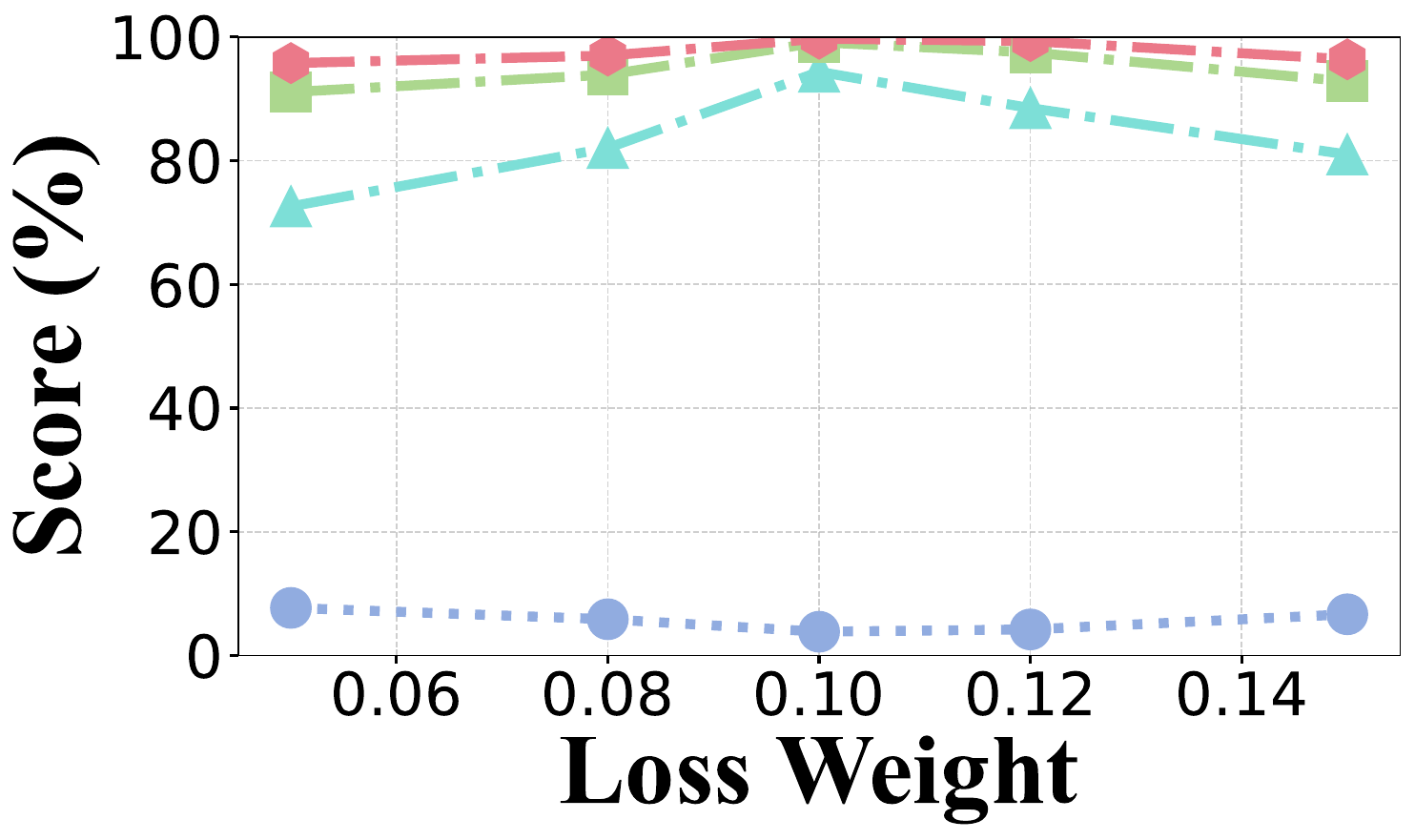}
    		\vspace{-0.1cm}
    	\end{minipage}
    }
    \subfigure[BNE]{
    	\begin{minipage}[b]{0.23\linewidth}
    		\centering
    		\includegraphics[trim=0mm 0mm 0mm 0mm, clip,width=0.95\textwidth]{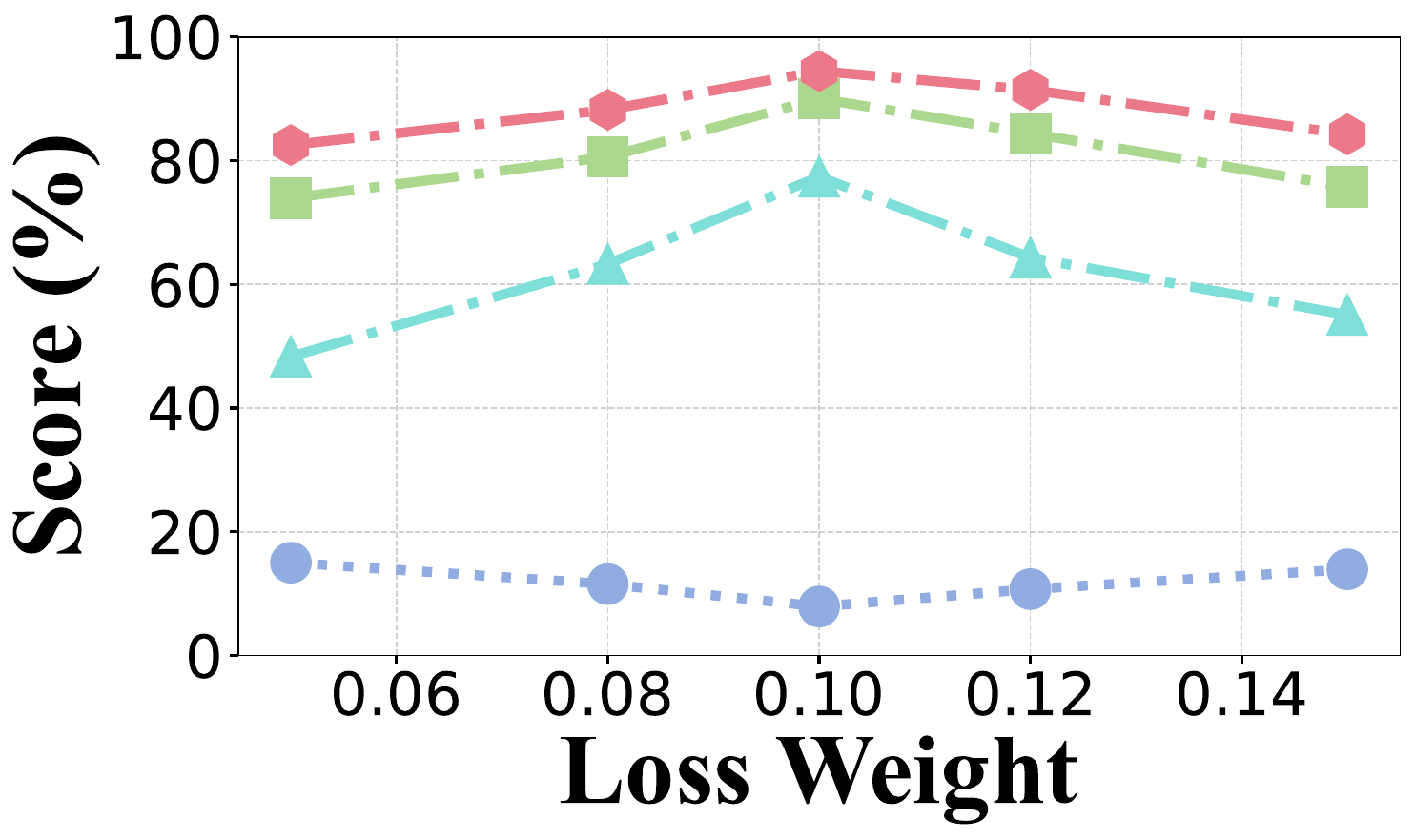}
    		\vspace{-0.1cm}
    	\end{minipage}
    }
    \subfigure[FreeVC]{
    	\begin{minipage}[b]{0.23\linewidth}
    		\centering
    		\includegraphics[trim=0mm 0mm 0mm 0mm, clip,width=0.95\textwidth]{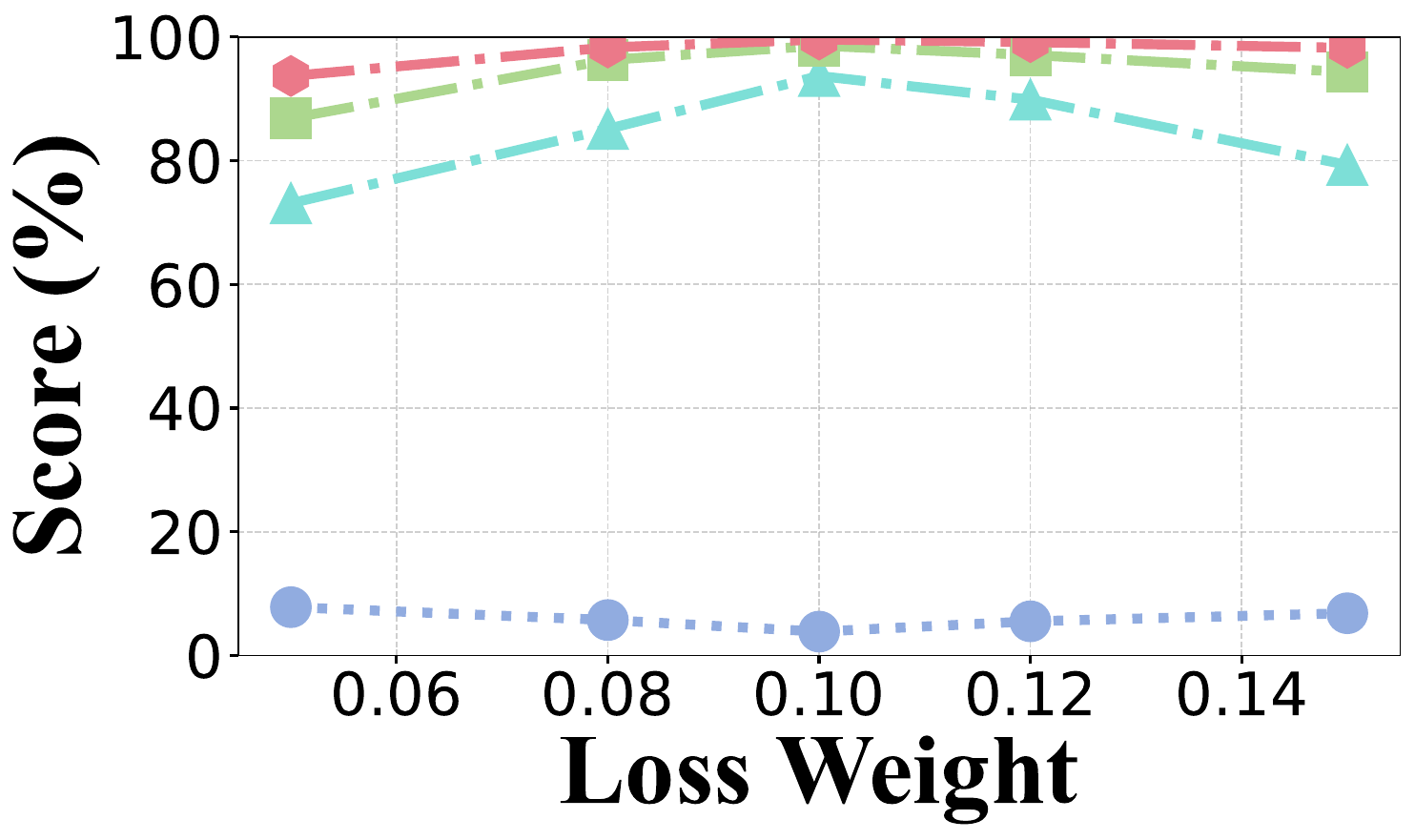}
    		\vspace{-0.1cm}
    	\end{minipage}
    }
    \subfigure[Diff]{
    	\begin{minipage}[b]{0.23\linewidth}
    		\centering
    		\includegraphics[trim=0mm 0mm 0mm 0mm, clip,width=0.95\textwidth]{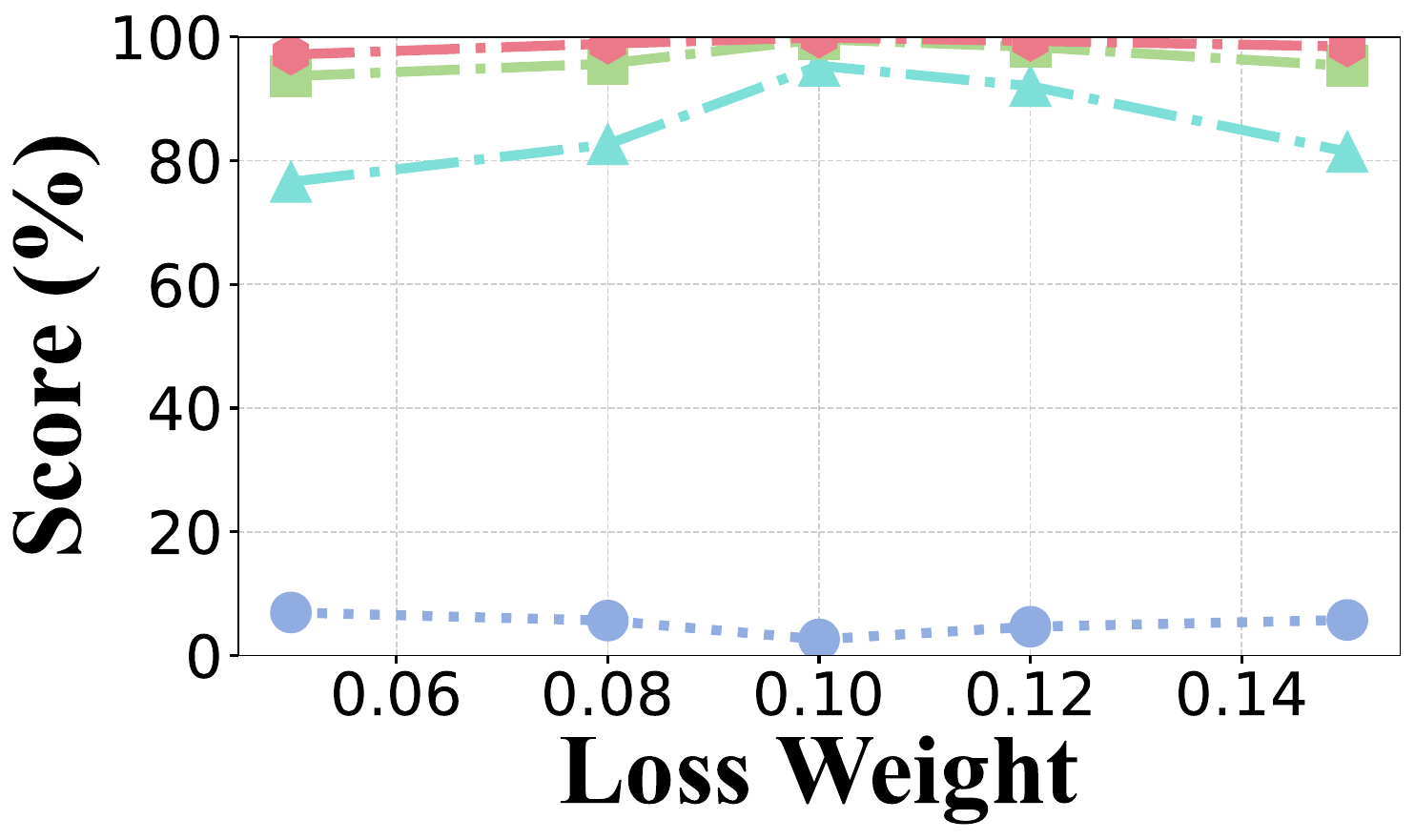}
    		\vspace{-0.1cm}
    	\end{minipage}
    }
    \subfigure[DDDM]{
    	\begin{minipage}[b]{0.23\linewidth}
    		\centering
    		\includegraphics[trim=0mm 0mm 0mm 0mm, clip,width=0.95\textwidth]{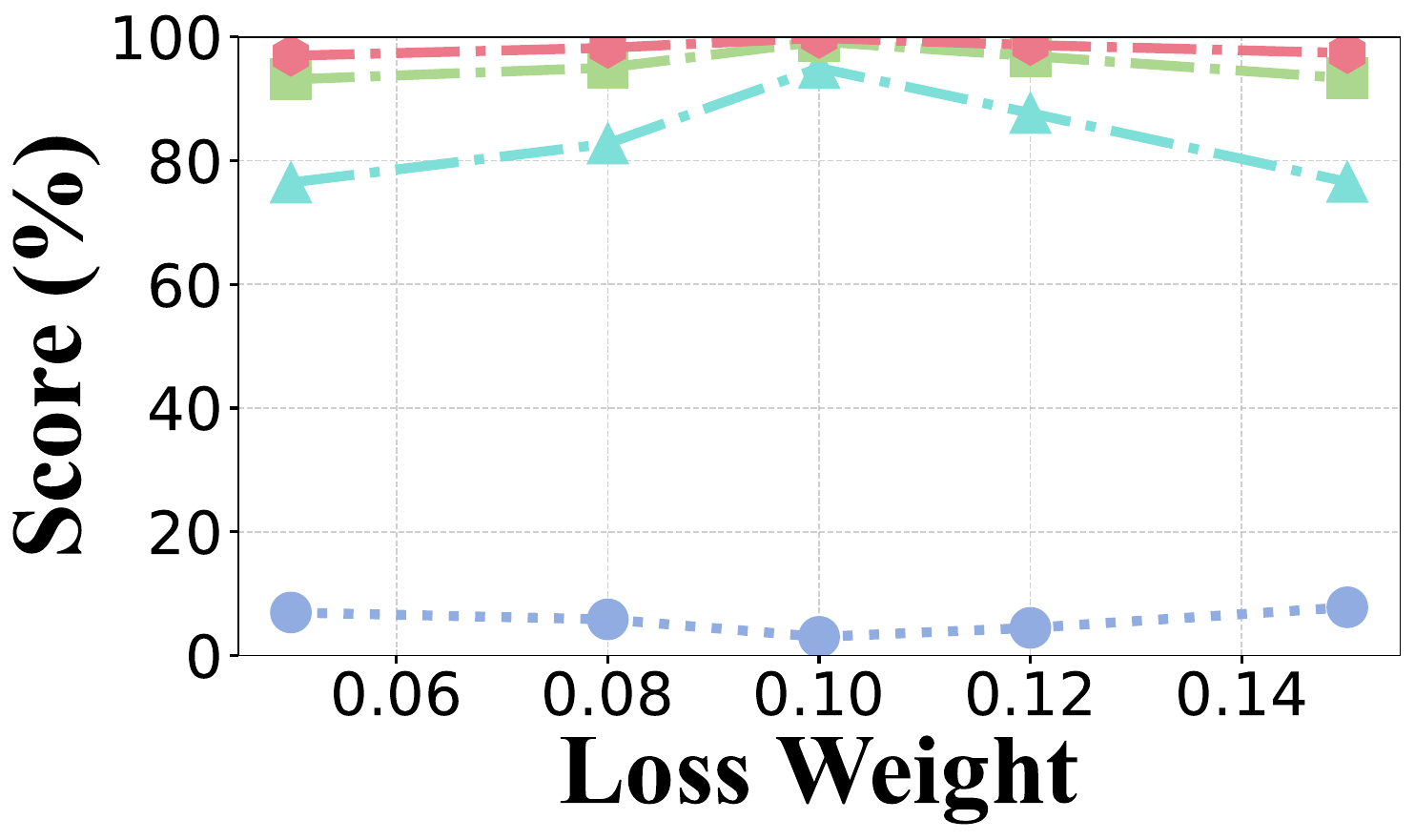}
    		\vspace{-0.1cm}
    	\end{minipage}
    }
 \vspace{-0.3cm}
	\caption{Impact of loss weight.}
	\label{fig:line_lw}
\end{figure*}

\end{document}